\documentclass[10pt]{article}
\usepackage{preamble}
\usepackage{xurl}

\title{The Network Structure of \texttt{Mathlib}}
\author{%
  Xinze Li$^{1}$
  \and
  Nanyun Peng$^{2,4}$
  \and
  Simone Severini$^{3,4}$
  \and
  Patrick Shafto$^{5}$
}
\date{}

\begin{document}

\maketitle
\footnotetext[1]{Department of Mathematics, University of Toronto. Corresponding author: \texttt{lixinze@math.utoronto.ca}}
\footnotetext[2]{Computer Science Department, University of California, Los Angeles}
\footnotetext[3]{Department of Computer Science, University College London}
\footnotetext[4]{Google}
\footnotetext[5]{Department of Mathematics and Computer Science, Rutgers University -- Newark}
\setcounter{footnote}{5}

\begin{abstract}
The ongoing development of Lean 4's \texttt{Mathlib} has produced a macroscopic structural complexity that interweaves logical, mathematical, and infrastructural dependencies. We present a network analysis of this library, extracting its dependency structure into a multilayer graph of 308,129 declarations, 8.4 million edges, and 7,563 modules. By introducing graph decompositions that isolate explicit edges from those synthesized by the compiler or driven by proofs, we quantify the structural properties of formalized mathematics. Our analysis reveals three findings. First, taxonomies designed by humans diverge from logical structures, exhibiting a 50.9\% coupling across namespaces. Second, developers utilize a median of 1.6\% of the imported scope. Third, formalization compresses semantic hierarchies, with network centrality capturing language infrastructure rather than mathematical relevance.
\end{abstract}

\addtocontents{toc}{\protect\setcounter{tocdepth}{-10}}
\section{Introduction}
Mathematical knowledge forms a vast network of logical dependencies that historically remained unrecorded: traditional writing preserves final results while compressing the underlying deductions and foundations, leaving the network as a latent construct obscured by coarse-grained citations. The advent of proof assistants offers the opportunity to map these dependencies, turning mathematics into explicit machine-analyzable networks. When a theorem is formalized, the compiler records every premise that its proof invokes, producing a machine-readable dependency graph that no amount of informal exposition could replicate. Recent large-scale formalization efforts illustrate the scale at which these networks now operate: the Equational Theories Project~\cite{bolan2025equationaltheoriesprojectadvancing} crowdsourced the classification of $4{,}694$ equational laws in Lean~4, and the ongoing Fermat's Last Theorem formalization~\cite{flt_lean} builds on thousands of \module{Mathlib}~\cite{The_mathlib_Community_2020} declarations.

\module{Mathlib} itself, the largest mathematical library for Lean~4, has reached a scale where its statistical structure begins to reflect the organizational patterns of mathematics itself. Its value lies not only in aggregating verified truths, but in transforming scattered results into a unified, queryable dependency network~\cite{Feng2026SemiAutonomousMD}. Yet the dependency network the compiler records is not the one mathematicians work with.

Figure~\ref{fig:blueprint-vs-decl} illustrates this divergence on a concrete example. The \texttt{leanblueprint}~\cite{massot_leanblueprint} graph on the left captures the mathematician's plan: four nodes connected by ``uses'' edges. The declaration graph on the right, extracted from the same formalized lemmas, reveals seven nodes and additional implicit edges, including infrastructure declarations (\texttt{List.rec}, \texttt{Nat.add}, \texttt{List.map}) that have no counterpart in the human plan. This paper systematically analyzes such gaps at the scale of the entire library.

Although the kernel deterministically enforces the validity of each theorem, the macroscopic organization of the library arises from human cognitive habits and collaborative workflows, encompassing naming conventions, file hierarchies, and namespace boundaries. Moreover, because the proof assistant also functions as a programming language, the type system of Lean introduces several structural layers absent from traditional mathematics. Each layer represents a necessary compromise between implicit human intuition and the strict demand of the compiler for explicitness. The typeclass mechanism, in particular, encodes mathematical knowledge that exists as implicit shared understanding among human practitioners (e.g., ``the integers form a ring'') as explicit \emph{instance declarations} for automated reasoning. This creates an entire stratum of nodes in the dependency graph that has no traditional counterpart (Definition~\ref{def:tc-graph}). More broadly, Lean's type system introduces distinct such compromises at every level:
\begin{itemize}
\item \emph{Synthesized edges} (Definition~\ref{def:synth-partition}). When a mathematician cites a lemma, the citation is visible on the page; when Lean resolves \texttt{a + 0 = a}, the compiler silently inserts references to \decl{HAdd}, \decl{OfNat}, and the relevant instances. In total, $74.2\%$ of all dependency edges are invisible in the source code.
\begin{center}
\begin{tikzpicture}[scale=0.95, every node/.style={transform shape},
  dot/.style={circle, fill=red!70!black, minimum size=5pt, inner sep=0pt},
  sdot/.style={circle, fill=gray!50, minimum size=5pt, inner sep=0pt},
  elink/.style={->, >=Stealth, red!70!black, semithick},
  slink/.style={->, >=Stealth, gray!50, semithick, dashed},
]
  \fill[red!6, rounded corners=4pt] (-0.8, -0.6) rectangle (1.8, 1.1);
  \node[font=\scriptsize\sffamily\itshape, red!40!black] at (0.5, 0.9) {math};
  \fill[gray!8, rounded corners=4pt] (2.1, -0.6) rectangle (5.2, 1.1);
  \node[font=\scriptsize\sffamily\itshape, gray!50] at (3.65, 0.9) {infrastructure};
  \node[dot, label={[font=\scriptsize\ttfamily, red!70!black]above:add\_zero}]
    (thm) at (2.5, 2.8) {};
  \node[dot, label={[font=\scriptsize\ttfamily, red!70!black]below:AddMonoid.add\_zero}]
    (axiom) at (0.5, 0.2) {};
  \node[sdot, label={[font=\scriptsize\ttfamily, gray!60]below:HAdd.hAdd}]
    (hadd) at (3.0, 0.2) {};
  \node[sdot, label={[font=\scriptsize\ttfamily, gray!60]below:OfNat.ofNat}]
    (ofnat) at (4.5, 0.2) {};
  \draw[elink] (thm) -- (axiom);
  \draw[slink] (thm) -- (hadd);
  \draw[slink] (thm) -- (ofnat);
\end{tikzpicture}
\end{center}

\item \emph{Coercions} (Definition~\ref{def:coercion-graph}). A mathematician writes ``let $n \in \mathbb{R}$'' when $n$ is a natural number and the embedding is universally understood; Lean must register an explicit chain of coercion instances $\mathbb{N} \hookrightarrow \mathbb{Z} \hookrightarrow \mathbb{Q} \hookrightarrow \mathbb{R}$, each silently inserting dependency edges.
\begin{center}
\begin{tikzpicture}[scale=0.95, every node/.style={transform shape},
  tp/.style={circle, draw=red!60!black, fill=red!6,
    font=\small, text=red!70!black, inner sep=3pt, minimum size=20pt},
  coe/.style={->, >=Stealth, red!60!black, semithick},
]
  \node[tp] (N) at (0, 0) {$\mathbb{N}$};
  \node[tp] (Z) at (1.8, 0) {$\mathbb{Z}$};
  \node[tp] (Q) at (3.6, 0) {$\mathbb{Q}$};
  \node[tp] (R) at (5.4, 0) {$\mathbb{R}$};
  \draw[coe] (N) -- node[above, font=\scriptsize\sffamily, text=gray!60] {ofNat} (Z);
  \draw[coe] (Z) -- node[above, font=\scriptsize\sffamily, text=gray!60] {ofInt} (Q);
  \draw[coe] (Q) -- node[above, font=\scriptsize\sffamily, text=gray!60] {ofRat} (R);
\end{tikzpicture}
\end{center}

\item \emph{Structure inheritance} (Definition~\ref{def:extends-graph}). The \texttt{extends} mechanism encodes ``every group is a monoid'' not as a logical implication but as a field-packing directive that auto-generates forgetful instances, building the algebraic hierarchy's inheritance lattice.
\begin{center}
\begin{tikzpicture}[scale=0.9, every node/.style={transform shape},
  nd/.style={rounded corners=2pt, draw=red!60!black, fill=red!6,
    font=\scriptsize\ttfamily, text=red!70!black, inner sep=3pt, minimum height=15pt},
  lf/.style={rounded corners=2pt, draw=teal!60!black, fill=teal!6,
    font=\scriptsize\ttfamily, text=teal!70!black, inner sep=3pt, minimum height=15pt},
  edg/.style={->, >=Stealth, red!60!black, semithick},
]
  \node[nd] (cm) at (1.6, 4.5) {CommMonoid};
  \node[nd] (mon) at (0.2, 3.2) {Monoid};
  \node[nd] (cs) at (3.0, 3.2) {CommSemi.};
  \node[nd] (sg) at (0.0, 1.9) {Semigroup};
  \node[nd] (moc) at (2.2, 1.9) {MulOneCl.};
  \node[lf] (mul) at (0.5, 0.6) {Mul};
  \node[lf] (one) at (2.6, 0.6) {One};
  \draw[edg] (cm) -- (mon);
  \draw[edg] (cm) -- (cs);
  \draw[edg] (mon) -- (sg);
  \draw[edg] (mon) -- (moc);
  \draw[edg] (cs) -- (sg);
  \draw[edg] (sg) -- (mul);
  \draw[edg] (moc) -- (mul);
  \draw[edg] (moc) -- (one);
\end{tikzpicture}
\end{center}

\item \emph{Additive mirroring} (Definition~\ref{def:decl-attributes}). A textbook proves a theorem for groups and writes ``similarly for the additive case''; Lean requires a separate declaration with a distinct proof term, auto-generated by the \texttt{to\_additive} metaprogram that duplicates every marked multiplicative result.

\item \emph{Definitional height} (Definition~\ref{def:def-height}). In traditional mathematics, the number of definitions that must be unfolded to verify a statement is rarely considered; Lean's kernel tracks this depth, which governs compilation cost and constitutes a form of implicit coupling invisible in the dependency graph.
\begin{center}
\begin{tikzpicture}[scale=0.95, every node/.style={transform shape},
  def/.style={rounded corners=2pt, draw=red!60!black, fill=red!6,
    font=\scriptsize\ttfamily, text=red!70!black, inner sep=3pt, minimum height=15pt},
  unfold/.style={->, >=Stealth, red!60!black, semithick, dashed},
  ht/.style={font=\scriptsize\sffamily\bfseries, text=blue!60!black},
]
  \node[def] (pow) at (0, 0) {Nat.pow};
  \node[def] (mul) at (3.5, 0) {Nat.mul};
  \node[def] (add) at (7.0, 0) {Nat.add};
  \draw[unfold] (pow) -- node[above, font=\scriptsize\sffamily, text=gray!60] {unfolds} (mul);
  \draw[unfold] (mul) -- node[above, font=\scriptsize\sffamily, text=gray!60] {unfolds} (add);
\end{tikzpicture}
\end{center}

\item \emph{Tactic usage} (Definition~\ref{def:tactic-usage}). Traditional proofs are written in natural language; Lean proofs are constructed via \emph{tactics}, automated proof strategies such as \texttt{simp}, \texttt{rw}, and \texttt{omega}, whose frequency and distribution vary across mathematical domains and encode the proof methodology favored by each subfield.
\begin{center}
\begin{tikzpicture}[scale=0.95, every node/.style={transform shape},
  dot/.style={circle, fill=red!70!black, minimum size=4pt, inner sep=0pt},
  tac/.style={rounded corners=2pt, draw=violet!60, fill=violet!8,
    font=\scriptsize\ttfamily, text=violet!70!black, inner sep=3pt, minimum height=14pt},
  dep/.style={->, >=Stealth, red!70!black, semithick},
]
  \node[dot, label={[font=\scriptsize\ttfamily, red!70!black]left:mul\_div\_assoc}]
    (thm) at (2.2, 3.2) {};
  \node[tac, anchor=west] at (2.5, 3.2) {rw};
  \node[dot, label={[font=\scriptsize\ttfamily, red!70!black]below:div\_eq\_mul\_inv}]
    (dm) at (1.0, 1.2) {};
  \node[dot, label={[font=\scriptsize\ttfamily, red!70!black]below:mul\_assoc}]
    (ma) at (3.8, 1.2) {};
  \draw[dep] (thm) -- (dm);
  \draw[dep] (thm) -- (ma);
  \fill[violet!30] (5.5, 2.7) rectangle (8.3, 3.0);
  \node[font=\scriptsize\sffamily, text=violet!70!black, anchor=west] at (5.6, 2.85)
    {\texttt{rw}\hspace{4pt}16.9\%};
  \fill[violet!25] (5.5, 2.2) rectangle (7.7, 2.5);
  \node[font=\scriptsize\sffamily, text=violet!70!black, anchor=west] at (5.6, 2.35)
    {\texttt{simp}\hspace{4pt}11.2\%};
  \fill[violet!20] (5.5, 1.7) rectangle (7.6, 2.0);
  \node[font=\scriptsize\sffamily, text=violet!70!black, anchor=west] at (5.6, 1.85)
    {\texttt{exact}\hspace{4pt}10.4\%};
\end{tikzpicture}
\end{center}

\item \emph{Automatic deriving} (Definition~\ref{def:deriving-edges}). Properties like decidable equality or printable representation are meta-level common sense in traditional mathematics; Lean requires explicit instance witnesses, which \texttt{deriving} handlers generate automatically.
\begin{center}
\begin{tikzpicture}[scale=0.95, every node/.style={transform shape},
  human/.style={circle, fill=red!70!black, minimum size=4pt, inner sep=0pt},
  auto/.style={circle, fill=gray!40, minimum size=4pt, inner sep=0pt},
  humanedge/.style={->, >=Stealth, red!70!black, semithick},
  autoedge/.style={->, >=Stealth, gray!40, semithick, dashed},
]
  \fill[gray!8, rounded corners=4pt] (-0.8, -1.2) rectangle (4.2, 0.4);
  \node[font=\scriptsize\sffamily\itshape, gray!50] at (1.7, 0.2) {$E_{\mathrm{auto}}$};
  \node[auto, label={[font=\scriptsize\ttfamily, gray!50]below:instRepr}]
    (r) at (0.0, -0.4) {};
  \node[auto, label={[font=\scriptsize\ttfamily, gray!50]below:instDecEq}]
    (d) at (1.7, -0.4) {};
  \node[auto, label={[font=\scriptsize\ttfamily, gray!50]below:instHash.}]
    (h) at (3.4, -0.4) {};
  \node[human, label={[font=\scriptsize\ttfamily, red!70!black]above:Stained}]
    (st) at (1.7, 2.8) {};
  \node[human, label={[font=\scriptsize\ttfamily, red!70!black]left:name}]
    (sn) at (0.0, 1.6) {};
  \node[human, label={[font=\scriptsize\ttfamily, red!70!black]right:goal}]
    (sg) at (3.4, 1.6) {};
  \draw[autoedge] (r) -- (st);
  \draw[autoedge] (d) -- (st);
  \draw[autoedge] (h) -- (st);
  \draw[humanedge] (sn) -- (st);
  \draw[humanedge] (sg) -- (st);
\end{tikzpicture}
\end{center}

\end{itemize}
Each mechanism deposits its own signature in the dependency graph (\S\ref{sec:declaration-graph}), and collectively, they constitute the \emph{tool-infrastructure layer}. Consequently, we hypothesize that these network properties extend beyond the mere documentation of logical dependencies. They instead provide measurable traces of mathematical production, reflecting community behavior, tooling mechanisms, and the subtle nuances of language design.

The present paper analyzes these structural patterns to establish a quantitative baseline for formal mathematics produced by humans. This baseline serves two purposes: it surfaces actionable engineering targets (e.g., the 1.6\% median import utilization that motivates finer-grained imports, or the 153-layer critical path that constrains parallel compilation), and it provides a reference point against which future changes, particularly those driven by AI-assisted formalization~\cite{Hubert_2026,achim2025aristotleimolevelautomatedtheorem,chen2025seedproverdeepbroadreasoning,xin2025deepseekproverv}, can be measured.

We focus on \module{Mathlib}: 308,129 declarations, 8,436,366 dependency edges, and 7,563 modules (Definitions~\ref{def:module}--\ref{def:module-tree}). All statistics refer to commit \texttt{534cf0b} (2 Feb 2026), extracted using \texttt{importGraph} for the module graph (Definition~\ref{def:module-graph}) and premise-extraction tooling for the declaration graph (Definition~\ref{def:thm-graph}). Our module import graph reflects a post-\texttt{shake} state (the linter removes certain transitively redundant imports in CI). Moreover, Lean~4's module system now distinguishes \texttt{public import} from \texttt{import} (Definition~\ref{def:visibility-graph}); our snapshot captures a transitional period where most imports remain \texttt{public}.

The remainder of the paper is structured as follows. \S\ref{sec:contribution} details our primary findings (summarized quantitatively in Table~\ref{tab:summary}). Following a review of prior literature, \S\ref{sec:conceptual-framework} articulates our conceptual framework. This section illustrates how the type system of Lean mediates this dynamic. \S\ref{sec:methodology} outlines our research workflow and analytical tools. The appendices provide formal graph definitions and structural decompositions (Appendix~\ref{sec:graph-definitions} and Table~\ref{tab:decompositions}), alongside extended statistical analyses for each dependency layer that include cross level comparisons (Appendices~\ref{app:module-detail} through \ref{sec:cross_level}). Finally, \S\ref{sec:conclusion} discusses the implications of our results for practical application and suggests directions for future research.
\paragraph{Data and code availability.}
To support reproducibility and downstream research, we release the complete extraction pipeline and graph datasets. The raw graph data, including the module import graph~$G_{\mathrm{module}}$, the declaration dependency graph~$G_{\mathrm{thm}}$, and all namespace-level aggregations~$G_{\mathrm{ns}}^{(k)}$ (Definition~\ref{def:ns-graph}), are published as a HuggingFace dataset at \url{https://huggingface.co/datasets/MathNetwork/MathlibGraph}, enabling independent analysis without requiring a local Lean~4 installation. The extraction and analysis code is available at \url{https://github.com/MathNetwork/mathlib-network}, with the v1.0.0 release archived on Zenodo at \url{https://doi.org/10.5281/zenodo.19746468}.

\section{Our Contribution}
\label{sec:contribution}

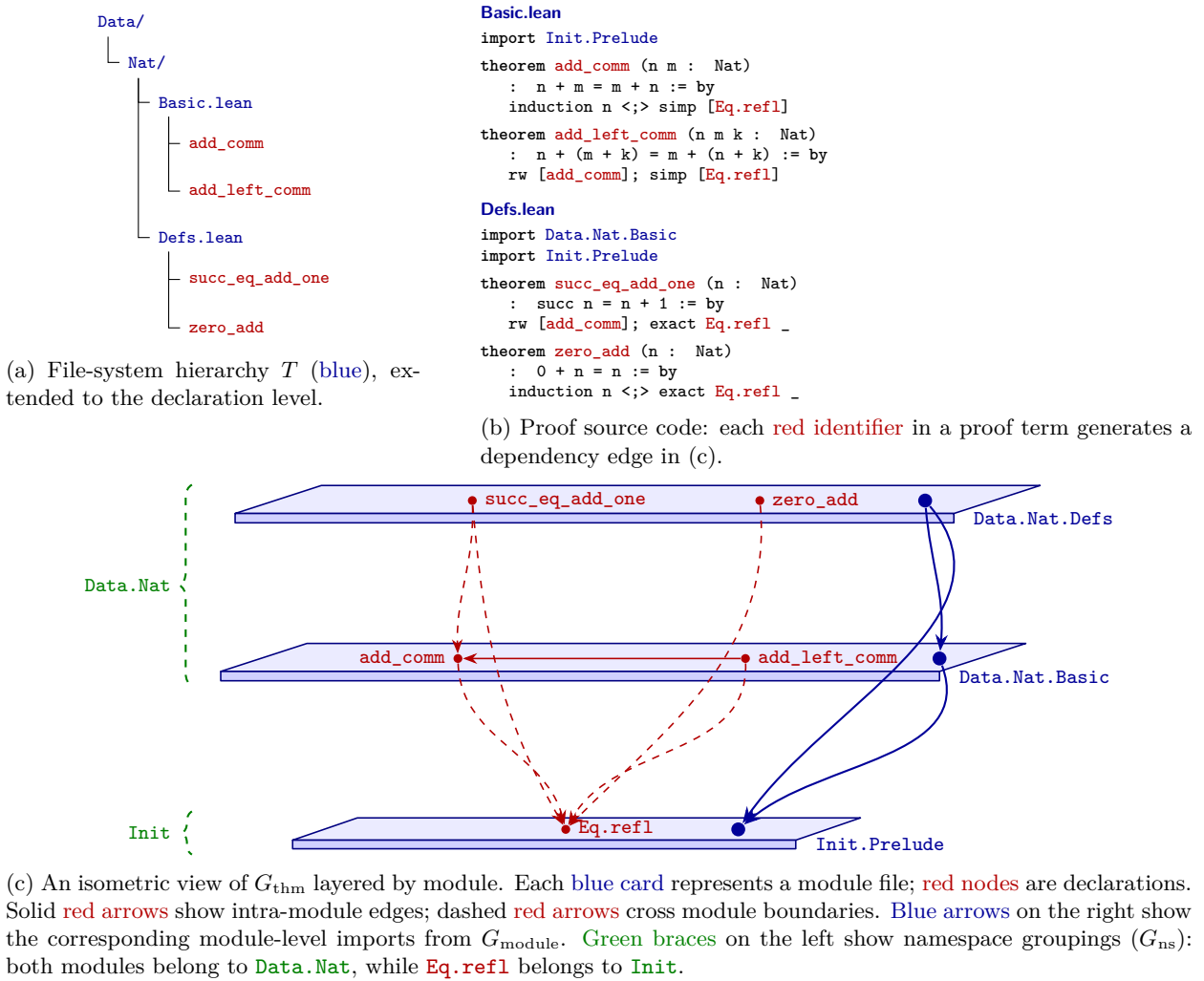
\begin{figure}[ht]
\centering
\begin{subfigure}[t]{0.35\textwidth}
\centering
\vspace{0pt}
\begin{forest}
  for tree={
    font=\scriptsize\ttfamily,
    text=blue!60!black,
    grow'=0,
    child anchor=west,
    parent anchor=south,
    anchor=west,
    calign=first,
    edge path={
      \noexpand\path [draw, black, \forestoption{edge}]
        (!u.south west) +(7.5pt,0) |- (.child anchor)\forestoption{edge label};
    },
    before typesetting nodes={
      if n=1 {insert before={[,phantom]}} {}
    },
    fit=band,
    before computing xy={l=12pt},
  }
[Data/
  [Nat/
    [Basic.lean
      [\textcolor{red!70!black}{add\_comm}]
      [\textcolor{red!70!black}{add\_left\_comm}]
    ]
    [Defs.lean
      [\textcolor{red!70!black}{succ\_eq\_add\_one}]
      [\textcolor{red!70!black}{zero\_add}]
    ]
  ]
]
\end{forest}
\caption{File-system hierarchy~$T$ (\textcolor{blue!60!black}{blue}), extended to the declaration level.}
\label{fig:tree-view-main}
\end{subfigure}%
\hfill
\begin{subfigure}[t]{0.60\textwidth}
\centering
\vspace{0pt}
\raggedright\scriptsize
\textsf{\bfseries\color{blue!60!black}Basic.lean}\\[2pt]
\ttfamily
\textbf{import} \textcolor{blue!60!black}{Init.Prelude}\\[3pt]
\textbf{theorem} \textcolor{red!70!black}{add\_comm} (n m : Nat)\\
\hspace{1.5em}: n + m = m + n := \textbf{by}\\
\hspace{1.5em}induction n <;> simp [\textcolor{red!70!black}{Eq.refl}]\\[3pt]
\textbf{theorem} \textcolor{red!70!black}{add\_left\_comm} (n m k : Nat)\\
\hspace{1.5em}: n + (m + k) = m + (n + k) := \textbf{by}\\
\hspace{1.5em}rw [\textcolor{red!70!black}{add\_comm}]; simp [\textcolor{red!70!black}{Eq.refl}]\\[6pt]
\textsf{\bfseries\color{blue!60!black}Defs.lean}\\[2pt]
\ttfamily
\textbf{import} \textcolor{blue!60!black}{Data.Nat.Basic}\\
\textbf{import} \textcolor{blue!60!black}{Init.Prelude}\\[3pt]
\textbf{theorem} \textcolor{red!70!black}{succ\_eq\_add\_one} (n : Nat)\\
\hspace{1.5em}: succ n = n + 1 := \textbf{by}\\
\hspace{1.5em}rw [\textcolor{red!70!black}{add\_comm}]; exact \textcolor{red!70!black}{Eq.refl} \_\\[3pt]
\textbf{theorem} \textcolor{red!70!black}{zero\_add} (n : Nat)\\
\hspace{1.5em}: 0 + n = n := \textbf{by}\\
\hspace{1.5em}induction n <;> exact \textcolor{red!70!black}{Eq.refl} \_
\caption{Proof source code: each \textcolor{red!70!black}{red identifier} in a proof term generates a dependency edge in~(c).}
\label{fig:source-view-main}
\end{subfigure}

\par\vspace{4pt}

\begin{subfigure}[b]{\textwidth}
\centering
\begin{tikzpicture}[
  every node/.style={font=\scriptsize\ttfamily, inner sep=2pt},
  dot/.style={circle, fill=red!70!black, minimum size=3.5pt, inner sep=0pt},
  dep/.style={->, >=Stealth, red!70!black, semithick},
  depext/.style={->, >=Stealth, red!70!black, semithick, dashed},
  modimp/.style={->, >=Stealth, blue!60!black, thick},
]
  \filldraw[fill=blue!8, draw=blue!50!black, semithick, line join=round]
    (0.2,4.7) -- (10.2,4.7) -- (11.4,5.09) -- (1.4,5.09) -- cycle;
  \filldraw[fill=blue!18, draw=blue!50!black, semithick, line join=round]
    (0.2,4.7) -- (10.2,4.7) -- (10.2,4.57) -- (0.2,4.57) -- cycle;
  \node[font=\footnotesize\sffamily, blue!60!black, anchor=west] at (10.4,4.63)
    {\textnormal{\module{Data.Nat.Defs}}};
  \node[dot] (sea) at (3.5,4.88) {};
  \node[font=\footnotesize\ttfamily, red!70!black, anchor=west] at (3.6,4.88) {succ\_eq\_add\_one};
  \node[dot] (za)  at (7.5,4.88) {};
  \node[font=\footnotesize\ttfamily, red!70!black, anchor=west] at (7.6,4.88) {zero\_add};

  \filldraw[fill=blue!8, draw=blue!50!black, semithick, line join=round]
    (0,2.5) -- (10.0,2.5) -- (11.2,2.89) -- (1.2,2.89) -- cycle;
  \filldraw[fill=blue!18, draw=blue!50!black, semithick, line join=round]
    (0,2.5) -- (10.0,2.5) -- (10.0,2.37) -- (0,2.37) -- cycle;
  \node[font=\footnotesize\sffamily, blue!60!black, anchor=west] at (10.2,2.43)
    {\textnormal{\module{Data.Nat.Basic}}};
  \node[dot] (ac)  at (3.3,2.68) {};
  \node[font=\footnotesize\ttfamily, red!70!black, anchor=east] at (3.2,2.68) {add\_comm};
  \node[dot] (alc) at (7.3,2.68) {};
  \node[font=\footnotesize\ttfamily, red!70!black, anchor=west] at (7.4,2.68) {add\_left\_comm};

  \filldraw[fill=blue!8, draw=blue!50!black, semithick, line join=round]
    (1.0,0.15) -- (8.0,0.15) -- (8.9,0.46) -- (1.9,0.46) -- cycle;
  \filldraw[fill=blue!18, draw=blue!50!black, semithick, line join=round]
    (1.0,0.15) -- (8.0,0.15) -- (8.0,0.03) -- (1.0,0.03) -- cycle;
  \node[font=\footnotesize\sffamily, blue!60!black, anchor=west] at (8.2,0.09)
    {\textnormal{\module{Init.Prelude}}};
  \node[dot] (er)  at (4.8,0.3) {};
  \node[font=\footnotesize\ttfamily, red!70!black, anchor=west] at (4.9,0.3) {Eq.refl};

  \node[circle, fill=blue!60!black, minimum size=5.5pt, inner sep=0pt] (md) at (9.8, 4.88) {};
  \node[circle, fill=blue!60!black, minimum size=5.5pt, inner sep=0pt] (mb) at (10.0, 2.68) {};
  \node[circle, fill=blue!60!black, minimum size=5.5pt, inner sep=0pt] (mo) at (7.2, 0.3) {};
  \draw[modimp] (md) to[out=-85,in=85] (mb);
  \draw[modimp] (mb) to[out=-70,in=40] (mo);
  \draw[modimp] (md) to[out=-50,in=55] (mo);

  \draw[dep] (alc) -- (ac);
  \draw[depext] (sea) to[out=-88,in=92] (ac);
  \draw[depext] (ac)  to[out=-88,in=105] (er);
  \draw[depext] (alc) to[out=-88,in=60] (er);
  \draw[depext] (sea) to[out=-85,in=120] (er);
  \draw[depext] (za)  to[out=-85,in=45] (er);

  \draw[green!50!black, thick, dashed, decorate, decoration={brace, amplitude=5pt, mirror}]
    (-0.4, 5.1) -- (-0.4, 2.3)
    node[midway, left=6pt, font=\footnotesize\sffamily, green!50!black, align=right] {\ns{Data.Nat}};
  \draw[green!50!black, thick, dashed, decorate, decoration={brace, amplitude=4pt, mirror}]
    (-0.4, 0.55) -- (-0.4, -0.05)
    node[midway, left=6pt, font=\footnotesize\sffamily, green!50!black, align=right] {\ns{Init}};

\end{tikzpicture}
\caption{An isometric view of $G_{\mathrm{thm}}$ layered by module. Each \textcolor{blue!60!black}{blue card} represents a module file; \textcolor{red!70!black}{red nodes} are declarations. Solid \textcolor{red!70!black}{red arrows} show intra-module edges; dashed \textcolor{red!70!black}{red arrows} cross module boundaries. \textcolor{blue!60!black}{Blue arrows} on the right show the corresponding module-level imports from $G_{\mathrm{module}}$. \textcolor{green!50!black}{Green braces} on the left show namespace groupings ($G_{\mathrm{ns}}$): both modules belong to \ns{Data.Nat}, while \decl{Eq.refl} belongs to \ns{Init}.}
\label{fig:dep-view-main}
\end{subfigure}
\caption{Three views of the same declarations. (a)~The file-system hierarchy~$T$ (\textcolor{blue!60!black}{blue}), extended to the declaration level. (b)~Proof source code: each \textcolor{red!70!black}{red identifier} creates a dependency edge in~(c). (c)~An isometric view of~$G_{\mathrm{thm}}$ layered by module; of six dependency edges, only one stays within its module. \textcolor{blue!60!black}{Blue arrows} on the right show the corresponding module-level imports from $G_{\mathrm{module}}$. All names have the \texttt{Nat.}\ prefix removed.}
\label{fig:three-views-main}
\end{figure}

We define a family of dependency graphs from \module{Mathlib} and analyze them as a multi-layer network. Figure~\ref{fig:three-views-main} illustrates the three primary layers on a concrete example. When a proof cites a premise, the kernel records a directed edge between the two declarations (Figure~\ref{fig:three-views-main}b, red identifiers). The collection of all such edges forms the \emph{declaration dependency graph}~$G_{\mathrm{thm}}$ (Definition~\ref{def:thm-graph}), whose 308,129 nodes and 8.4 million edges encode the logical content of the library. Declarations reside in source files (Figure~\ref{fig:three-views-main}a); each file constitutes a module, and each \texttt{import} statement between files produces an edge in the \emph{module import graph}~$G_{\mathrm{module}}$ (Definition~\ref{def:module-graph}). In the isometric view (Figure~\ref{fig:three-views-main}c), the blue cards are modules: of six declaration-level edges, only one stays within its module, while the remaining five cross file boundaries and collapse into blue import arrows at the module level. A third organizational layer, orthogonal to the file hierarchy, is the human-chosen naming taxonomy: the \emph{namespace graphs}~$G_{\mathrm{ns}}^{(k)}$ (Definition~\ref{def:ns-graph}) aggregate declarations by their dotted-name prefix (e.g., all declarations under \ns{Mathlib.Topology}). Together with the module tree~$T$ (Definition~\ref{def:module-tree}), these layers range from fine-grained mathematical logic to coarse-grained human organization.

We apply standard network metrics to each layer and introduce graph decompositions that isolate individual structural dimensions (Appendix~\ref{sec:graph-definitions}, Table~\ref{tab:decompositions}). Three findings emerge, which we interpret through the product/process framework of \S\ref{sec:conceptual-framework}.

\begin{finding}[Finding 1:]
Human file folders and naming conventions diverge from logical dependencies. Formal libraries encode human convenience and tooling infrastructure as much as mathematics itself.
\end{finding}

\begin{figure}[ht]
\centering
\begin{tikzpicture}[
  bp/.style={draw, rounded corners=2pt, fill=green!10, font=\scriptsize, inner sep=3pt, minimum width=1.4cm, minimum height=0.5cm},
  decl/.style={circle, fill, inner sep=1.5pt},
  arr/.style={->, >=Stealth, thin},
  darr/.style={->, >=Stealth, thin, red!60!black}
]
\node[font=\small\bfseries] at (1.5, 2.8) {Blueprint};
\node[bp] (la) at (1.5, 2) {\texttt{length\_append}};
\node[bp] (lm) at (0, 0.5) {\texttt{length\_map}};
\node[bp] (mm) at (3, 0.5) {\texttt{map\_map}};
\node[bp] (len) at (1.5, -0.8) {\texttt{List.length}};
\draw[arr] (la) -- node[left, font=\tiny] {uses} (lm);
\draw[arr] (la) -- node[right, font=\tiny] {uses} (len);
\draw[arr] (lm) -- node[left, font=\tiny] {uses} (len);
\draw[arr] (mm) -- node[right, font=\tiny] {uses} (len);

\node[font=\small\bfseries] at (7.5, 2.8) {Declaration graph};
\node[decl, label={[font=\scriptsize]above:\texttt{length\_append}}] (a) at (7.5, 2) {};
\node[decl, label={[font=\scriptsize]left:\texttt{length\_map}}] (b) at (5.8, 0.8) {};
\node[decl, label={[font=\scriptsize]right:\texttt{map\_map}}] (c) at (9.2, 0.8) {};
\node[decl, label={[font=\scriptsize]below:\texttt{List.length}}] (d) at (7.5, -0.3) {};
\node[decl, label={[font=\scriptsize]left:\texttt{List.rec}}] (rec) at (5.5, -0.5) {};
\node[decl, label={[font=\scriptsize]right:\texttt{Nat.add}}] (add) at (9.5, -0.5) {};
\node[decl, label={[font=\scriptsize]below:\texttt{List.map}}] (map) at (7.5, -1.3) {};
\draw[darr] (a) -- (b);
\draw[darr] (a) -- (d);
\draw[darr] (a) -- (add);
\draw[darr] (b) -- (d);
\draw[darr] (b) -- (map);
\draw[darr] (c) -- (map);
\draw[darr] (c) -- (rec);
\draw[darr] (d) -- (rec);
\draw[darr, dashed, gray] (a) to[bend right=15] (rec);
\draw[darr, dashed, gray] (a) to[bend left=15] (map);
\end{tikzpicture}
\caption{Blueprint (left) versus compiler-extracted declaration graph (right) for basic list lemmas. The compiler graph introduces additional nodes and implicit edges (dashed) absent from the human plan.}
\label{fig:blueprint-vs-decl}
\end{figure}
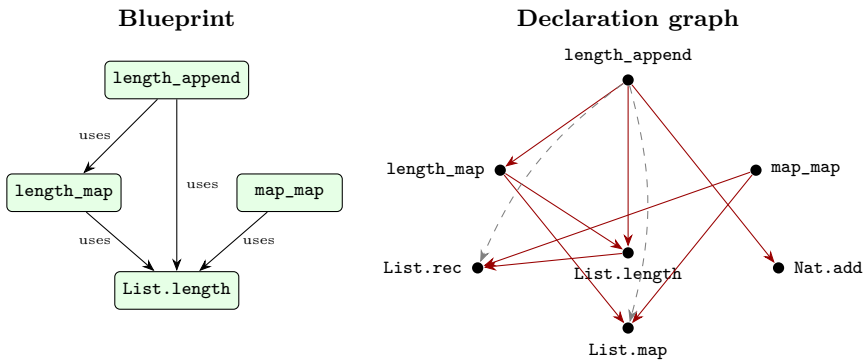

File layout and namespace naming agree with each other (NMI $= 0.71$) but diverge from the logical dependency structure (NMI $= 0.34$ against dependency-based communities; \S\ref{sec:sa-community}). Cross-namespace edges reach $50.9\%$ at the declaration level (\S\ref{sec:thm-cross-namespace}), and structural containment decays rapidly with hierarchy depth (\S\ref{sec:containment-decay}, \S\ref{sec:module-containment}). Meanwhile, Lean's elaborator inserts $74.2\%$ of all dependency edges automatically (Definition~\ref{def:synth-partition}), and the hub structure bifurcates into language infrastructure (coercions, equality) and mathematical infrastructure (\decl{CategoryTheory.Category}, \decl{Real}, \decl{TopologicalSpace}; \S\ref{sec:thm-types}). Figure~\ref{fig:blueprint-vs-decl} illustrates the gap concretely: a \texttt{leanblueprint}~\cite{massot_leanblueprint} graph for basic list lemmas has four nodes; the compiler-extracted declaration graph for the same lemmas has seven, with additional implicit edges that have no counterpart in the human plan.

\begin{finding}[Finding 2:]
Developers over-import dependencies, using only a fraction of the declarations they pull into scope. This creates compilation bottlenecks. The network-theoretic approach helps quantify the resulting overhead.
\end{finding}

Lean's module system operates at file granularity, so each \texttt{import} pulls in an entire source file. The median import utilization is only $1.6\%$ (Definition~\ref{def:import-util}), and the module graph contains $17.5\%$ transitive redundancy that the \texttt{shake} linter~\cite{baanen2025growing} intentionally preserves (\S\ref{sec:transitive-reduction}). This redundancy constrains build performance: the critical path spans $161$ modules with a parallelism ratio of $22.4\times$ (Definition~\ref{def:build-graph}).

\begin{finding}[Finding 3:]
Formalization preserves mathematical labels while erasing their conceptual hierarchy. Network centrality measures technical utility rather than mathematical depth.
\end{finding}

The DAG depths of the three layers are $153$, $84$, and $8$ (\S\ref{sec:dag-depth}, \S\ref{sec:thm-dag-depth}, \S\ref{sec:ns-dag-depth}): human-organized structures are deeper than logical ones. Edge classification (Definition~\ref{def:stmt-proof-dep}) shows that $43.9\%$ of dependencies arise exclusively from proofs versus only $8.1\%$ from statements, confirming that the dependency structure is shaped by the process of proving rather than the content of assertions. Centrality rankings (\S\ref{sec:thm-types}) make the flattening concrete: \decl{Eq.refl}, a logically trivial axiom (reflexivity of equality), ranks second by in-degree with $69{,}580$ citations, while results such as the Chinese Remainder Theorem do not appear in the top~$100$.
\section{Previous and related work}
\paragraph{Software dependency networks.}
Dependency graphs arise naturally in software ecosystems: package managers such as npm, PyPI, and Maven define directed acyclic graphs in which each package declares its requirements~\cite{Decan_2019,labelle2004interpackagedependencynetworksopensource}. In the formal mathematics setting, Blanchette et al.~\cite{Blanchette_2015} mine structural properties of the Isabelle AFP, making it the closest methodological predecessor to our work, and Bancerek et al.~\cite{Bancerek_2018} document the MML's role and evolution. Studies of these ecosystems have revealed heavy-tailed degree distributions, small-world connectivity, and fragility under targeted removal of hub packages. Our analysis applies a similar methodology to a formal mathematics library, a setting in which every edge carries a stronger guarantee than ``package~$A$ requires package~$B$ at build time.'' In a type-checked library, premise-level dependencies are mechanically certified: if a constant appears in a proof term, the kernel enforces that dependency. Module-level imports, by contrast, are developer-declared and can be redundant relative to reachability ($17.5\%$ are transitively implied, measured after Mathlib's \texttt{shake} linter has already pruned certain redundant imports in CI). The graph thus records not merely a build requirement, but a semantic relationship.

\paragraph{\module{Mathlib} growth and maintenance.}
Van Doorn, Ebner, and Lewis~\cite{vandoorn2020maintaining} described the engineering practices that sustain \module{Mathlib} as a collaboratively maintained library, addressing linting, documentation, and refactoring workflows. Ringer et al.~\cite{Ringer_2019} provide a canonical survey of proof engineering, noting that ``there is still relatively little work describing tools and best practices''; our contribution fills part of this gap. Commelin and Topaz~\cite{commelin2023abstractionboundariesspecdriven} proposed abstraction boundaries and specification-driven development as organizing principles for formal mathematics. More recently, Commelin et al.~\cite{baanen2025growing} documented the challenges of scaling \module{Mathlib}, including build times, namespace reorganization, and technical debt. On the tooling side, Massot's \textsc{leanblueprint} system~\cite{massot_leanblueprint} provides infrastructure for managing large formalization projects: mathematicians manually annotate dependency relations between statements in \LaTeX{} using \verb|\uses{}| tags, producing a dependency graph that tracks formalization progress. This approach has been adopted by several major projects, including the Liquid Tensor Experiment~\cite{lean_liquid} and the ongoing formalization of Fermat's Last Theorem~\cite{flt_lean}. Zhu et al.~\cite{zhu2026leanarchitectautomatingblueprintgeneration} complement this with \textsc{LeanArchitect}, which automatically infers dependency information from Lean source code, eliminating the manual duplication between informal and formal representations. Notably, their work includes AI integration: in a case study, the automatically extracted dependency graph guides an AI prover (Achim et al.'s Aristotle system~\cite{achim2025aristotleimolevelautomatedtheorem}) to formalize theorems in topological order. Our work differs from both in scope and purpose: \textsc{leanblueprint} and \textsc{LeanArchitect} extract dependency graphs to manage individual formalization projects; we analyze the global dependency structure of the entire \module{Mathlib} library as a network, applying tools from network science (degree distributions, community detection, robustness analysis) to characterize structural patterns that no single project would reveal. Avigad et al.~\cite{avigad2024anatomyformalproof} describe the practitioner experience of using proof assistants; Klein et al.~\cite{Klein_2012} address managing large-scale proofs in Isabelle, including tools for moving lemmas across modules. These works address \module{Mathlib}'s maintenance from a software engineering perspective. Our analysis complements theirs by providing a network-theoretic diagnostic of the same challenges they address qualitatively: we show, for example, that the $17.5\%$ import redundancy rate (Definition~\ref{def:redundancy-rate}, measured post-\texttt{shake}) and $0.107$ module cohesion quantify the development ergonomics overhead that maintenance practices must manage.

\paragraph{Machine learning for formal mathematics.}
The graph structure of \module{Mathlib} has been exploited in machine learning, particularly for premise selection. Yang et al.~\cite{yang2023leandojo} introduced the LeanDojo benchmark, extracting proof traces from \module{Mathlib} that implicitly define a dependency graph. Petrov\v{c}i\v{c} et al.~\cite{petrovcic2025combining} constructed a heterogeneous dependency graph from this benchmark and applied relational graph convolutional networks for premise selection, achieving a $25\%$ improvement over text-only baselines. Lu et al.~\cite{lu2025lean} built a directed acyclic graph of Lean~4 statements with dependency edges to support semantic search. On the knowledge-representation side, Uskuplu et al.~\cite{uskuplu2025knowtexvisualizingmathematicaldependencies} argued that dependency graphs should become a standard feature of mathematical writing. On the extraction side, our analysis draws on several complementary tools: \texttt{lean4export}~\cite{lean4export} dumps every declaration in the Lean environment; \texttt{lean-training-data}~\cite{lean_training_data} extracts the premises invoked in each proof; \texttt{importGraph}~\cite{importgraph} produces the module-level import graph; and \texttt{jixia}~\cite{jixia2024} exposes per-declaration metadata such as tactic usage and definitional heights. These systems consume the dependency graph as training data or as a retrieval index. We study the graph itself, not as input to a downstream task, but as an object of independent structural interest. Our structural analysis (two-layer hub structure, community detection) could inform premise selection by quantifying how much of the search space is language infrastructure versus mathematical content.

\paragraph{Complex networks.}
The analytical tools we employ are basic and standard in network science. Heavy-tailed network theory~\cite{Baraba_si_1999}, small-world analysis~\cite{Watts_1998}, and modularity-based community detection~\cite{Newman_2004,Blondel_2008} have been applied to biological, social, and technological networks, but not, to our knowledge, to a large formal mathematics library at the multi-layer depth we pursue here. (Blanchette et al.~\cite{Blanchette_2015} apply mining techniques to the Isabelle AFP, and Huch and Wenzel~\cite{huch_et_al:LIPIcs.ITP.2024.22} analyze the AFP's theory DAG for parallel build scheduling, but neither pursues the three-layer network analysis we develop.) In a related setting, Gonzaga et al.~\cite{GONZAGA_2014} studied the network structure of three online mathematical libraries (Wikipedia restricted to mathematics, MathWorld, and DLMF), finding large strongly connected components, the absence of clear power laws, and the effectiveness of stress centrality for navigating mathematical content. Their networks, however, are hyperlinked encyclopedias with informal content, while here we consider a machine-verified library in which every dependency is logically certified.

\section{Conceptual Framework}
\label{sec:conceptual-framework}
Our premise is that the evolution of mathematical practice is driven by a continuous feedback loop between two dimensions: inherited mathematical structure and human organizational process~\cite{Avigad2023MathematicsAT,commelin2023abstractionboundariesspecdriven}. \emph{Inherited structure} encompasses definitions, theorems, canonical proof pathways, and disciplinary boundaries. It represents the enduring record of successive generations of mathematical labor, transmitted through formal literature, curricula, and computational libraries. Conversely, \emph{human organizational process} involves the active reconfiguration of this inheritance through contemporary tooling. It includes both the systematic refactoring of the existing corpus and the generative synthesis of novel mathematics. Ultimately, the organizational practice of one era crystallizes into the foundational inheritance of the next.

Each revolution in mathematical tooling, from writing to print and from print to computation, has reshaped the balance between these two dimensions. New tools do not render prior mathematics irrelevant; rather, their utility depends on the underlying mathematical framework. The central insight of our analytical framework is that \module{Mathlib} functions as a \emph{transitional artifact}. It represents a large-scale, systematic migration of traditional mathematical knowledge, which was historically bound by the analog constraints of print media and physical locality, into a digital sociotechnical environment characterized by decentralized collaboration and version control. Consequently, its architecture simultaneously embeds the structural signatures of both epochs.

Crucially, the proof assistant mediating this migration is itself a programming language. Lean operates simultaneously as a proof checker and a general-purpose language, and this dual nature directly shapes the resulting graph topology. Because Lean is programmable, mechanisms such as typeclasses, coercions, structure inheritance, compiler-synthesized edges, metaprograms like \texttt{to\_additive}, and \texttt{deriving} handlers all deposit distinct structural signatures into the dependency graph. Together, they constitute the \emph{tooling infrastructure layer}, a complete stratum of nodes and edges that encodes knowledge traditionally left implicit by human practitioners (Appendix~\ref{sec:graph-definitions} defines the corresponding graph decompositions). The premise dependency network is therefore neither an exact transcription of traditional mathematics nor a completely novel creation. Instead, it acts as a \emph{tool-mediated migration product}. The inherited structure dictates the semantic content, keeping canonical theorems and macroscopic dependencies invariant. Concurrently, the contemporary medium dictates the syntactic form, requiring microscopic dependency chains, proof architectures, and type-theoretic representations to be completely reconfigured to satisfy the compiler. Accordingly, our metrics provide a snapshot of a specific state of production rather than an eternal structural law. We acknowledge the temporal contingency of these measurements in \S\ref{sec:limitations}.

However, each raw dependency graph is not a pure embodiment of a single dimension, but rather a \emph{composite} of both. The graph decompositions defined in Appendix~\ref{sec:graph-definitions} serve as systematic tools for decoupling these elements, acting as projections that isolate individual dimensions from the combined structure. At the declaration level, partitions between statements and proofs, as well as between explicit and synthesized edges, separate inherited content from tool-mediated form. Similarly, the typeclass instance graph and its finer projections isolate the tooling infrastructure layer by its underlying mechanism. At the module level, visibility and build graphs distinguish intentional API design from engineering overhead. Finally, at the node level, declaration attributes enrich nodes with metadata spanning both dimensions.
The macro-organizational structure of the library, including its module partitioning and namespace taxonomies, further instantiates this tension. Top-level directories such as \module{Mathlib.Analysis}, \module{Mathlib.Topology}, and \module{Mathlib.Algebra} map directly onto classical disciplinary boundaries. This alignment demonstrates that legacy cognitive classifications persist across the shift in medium. At the same time, the demands of the proof assistant have catalyzed the emergence of entirely novel, medium-specific categories, such as \module{Mathlib.Tactic} and \module{Mathlib.Lean}, which possess no analogue in traditional mathematics. The gap between what is formalizable in principle and what is feasible in practice is precisely where proof assistant design operates. Our measurements of $17.5\%$ import redundancy, a cohesion score of $0.107$, and a graph depth of $153$ layers explicitly quantify this gap, representing the structural distance between what is logically necessary and what is practically maintained.

The primary empirical signal for our research is the structural friction generated by this transition. For instance, tightly interdependent declarations are routinely distributed across distinct modules to improve compilation efficiency, while logically orthogonal content is aggregated under single namespaces due to legacy conventions. This friction is amplified by the decentralized nature of \module{Mathlib}. Because contributors lack a global view of the library's dependency structure, local pragmatic decisions accumulate unseen, making macroscopic network diagnostics an absolute necessity. Furthermore, this evolving ecosystem introduces a self-reinforcing flywheel effect: well-structured proofs provide training data for artificial intelligence, and AI-generated proofs subsequently reshape the library. (We discuss the consequences of this dynamic for library trust and maintenance in \S\ref{sec:interplay} and \S\ref{sec:practice}.)


\section{Methodology}
\label{sec:methodology}

This section describes how the study was conducted. The formal definitions of the multi-layer dependency graphs are presented separately in Appendix~\ref{sec:graph-definitions}. \label{sec:workflow} This paper is a human-AI collaborative open-source research project. We invite readers to regard it as a continuously evolving effort rather than a traditional, self-contained academic publication. All analyses were conducted in a command-line environment; the complete operation history, commit log, and revision trail are publicly available in the GitHub repository for inspection, reproduction, and critique. 
Our methodological starting point is to follow the internal logic of \module{Mathlib} and to deploy a toolchain for systematic extraction and analysis. Within this framework, humans drive the research questions, interpret results, and decide what to include or discard, while AI orchestrates and executes open-source tools (managing dependency installation, running extraction scripts, generating analysis code, and coordinating data flow between tools).

Raw data extraction relies on three complementary open-source tools: \texttt{importGraph}~\cite{importgraph} (module-level import edges), \texttt{jixia}~\cite{jixia2024} (declaration-level dependency edges and tactic usage profiles), and \texttt{lean-training-data}~\cite{lean_training_data} (declaration metadata). The resulting dataset comprises $317{,}655$ nodes, $8.4$M edges across $7{,}563$ modules, and is publicly available on HuggingFace. All analyses are implemented in Python using \texttt{NetworkX}~\cite{hagberg2008} (graph construction and centrality), \texttt{python-louvain}~\cite{python_louvain} (community detection), \texttt{powerlaw}~\cite{Alstott_2014} (heavy-tailed fitting), \texttt{scikit-learn}~\cite{JMLR:v12:pedregosa11a} (NMI/ARI evaluation), \texttt{Pandas}~\cite{mckinney2010}, \texttt{NumPy}~\cite{Harris_2020}, and \texttt{Matplotlib}~\cite{4160265}. Analysis scripts follow a test-driven development cycle: write the test first, then implement, then verify incrementally.

The three extraction tools are maintained as part of the \module{Mathlib} ecosystem and are routinely used by the community for dependency analysis and documentation generation; we treat their output as authoritative.

\section{Conclusion and Future Work}
\label{sec:conclusion}


\begin{table}[ht]
\centering
\caption{Cross-level summary of structural findings.}
\label{tab:summary}
\small
\begin{tabular}{lrrrr}
\toprule
Metric & \S\ref{sec:module-import} Module & \S\ref{sec:theorem-premise} Declaration & \S\ref{sec:namespace-graph} Namespace & \S\ref{sec:module-declaration} Cross-Level \\
\midrule
\multicolumn{5}{@{}l}{\textit{Scale and topology}} \\[2pt]
Nodes                      & $7{,}563$        & $308{,}129$      & $10{,}097$       & --- \\
Edges                      & $23{,}570$       & $8{,}436{,}366$  & $332{,}081$      & --- \\
DAG depth                  & $153$            & $84$             & $8$              & --- \\
Cross-namespace edges       & $37.1\%$         & $50.9\%$         & $85.8\%$         & --- \\
Intra-module edges          & ---              & $9.6\%$          & ---              & --- \\
Terminal nodes (zero out-citation) & ---       & $44.8\%$         & ---              & --- \\[4pt]
\multicolumn{5}{@{}l}{\textit{Decomposition metrics}} \\[2pt]
Synthesized edges ($\sigma$) & ---             & $74.2\%$         & ---              & --- \\
Statement-only / proof-only / both & ---       & $8.1/43.9/48.0\%$ & ---            & --- \\
Transitive redundancy       & $17.5\%$         & ---              & ---              & --- \\
Definitional height         & ---              & med.\,$7$, max\,$60$ & ---          & --- \\
Parallelism ratio           & $22.4\times$     & ---              & ---              & --- \\[4pt]
\multicolumn{5}{@{}l}{\textit{Community and alignment}} \\[2pt]
Centrality separation       & partial          & complete         & complete         & --- \\
Community--namespace NMI   & ---              & $0.34$           & $0.26$           & --- \\
Modularity                 & ---              & $0.48$           & $0.27$           & --- \\
Single-node removal impact & $29$ components  & $\le 6$ nodes    & GCC $20.8\%$ at $20\%$ & --- \\
Containment (depth $1$)    & ---              & ---              & $22.2\%$         & --- \\
Theorem/lemma in-degree ratio & ---            & $1.47\times$     & ---              & --- \\[4pt]
\multicolumn{5}{@{}l}{\textit{Cross-level}} \\[2pt]
Module cohesion (mean)     & ---              & ---              & ---              & $0.107$ \\
Zero-cohesion modules      & ---              & ---              & ---              & $8.4\%$ \\
$G_{\mathrm{file}}/G_{\mathrm{module}}$ edge ratio & ---  & ---  & ---              & $9.1\times$ \\
Active imports             & ---              & ---              & ---              & $72.1\%$ \\
Indirect declaration deps  & ---              & ---              & ---              & $92.2\%$ \\
Import utilization (median) & ---              & ---              & ---              & $1.6\%$ \\
\bottomrule
\end{tabular}
\end{table}

Table~\ref{tab:summary} collects the key quantitative findings. Rather than recapitulating per-level findings (detailed in Appendix~\ref{app:supplementary}), we organize the discussion around the product and process lens introduced in \S\ref{sec:conceptual-framework}. 

\label{sec:summary}
\label{sec:interplay} 
Our data identify three distinct process traces embedded within the product, corresponding to the findings of \S\ref{sec:contribution}:

\begin{enumerate}
\item At every granularity level, naming conventions and dependency-based community structures diverge. Contributors impose a coarser taxonomy than the dependency graph logically warrants, trading structural precision for human navigability. The cross-file declaration dependency graph contains $9.1\times$ more edges than the module import graph that abstracts it (\S\ref{sec:import_vs_declaration}). Furthermore, $92.2\%$ of cross-file declaration dependencies are reached through transitive import chains rather than direct imports. An additional asymmetry separates product from process: in-degree is heavy-tailed and open-ended because mathematical content dictates how widely a result is reused, whereas out-degree remains bounded because human cognitive complexity limits the number of premises a single proof can practically invoke.

\item The $17.5\%$ post-\texttt{shake} redundancy rate (\S\ref{sec:transitive-reduction}) explicitly quantifies the gap between compiler requirements and developer workflows. This redundancy is directional, inflating out-degree without affecting in-degree. This asymmetry strongly aligns with an ergonomic motivation: developers add redundant imports for local convenience, rather than to satisfy the global demands of the imported module's consumers.

\item The DAG depths span $153$, $84$, and $8$ layers across the module, declaration, and namespace graphs, respectively. The fact that the coarser module graph is significantly deeper than the underlying declaration graph highlights how a single module import pulls in an entire file, artificially chaining import layers even when only a few declarations are actually required. Consequently, incremental organizational layering shapes the file hierarchy, whereas the underlying mathematical logic resists such deep stratification. Finally, while formalization preserves traditional mathematical labels, it often flattens their semantic meaning; for instance, the distinction between a \texttt{theorem} and a \texttt{lemma} is partially validated in aggregate but frequently contradicted in individual cases (\S\ref{sec:thm-vs-lemma}).
\end{enumerate}

The relationship between product and process is not unidirectional. The $37.1\%$ cross-directory edge rate in $G_{\mathrm{module}}^{-}$ (\S\ref{sec:cross-namespace}) exerts substantial reorganization pressure on the directory hierarchy, as evidenced by the ongoing absorption of \module{RingTheory} into \module{Algebra} within \module{Mathlib}. Additionally, the deep dependency chain of $153$ layers directly constrains the development process. Modifications to high-PageRank modules trigger recompilation cascades, which strongly incentivize stability at the core and isolate experimentation to the periphery. Formalization thus initiates a continuous feedback cycle: the directory tree dictates which imports are convenient (as co-located files are easier to discover), thereby shaping the module graph. In turn, the module graph reshapes the directory tree whenever cross-directory coupling becomes untenable.

\label{sec:practice}
\paragraph{Implications for practice.}
These findings are not merely descriptive; several metrics admit direct operational use. For instance, module cohesion (Definition~\ref{def:cohesion}) can actively flag scope drift when computed incrementally, while the zero-citation rate (\S\ref{sec:thm-leaves}) identifies potentially obsolete or redundant API surface. PageRank and betweenness rankings pinpoint modules whose modification triggers the widest recompilation cascades. These metrics could immediately inform continuous integration systems by assigning higher test priority to pull requests that touch high-centrality modules. Prior work further supports this operational utility: Huch and Wenzel~\cite{huch_et_al:LIPIcs.ITP.2024.22} demonstrate a ${>}100\times$ speedup from DAG-based parallel builds, and Baanen et al.~\cite{baanen2025growing} report speedups of $6\%$ to $33\%$ resulting from structural refactors.

For language design, the two-layer hub structure (\S\ref{sec:thm-types}) suggests that creating an explicit boundary between language infrastructure and mathematical content could enable entirely independent dependency analysis and compilation. Furthermore, the $92.2\%$ indirect dependency rate (\S\ref{sec:import_vs_declaration}) motivates the need for finer-grained import mechanisms. Lean's updated module system (introduced in November 2025), which structurally distinguishes \texttt{public import} from standard \texttt{import}, represents a significant step in this direction.

For artificial intelligence systems tasked with premise selection or proof generation, current datasets typically present each theorem as an isolated data point with its statement and proof, discarding the rich structural context that our network analysis reveals. Community membership, hub and bridge roles, the explicit/synthesized edge distinction, and cross-namespace coupling all carry information that flat representations lose. Our decompositions provide a vocabulary for encoding this structural context as training metadata: for instance, Louvain community labels as disciplinary tags, declaration types as role indicators, and the explicit subgraph (Definition~\ref{def:synth-partition}) as a closer proxy for human-intended dependencies. As AI-generated proofs become a larger fraction of community contributions, the metrics established here provide a baseline for detecting future structural drift.

\label{sec:limitations}
\paragraph{Limitations.}
Several limitations constrain these conclusions. First, we work with a single commit (\texttt{534cf0b}, February 2026). The cross-version comparison in \S\ref{sec:contribution} covers only a handful of aggregate indicators across four snapshots. It follows that we cannot definitively determine whether the finer structural properties we observe are stable, growing, or approaching critical thresholds. Furthermore, the evolutionary hypotheses advanced in \S\ref{sec:cross-level-summary} currently remain untested predictions. The snapshot also captures the library in a transitional state where the vast majority of imports are still public; as the community increasingly adopts private imports, redundancy rates and transitive visibility will inevitably shift. Additionally, the organizational ``process'' component is inferred from structural traces embedded in the product, rather than observed directly through pull request workflows or commit histories. Finally, while our data strongly suggest a bidirectional influence between the module tree and the dependency graph, establishing a rigorous causal account will require future longitudinal or experimental evidence. Additionally, community detection uses the Louvain algorithm, which has a known resolution limit: communities smaller than a threshold determined by total edge weight may be merged into larger clusters. With $8.4$M edges, genuine small-scale mathematical subcommunities could go undetected; running Leiden or varying the resolution parameter would serve as a robustness check.

\label{sec:future}
\paragraph{Open problems.}
Several graph definitions in \S\ref{sec:graph-definitions} remain empirically unexplored, and \S\ref{sec:temporal-graphs} outlines a program for investigating temporal, visibility, and co-modification graphs. Beyond these immediate extensions, we highlight several broader research directions. The growing ecosystem of downstream projects (e.g., FLT~\cite{flt_lean}, PFR~\cite{Anderson_Formalization_of_the_2023}, and Sphere Eversion~\cite{vandoorn2023sphereeversion}) suggests the emergence of a ``core and satellite'' architecture whose structural properties merit dedicated investigation. The federated-tiers hypothesis (\S\ref{app:module-analysis}) can thus be reframed to view \module{Mathlib} as consolidating into a universal foundational tier. Moreover, applying this network analysis framework to other formal libraries, such as Isabelle/AFP~\cite{Blanchette_2015,Klein_2012}, Coq/MathComp, and Mizar/MML~\cite{Bancerek_2018}, would help distinguish structural features that are intrinsic to formalized mathematics from those that are contingent upon a specific proof assistant community.

Network analysis also exposes the structural vulnerabilities of \module{Mathlib}'s monolithic design, providing a quantitative basis for future decentralization. Currently, compilation costs grow with every contribution due to a massive concentration of in-degree on a few infrastructure hubs, a $153$-layer critical path, and a $17.5\%$ transitive redundancy rate. As the library expands, especially in cross-cutting theories where the $50.9\%$ cross-namespace density will likely increase, clean modular boundaries will become harder to maintain. In principle, community detection and minimum-cut analysis on $G_{\mathrm{module}}$ could identify natural package boundaries, clustering modules with dense internal coupling while preserving the inherently narrow cross-package interfaces suggested by the $1.6\%$ median import utilization. Until such a split is feasible, the network metrics developed here can serve as a diagnostic dashboard. By tracking PageRank thresholds for hub declarations, monitoring critical-path lengths across releases, and enforcing strict import utilization targets, maintainers can proactively manage the engineering pressures of a rapidly scaling mathematical corpus.

The dependency graphs studied here capture only the strict logical skeleton of mathematics. Traditional mathematical texts encode a far richer network of relations that carry essential structural information, despite not being formal logical dependencies. These include motivating insights, instructive counterexamples, clarifying remarks, historical context, and heuristic analogies between subfields. These ``soft'' edges have no native representation in the environment of a proof assistant, yet they fundamentally guide how mathematicians navigate and extend a body of knowledge. Tools such as Leanblueprint~\cite{massot_leanblueprint} already extract the strict dependency DAG of a formalization project to serve as a scaffolding for coordinating contributions. Extending this scaffolding with soft-dependency layers (e.g., extracted from the natural-language prose surrounding formal statements in textbooks and papers) could yield a vastly more faithful map of mathematical knowledge. Such an extended network would have direct practical value. In a large formalization campaign, the order in which results are formalized is currently determined by the logical DAG alone. However, integrating soft dependencies, such as noting that a proof is best understood after reviewing a specific counterexample, could inform a much more cognitively efficient sequencing.

The core question is deeper: is the dependency topology we observe an inherent property of \emph{mathematics}, or is it an artifact of \emph{how mathematics is currently produced}? The metrics and theoretical framework developed in this paper offer a preliminary quantitative vocabulary for exploring this distinction.

\paragraph{Acknowledgements.}
We thank Leonardo de Moura for facilitating the collaboration that led to this work, and for valuable feedback and discussions. We are grateful to Ginestra Bianconi for insightful feedback and for drawing the analogy between cross-namespace dependencies and horizontal gene transfer in phylogenetic networks. We thank Johan Commelin for connecting us with the Mathlib maintainer community, Oliver Nash for valuable comments on an earlier draft and for clarifications regarding \texttt{Mathlib}'s tooling and module system, and Franz Josef Och and Jennifer Bennett for valuable discussions. Xinze Li thanks his doctoral advisor Yevgeny Liokumovich and Luis Seco for their support, and Kasra Rafi, Bennett Chow, Bin Dong, Ronghui Gu, Zaiwen Wen, Tianyu Wang, and Shuangjian Zhang for helpful discussions on AI for mathematics.

\begingroup
\footnotesize
\let\oldbibliography\thebibliography
\renewcommand{\thebibliography}[1]{%
  \oldbibliography{#1}%
  \setlength{\itemsep}{0pt}%
  \setlength{\parsep}{0pt}%
  \setlength{\parskip}{0pt}%
}
\bibliographystyle{alphaurl}
\bibliography{references}
\endgroup

\newpage
\addtocontents{toc}{\protect\setcounter{tocdepth}{2}}
\appendix
\begin{center}
\vspace*{2cm}
{\Huge\bfseries Appendices}
\vspace{1cm}

\end{center}
\vfill
\setcounter{tocdepth}{2}
\tableofcontents
\newpage
\section{Mathematical Preliminaries}
\label{sec:preliminaries}

We collect the standard graph-theoretic and network-analytic definitions used throughout the paper. Readers familiar with these notions may skip to \S\ref{sec:module-import}.

\subsection{Graph-Theoretic Notation}
\label{sec:prelim-graph}

A \emph{directed graph} is a pair $G = (V, E)$ where $V$ is a set (the \emph{vertex set}) and $E \subseteq V \times V$ (the \emph{edge set}). A \emph{path} in~$G$ is a sequence of vertices $(v_0, v_1, \ldots, v_k)$ such that $(v_i, v_{i+1}) \in E$ for all $0 \le i < k$; its \emph{length} is~$k$. A \emph{cycle} is a path with $k \ge 1$ and $v_k = v_0$. A directed graph is a \emph{directed acyclic graph (DAG)} if it contains no cycle. A \emph{rooted tree} is a DAG $T = (V, E)$ with a distinguished vertex $r \in V$ (the \emph{root}) such that for every $v \in V$ there is a unique directed path from~$r$ to~$v$.

For a vertex~$v$ in a directed graph $G = (V, E)$, the \emph{in-degree} is $\deg^{-}(v) = |\{u \in V : (u, v) \in E\}|$ and the \emph{out-degree} is $\deg^{+}(v) = |\{w \in V : (v, w) \in E\}|$. A vertex~$v$ is a \emph{source} if $\deg^{-}(v) = 0$ and a \emph{sink} if $\deg^{+}(v) = 0$.

The \emph{depth} of a DAG is the length of its longest directed path. The \emph{topological level} $\ell(v)$ of a vertex~$v$ is the length of the longest directed path ending at~$v$; in particular, all sources sit at level~$0$.

A directed graph~$G$ is \emph{weakly connected} if the underlying undirected graph (obtained by ignoring edge orientations) is connected. A \emph{weakly connected component (WCC)} is a maximal weakly connected subgraph.

A directed graph~$G$ is \emph{strongly connected} if for every pair of vertices~$u, v$ there exists a directed path from~$u$ to~$v$ and a directed path from~$v$ to~$u$. A \emph{strongly connected component (SCC)} is a maximal strongly connected subgraph. Every DAG has only trivial SCCs (single vertices). The \emph{condensation} of a directed graph replaces each SCC by a single super-node and each inter-SCC edge by an edge between the corresponding super-nodes; the resulting graph is always a DAG.

\subsection{Network-Analytic Methods}
\label{sec:prelim-network}

We employ three families of network-analytic tools: \emph{centrality measures} (\S\ref{sec:prelim-centrality}) that quantify the structural importance of individual nodes, \emph{community detection and partition comparison} (\S\ref{sec:prelim-community}) that reveal and evaluate mesoscale structure, and \emph{heavy-tailed distribution fitting} (\S\ref{sec:prelim-powerlaw}) that characterizes degree heterogeneity.

\subsubsection{Centrality Measures}
\label{sec:prelim-centrality}

We use three centrality measures, each capturing a different notion of structural importance. For textbook treatments, see Newman~\cite[\S\S7.1, 7.4, 7.7]{Newman_2010}.

\emph{In-degree} $\deg^{-}(v)$ counts how many other vertices point to~$v$. It is the simplest measure of direct importance~\cite[\S7.1]{Newman_2010}.

\begin{definition}[PageRank~\cite{334}; see also {\cite[\S7.4]{Newman_2010}}]
\label{def:pagerank}
The \emph{PageRank} of a vertex~$v$ in a directed graph is the stationary probability of~$v$ under a random walk that, at each step, follows an outgoing edge uniformly at random with probability~$\alpha$ (the \emph{damping factor}, typically $\alpha = 0.85$) and jumps to a uniformly random vertex with probability $1 - \alpha$.  Equivalently, PageRank is the unique solution to
\[
  \mathrm{PR}(v) \;=\; \frac{1-\alpha}{N} \;+\; \alpha \sum_{u \to v} \frac{\mathrm{PR}(u)}{\deg^{+}(u)},
\]
where $N$ is the number of vertices and $\deg^{+}(u)$ is the out-degree of~$u$.
\end{definition}

PageRank measures \emph{recursive} importance: a vertex ranks high not merely because many others point to it, but because \emph{important} others point to it. It identifies the deep foundations on which large subgraphs transitively rest.

\begin{definition}[Betweenness centrality~\cite{Fre77}; see also {\cite[\S7.7, Eq.~(7.36)]{Newman_2010}}]
\label{def:betweenness}
The \emph{betweenness centrality} of a vertex~$v$ is
\[
  c_B(v) = \sum_{\substack{s, t \in V \\ s \ne v \ne t}} \frac{\sigma_{st}(v)}{\sigma_{st}},
\]
where $\sigma_{st}$ is the number of shortest paths from~$s$ to~$t$, and $\sigma_{st}(v)$ is the number of those paths passing through~$v$.
\end{definition}

Betweenness identifies \emph{bridge} vertices: those that sit on many shortest paths between other pairs. Removing a high-betweenness vertex forces information flow to take long detours.

\subsubsection{Community Detection and Partition Comparison}
\label{sec:prelim-community}

\begin{definition}[Modularity~\cite{Newman_2004}; see also {\cite[\S11.6]{Newman_2010}}]
\label{def:modularity}
Given an undirected graph with adjacency matrix~$A$, total edge weight $m = \frac{1}{2}\sum_{ij} A_{ij}$, and a partition assigning each vertex~$i$ to community~$c_i$, the \emph{modularity} is
\[
  Q = \frac{1}{2m} \sum_{ij} \left[ A_{ij} - \frac{k_i k_j}{2m} \right] \delta(c_i, c_j),
\]
where $k_i = \sum_j A_{ij}$ is the degree (or weighted degree) of vertex~$i$ and $\delta(c_i, c_j) = 1$ if $i$ and $j$ belong to the same community. The term $k_i k_j / 2m$ is the expected number of edges between $i$ and $j$ under a random null model preserving the degree sequence. Modularity ranges from $-1/2$ to $1$; values above ${\sim}0.3$ indicate significant community structure.
\end{definition}

The \emph{Louvain algorithm}~\cite{Blondel_2008} is a greedy, hierarchical method that maximizes modularity through two alternating phases:

\begin{enumerate}
\item \textbf{Local move phase.} Each vertex is initially placed in its own community. The algorithm then iterates over all vertices in random order: for each vertex~$v$, it computes the change in modularity $\Delta Q$ that would result from moving $v$ to each of its neighbors' communities, and moves $v$ to the community yielding the largest positive $\Delta Q$. This sweep repeats until no move produces a positive gain.

\item \textbf{Aggregation phase.} The communities found in Phase~1 are contracted into single super-nodes, with edge weights between super-nodes equal to the sum of edge weights between their constituent vertices (self-loops encode intra-community edges). The algorithm then returns to Phase~1 on this coarsened graph.
\end{enumerate}

\noindent
The two phases alternate until modularity converges. The result is a hierarchical dendrogram of community merges; in practice, we use the partition at the level that maximizes~$Q$. The algorithm runs in $O(n \log n)$ time on sparse graphs, making it tractable for our largest graph ($308{,}054$ nodes, $8.4$M edges).

To compare two partitions of the same vertex set (for example, a community assignment and a namespace labeling), we use two standard measures.

\begin{definition}[Shannon Entropy~{\cite[Ch.~2]{cover1991}}]
\label{def:shannon-entropy}
Given a partition $\mathcal{A} = \{A_1, \ldots, A_k\}$ of a finite set~$V$, the \emph{Shannon entropy} is
\[
  H(\mathcal{A}) = -\sum_{i=1}^{k} \frac{|A_i|}{|V|} \log \frac{|A_i|}{|V|}.
\]
It measures the uncertainty in a randomly chosen element's cluster assignment: $H = 0$ when all elements belong to a single cluster, and $H = \log k$ when elements are uniformly distributed.
\end{definition}

\begin{definition}[Mutual Information~{\cite[Ch.~2]{cover1991}}]
\label{def:mutual-information}
Given two partitions $\mathcal{A} = \{A_1, \ldots, A_k\}$ and $\mathcal{B} = \{B_1, \ldots, B_l\}$ of~$V$, the \emph{mutual information} is
\[
  I(\mathcal{A}; \mathcal{B}) = \sum_{i=1}^{k} \sum_{j=1}^{l} \frac{|A_i \cap B_j|}{|V|} \log \frac{|V| \cdot |A_i \cap B_j|}{|A_i| \cdot |B_j|},
\]
with the convention $0 \log 0 = 0$. It quantifies how much knowing one partition reduces uncertainty about the other.
\end{definition}

\begin{definition}[Normalized Mutual Information (NMI)~{\cite[\S3, Eq.~(2)]{Danon_2005}}]
\label{def:nmi}
Given two partitions $\mathcal{A}$ and $\mathcal{B}$ of a set~$V$, the \emph{normalized mutual information} is
\[
  \mathrm{NMI}(\mathcal{A}, \mathcal{B}) = \frac{2\, I(\mathcal{A}; \mathcal{B})}{H(\mathcal{A}) + H(\mathcal{B})},
\]
where $I$ and $H$ are defined above. NMI ranges from~$0$ (independent) to~$1$ (identical partitions).
\end{definition}

\begin{definition}[Adjusted Rand Index (ARI)~{\cite[\S2.1, Eq.~(4)--(5)]{Hubert_1985}}]
\label{def:ari}
Given two partitions $\mathcal{A}$ and $\mathcal{B}$ of~$V$, consider all $\binom{|V|}{2}$ pairs of elements. Each pair is classified as:
\begin{itemize}
\item $a$: same cluster in both $\mathcal{A}$ and $\mathcal{B}$ (concordant---together),
\item $b$: same in $\mathcal{A}$, different in $\mathcal{B}$ (discordant),
\item $c$: different in $\mathcal{A}$, same in $\mathcal{B}$ (discordant),
\item $d$: different in both (concordant---apart).
\end{itemize}
The \emph{Rand Index} is $\mathrm{RI} = (a + d) / \binom{|V|}{2}$. Since even random partitions yield $\mathrm{RI} > 0$, the \emph{Adjusted Rand Index} corrects for chance:
\[
  \mathrm{ARI}(\mathcal{A}, \mathcal{B}) = \frac{\mathrm{RI} - \mathbb{E}[\mathrm{RI}]}{\max(\mathrm{RI}) - \mathbb{E}[\mathrm{RI}]},
\]
where the expectation is over random partitions with the same cluster-size distributions as $\mathcal{A}$ and $\mathcal{B}$. ARI equals~$1$ for identical partitions, $0$ for agreement at the level expected by chance, and can be negative for agreement worse than random.
\end{definition}

\subsubsection{Heavy-Tailed Distributions}
\label{sec:prelim-powerlaw}

\begin{definition}[Degree Distribution~{\cite[\S8.3]{Newman_2010}}]
\label{def:degree-distribution}
The \emph{degree distribution} of a graph $G = (V, E)$ is the function
\[
  P(k) = \frac{|\{v \in V : \deg(v) = k\}|}{|V|},
\]
giving the fraction of nodes with degree exactly~$k$. For directed graphs, one defines $P^{+}(k)$ and $P^{-}(k)$ separately for out-degree and in-degree.
\end{definition}

A distribution follows a \emph{power law} if $P(k) \propto k^{-\alpha}$ for some exponent $\alpha > 1$. On a log--log plot, a power law appears as a straight line with slope~$-\alpha$. We apply the Clauset--Shalizi--Newman method~\cite{Clauset_2009} to test power-law hypotheses: it fits a discrete power law $P(k) \propto k^{-\alpha}$ for $k \ge x_{\min}$ and compares the fit against alternatives (lognormal, exponential, truncated power law, stretched exponential) using likelihood ratio tests.

\section{Dependency Graphs}
\label{sec:graph-definitions}

The dependency graphs defined in this paper are not pure embodiments of mathematical logic; each is an entanglement of inherited mathematical structure and the organizational imprint of Lean's language mechanisms. Table~\ref{tab:decompositions} catalogs the complete family of decompositions, organized by the Lean mechanism they isolate.

\begin{table}[!htb]
\centering
\caption{Tool-mediated decompositions of the dependency graphs. Each row identifies a Lean language mechanism, the corresponding graph decomposition or projection, and the dimension it isolates. Rows above the mid-rule have been extracted and analyzed; rows below are deferred to future work.}
\label{tab:decompositions}
\footnotesize
\begin{tabular}{@{}lllrl@{}}
\toprule
\textbf{Lean mechanism} & \textbf{Decomposition} & \textbf{Isolates} & \textbf{Scale} & \textbf{Ref.} \\
\midrule
\multicolumn{5}{@{}l}{\textit{Edge-level decompositions of $G_{\mathrm{thm}}$}} \\[3pt]
Statement vs.\ proof & $E_S \sqcup E_P \sqcup E_{SP}$ & Assertion vs.\ proof strategy & $8.1/43.9/48.0\%$ & Def.~\ref{def:stmt-proof-dep} \\
Explicit vs.\ synthesized & $E_{\mathrm{explicit}} \sqcup E_{\mathrm{synth}}$ & Human-written vs.\ compiler-generated & $\sigma = 74.2\%$ & Def.~\ref{def:synth-partition} \\[6pt]
\multicolumn{5}{@{}l}{\textit{Subgraphs of $G_{\mathrm{thm}}$}} \\[3pt]
Typeclass synthesis & $G_{\mathrm{tc}}$ & Algebraic hierarchy infrastructure & $1{,}817$ / $29{,}952$ & Def.~\ref{def:tc-graph} \\[6pt]
\multicolumn{5}{@{}l}{\textit{Node-level enrichment of $G_{\mathrm{thm}}$}} \\[3pt]
Attributes \& metadata & $\alpha(d)$, $\kappa(d)$, $\upsilon(d)$ & Proof complexity, polymorphism & $165$ types & Def.~\ref{def:decl-attributes} \\
\texttt{to\_additive} & $\sim_{\mathrm{add}}$ mirror pairs & Metaprogrammed duplication & $18{,}659$ pairs & Def.~\ref{def:decl-attributes} \\[6pt]
\multicolumn{5}{@{}l}{\textit{Refinements of $G_{\mathrm{module}}$}} \\[3pt]
Public/private imports & $G_{\mathrm{module}}^{\mathrm{pub}}$ & Intentional API vs.\ all imports & $45{,}289$ / $1{,}912$ & Def.~\ref{def:visibility-graph} \\
Compilation cost & $G_{\mathrm{build}}$ & Engineering bottlenecks & $47{,}201$ edges & Def.~\ref{def:build-graph} \\
Import utilization & $\mathrm{util}(m_i, m_j)$ & Module granularity & median $1.6\%$ & Def.~\ref{def:import-util} \\[6pt]
\multicolumn{5}{@{}l}{\textit{Additional subgraphs of $G_{\mathrm{thm}}$}} \\[3pt]
Coercion insertion & $G_{\mathrm{coe}}$ & Implicit type conversion paths & $256$ coercions & Def.~\ref{def:coercion-graph} \\
Structure inheritance & $G_{\mathrm{ext}}$ & \texttt{extends} hierarchy & $1{,}535$ edges & Def.~\ref{def:extends-graph} \\
\texttt{deriving} handlers & $E_{\mathrm{auto}}$ & Auto-generated instances & $15$ handlers & Def.~\ref{def:deriving-edges} \\[6pt]
\multicolumn{5}{@{}l}{\textit{Additional node-level enrichment of $G_{\mathrm{thm}}$}} \\[3pt]
Tactic usage & $\tau(d)$ & Proof strategy selection & top: \texttt{rw}, \texttt{exact}, \texttt{simp} & Def.~\ref{def:tactic-usage} \\
Definitional height & $\delta(d)$ & Kernel-level implicit coupling & median $7$, max $60$ & Def.~\ref{def:def-height} \\
\midrule
\multicolumn{5}{@{}l}{\textit{Planned decompositions (future work)}} \\[3pt]
File co-modification & $G_{\mathrm{co}}$ & Social/development coupling & --- & \S\ref{sec:co-edit-graph} \\
Temporal evolution & $G^{(t)}$ & Structural stability over time & --- & \S\ref{sec:temporal-graphs} \\
\bottomrule
\end{tabular}
\end{table}

\noindent Each row acts as a \emph{projection} that isolates one dimension---inherited mathematical structure or human organizational process---from the entangled raw graph. The ``Scale'' column reports the footprint of each mechanism at our snapshot, extracted programmatically from Lean's \texttt{Environment} API across all $499{,}732$~Mathlib constants. All decompositions above the mid-rule have been extracted and empirically analyzed in this paper; the two remaining rows (file co-modification and temporal evolution) require additional data sources and are deferred to future work (\S\ref{sec:future-graphs}).

\subsection{The Module Graph: Definitions and Examples}
\label{sec:module-import}

This appendix collects the formal definitions, worked examples, and illustrative figures for the module tree and module graph. Extended statistical analyses (degree distributions, centrality rankings, robustness curves) appear in Appendix~\ref{app:module-detail}.

\subsubsection{Modules and the Module Tree}

\begin{definition}[Module]\label{def:module}
A \emph{module} in \module{Mathlib} is identified by a dot-separated sequence of identifiers, such as $\module{Mathlib.Algebra.Group.Defs}$. Each module corresponds to a \texttt{.lean} source file, with dots replacing path separators. We write $\mathcal{M}$ for the set of all modules in \module{Mathlib}; at our snapshot, $|\mathcal{M}| = 7{,}563$.
The \emph{depth} of a module $c_1.c_2.\cdots.c_k$ is~$k$, the number of dot-separated components.
\end{definition}

\begin{example}
\leavevmode
\begin{itemize}
  \item $\module{Mathlib}$ has depth~$1$.
  \item $\module{Mathlib.Algebra}$ has depth~$2$.
  \item $\module{Mathlib.Algebra.Group.Defs}$ has depth~$4$.
\end{itemize}
The maximum depth in \module{Mathlib} is~$7$.\footnote{Attained by, e.g., \module{Mathlib.CategoryTheory.Limits.Shapes.Pullback.IsPullback.Kernels}.}
\end{example}

We say $m$ is a \emph{prefix} of $m'$ if $m'$ extends $m$ by additional components. For example, \module{Mathlib.Algebra} is a prefix of \module{Mathlib.Algebra.Group.Defs}.

\begin{definition}[Module tree]\label{def:module-tree}
The prefix relation induces a rooted tree $T = (\mathcal{M}, E_T)$ with root $\module{Mathlib}$, which we call the \emph{module tree}, where
\[
  E_T = \{(m, m') \in \mathcal{M} \times \mathcal{M} \mid m \text{ is a prefix of } m' \text{ and } \mathrm{depth}(m') = \mathrm{depth}(m) + 1\}.
\]
Its leaves are modules (source files); its internal nodes are shared prefixes of modules, corresponding to directories in the source repository.
\end{definition}

\begin{figure}[H]
\centering
\begin{subfigure}[t]{0.42\textwidth}
\centering
\begin{forest}
  for tree={
    font=\scriptsize\ttfamily,
    grow'=0,
    child anchor=west,
    parent anchor=south,
    anchor=west,
    calign=first,
    edge path={
      \noexpand\path [draw, \forestoption{edge}]
        (!u.south west) +(7.5pt,0) |- (.child anchor)\forestoption{edge label};
    },
    before typesetting nodes={
      if n=1 {insert before={[,phantom]}} {}
    },
    fit=band,
    before computing xy={l=13pt},
  }
[Mathlib/
  [Algebra/
    [Group/
      [Defs.lean]
    ]
  ]
  [Data/
    [Nat/
      [Notation.lean]
    ]
  ]
  [\ldots, edge={draw, densely dotted}]
]
\end{forest}
\caption{Directory structure}
\label{fig:dir-tree-app}
\end{subfigure}%
\hfill
\begin{subfigure}[t]{0.52\textwidth}
\centering
\begin{tikzpicture}[
  every node/.style={font=\small\ttfamily, inner sep=2pt, text=blue!60!black},
  dot/.style={circle, fill=blue!60!black, minimum size=4pt, inner sep=0pt},
  dep/.style={->, >=Stealth, thin, blue!60!black},
]
  \node[dot, label=above:{Mathlib}]              (root) at (1.8,0) {};
  \node[dot, label=left:{Algebra}]               (alg)  at (0,-1.5) {};
  \node[dot, label=right:{Data}]                 (dat)  at (3.6,-1.5) {};
  \node[dot, label=left:{Algebra.Group}]         (grp)  at (0,-3.0) {};
  \node[dot, label=right:{Data.Nat}]             (nat)  at (3.6,-3.0) {};
  \node[dot, label=below:{Algebra.Group.Defs}]   (gd)   at (0,-4.5) {};
  \node[dot, label=below:{Data.Nat.Notation}]    (nn)   at (3.6,-4.5) {};
  \draw[dep] (root) -- (alg);
  \draw[dep] (root) -- (dat);
  \draw[dep] (alg)  -- (grp);
  \draw[dep] (dat)  -- (nat);
  \draw[dep] (grp)  -- (gd);
  \draw[dep] (nat)  -- (nn);
  \node[font=\small, text=gray] at (1.8,-1.5) {$\cdots$};
\end{tikzpicture}
\caption{Module tree $T$: arrows point parent $\to$ child ($\in E_T$)}
\label{fig:namespace-tree-app}
\end{subfigure}
\caption{A fragment of the \module{Mathlib} source tree (commit \texttt{534cf0b}). Left: the directory layout. Right: the corresponding rooted tree~$T$. Each directory path maps to a module with dots replacing separators, e.g., \module{Mathlib/\allowbreak Algebra/\allowbreak Group/\allowbreak Defs.lean} $\leftrightarrow$ \module{Mathlib.\allowbreak Algebra.\allowbreak Group.\allowbreak Defs}.}
\label{fig:prefix-tree-app}
\end{figure}
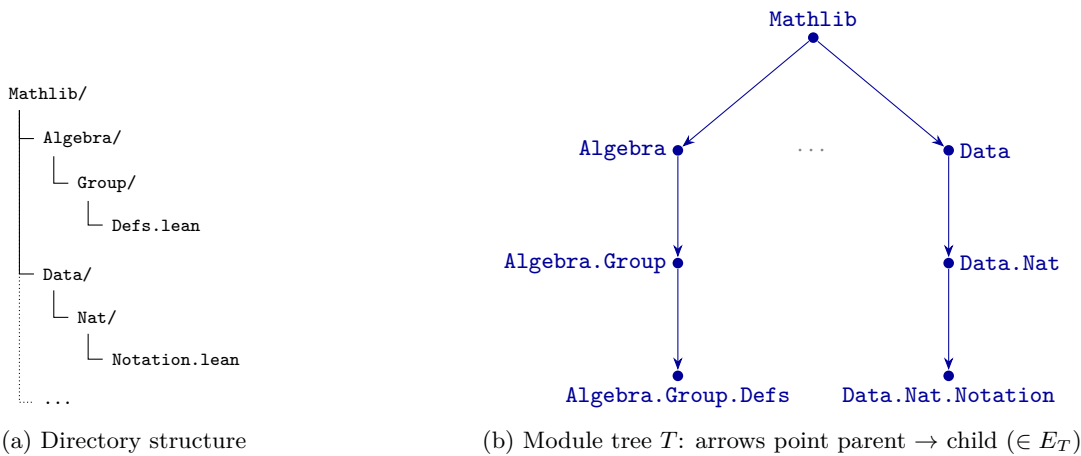

\subsubsection{The Module Graph $G_{\mathrm{module}}$}
\label{sec:import-graph}

Each Lean source file begins with \texttt{import} statements that declare its dependencies. Lean~4 supports both \texttt{import} and \texttt{public import}; for the purposes of the module graph, both forms create an edge. After defining the base graph, we introduce three refinements: the visibility graph $G_{\mathrm{module}}^{\mathrm{pub}}$ (which distinguishes public from private imports), import utilization $\mathrm{util}(m_i, m_j)$ (which measures how much of an imported module is actually used), and the build graph $G_{\mathrm{build}}$ (which augments the module graph with compilation cost).

\begin{definition}[Import set]
\label{def:import-set}
For each $m \in \mathcal{M}$, let $\mathrm{imports}(m) \subseteq \mathcal{M}$ denote the set of modules directly imported by $m$, regardless of visibility modifier.
\end{definition}

\begin{definition}[Module graph]\label{def:module-graph}
The \emph{module graph} of \module{Mathlib} is $G_{\mathrm{module}} = (\mathcal{M}, E_{\mathrm{module}})$ where
\[
  E_{\mathrm{module}} = \{(m_1, m_2) \in \mathcal{M} \times \mathcal{M} \mid m_2 \in \mathrm{imports}(m_1)\}.
\]
\end{definition}

\noindent At our snapshot, $|E_{\mathrm{module}}| = 23{,}570$.

For example, the file \module{Mathlib/Data/Nat/Notation.lean} contains:

\begin{codebox}
\textsf{\bfseries\color{blue!60!black}Data/Nat/Notation.lean}\\[2pt]
\textbf{public import} \textcolor{blue!60!black}{Mathlib.Init}
\end{codebox}

\noindent
This file depends on exactly one other module: \module{Mathlib.Init}.
A file with more dependencies is \module{Mathlib/Algebra/Group/Defs.lean}, which imports nine modules (shown here in abbreviated form):

\begin{codebox}
\textsf{\bfseries\color{blue!60!black}Algebra/Group/Defs.lean}\\[2pt]
\textbf{public import} \textcolor{blue!60!black}{Batteries.Logic}\\
\textbf{public import} \textcolor{blue!60!black}{Mathlib.Algebra.Notation.Defs}\\
\textbf{public import} \textcolor{blue!60!black}{Mathlib.Algebra.Regular.Defs}\\
\hspace{1.5em}\textit{\color{gray}-- \ldots\ 6 more imports}
\end{codebox}

\noindent
Each \texttt{import} statement is a directed edge from the importing module to the imported module.
Since \module{Mathlib/Data/Nat/Notation.lean} contains \texttt{public import Mathlib.Init}, we have the edge $(\module{Data.Nat.Notation},\; \module{Init}) \in E_{\mathrm{module}}$:

\begin{figure}[H]
\centering
\begin{tikzpicture}[
  every node/.style={font=\small\ttfamily, inner sep=2pt, text=blue!60!black},
  dot/.style={circle, fill=blue!60!black, minimum size=3pt, inner sep=0pt},
  dep/.style={->, >=Stealth, thin, blue!60!black},
]
  \node[dot, label=above:{Init}]              (b) at (0,0) {};
  \node[dot, label=above:{Data.Nat.Notation}] (a) at (4,0) {};
  \draw[dep] (a) -- (b);
\end{tikzpicture}
\caption{The simplest import edge: \module{Data.Nat.Notation} imports \module{Init}. The arrow points from the importing module to the imported module. All names have the \module{Mathlib.}\ prefix removed.}
\label{fig:single-import-edge}
\end{figure}
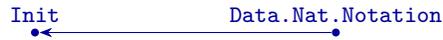

Since November 2025, Lean~4's module system distinguishes \texttt{public import} from \texttt{import} (private), allowing modules to selectively limit transitive visibility. This distinction refines the module graph:

\begin{definition}[Visibility graph]\label{def:visibility-graph}
For each import edge $(m_i, m_j) \in E_{\mathrm{module}}$, let
\[
  \mathrm{vis}(m_i, m_j) \in \{\texttt{public},\, \texttt{private}\}.
\]
The \emph{public import subgraph} is
\[
  G_{\mathrm{module}}^{\mathrm{pub}} = (\mathcal{M},\, \{(m_i, m_j) \in E_{\mathrm{module}} : \mathrm{vis}(m_i, m_j) = \texttt{public}\}).
\]
The transitive closure of $G_{\mathrm{module}}^{\mathrm{pub}}$ determines the actual namespace visibility scope: which declarations are transitively accessible from which modules.
\end{definition}

\noindent Our snapshot captures a transitional state in which most imports are still \texttt{public}; as private imports are adopted, $G_{\mathrm{module}}^{\mathrm{pub}}$ will become a proper subgraph of $G_{\mathrm{module}}$, reducing transitive visibility and reshaping redundancy (\S\ref{sec:transitive-reduction}).

\begin{figure}[!htb]
\centering
\begin{subfigure}[t]{0.48\textwidth}
\centering
\vspace{0pt}
\raggedright\small\ttfamily
\textsf{\bfseries\color{blue!60!black}Data/Nat/Notation.lean}\\[2pt]
\textbf{public import} \textcolor{blue!60!black}{Init}%
\hfill\textnormal{\sffamily\footnotesize\textcolor{gray!60}{re-exports Init's declarations}}\\[10pt]
\textsf{\itshape\normalfont\sffamily vs.}\\[10pt]
\textsf{\bfseries\color{blue!60!black}Data/Nat/Notation.lean}\\[2pt]
\textbf{import} \textcolor{blue!60!black}{Init}%
\hfill\textnormal{\sffamily\footnotesize\textcolor{gray!60}{Init hidden from importers}}
\caption{\texttt{public import} re-exports; plain \texttt{import} keeps declarations private.}
\label{fig:vis-source}
\end{subfigure}%
\hfill
\begin{subfigure}[t]{0.48\textwidth}
\centering
\vspace{0pt}
\begin{tikzpicture}[scale=1.25, every node/.style={transform shape, font=\footnotesize, inner sep=2pt},
  dot/.style={circle, fill=blue!60!black, minimum size=5pt, inner sep=0pt},
  pub/.style={->, >=Stealth, blue!60!black, thick},
  priv/.style={->, >=Stealth, blue!60!black, thick, dashed},
]
  \node[font=\footnotesize\sffamily\bfseries, blue!60!black] at (0.7, 3.8)
    {$G_{\mathrm{module}}^{\mathrm{pub}} = G_{\mathrm{module}}$};
  \fill[blue!6, rounded corners=4pt] (-0.2, -0.3) rectangle (1.6, 3.5);
  \node[font=\footnotesize\sffamily\itshape, blue!40!black] at (0.7, -0.6)
    {A sees \{B, C\}};
  \node[dot, label={[font=\footnotesize\ttfamily, blue!60!black]right:A}]
    (a1) at (0.7, 3.0) {};
  \node[dot, label={[font=\footnotesize\ttfamily, blue!60!black]right:B}]
    (b1) at (0.7, 1.5) {};
  \node[dot, label={[font=\footnotesize\ttfamily, blue!60!black]right:C}]
    (c1) at (0.7, 0.0) {};
  \draw[pub] (a1) -- node[left, font=\footnotesize\sffamily, text=blue!50!black] {pub} (b1);
  \draw[pub] (b1) -- node[left, font=\footnotesize\sffamily, text=blue!50!black] {pub} (c1);
  \node[font=\footnotesize\sffamily\bfseries, blue!60!black] at (3.7, 3.8)
    {$G_{\mathrm{module}}^{\mathrm{pub}} \subset G_{\mathrm{module}}$};
  \fill[blue!6, rounded corners=4pt] (2.8, 0.9) rectangle (4.6, 3.5);
  \node[font=\footnotesize\sffamily\itshape, blue!40!black] at (3.7, -0.6)
    {A sees \{B\} only};
  \node[dot, label={[font=\footnotesize\ttfamily, blue!60!black]right:A}]
    (a2) at (3.7, 3.0) {};
  \node[dot, label={[font=\footnotesize\ttfamily, blue!60!black]right:B}]
    (b2) at (3.7, 1.5) {};
  \node[dot, fill=gray!40, label={[font=\footnotesize\ttfamily, gray!50]right:C}]
    (c2) at (3.7, 0.0) {};
  \draw[pub] (a2) -- node[left, font=\footnotesize\sffamily, text=blue!50!black] {pub} (b2);
  \draw[priv] (b2) -- node[left, font=\footnotesize\sffamily, text=gray!50] {priv} (c2);
\end{tikzpicture}
\caption{Transitive visibility scope (blue shading) of module~A.}
\label{fig:vis-graph}
\end{subfigure}
\caption{Visibility graph (Definition~\ref{def:visibility-graph}). (a)~\texttt{public import} re-exports the imported module's declarations; plain \texttt{import} keeps them private. (b)~Left: when~B publicly imports~C, A's transitive visibility includes~\{B,\,C\}. Right: when~B privately imports~C, A can see only~\{B\}, because the private edge breaks the transitive chain, removing~C from $G_{\mathrm{module}}^{\mathrm{pub}}$.}
\label{fig:visibility}
\end{figure}
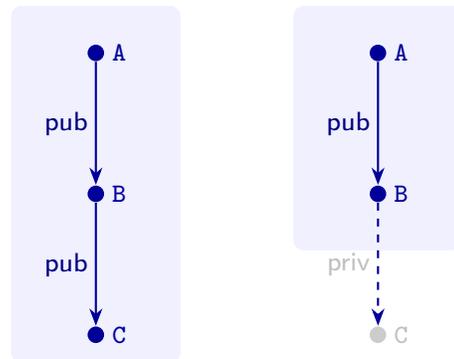
\begin{figure}[H]
\centering
\begin{tikzpicture}[every node/.style={font=\small\ttfamily, inner sep=2pt, text=blue!60!black},
  dot/.style={circle, fill=blue!60!black, minimum size=4pt, inner sep=0pt},
  dep/.style={->, >=Stealth, thin, blue!60!black},
]
  \node[dot, label=above:{Init}]                (d) at (0,0) {};
  \node[dot, label=left:{Data.Int.Notation}]    (b) at (-1.5,-1.5) {};
  \node[dot, label=right:{Data.Nat.Notation}]   (c) at (1.5,-1.5) {};
  \node[dot, label=below:{Algebra.Group.Defs}]  (a) at (0,-3) {};

  \draw[dep] (a) -- (b);
  \draw[dep] (a) -- (c);
  \draw[dep] (b) -- (d);
  \draw[dep] (c) -- (d);
\end{tikzpicture}
\caption{A fragment of $G_{\mathrm{module}}$: \module{Algebra.Group.Defs} imports two modules that both depend on \module{Init}, forming a diamond. All names have the \module{Mathlib.}\ prefix removed.}
\label{fig:import-subgraph}
\end{figure}
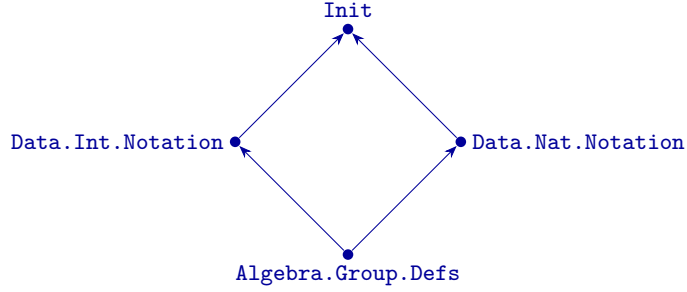

\subsubsection{Transitive Reduction and Redundant Imports}
\label{sec:transitive-reduction}

Since $G_{\mathrm{module}}$ is a DAG, it has a unique \emph{transitive reduction}~\cite{aho1972transitive}: the smallest subgraph $G_{\mathrm{module}}^{-} \subseteq G_{\mathrm{module}}$ with the same reachability relation. An edge $(m_1, m_2) \in E_{\mathrm{module}}$ is \emph{transitively redundant} if $m_2$ is reachable from $m_1$ via a path of length $\ge 2$ in $G_{\mathrm{module}}$; that is, the dependency is already implied by other imports.

\begin{definition}[Transitive reduction]\label{def:transitive-reduction}
The \emph{transitive reduction} of $G_{\mathrm{module}}$ is $G_{\mathrm{module}}^{-} = (\mathcal{M}, E_{\mathrm{module}}^{-})$ where
\[
  E_{\mathrm{module}}^{-} = E_{\mathrm{module}} \setminus \{(m_1, m_2)\in E_{\mathrm{module}} \mid \exists\, \text{path } m_1 \to \cdots \to m_2 \text{ in } G_{\mathrm{module}} \text{ of length} \ge 2\}.
\]
\end{definition}

\begin{definition}[Redundancy rate]
\label{def:redundancy-rate}
The \emph{redundancy rate} of $G_{\mathrm{module}}$ is the fraction of edges removed by transitive reduction:
\[
  r \;=\; \frac{|E_{\mathrm{module}} \setminus E_{\mathrm{module}}^{-}|}{|E_{\mathrm{module}}|}.
\]
\end{definition}

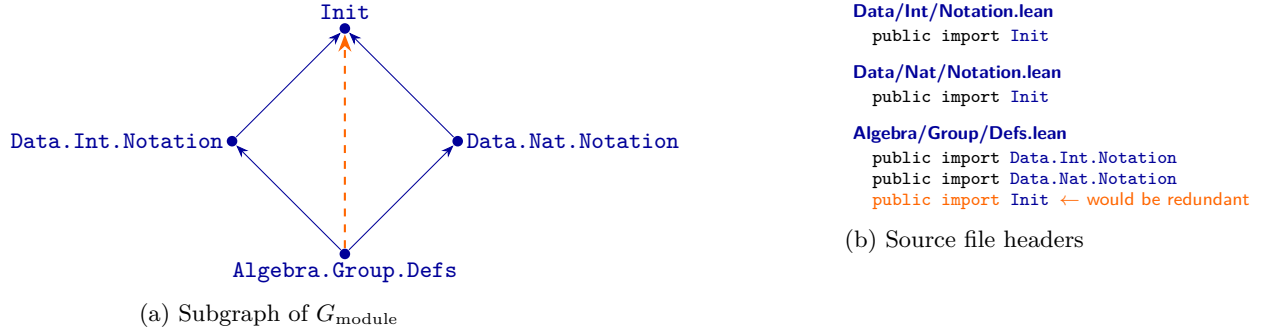
\begin{figure}[H]
\centering
\begin{subfigure}[t]{0.42\textwidth}
\centering
\vspace{0pt}
\begin{tikzpicture}[every node/.style={font=\small\ttfamily, inner sep=1pt, text=blue!60!black},
  dot/.style={circle, fill=blue!60!black, minimum size=4pt, inner sep=0pt},
  dep/.style={->, >=Stealth, thin, blue!60!black},
  redundant/.style={->, >=Stealth, orange!80!red, dashed, thick},
]
  \node[dot, label=above:{Init}]                (d) at (0,0) {};
  \node[dot, label=left:{Data.Int.Notation}]    (b) at (-1.5,-1.5) {};
  \node[dot, label=right:{Data.Nat.Notation}]   (c) at (1.5,-1.5) {};
  \node[dot, label=below:{Algebra.Group.Defs}]  (a) at (0,-3) {};

  \draw[dep] (a) -- (b);
  \draw[dep] (a) -- (c);
  \draw[dep] (b) -- (d);
  \draw[dep] (c) -- (d);
  \draw[redundant] (a) -- (d);
\end{tikzpicture}
\caption{Subgraph of $G_{\mathrm{module}}$}
\label{fig:tr-graph-app}
\end{subfigure}%
\hfill
\begin{subfigure}[t]{0.46\textwidth}
\vspace{0pt}\raggedleft
\scriptsize\ttfamily
\begin{tabular}[t]{@{}l@{}}
  \textnormal{\sffamily\bfseries\color{blue!60!black} Data/Int/Notation.lean} \\[1pt]
  \hspace{1em}public import \textcolor{blue!60!black}{Init} \\[6pt]
  \textnormal{\sffamily\bfseries\color{blue!60!black} Data/Nat/Notation.lean} \\[1pt]
  \hspace{1em}public import \textcolor{blue!60!black}{Init} \\[6pt]
  \textnormal{\sffamily\bfseries\color{blue!60!black} Algebra/Group/Defs.lean} \\[1pt]
  \hspace{1em}public import \textcolor{blue!60!black}{Data.Int.Notation} \\
  \hspace{1em}public import \textcolor{blue!60!black}{Data.Nat.Notation} \\
  \hspace{1em}\textcolor{orange!80!red}{public import \textcolor{blue!60!black}{Init}}
    \textnormal{\sffamily\color{orange!80!red} $\leftarrow$ would be redundant} \\
\end{tabular}
\caption{Source file headers}
\label{fig:tr-source-app}
\end{subfigure}
\caption{Illustrating transitive redundancy on the diamond from Figure~\ref{fig:import-subgraph}. If \module{Algebra.Group.Defs} were to import \module{Init} directly (\textcolor{orange!80!red}{dashed orange} edge), that edge would be transitively redundant: \module{Init} is already reachable via both \module{Data.Int.Notation} and \module{Data.Nat.Notation}. All names have the \module{Mathlib.}\ prefix removed.}
\label{fig:transitive-reduction-app}
\end{figure}
Transitive reduction removes $4{,}122$ of $23{,}570$ edges ($17.5\%$), yielding $|E_{\mathrm{module}}^{-}| = 19{,}448$. This residual redundancy persists \emph{after} the \texttt{shake} linter and reflects development ergonomics rather than careless coding; a detailed discussion appears in Appendix~\ref{app:module-analysis}.

\subsubsection{Build Graph}

A complementary refinement augments the module graph with compilation cost:

\begin{definition}[Build graph]\label{def:build-graph}
The \emph{build graph} is a weighted DAG
\[
  G_{\mathrm{build}} = (\mathcal{M},\, E_{\mathrm{module}},\, w),
\]
where $w : \mathcal{M} \to \mathbb{R}_{\ge 0}$ assigns to each module~$m$ its compilation time (in seconds). The \emph{critical path} is the longest path in $G_{\mathrm{build}}$ under the weight function
\[
  W(\gamma) = \sum_{m \in \gamma} w(m),
\]
maximized over all directed paths~$\gamma$ from a source to a sink. The modules on this path are the \emph{compilation bottlenecks}: no amount of parallelism can reduce build time below~$W(\gamma^*)$.
\end{definition}

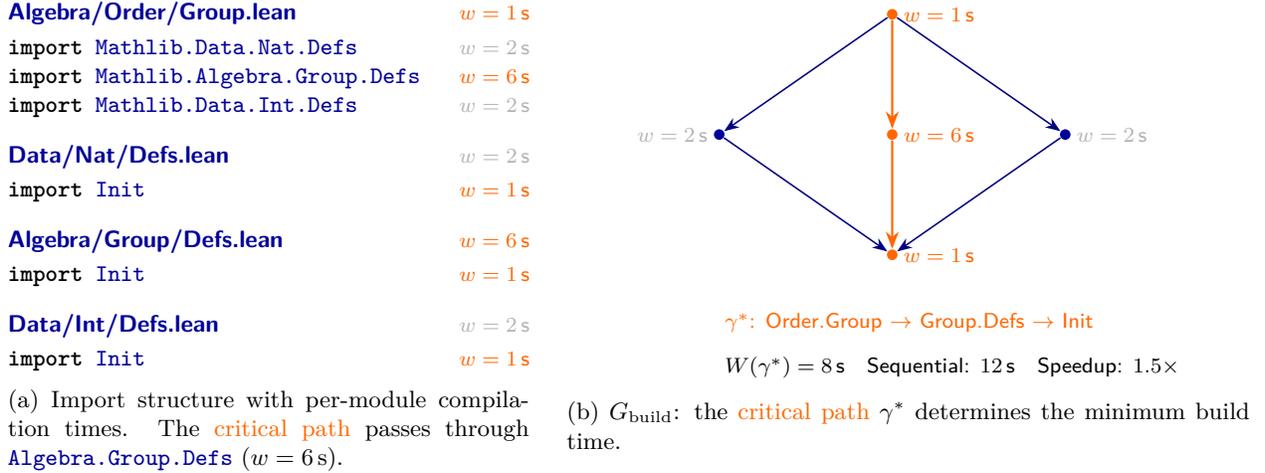
\begin{figure}[!htb]
\centering
\begin{subfigure}[t]{0.42\textwidth}
\centering
\vspace{0pt}
\raggedright\small\ttfamily
\textsf{\bfseries\color{blue!60!black}Algebra/Order/Group.lean}%
\hfill\textnormal{\sffamily\footnotesize\textcolor{orange!80!red}{$w = 1$\,s}}\\[2pt]
\textbf{import} \textcolor{blue!60!black}{Mathlib.Data.Nat.Defs}%
\hfill\textnormal{\sffamily\footnotesize\textcolor{gray!60}{$w = 2$\,s}}\\
\textbf{import} \textcolor{blue!60!black}{Mathlib.Algebra.Group.Defs}%
\hfill\textnormal{\sffamily\footnotesize\textcolor{orange!80!red}{$w = 6$\,s}}\\
\textbf{import} \textcolor{blue!60!black}{Mathlib.Data.Int.Defs}%
\hfill\textnormal{\sffamily\footnotesize\textcolor{gray!60}{$w = 2$\,s}}\\[8pt]
\textsf{\bfseries\color{blue!60!black}Data/Nat/Defs.lean}%
\hfill\textnormal{\sffamily\footnotesize\textcolor{gray!60}{$w = 2$\,s}}\\[2pt]
\textbf{import} \textcolor{blue!60!black}{Init}%
\hfill\textnormal{\sffamily\footnotesize\textcolor{orange!80!red}{$w = 1$\,s}}\\[8pt]
\textsf{\bfseries\color{blue!60!black}Algebra/Group/Defs.lean}%
\hfill\textnormal{\sffamily\footnotesize\textcolor{orange!80!red}{$w = 6$\,s}}\\[2pt]
\textbf{import} \textcolor{blue!60!black}{Init}%
\hfill\textnormal{\sffamily\footnotesize\textcolor{orange!80!red}{$w = 1$\,s}}\\[8pt]
\textsf{\bfseries\color{blue!60!black}Data/Int/Defs.lean}%
\hfill\textnormal{\sffamily\footnotesize\textcolor{gray!60}{$w = 2$\,s}}\\[2pt]
\textbf{import} \textcolor{blue!60!black}{Init}%
\hfill\textnormal{\sffamily\footnotesize\textcolor{orange!80!red}{$w = 1$\,s}}
\caption{Import structure with per-module compilation times. The \textcolor{orange!80!red}{critical path} passes through \module{Algebra.Group.Defs} ($w=6$\,s).}
\label{fig:build-source}
\end{subfigure}%
\hfill
\begin{subfigure}[t]{0.55\textwidth}
\centering
\vspace{0pt}
\begin{tikzpicture}[every node/.style={font=\footnotesize, inner sep=2pt},
  dot/.style={circle, fill=blue!60!black, minimum size=4pt, inner sep=0pt},
  critdot/.style={circle, fill=orange!80!red, minimum size=4pt, inner sep=0pt},
  edge/.style={->, >=Stealth, blue!50!black, semithick},
  critedge/.style={->, >=Stealth, orange!80!red, thick},
]
  \node[critdot, label={[font=\footnotesize\sffamily, text=orange!80!red]right:{$w = 1$\,s}}] (top) at (2, 3.2) {};
  \node[dot, label={[font=\footnotesize\sffamily, text=gray!60]left:{$w = 2$\,s}}] (nat) at (-0.3, 1.6) {};
  \node[critdot, label={[font=\footnotesize\sffamily, text=orange!80!red]right:{$w = 6$\,s}}] (grp) at (2, 1.6) {};
  \node[dot, label={[font=\footnotesize\sffamily, text=gray!60]right:{$w = 2$\,s}}] (int) at (4.3, 1.6) {};
  \node[critdot, label={[font=\footnotesize\sffamily, text=orange!80!red]right:{$w = 1$\,s}}] (init) at (2, 0) {};
  \draw[edge] (top) -- (nat);
  \draw[edge] (top) -- (int);
  \draw[edge] (nat) -- (init);
  \draw[edge] (int) -- (init);
  \draw[critedge] (top) -- (grp);
  \draw[critedge] (grp) -- (init);
  \node[font=\footnotesize\sffamily, text=orange!80!red, anchor=west]
    at (-0.3, -0.9) {$\gamma^*$: Order.Group $\to$ Group.Defs $\to$ Init};
  \node[font=\footnotesize\sffamily, anchor=west]
    at (-0.3, -1.5) {$W(\gamma^*) = 8$\,s\quad Sequential: $12$\,s\quad Speedup: $1.5\times$};
\end{tikzpicture}
\caption{$G_{\mathrm{build}}$: the \textcolor{orange!80!red}{critical path} $\gamma^*$ determines the minimum build time.}
\label{fig:build-graph}
\end{subfigure}
\caption{Build graph (Definition~\ref{def:build-graph}). (a)~Import declarations with per-module compilation times $w(m)$. (b)~The weighted DAG: the \textcolor{orange!80!red}{critical path}~$\gamma^*$ (orange boxes and edges) yields $W(\gamma^*)=1{+}6{+}1=8$\,s; \module{Data.Nat.Defs} and \module{Data.Int.Defs} compile in parallel but cannot reduce the total below~$W(\gamma^*)$. All names have the \module{Mathlib.}\ prefix removed.}
\label{fig:build}
\end{figure}

In the current snapshot, the critical path of $G_{\mathrm{build}}$ traverses $161$ modules; with unlimited parallelism, the theoretical speedup is $7{,}563 / 161 \approx 22.4\times$ relative to sequential compilation. This parallelism ratio quantifies how much of the build time is inherently sequential due to dependency ordering.

\subsubsection{Module Containment Decay}
\label{sec:module-containment}

The module graph's relationship to the module hierarchy $T$ can be quantified at every depth of the hierarchy, not only at the top level. At each depth~$k$, we truncate both endpoints of every import edge to their first~$k$ dot-separated components and measure the proportion of edges whose truncated endpoints match.

\begin{definition}[Module containment at depth~$k$]
\label{def:module-containment}
For each depth $k \ge 1$, let $\mathrm{trunc}_k(m)$ denote the first~$k$ components of module~$m$. For example,
\[
  \mathrm{trunc}_2(\module{Mathlib.Algebra.Group.Defs}) = \module{Mathlib.Algebra}.
\]
The \emph{module containment ratio at depth~$k$} is
\[
  \mathrm{contain}_{\mathrm{mod}}(k) \;=\; \frac{|\{(m_1,m_2) \in E_{\mathrm{module}} \mid \mathrm{trunc}_k(m_1) = \mathrm{trunc}_k(m_2)\}|}{|E_{\mathrm{module}}|}.
\]
\end{definition}

\subsubsection{Cross-Directory Dependencies}
\label{sec:cross-namespace}

\begin{definition}[Top-level directory]
For a module $m = \module{Mathlib.}X.Y.\cdots$, the \emph{top-level directory} is $\mathrm{dir}(m) = X$, the first directory component under the \module{Mathlib/} root.
\end{definition}

\begin{definition}[Intra-/inter-directory edge]
An edge $(m_1, m_2) \in E_{\mathrm{module}}$ is \emph{intra-directory} if $\mathrm{dir}(m_1) = \mathrm{dir}(m_2)$ and \emph{inter-directory} otherwise. For example, the edge
\[
  (\module{Mathlib.Algebra.Group.Defs},\;\; \module{Mathlib.Algebra.Notation.Defs})
\]
is intra-directory (both have $\mathrm{dir} = \module{Algebra}$), while
\[
  (\module{Mathlib.Algebra.Group.Defs},\;\; \module{Mathlib.Order.Defs})
\]
is inter-directory.
\end{definition}

\subsection{The Declaration Graph: Definitions and Decompositions}
\label{sec:theorem-premise}
\label{sec:declaration-graph}

This appendix collects the formal definitions, edge decompositions, mechanism-specific subgraphs, node-level metadata, and all illustrative figures for the declaration dependency graph. Extended statistical analysis (degree distributions, centrality rankings, community detection, robustness curves) appears in Appendix~\ref{app:decl-detail}.

\subsubsection{The declaration graph $G_{\mathrm{thm}}$}
When a mathematician writes a proof in Lean, the tactic script is compiled into a \emph{proof term}, a fully explicit expression in Lean's dependent type theory. Every named constant from the environment that appears in this term is called a \textbf{premise} of the declaration. Premises include both the results a mathematician explicitly invokes and the infrastructure (typeclass instances, coercions, notation) that Lean's \emph{elaborator} (the compiler component responsible for resolving implicit arguments, typeclass instances, and coercions) inserts automatically.

\begin{definition}[Declaration graph]
\label{def:thm-graph}
The \emph{declaration graph} is a directed graph $G_{\mathrm{thm}} = (\mathcal{D}, E_{\mathrm{thm}})$, where $\mathcal{D}$ is the set of all declarations (theorems, definitions, abbreviations, inductive types, constructors, opaques, quotients, and axioms) in \module{Mathlib}, and
\[
  E_{\mathrm{thm}} = \{(d_1, d_2) \in \mathcal{D} \times \mathcal{D} \mid d_2 \text{ appears as a premise in the proof term of } d_1\}.
\]
Each node carries two attributes: its \emph{kind} (one of eight declaration types) and its \emph{module} (the source file in which it is defined).
\end{definition}

\noindent At our snapshot, $|\mathcal{D}| = 308{,}129$ and $|E_{\mathrm{thm}}| = 8{,}436{,}366$.
Where the module graph captures the \emph{architecture} of the library, the declaration graph captures the \emph{reasoning structure} of mathematics itself.

\paragraph{What is a premise?}
Consider a simple example:

\begin{codebox}
\textit{\color{gray}-- Tactic proof (what the mathematician writes):}\\
\textbf{theorem} \textcolor{red!70!black}{Nat.succ\_pos} (n : Nat) : 0 < n + 1 :=\\
\hspace{1.5em}\textcolor{red!70!black}{Nat.zero\_lt\_succ} n\\[3pt]
\textit{\color{gray}-- The proof term contains:}\\
\textit{\color{gray}--\hspace{0.5em}Nat.zero\_lt\_succ\hspace{1em}(named constant from environment $\to$ premise)}\\
\textit{\color{gray}--\hspace{0.5em}n\hspace{6.3em}(local variable, bound by theorem $\to$ NOT a premise)}
\end{codebox}

\noindent
The distinction is precise: a premise is an \emph{external} declaration (a theorem, definition, constructor, or other named constant already registered in the environment), not a local variable or bound parameter. In the example above, \decl{Nat.zero\_lt\_succ} is a premise because it is a theorem proved elsewhere; \texttt{n} is not, because it is a locally bound variable. In practice, the kernel records additional infrastructure premises (typeclass instances, notation abbreviations) that are invisible in the tactic script but present in the compiled term. It is these mechanically extracted premises that form the edges of $G_{\mathrm{thm}}$.

The eight declaration kinds are Lean~4's fundamental building blocks: theorems ($79.1\%$), definitions ($15.8\%$), abbreviations ($2.2\%$), constructors ($1.5\%$), inductive types ($1.2\%$), and three minor kinds (Table~\ref{tab:kind-overview} in Appendix~\ref{app:decl-detail}). We highlight two structural extremes below.

\paragraph{Theorems as edge producers; axioms as pure sinks.}
In $G_{\mathrm{thm}}$, a declaration is an \emph{edge producer} if it has high out-degree (its proof invokes many premises) and a \emph{pure sink} if it has zero or near-zero out-degree but is cited by many others (high in-degree). Theorems constitute $79.1\%$ of all declarations and are the primary edge producers. A theorem's out-edges point to the premises invoked in its proof, while its in-edges come from later theorems that cite it. At the opposite extreme, \module{Mathlib} rests on exactly three axioms (\decl{propext} (propositional extensionality), \decl{Classical.choice} (the axiom of choice), and \decl{Quot.sound}), which have near-zero out-degree but are transitively required by all $308{,}129$ declarations. Among the three, \decl{propext} alone accounts for $23{,}226$ incoming edges.

\paragraph{Abbreviations as invisible hubs.}
Abbreviations (\texttt{abbrev}) are definitionally transparent shorthands that the elaborator always unfolds, acting as notational glue, typically wrapping typeclass projections or operator notation:

\begin{codebox}
\textbf{abbrev} \textcolor{red!70!black}{OfNat.ofNat} (n : Nat) [inst : OfNat a n] : a := inst.ofNat
\end{codebox}

\noindent
The most-cited node in the entire graph, \decl{OfNat.ofNat} (in-degree $89{,}936$), is an abbreviation: every declaration that mentions a numeric literal depends on it. Because the elaborator inserts these references automatically, abbreviations have the most extreme in-degree distribution among non-axiom types; their citation counts reflect the machinery of the proof assistant, not the structure of mathematical reasoning. By contrast, the most-cited \emph{inductive types} (\decl{CategoryTheory.Category} ($44{,}487$), \decl{Real} ($26{,}637$), \decl{Module} ($23{,}404$)) form a layer of mathematical infrastructure whose centrality reflects genuine mathematical importance.

\paragraph{Edge directionality.}
Not all declaration kinds participate equally in $G_{\mathrm{thm}}$. An edge $(d_1, d_2)$ means that $d_1$'s proof or body \emph{cites} $d_2$; it does not imply an edge in the reverse direction. For example, the theorem \decl{Nat.lt\_irrefl} proves the irreflexivity of strict ordering on natural numbers: no $n$ satisfies $n < n$.

\begin{codebox}
\textit{\color{gray}-- Irreflexivity of < on natural numbers}\\
\textbf{theorem} \textcolor{red!70!black}{Nat.lt\_irrefl} (n : Nat) : Not (n < n) :=\\
\hspace{1.5em}\textcolor{red!70!black}{Nat.not\_succ\_le\_self} n
\end{codebox}

\noindent
The proof invokes \decl{Nat.not\_succ\_le\_self}, which establishes that $n + 1 \le n$ is impossible. The kernel also records two infrastructure premises: \decl{LT.lt}, the abbreviation that provides the~$<$ notation, and \decl{instLTNat}, the typeclass instance that connects~$<$ to \decl{Nat}. This gives three edges:

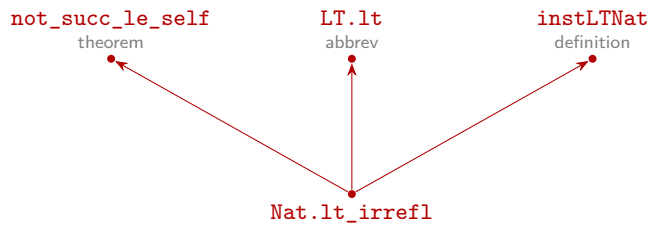
\begin{figure}[H]
\centering
\begin{tikzpicture}[
  every node/.style={font=\small\ttfamily, inner sep=2pt, text=red!70!black},
  dot/.style={circle, fill=red!70!black, minimum size=3pt, inner sep=0pt},
  dep/.style={->, >=Stealth, thin, red!70!black},
]
  \node[dot, label={[align=center]above:{not\_succ\_le\_self\\[-2pt]{\normalfont\sffamily\scriptsize\color{gray}theorem}}}] (thm) at (-3.2,0) {};
  \node[dot, label={[align=center]above:{LT.lt\\[-2pt]{\normalfont\sffamily\scriptsize\color{gray}abbrev}}}]              (ab)  at (0,0) {};
  \node[dot, label={[align=center]above:{instLTNat\\[-2pt]{\normalfont\sffamily\scriptsize\color{gray}definition}}}]       (def) at (3.2,0) {};
  \node[dot, label=below:{Nat.lt\_irrefl}] (t) at (0,-1.8) {};
  \draw[dep] (t) -- (thm);
  \draw[dep] (t) -- (ab);
  \draw[dep] (t) -- (def);
\end{tikzpicture}
\caption{Premise edges for the theorem \decl{Nat.lt\_irrefl}: the mathematical premise \decl{not\_succ\_le\_self} (a theorem) and two infrastructure premises \decl{LT.lt} (an abbreviation providing $<$ notation) and \decl{instLTNat} (a typeclass instance). The edges are directed: none of the three premises cites \decl{lt\_irrefl} in return.}
\label{fig:thm-single-edge}
\end{figure}

\noindent
The eight kinds arrange into a clear hierarchy of edge production and consumption (Table~\ref{tab:inter-kind-flow} in Appendix~\ref{app:decl-detail}). Theorems produce $89\%$ of all edges ($7.50$M of $8.44$M), primarily citing definitions and abbreviations, the computational and notational infrastructure on which proofs are built. Axioms and quotients are near-pure sinks (combined out-degree $= 3$). The graph is nearly acyclic; the only bidirectional edges ($4{,}713$ pairs) link inductive types to their own constructors (e.g., \decl{Nat}~$\leftrightarrow$~\decl{Nat.succ}).

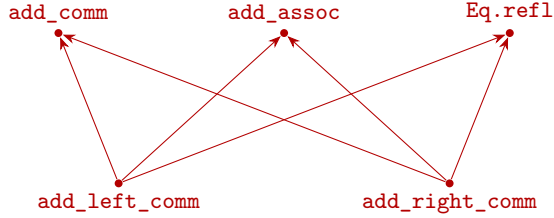
\begin{figure}[H]
\centering
\begin{tikzpicture}[
  every node/.style={font=\small\ttfamily, inner sep=2pt, text=red!70!black},
  dot/.style={circle, fill=red!70!black, minimum size=3pt, inner sep=0pt},
  dep/.style={->, >=Stealth, thin, red!70!black},
]
  \node[dot, label=above:{add\_comm}]  (ac) at (-3,0) {};
  \node[dot, label=above:{add\_assoc}] (aa) at (0,0)  {};
  \node[dot, label=above:{Eq.refl}]    (er) at (3,0)  {};

  \node[dot, label=below:{add\_left\_comm}]  (alc) at (-2.2,-2) {};
  \node[dot, label=below:{add\_right\_comm}] (arc) at (2.2,-2)  {};

  \draw[dep] (alc) -- (ac);
  \draw[dep] (alc) -- (aa);
  \draw[dep] (alc) -- (er);

  \draw[dep] (arc) -- (ac);
  \draw[dep] (arc) -- (aa);
  \draw[dep] (arc) -- (er);
\end{tikzpicture}
\caption{A fragment of $G_{\mathrm{thm}}$: the theorems \decl{add\_left\_comm} and \decl{add\_right\_comm} both cite the premises \decl{add\_comm}, \decl{add\_assoc}, and \decl{Eq.refl}. The shared node \decl{Eq.refl} (in-degree $69{,}580$) exemplifies the hub phenomenon. All names have the \texttt{Nat.}\ prefix removed except \decl{Eq.refl}.}
\label{fig:thm-fragment}
\end{figure}

\medskip
\noindent
The \decl{Nat.lt\_irrefl} example above showed a single theorem's premises; we now turn to a \emph{pair} of theorems to illustrate how shared premises create the hub structure that dominates $G_{\mathrm{thm}}$. Consider \decl{Nat.add\_left\_comm}, which proves $a + (b + c) = b + (a + c)$ for natural numbers. The extracted premises are:

\smallskip
\begin{center}
\small
\begin{tabular}{ll}
\textit{Mathematical premises:} & \decl{Nat.add\_comm}, \decl{Nat.add\_assoc} \\
\textit{Equality infrastructure:} & \decl{Eq.refl}, \decl{Eq.symm}, \decl{Eq.mpr} \\
\textit{Typeclass instances:} & \decl{instHAdd}, \decl{HAdd.hAdd}, \decl{instAddNat}
\end{tabular}
\end{center}
\smallskip

\noindent
A closely related theorem, \decl{Nat.add\_right\_comm} ($a + b + c = a + c + b$), depends on \emph{exactly the same} eight premises. The resulting fragment of $G_{\mathrm{thm}}$ is shown in Figure~\ref{fig:thm-fragment}: both theorems share all three displayed premises, and \decl{Eq.refl} (a constructor with in-degree $69{,}580$) is the single most-cited declaration in the entire graph. This shared-hub pattern is the mechanism by which a small number of foundational declarations accumulate enormous in-degree.

\noindent\textbf{Note.}
The module \module{Algebra.Group.Defs} contains dozens of declarations. In $G_{\mathrm{module}}$, this entire module is a single vertex; in $G_{\mathrm{thm}}$, each declaration within it becomes a separate vertex, and each premise reference becomes a separate edge. A single import edge in $G_{\mathrm{module}}$ thus unfolds into potentially hundreds of premise edges in $G_{\mathrm{thm}}$; this is why $|E_{\mathrm{thm}}|$ exceeds $|E_{\mathrm{module}}|$ by a factor of $360$.

The graph is almost entirely connected: $308{,}054$ of $308{,}129$ nodes ($99.98\%$) belong to a single weakly connected component. The remaining $75$ nodes distribute across $71$ tiny components, of which $69$ are singletons.

\paragraph{Empirical observations.}
The degree statistics by declaration kind reveal that each type occupies a distinct structural niche: axioms and inductive types are heavily cited infrastructure, while theorems are dependency consumers (\S\ref{sec:thm-types}). The hub structure decomposes into a \emph{language infrastructure} layer and a \emph{mathematical infrastructure} layer (Remark~\ref{rem:two-layers}). Of all theorems, $44.8\%$ are never cited, leaf nodes whose consumers are necessarily external (\S\ref{sec:thm-leaves}, Remark~\ref{rem:leaf-theorems}). Cross-namespace edges account for $50.9\%$ of all declaration-level dependencies, far exceeding the module-level figure (\S\ref{sec:thm-cross-namespace}). Full analysis is in Appendix~\ref{app:decl-detail}.

\medskip
The declaration graph admits several decompositions that disentangle inherited mathematical structure from tool-mediated organizational choices. We organize these into three groups: edge-level decompositions that partition $E_{\mathrm{thm}}$ by the origin of each dependency (\S\ref{sec:stmt-proof}--\S\ref{sec:typeclass-synth}), subgraph constructions that isolate specific language mechanisms such as typeclass synthesis and coercions (\S\ref{sec:typeclass-synth}), and node-level enrichments that attach metadata (attributes, definitional height, tactic usage) to each declaration (\S\ref{sec:decl-attributes}).

\subsubsection{Statement and proof decomposition}
\label{sec:stmt-proof}
Every declaration $d$ has a type~$\tau$ (the statement) and optionally a proof term~$\pi$; an edge may originate from either or both.

\begin{definition}[Statement and proof dependency]\label{def:stmt-proof-dep}
For declarations $a, b \in \mathcal{D}$, define:
\begin{align*}
  a \xrightarrow{S} b &\iff b \in \mathrm{refs}(a.\tau), \\
  a \xrightarrow{P} b &\iff b \in \mathrm{refs}(a.\pi) \setminus \mathrm{refs}(a.\tau).
\end{align*}
The \emph{statement dependency graph} is $G_S = (\mathcal{D}, E_S)$ where $E_S = \{(a,b) : a \xrightarrow{S} b\}$, and the \emph{pure proof dependency graph} is $G_P = (\mathcal{D}, E_P)$ where $E_P = \{(a,b) : a \xrightarrow{P} b\}$. An edge $(a,b) \in E_{\mathrm{thm}}$ that belongs to both $\mathrm{refs}(a.\tau)$ and $\mathrm{refs}(a.\pi)$ is classified as a \emph{mixed edge} $E_{SP}$. This yields a partition $E_{\mathrm{thm}} = E_S \sqcup E_P \sqcup E_{SP}$ (where $E_S$ excludes mixed edges for disjointness, i.e., $E_S$ consists of edges in $\mathrm{refs}(a.\tau) \setminus \mathrm{refs}(a.\pi)$, $E_{SP} = \mathrm{refs}(a.\tau) \cap \mathrm{refs}(a.\pi)$, and $E_P$ as above).
\end{definition}

\begin{figure}[!htb]
\centering
\begin{subfigure}[t]{0.52\textwidth}
\centering
\vspace{0pt}
\raggedright\scriptsize\ttfamily
\textbf{theorem} \textcolor{red!70!black}{length\_nil}\\
\hspace{1em}: [].{\textcolor{teal}{length}} = 0 := rfl\\[6pt]
\textbf{theorem} \textcolor{red!70!black}{length\_cons} (a : $\alpha$) (l : List $\alpha$)\\
\hspace{1em}: (a :: l).{\textcolor{teal}{length}} = l.{\textcolor{teal}{length}} + 1 := rfl\\[6pt]
\textbf{theorem} \textcolor{red!70!black}{length\_append} (l1 l2 : List $\alpha$)\\
\hspace{1em}: (l1 ++ l2).{\textcolor{violet}{length}} = l1.{\textcolor{violet}{length}} + l2.{\textcolor{violet}{length}}\\
\hspace{1em}:= \textbf{by} induction l1 \textbf{with}\\
\hspace{1em}| nil => simp [\textcolor{orange!80!red}{length\_nil}]\\
\hspace{1em}| cons a l ih => simp [\textcolor{orange!80!red}{length\_cons}, ih]
\caption{Source code: references colored by edge type.}
\label{fig:sp-source}
\end{subfigure}%
\hfill
\begin{subfigure}[t]{0.44\textwidth}
\centering
\vspace{0pt}
\begin{tikzpicture}[
  every node/.style={font=\scriptsize, inner sep=2pt},
  dot/.style={circle, fill=red!70!black, minimum size=3.5pt, inner sep=0pt},
  sedge/.style={->, >=Stealth, teal, semithick},
  pedge/.style={->, >=Stealth, orange!80!red, semithick, dashed},
  spedge/.style={->, >=Stealth, violet, thick},
]
  \node[dot, label={[font=\footnotesize\ttfamily, red!70!black]above:length\_append}]
    (la) at (2, 3.2) {};
  \node[dot, label={[font=\footnotesize\ttfamily, red!70!black]left:length\_nil}]
    (ln) at (0.3, 1.6) {};
  \node[dot, label={[font=\footnotesize\ttfamily, red!70!black]right:length\_cons}]
    (lc) at (3.7, 1.6) {};
  \node[dot, label={[font=\footnotesize\ttfamily, red!70!black]below:length}]
    (ll) at (2, 0) {};
  \draw[pedge] (la) -- (ln);
  \draw[pedge] (la) -- (lc);
  \draw[spedge] (la) -- (ll);
  \draw[sedge] (ln) -- (ll);
  \draw[sedge] (lc) -- (ll);
\end{tikzpicture}
\caption{$E_{\mathrm{thm}} = \textcolor{teal}{E_S} \sqcup \textcolor{orange!80!red}{E_P} \sqcup \textcolor{violet}{E_{SP}}$.}
\label{fig:sp-graph}
\end{subfigure}
\caption{Statement vs.\ proof decomposition (Definition~\ref{def:stmt-proof-dep}). (a)~Source code for three \texttt{List.length} theorems from \module{Init.Data.List.Lemmas}: \textcolor{teal}{teal} = reference only in the type~$\tau$ ($E_S$), \textcolor{orange!80!red}{orange dashed} = only in the proof~$\pi$ ($E_P$), \textcolor{violet}{violet} = in both ($E_{SP}$). For \decl{length\_nil} and \decl{length\_cons}, the definition \decl{length} appears only in~$\tau$; for \decl{length\_append}, it appears in both~$\tau$ and~$\pi$. (b)~The edge-classified subgraph. All names have the \texttt{List.}\ prefix removed.}
\label{fig:stmt-proof-decomp}
\end{figure}
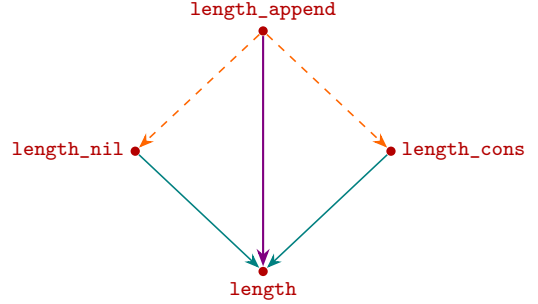

\subsubsection{Typeclass instance graph, coercions, and synthesized edges}
\label{sec:typeclass-synth}

Lean's \emph{typeclass mechanism} encodes mathematical knowledge that human practitioners share implicitly (e.g., which addition is meant by $a + b$) as explicit, machine-searchable declarations.  A \texttt{class} defines an abstract interface; an \texttt{instance} registers a concrete witness that a given type satisfies it.  The mechanism enables readable formal statements at the cost of creating a large infrastructure layer: $24{,}947$ instance declarations ($10.6\%$ of $\mathcal{D}$) form this layer, dominating the dependency graph's in-degree distribution.

This subsection defines three related constructions: the typeclass instance graph $G_{\mathrm{tc}}$, the synthesized/explicit edge partition, and the coercion graph $G_{\mathrm{coe}}$ and structure inheritance graph $G_{\mathrm{ext}}$.

\begin{definition}[Typeclass instance graph]\label{def:tc-graph}
Let $\mathcal{C} \subset \mathcal{D}$ be the set of all typeclass declarations and $\mathcal{I} \subset \mathcal{D}$ the set of all instance declarations. For each instance $i \in \mathcal{I}$, let $\mathrm{target}(i) \in \mathcal{C}$ be the class it provides and $\mathrm{requires}(i) \subseteq \mathcal{C}$ the classes it depends on as preconditions. The \emph{typeclass instance graph} is
\[
  G_{\mathrm{tc}} = (\mathcal{C},\, E_{\mathrm{tc}}), \quad (C_1, C_2) \in E_{\mathrm{tc}} \iff \exists\, i \in \mathcal{I} : \mathrm{target}(i) = C_2 \land C_1 \in \mathrm{requires}(i).
\]
\end{definition}

\begin{figure}[!htb]
\centering
\begin{subfigure}[t]{0.30\textwidth}
\centering
\vspace{0pt}
\raggedright\scriptsize\ttfamily
\textsf{\bfseries\color{blue!60!black}Algebra/Group/Defs.lean}\\[2pt]
\textbf{class} \textcolor{red!70!black}{Semigroup} ($\alpha$)\\
\hspace{1em}\textbf{extends} \textcolor{teal}{Mul} $\alpha$\\[3pt]
\textbf{class} \textcolor{red!70!black}{MulOneClass} ($\alpha$)\\
\hspace{1em}\textbf{extends} \textcolor{teal}{Mul} $\alpha$, \textcolor{teal}{One} $\alpha$\\[3pt]
\textbf{class} \textcolor{red!70!black}{Monoid} ($\alpha$)\\
\hspace{1em}\textbf{extends} \textcolor{red!70!black}{Semigroup} $\alpha$,\\
\hspace{1em}\phantom{\textbf{extends}} \textcolor{red!70!black}{MulOneClass} $\alpha$\\[3pt]
\textbf{class} \textcolor{red!70!black}{CommMonoid} ($\alpha$)\\
\hspace{1em}\textbf{extends} \textcolor{red!70!black}{Monoid} $\alpha$,\\
\hspace{1em}\phantom{\textbf{extends}} \textcolor{red!70!black}{CommSemigroup} $\alpha$
\caption{Source: each \texttt{extends} creates an edge in $G_{\mathrm{ext}}$ and auto-generates a forgetful instance in~$G_{\mathrm{tc}}$.}
\label{fig:tc-ext-source}
\end{subfigure}%
\hfill
\begin{subfigure}[t]{0.33\textwidth}
\centering
\vspace{0pt}
\begin{tikzpicture}[
  scale=0.82, every node/.style={transform shape},
  nd/.style={rounded corners=2pt, draw=red!60!black, fill=red!6,
    font=\footnotesize\ttfamily, text=red!70!black, inner sep=2.5pt, minimum height=13pt},
  lf/.style={rounded corners=2pt, draw=teal!60!black, fill=teal!6,
    font=\footnotesize\ttfamily, text=teal!70!black, inner sep=2.5pt, minimum height=13pt},
  edg/.style={->, >=Stealth, red!60!black, semithick},
]
  \node[nd] (cm) at (1.6, 4.5) {CommMonoid};
  \node[nd] (mon) at (0.2, 3.2) {Monoid};
  \node[nd] (cs) at (3.0, 3.2) {CommSemi.};
  \node[nd] (sg) at (0.0, 1.9) {Semigroup};
  \node[nd] (moc) at (2.2, 1.9) {MulOneCl.};
  \node[lf] (mul) at (0.5, 0.6) {Mul};
  \node[lf] (one) at (2.6, 0.6) {One};
  \draw[edg] (cm) -- (mon);
  \draw[edg] (cm) -- (cs);
  \draw[edg] (mon) -- (sg);
  \draw[edg] (mon) -- (moc);
  \draw[edg] (cs) -- (sg);
  \draw[edg] (sg) -- (mul);
  \draw[edg] (moc) -- (mul);
  \draw[edg] (moc) -- (one);
\end{tikzpicture}
\caption{$G_{\mathrm{ext}}$: \texttt{extends} edges only (8 edges).}
\label{fig:ext-graph}
\end{subfigure}%
\hfill
\begin{subfigure}[t]{0.33\textwidth}
\centering
\vspace{0pt}
\begin{tikzpicture}[
  scale=0.82, every node/.style={transform shape},
  nd/.style={rounded corners=2pt, draw=red!60!black, fill=red!6,
    font=\footnotesize\ttfamily, text=red!70!black, inner sep=2.5pt, minimum height=13pt},
  lf/.style={rounded corners=2pt, draw=teal!60!black, fill=teal!6,
    font=\footnotesize\ttfamily, text=teal!70!black, inner sep=2.5pt, minimum height=13pt},
  edg/.style={->, >=Stealth, red!60!black, semithick},
  extra/.style={->, >=Stealth, blue!60!black, thick},
]
  \node[nd] (cm) at (1.6, 4.5) {CommMonoid};
  \node[nd] (mon) at (0.2, 3.2) {Monoid};
  \node[nd] (cs) at (3.0, 3.2) {CommSemi.};
  \node[nd] (sg) at (0.0, 1.9) {Semigroup};
  \node[nd] (moc) at (2.2, 1.9) {MulOneCl.};
  \node[lf] (mul) at (0.5, 0.6) {Mul};
  \node[lf] (one) at (2.6, 0.6) {One};
  \draw[edg, thin, red!30] (cm) -- (mon);
  \draw[edg, thin, red!30] (cm) -- (cs);
  \draw[edg, thin, red!30] (mon) -- (sg);
  \draw[edg, thin, red!30] (mon) -- (moc);
  \draw[edg, thin, red!30] (cs) -- (sg);
  \draw[edg, thin, red!30] (sg) -- (mul);
  \draw[edg, thin, red!30] (moc) -- (mul);
  \draw[edg, thin, red!30] (moc) -- (one);
  \draw[extra, bend right=25] (cm) to (sg);
  \draw[extra, bend left=20] (cm) to (moc);
  \draw[extra, bend left=25] (cs) to (mul);
  \draw[extra, bend right=15] (cm) to (mul);
  \draw[extra, bend left=15] (cm) to (one);
\end{tikzpicture}
\caption{$G_{\mathrm{tc}}$: adds \textcolor{blue!60!black}{5 transitive} forgetful instance edges.}
\label{fig:tc-graph}
\end{subfigure}
\caption{$G_{\mathrm{ext}}$ vs.\ $G_{\mathrm{tc}}$ (Definitions~\ref{def:tc-graph} and~\ref{def:extends-graph}), from \module{Mathlib.Algebra.Group.Defs} (lines 173--759). (a)~Source: each \texttt{extends} creates a field inheritance edge. (b)~$G_{\mathrm{ext}}$: edges from \texttt{extends} only. (c)~$G_{\mathrm{tc}}$: the same nodes, but Lean auto-generates transitive forgetful instances (\textcolor{blue!60!black}{blue}), e.g.\ \texttt{CommMonoid} directly provides \texttt{Semigroup} without going through \texttt{Monoid}. Thus $G_{\mathrm{ext}} \subset G_{\mathrm{tc}}$.}
\label{fig:tc-ext}
\end{figure}

In a traditional mathematical paper, a proof's dependencies are exactly what the author explicitly cites: if a proof invokes the intermediate value theorem, that citation is visible on the page.  In Lean, the situation is radically different.  When a mathematician writes \texttt{a + 0 = a}, the elaborator silently resolves \texttt{+} to a specific addition operation (via \texttt{HAdd}), interprets \texttt{0} as \texttt{OfNat.ofNat 0} for the relevant type, and synthesizes the required algebraic instances, each resolution creating a dependency edge that appears nowhere in the source code.  In Mathlib, $74.2\%$ of all dependency edges are synthesized in this way, meaning that the ``visible'' mathematical content accounts for barely a quarter of the actual dependency graph.

\begin{definition}[Synthesized vs.\ explicit edge partition]\label{def:synth-partition}
For each edge $(a, b) \in E_{\mathrm{thm}}$, define
\[
  \mathrm{synth}(a, b) =
  \begin{cases}
    1 & \text{if } b \text{ was inserted by typeclass instance synthesis during elaboration}, \\
    0 & \text{if } b \text{ was explicitly referenced in the source}.
  \end{cases}
\]
This partitions $E_{\mathrm{thm}} = E_{\mathrm{explicit}} \sqcup E_{\mathrm{synth}}$. The \emph{synthesis ratio} $\sigma = |E_{\mathrm{synth}}| / |E_{\mathrm{thm}}|$ measures the fraction of the observed dependency structure attributable to the proof assistant's type system rather than to mathematical logic.
\end{definition}

\noindent At our snapshot, $\sigma = 74.2\%$: $6{,}257{,}832$ edges are synthesized and $2{,}178{,}534$ are explicit. The ``visible'' mathematical reasoning thus accounts for barely a quarter of the actual dependency graph.

\begin{definition}[Coercion graph]\label{def:coercion-graph}
Lean allows automatic type coercions (e.g., $\mathbb{N} \hookrightarrow \mathbb{Z} \hookrightarrow \mathbb{Q}$), which the elaborator chains silently. Let $\mathcal{K} \subset \mathcal{D}$ be the set of all registered coercion instances. The \emph{coercion graph} is
\[
  G_{\mathrm{coe}} = (\mathcal{T},\, E_{\mathrm{coe}}), \quad (T_1, T_2) \in E_{\mathrm{coe}} \iff \exists\, k \in \mathcal{K} : k \text{ coerces } T_1 \to T_2.
\]
Long coercion chains (paths of length $\ge 3$) degrade compilation performance; the diameter of $G_{\mathrm{coe}}$ measures this fragility. Together with $G_{\mathrm{tc}}$, the coercion graph captures the two main mechanisms by which Lean's type system silently inflates the dependency graph.
\end{definition}

\begin{figure}[!htb]
\centering
\begin{subfigure}[t]{0.48\textwidth}
\centering
\vspace{0pt}
\raggedright\scriptsize\ttfamily
\textsf{\bfseries\color{blue!60!black}Lean number type coercions}\\[2pt]
\textnormal{\sffamily The elaborator silently chains coercions:}\\[3pt]
\textbf{example} (n : \textcolor{teal}{$\mathbb{N}$}) :
  \textcolor{teal}{$\mathbb{R}$} := n\\
\textnormal{\sffamily\scriptsize\textcolor{gray!60}{\hspace{1em}%
--- inserts $\mathbb{N} \xrightarrow{\text{ofNat}}
  \mathbb{Z} \xrightarrow{\text{ofInt}}
  \mathbb{Q} \xrightarrow{\text{ofRat}} \mathbb{R}$}}\\[6pt]
\textnormal{\sffamily Registered coercion instances:}\\[2pt]
\hspace{1em}\textcolor{red!70!black}{Int.ofNat}%
  \textnormal{\sffamily\scriptsize\textcolor{gray!60}{\hspace{4pt}: $\mathbb{N} \to \mathbb{Z}$}}\\
\hspace{1em}\textcolor{red!70!black}{Rat.ofInt}%
  \textnormal{\sffamily\scriptsize\textcolor{gray!60}{\hspace{4pt}: $\mathbb{Z} \to \mathbb{Q}$}}\\
\hspace{1em}\textcolor{red!70!black}{Real.ofRat}%
  \textnormal{\sffamily\scriptsize\textcolor{gray!60}{\hspace{4pt}: $\mathbb{Q} \to \mathbb{R}$}}\\
\hspace{1em}\textcolor{red!70!black}{Complex.ofReal}%
  \textnormal{\sffamily\scriptsize\textcolor{gray!60}{\hspace{4pt}: $\mathbb{R} \to \mathbb{C}$}}
\caption{Source: a single coercion \texttt{n} $\to$ $\mathbb{R}$ triggers a chain of length~$3$.}
\label{fig:coe-source}
\end{subfigure}%
\hfill
\begin{subfigure}[t]{0.48\textwidth}
\centering
\vspace{0pt}
\begin{tikzpicture}[
  every node/.style={font=\scriptsize, inner sep=2pt},
  tp/.style={circle, draw=red!60!black, fill=red!6,
    font=\small, text=red!70!black, inner sep=3pt, minimum size=20pt},
  coe/.style={->, >=Stealth, red!60!black, semithick},
]
  \node[tp] (N) at (0, 0) {$\mathbb{N}$};
  \node[tp] (Z) at (1.8, 0) {$\mathbb{Z}$};
  \node[tp] (Q) at (3.6, 0) {$\mathbb{Q}$};
  \node[tp] (R) at (5.4, 0) {$\mathbb{R}$};
  \node[tp] (C) at (7.2, 0) {$\mathbb{C}$};
  \draw[coe] (N) -- node[above, font=\footnotesize\sffamily, text=gray!60] {ofNat} (Z);
  \draw[coe] (Z) -- node[above, font=\footnotesize\sffamily, text=gray!60] {ofInt} (Q);
  \draw[coe] (Q) -- node[above, font=\footnotesize\sffamily, text=gray!60] {ofRat} (R);
  \draw[coe] (R) -- node[above, font=\footnotesize\sffamily, text=gray!60] {ofReal} (C);
  \node[font=\footnotesize\sffamily, text=gray!60] at (3.6, -0.7)
    {$\underbrace{\hspace{5.2cm}}_{\text{chain length} = 3}$};
\end{tikzpicture}
\caption{$G_{\mathrm{coe}}$: the number type coercion chain. Diameter~$= 4$.}
\label{fig:coe-graph}
\end{subfigure}
\caption{Coercion graph (Definition~\ref{def:coercion-graph}). (a)~Writing \texttt{(n :~$\mathbb{R}$)} where \texttt{n~:~$\mathbb{N}$} triggers the elaborator to chain three coercions $\mathbb{N} \to \mathbb{Z} \to \mathbb{Q} \to \mathbb{R}$, each a registered instance in~$\mathcal{K}$. The individual coercions are defined in \module{Init.Data.Int.Basic}, \module{Mathlib.Data.Rat.Cast.Defs}, and \module{Mathlib.Topology.Algebra.Order.Field}. (b)~The subgraph of~$G_{\mathrm{coe}}$ for the number tower. Long chains (length~$\ge 3$) degrade elaboration performance and can produce opaque error messages.}
\label{fig:coercion}
\end{figure}

\noindent At our snapshot, $|\mathcal{K}| = 256$ registered coercions with diameter~$4$ (the number tower $\mathbb{N} \to \mathbb{Z} \to \mathbb{Q} \to \mathbb{R} \to \mathbb{C}$). The most-cited coercion node, \decl{DFunLike.coe} (in-degree $65{,}437$), mediates function-like coercions across the library; \decl{SetLike.coe} ($9{,}519$) is the second hub.

\begin{definition}[Structure inheritance graph]\label{def:extends-graph}
Lean structures may extend other structures via the \texttt{extends} keyword, inheriting their fields. The \emph{structure inheritance graph} is
\[
  G_{\mathrm{ext}} = (\mathcal{S},\, E_{\mathrm{ext}}), \quad (S_1, S_2) \in E_{\mathrm{ext}} \iff S_1 \texttt{ extends } S_2,
\]
where $\mathcal{S} \subset \mathcal{D}$ is the set of all \texttt{structure} declarations. This graph encodes the algebraic hierarchy's inheritance lattice, a tool-mediated organizational layer that shapes which instances need to be defined and which are inherited automatically.
\end{definition}

\noindent At our snapshot, $|E_{\mathrm{ext}}| = 1{,}535$ \texttt{extends} edges among $|\mathcal{S}|$ structures. The most-extended structures, at the top of the algebraic hierarchy, are the same inductive types that dominate $G_{\mathrm{thm}}$'s in-degree: \decl{CategoryTheory.Category} ($44{,}487$), \decl{Real} ($26{,}637$), \decl{TopologicalSpace} ($25{,}310$), \decl{Module} ($23{,}404$), \decl{CommRing} ($20{,}837$).


\begin{figure}[!htb]
\centering
\begin{subfigure}[t]{0.50\textwidth}
\centering
\vspace{0pt}
\raggedright\scriptsize\ttfamily
\textbf{theorem} \textcolor{red!70!black}{add\_zero} [AddMonoid $\alpha$] (a : $\alpha$)\\
\hspace{1em}: a + 0 = a := \textcolor{red!70!black}{AddMonoid.add\_zero} a\\[8pt]
\textnormal{\sffamily Elaborated term references:}\\[2pt]
\hspace{1em}\textcolor{red!70!black}{AddMonoid.add\_zero}%
  \textnormal{\sffamily\scriptsize\textcolor{gray!60}{\hspace{4pt}--- explicit}}\\
\hspace{1em}\textcolor{gray!50}{HAdd.hAdd}%
  \textnormal{\sffamily\scriptsize\textcolor{gray!60}{\hspace{4pt}--- synth.\ for \texttt{+}}}\\
\hspace{1em}\textcolor{gray!50}{OfNat.ofNat}%
  \textnormal{\sffamily\scriptsize\textcolor{gray!60}{\hspace{4pt}--- synth.\ for \texttt{0}}}
\caption{Source: only \decl{AddMonoid.add\_zero} is explicit; \texttt{+}~and~\texttt{0} trigger instance synthesis.}
\label{fig:tc-source}
\end{subfigure}%
\hfill
\begin{subfigure}[t]{0.46\textwidth}
\centering
\vspace{0pt}
\begin{tikzpicture}[
  every node/.style={font=\scriptsize, inner sep=2pt},
  dot/.style={circle, fill=red!70!black, minimum size=3.5pt, inner sep=0pt},
  sdot/.style={circle, fill=gray!50, minimum size=3.5pt, inner sep=0pt},
  elink/.style={->, >=Stealth, red!70!black, semithick},
  slink/.style={->, >=Stealth, gray!50, semithick, dashed},
]
  \fill[red!6, rounded corners=3pt] (-0.6, -0.5) rectangle (1.6, 1.0);
  \node[font=\footnotesize\sffamily\itshape, red!40!black] at (0.5, 0.85) {math content};
  \fill[gray!8, rounded corners=3pt] (1.9, -0.5) rectangle (5.0, 1.0);
  \node[font=\footnotesize\sffamily\itshape, gray!50] at (3.45, 0.85) {language infrastructure};
  \node[dot, label={[font=\footnotesize\ttfamily, red!70!black]above:add\_zero}]
    (thm) at (2.3, 2.8) {};
  \node[dot, label={[font=\footnotesize\ttfamily, red!70!black]below:AddMonoid.add\_zero}]
    (axiom) at (0.5, 0.2) {};
  \node[sdot, label={[font=\footnotesize\ttfamily, gray!60]below:HAdd.hAdd}]
    (hadd) at (2.8, 0.2) {};
  \node[sdot, label={[font=\footnotesize\ttfamily, gray!60]below:OfNat.ofNat}]
    (ofnat) at (4.3, 0.2) {};
  \draw[elink] (thm) -- (axiom);
  \draw[slink] (thm) -- (hadd);
  \draw[slink] (thm) -- (ofnat);
\end{tikzpicture}
\caption{$E_{\mathrm{thm}} = \textcolor{red!70!black}{E_{\mathrm{explicit}}} \sqcup \textcolor{gray!50}{E_{\mathrm{synth}}}$.}
\label{fig:synth-graph}
\end{subfigure}
\caption{Synthesized vs.\ explicit edge partition (Definition~\ref{def:synth-partition}). (a)~In the source for \decl{add\_zero} (from \module{Mathlib.Algebra.Group.Defs}, line 637), only \decl{AddMonoid.add\_zero} is an explicit reference; the notation~\texttt{+} and literal~\texttt{0} each trigger typeclass instance synthesis, silently inserting \textcolor{gray!50}{synthesized edges}. (b)~The subgraph of $G_{\mathrm{thm}}$: \textcolor{red!70!black}{solid} = explicit, \textcolor{gray!50}{dashed} = synthesized. The two zones separate mathematical content from language infrastructure (Remark~\ref{rem:two-layers}). Hub nodes such as \decl{OfNat.ofNat} (in-degree $89{,}936$) accumulate high centrality primarily through synthesized edges.}
\label{fig:synth-partition}
\end{figure}

\subsubsection{Declaration attributes and node-level metadata}
\label{sec:decl-attributes}
Beyond graph topology, each declaration carries metadata that enriches $G_{\mathrm{thm}}$ with node-level features. We define four groups: the attribute function $\alpha(d)$ with derived measures of proof complexity and universe polymorphism; the definitional height $\delta(d)$, which captures implicit coupling through the kernel's unfolding mechanism; the tactic usage profile $\tau(d)$, which records proof strategy choices; and the auto-derived instance edges $E_{\mathrm{auto}}$, which isolate the purely tool-generated fragment of the dependency graph.

\begin{definition}[Declaration attributes]\label{def:decl-attributes}
The \emph{attribute function} $\alpha : \mathcal{D} \to \mathcal{P}(\mathcal{A})$ maps each declaration to its set of Lean attributes, where
\[
  \mathcal{A} = \{\texttt{simp},\, \texttt{ext},\, \texttt{to\_additive},\, \texttt{instance},\, \texttt{reducible},\, \texttt{inline},\, \ldots\}.
\]
In particular:
\begin{itemize}
  \item The \emph{proof complexity} $\kappa(d) = |\{\text{Expr nodes in } d.\pi\}|$ measures proof term size.
  \item The \emph{universe polymorphism degree} $\upsilon(d) = |\text{universe parameters of } d|$.
  \item The \texttt{@[to\_additive]} attribute instructs a metaprogram to generate the additive mirror of each marked multiplicative declaration. Formally, $d_1 \sim_{\mathrm{add}} d_2$ when $\texttt{to\_additive} \in \alpha(d_1)$ and $d_2$ is the generated counterpart. The \emph{flattening ratio} $\phi = |\{d : \texttt{to\_additive} \in \alpha(d)\}| / |\mathcal{D}|$ quantifies the fraction of the library that exists as such mirrored pairs.
\end{itemize}
\end{definition}

\noindent At our snapshot, $32.9\%$ of declarations carry at least one attribute. The most prevalent are \texttt{@[simp]} ($40{,}223$ declarations, $17.1\%$), \texttt{@[to\_additive]} ($13{,}664$, $5.8\%$), and \texttt{@[simps]} ($6{,}830$, $2.9\%$). The flattening ratio is $\phi = 13{,}664 / 235{,}586 = 5.8\%$; including the generated mirrors, \texttt{@[to\_additive]} accounts for roughly $11.6\%$ of the library. A total of $105$ distinct attribute types are registered.

Lean's kernel decides definitional equality by unfolding chains of definitions; the depth of this chain (the \emph{definitional height}) governs compilation cost and constitutes a form of implicit coupling invisible in $G_{\mathrm{thm}}$.

\begin{definition}[Definitional height]\label{def:def-height}
For each definition $d \in \mathcal{D}$ with \texttt{DefinitionVal}, Lean computes a \emph{definitional height} $\delta(d)$ stored in the \texttt{ReducibilityHints} field:
\[
  \delta(d) =
  \begin{cases}
    h \in \mathbb{N} & \text{if } d \text{ is \emph{regular} (semireducible), where } h \text{ counts unfolding steps},\\
    \bot_{\mathrm{abbrev}} & \text{if } d \text{ is an \emph{abbreviation} (always unfolded)},\\
    \bot_{\mathrm{opaque}} & \text{if } d \text{ is \emph{opaque} (never unfolded)}.
  \end{cases}
\]
Declarations with high~$\delta$ are brittle: changes to any definition along the unfolding chain can break type-checking even if the explicit premise set is unchanged. This quantity is invisible in $G_{\mathrm{thm}}$ (it does not appear in $\mathrm{refs}(d.\pi)$), yet it constitutes a form of \emph{implicit coupling}.
\end{definition}

\begin{figure}[!htb]
\centering
\begin{subfigure}[t]{0.48\textwidth}
\centering
\vspace{0pt}
\raggedright\scriptsize\ttfamily
\textsf{\bfseries\color{blue!60!black}Init.Data.Nat.Basic}\\[2pt]
\textbf{def} \textcolor{red!70!black}{Nat.add} : Nat \textnormal{$\to$} Nat \textnormal{$\to$} Nat\\
\hspace{1em}| n, 0 => n\\
\hspace{1em}| n, succ m => succ (Nat.add n m)\\[4pt]
\textbf{def} \textcolor{red!70!black}{Nat.mul} : Nat \textnormal{$\to$} Nat \textnormal{$\to$} Nat\\
\hspace{1em}| \_, 0 => 0\\
\hspace{1em}| n, succ m => \textcolor{red!70!black}{Nat.add} (\textcolor{red!70!black}{Nat.mul} n m) n\\[4pt]
\textbf{def} \textcolor{red!70!black}{Nat.pow} : Nat \textnormal{$\to$} Nat \textnormal{$\to$} Nat\\
\hspace{1em}| \_, 0 => 1\\
\hspace{1em}| n, succ m => \textcolor{red!70!black}{Nat.mul} n (\textcolor{red!70!black}{Nat.pow} n m)
\caption{Source: each definition unfolds through the previous one, increasing~$\delta$.}
\label{fig:height-source}
\end{subfigure}%
\hfill
\begin{subfigure}[t]{0.48\textwidth}
\centering
\vspace{0pt}
\begin{tikzpicture}[
  every node/.style={font=\scriptsize, inner sep=2pt},
  def/.style={rounded corners=2pt, draw=red!60!black, fill=red!6,
    font=\footnotesize\ttfamily, text=red!70!black, inner sep=3pt, minimum height=14pt},
  unfold/.style={->, >=Stealth, red!60!black, semithick, dashed},
  ht/.style={font=\footnotesize\sffamily\bfseries, text=blue!60!black},
]
  \node[def] (pow) at (2.0, 3.6) {Nat.pow};
  \node[ht, right=3pt of pow] {$\delta = 3$};
  \node[def] (mul) at (2.0, 2.0) {Nat.mul};
  \node[ht, right=3pt of mul] {$\delta = 2$};
  \node[def] (add) at (2.0, 0.4) {Nat.add};
  \node[ht, right=3pt of add] {$\delta = 1$};
  \draw[unfold] (pow) -- node[left, font=\footnotesize\sffamily, text=gray!60] {unfolds} (mul);
  \draw[unfold] (mul) -- node[left, font=\footnotesize\sffamily, text=gray!60] {unfolds} (add);
  \node[font=\footnotesize\sffamily, text=gray!50, anchor=north west] at (-0.2, -0.3)
    {Higher $\delta$ = longer unfolding chain};
\end{tikzpicture}
\caption{Unfolding chain: $\delta$ increases with each layer.}
\label{fig:height-graph}
\end{subfigure}
\caption{Definitional height (Definition~\ref{def:def-height}). (a)~From \module{Init.Data.Nat.Basic}: \decl{Nat.pow} calls \decl{Nat.mul}, which calls \decl{Nat.add}. (b)~The unfolding chain: to type-check an expression involving \decl{Nat.pow}, the kernel may need to unfold through \decl{Nat.mul} and \decl{Nat.add}, yielding $\delta = 3$. Changes to any definition in the chain can break type-checking even if the explicit premises of \decl{Nat.pow} are unchanged.}
\label{fig:def-height}
\end{figure}

\noindent Of the $101{,}021$ definitions with reducibility hints, $53{,}962$ are regular (semireducible), $42{,}786$ are abbreviations (always unfolded), and $4{,}273$ are opaque (never unfolded). Among regular definitions, $\delta$ has median~$7$, mean~$10.1$, and maximum~$60$, with $90\%$ of values below~$23$. The distribution is concentrated at low depths ($27.8\%$ in the range $5$--$9$) but has a long tail: the ten highest-$\delta$ declarations all belong to \ns{MeasureTheory} (conditional expectation, Bochner integral, $L^p$ spaces), reflecting deep unfolding chains through the measure-theoretic hierarchy.

\begin{definition}[Tactic usage profile]\label{def:tactic-usage}
For each declaration $d \in \mathcal{D}$ whose proof is written in tactic mode (i.e., $d.\pi$ begins with \texttt{by}), define the \emph{tactic profile}
\[
  \tau(d) = (t_1, t_2, \ldots, t_k),
\]
where each $t_i \in \mathcal{T}_{\mathrm{tac}}$ is a tactic name invoked in the proof script, preserving multiplicity and order. The aggregate tactic frequency $\mathrm{freq}(t) = |\{(d, i) : t_i = t\}|$ over all tactic-mode proofs measures the prevalence of each proof strategy across the library. Per-namespace distributions enable Jensen--Shannon divergence analysis of proof methodology variation across mathematical domains.
\end{definition}

\begin{figure}[!htb]
\centering
\begin{subfigure}[t]{0.50\textwidth}
\centering
\vspace{0pt}
\raggedright\scriptsize\ttfamily
\textsf{\bfseries\color{blue!60!black}Algebra/Group/Defs.lean, line 1056}\\[2pt]
@[to\_additive]\\
\textbf{theorem} \textcolor{red!70!black}{mul\_div\_assoc}\\
\hspace{1em}(a b c : G) : a * b / c = a * (b / c) := \textbf{by}\\
\hspace{1em}\textcolor{violet}{rw} [\textcolor{red!70!black}{div\_eq\_mul\_inv},\\
\hspace{3em}\textcolor{red!70!black}{div\_eq\_mul\_inv},\\
\hspace{3em}\textcolor{red!70!black}{mul\_assoc} \_ \_ \_]\\[6pt]
\textnormal{\sffamily Extracted tactic profile:}\\[2pt]
\hspace{1em}\textnormal{$\tau(\text{mul\_div\_assoc}) = ($}\textcolor{violet}{rw}\textnormal{$)$}\\[2pt]
\textnormal{\sffamily Premises from proof:}\\[2pt]
\hspace{1em}\textnormal{\sffamily\scriptsize\textcolor{gray!60}{%
div\_eq\_mul\_inv (2$\times$), mul\_assoc (1$\times$)}}
\caption{Source: a tactic-mode proof. The profile $\tau(d)$ records tactic names.}
\label{fig:tactic-source}
\end{subfigure}%
\hfill
\begin{subfigure}[t]{0.46\textwidth}
\centering
\vspace{0pt}
\begin{tikzpicture}[
  every node/.style={font=\scriptsize, inner sep=2pt},
  dot/.style={circle, fill=red!70!black, minimum size=3.5pt, inner sep=0pt},
  tac/.style={rounded corners=2pt, draw=violet!60, fill=violet!8,
    font=\footnotesize\ttfamily, text=violet!70!black, inner sep=3pt, minimum height=12pt},
  dep/.style={->, >=Stealth, red!70!black, semithick},
]
  \node[dot, label={[font=\footnotesize\ttfamily, red!70!black]left:mul\_div\_assoc}]
    (thm) at (2.2, 3.2) {};
  \node[tac, anchor=west] at (2.5, 3.2) {rw};
  \node[dot, label={[font=\footnotesize\ttfamily, red!70!black]below:div\_eq\_mul\_inv}]
    (dm) at (1.0, 1.2) {};
  \node[dot, label={[font=\footnotesize\ttfamily, red!70!black]below:mul\_assoc}]
    (ma) at (3.8, 1.2) {};
  \draw[dep] (thm) -- (dm);
  \draw[dep] (thm) -- (ma);
  \node[font=\footnotesize\sffamily, text=gray!60, anchor=north west] at (0.2, -0.1)
    {Top tactics:};
  \fill[violet!30] (0.2, -0.65) rectangle (2.9, -0.45);
  \node[font=\footnotesize\sffamily, text=violet!70!black, anchor=west] at (0.3, -0.55)
    {\texttt{rw}\hspace{4pt}16.9\%};
  \fill[violet!25] (0.2, -1.0) rectangle (2.4, -0.8);
  \node[font=\footnotesize\sffamily, text=violet!70!black, anchor=west] at (0.3, -0.9)
    {\texttt{simp}\hspace{4pt}11.2\%};
  \fill[violet!20] (0.2, -1.35) rectangle (2.3, -1.15);
  \node[font=\footnotesize\sffamily, text=violet!70!black, anchor=west] at (0.3, -1.25)
    {\texttt{exact}\hspace{4pt}10.4\%};
\end{tikzpicture}
\caption{$G_{\mathrm{thm}}$ with \textcolor{violet!70!black}{tactic tags}; bottom: aggregate $\mathrm{freq}(t)$.}
\label{fig:tactic-graph}
\end{subfigure}
\caption{Tactic usage profile (Definition~\ref{def:tactic-usage}). (a)~From \module{Mathlib.Algebra.Group.Defs} (line 1056): \decl{mul\_div\_assoc} is proved by a single \textcolor{violet!70!black}{\texttt{rw}} invocation with three lemma arguments. The tactic profile $\tau(d)$ captures which tactics are used. (b)~The declaration node annotated with its tactic tag; the bar chart shows library-wide tactic frequencies over $79{,}910$ tactic-mode proofs (extracted via \texttt{jixia}~\cite{jixia2024}).}
\label{fig:tactic-usage}
\end{figure}

\noindent Across $79{,}910$ tactic-mode proofs ($325{,}454$ tactic invocations, $7{,}053$ distinct tactics), the top three tactics (\texttt{rw} ($16.9\%$), \texttt{simp} ($11.2\%$), \texttt{exact} ($10.4\%$)) account for $38.5\%$ of all invocations. Per-namespace profiles reveal systematic variation: \texttt{Algebra} proofs favor \texttt{rw} ($20\%$) while \texttt{CategoryTheory} proofs favor \texttt{simp} ($13\%$); \texttt{MeasureTheory} has the highest \texttt{exact} rate ($13\%$) and \texttt{have} rate ($10\%$), reflecting a more lemma-chaining proof style.

Lean's \texttt{deriving} mechanism automates the generation of meta-level typeclass instances (e.g., \texttt{DecidableEq}, \texttt{Repr}) that traditional mathematics takes for granted, creating declarations and dependency edges required for computational infrastructure.

\begin{definition}[Auto-derived instance edges]\label{def:deriving-edges}
Lean's \texttt{deriving} mechanism automatically generates typeclass instances (e.g., \texttt{Repr}, \texttt{DecidableEq}, \texttt{BEq}) for inductive types. Let $\mathcal{H} \subset \mathcal{A}$ be the set of registered \texttt{deriving} handlers. Each handler $h \in \mathcal{H}$ applied to a type $T$ produces a set of auto-generated declarations $\mathrm{derived}(h, T) \subset \mathcal{D}$, contributing edges in $G_{\mathrm{thm}}$ that were never written by a human. The set $E_{\mathrm{auto}} = \bigcup_{h, T} \{(d', d) : d \in \mathrm{derived}(h, T),\, d' \in \mathrm{refs}(d.\pi)\}$ captures the purely tool-generated fragment of the dependency graph.
\end{definition}

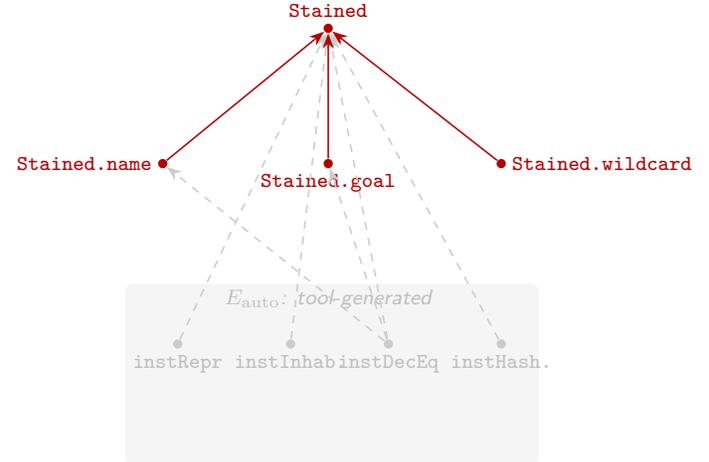
\begin{figure}[!htb]
\centering
\begin{subfigure}[t]{0.50\textwidth}
\centering
\vspace{0pt}
\raggedright\scriptsize\ttfamily
\textsf{\bfseries\color{blue!60!black}Tactic/Linter/FlexibleLinter.lean, line 190}\\[2pt]
\textbf{inductive} \textcolor{red!70!black}{Stained}\\
\hspace{1em}| name \hspace{4pt}: Name \textnormal{$\to$} Stained\\
\hspace{1em}| goal \hspace{4pt}: Stained\\
\hspace{1em}| wildcard : Stained\\
\hspace{1em}\textcolor{teal!70!black}{\textbf{deriving} Repr, Inhabited,}\\
\hspace{1em}\textcolor{teal!70!black}{\phantom{\textbf{deriving}} DecidableEq, Hashable}\\[6pt]
\textnormal{\sffamily Auto-generated declarations:}\\[2pt]
\hspace{1em}\textcolor{gray!50}{instReprStained}%
  \textnormal{\sffamily\scriptsize\textcolor{gray!60}{\hspace{4pt}--- by \texttt{Repr} handler}}\\
\hspace{1em}\textcolor{gray!50}{instInhabitedStained}%
  \textnormal{\sffamily\scriptsize\textcolor{gray!60}{\hspace{4pt}--- by \texttt{Inhabited} handler}}\\
\hspace{1em}\textcolor{gray!50}{instDecidableEqStained}%
  \textnormal{\sffamily\scriptsize\textcolor{gray!60}{\hspace{4pt}--- by \texttt{DecEq} handler}}\\
\hspace{1em}\textcolor{gray!50}{instHashableStained}%
  \textnormal{\sffamily\scriptsize\textcolor{gray!60}{\hspace{4pt}--- by \texttt{Hashable} handler}}
\caption{Source: \textcolor{teal!70!black}{\texttt{deriving}} auto-generates 4 instances.}
\label{fig:auto-source}
\end{subfigure}%
\hfill
\begin{subfigure}[t]{0.46\textwidth}
\centering
\vspace{0pt}
\begin{tikzpicture}[
  every node/.style={font=\scriptsize, inner sep=2pt},
  human/.style={circle, fill=red!70!black, minimum size=3.5pt, inner sep=0pt},
  auto/.style={circle, fill=gray!40, minimum size=3.5pt, inner sep=0pt},
  humanedge/.style={->, >=Stealth, red!70!black, semithick},
  autoedge/.style={->, >=Stealth, gray!40, semithick, dashed},
]
  \fill[gray!8, rounded corners=3pt] (-0.5, -1.8) rectangle (5.0, 0.6);
  \node[font=\footnotesize\sffamily\itshape, gray!50] at (2.2, 0.4) {$E_{\mathrm{auto}}$: tool-generated};
  \node[auto, label={[font=\footnotesize\ttfamily, gray!50]below:instRepr}]
    (r) at (0.2, -0.2) {};
  \node[auto, label={[font=\footnotesize\ttfamily, gray!50]below:instInhab.}]
    (i) at (1.7, -0.2) {};
  \node[auto, label={[font=\footnotesize\ttfamily, gray!50]below:instDecEq}]
    (d) at (3.0, -0.2) {};
  \node[auto, label={[font=\footnotesize\ttfamily, gray!50]below:instHash.}]
    (h) at (4.5, -0.2) {};
  \node[human, label={[font=\footnotesize\ttfamily, red!70!black]above:Stained}]
    (st) at (2.2, 4.0) {};
  \node[human, label={[font=\footnotesize\ttfamily, red!70!black]left:Stained.name}]
    (sn) at (0.0, 2.2) {};
  \node[human, label={[font=\footnotesize\ttfamily, red!70!black]below:Stained.goal}]
    (sg) at (2.2, 2.2) {};
  \node[human, label={[font=\footnotesize\ttfamily, red!70!black]right:Stained.wildcard}]
    (sw) at (4.5, 2.2) {};
  \draw[autoedge] (r) -- (st);
  \draw[autoedge] (i) -- (st);
  \draw[autoedge] (d) -- (st);
  \draw[autoedge] (h) -- (st);
  \draw[autoedge] (d) -- (sn);
  \draw[autoedge] (d) -- (sg);
  \draw[humanedge] (sn) -- (st);
  \draw[humanedge] (sg) -- (st);
  \draw[humanedge] (sw) -- (st);
\end{tikzpicture}
\caption{\textcolor{red!70!black}{$\bullet$}\,human, \textcolor{gray!40}{$\bullet$}\,auto: $E_{\mathrm{auto}}$ edges are \textcolor{gray!40}{dashed}.}
\label{fig:auto-graph}
\end{subfigure}
\caption{Auto-derived instance edges (Definition~\ref{def:deriving-edges}). (a)~From \module{Mathlib.Tactic.Linter.FlexibleLinter} (line 190): the \texttt{deriving} clause on \decl{Stained} instructs Lean to auto-generate four typeclass instances. (b)~In $G_{\mathrm{thm}}$, \textcolor{red!70!black}{solid} edges are human-written (constructors $\to$ type); \textcolor{gray!40}{dashed} edges belong to $E_{\mathrm{auto}}$: the auto-generated instances reference both the type and its constructors, but no human ever wrote these dependencies.}
\label{fig:auto-derived}
\end{figure}

\noindent At our snapshot, $|\mathcal{H}| = 15$ registered \texttt{deriving} handlers produce $1{,}515$ auto-derived declarations and $34{,}090$ edges in $G_{\mathrm{thm}}$ ($0.40\%$ of $|E_{\mathrm{thm}}|$). While this fraction is small, these declarations are structurally distinctive: they have uniform dependency patterns (each references the parent type and its constructors) and inflate the graph's surface without contributing mathematical content.

\begin{figure}[!htb]
\centering
\begin{subfigure}[t]{0.42\textwidth}
\centering
\vspace{0pt}
\raggedright\footnotesize\ttfamily
@[simp] \textbf{theorem} \textcolor{red!70!black}{mul\_one}\\
\hspace{1em}[Monoid $\alpha$] (a : $\alpha$) : a * 1 = a\\[5pt]
@[to\_additive] \textbf{theorem} \textcolor{red!70!black}{mul\_comm}\\
\hspace{1em}[CommMonoid $\alpha$] (a b : $\alpha$) : a * b = b * a\\[3pt]
\hspace{1em}\textnormal{\sffamily\scriptsize\textcolor{gray!50}{%
generates \decl{add\_comm} automatically}}\\[5pt]
@[simp, ext] \textbf{theorem} \textcolor{red!70!black}{Prod.mk.eta}\\
\hspace{1em}: (p.1, p.2) = p\\[8pt]
\textnormal{\sffamily Extracted metadata per node:}\\[2pt]
\hspace{1em}\textnormal{\sffamily\scriptsize%
$\alpha(d)$: attribute set \{simp, ext, \ldots\}}\\
\hspace{1em}\textnormal{\sffamily\scriptsize%
$\kappa(d)$: proof term size (Expr node count)}\\
\hspace{1em}\textnormal{\sffamily\scriptsize%
$\upsilon(d)$: universe polymorphism degree}
\caption{Declarations with Lean attributes and the metadata extracted from each.}
\label{fig:attr-source}
\end{subfigure}%
\hfill
\begin{subfigure}[t]{0.55\textwidth}
\centering
\vspace{0pt}
\begin{tikzpicture}[
  every node/.style={font=\scriptsize, inner sep=2pt},
  dot/.style={circle, fill=red!70!black, minimum size=4pt, inner sep=0pt},
  depedge/.style={->, >=Stealth, gray!50, semithick},
  tag/.style={rounded corners=1.5pt, font=\footnotesize\sffamily, inner sep=1.5pt},
  simptag/.style={tag, fill=teal!15, text=teal!70!black, draw=teal!40},
  exttag/.style={tag, fill=violet!15, text=violet!70!black, draw=violet!40},
  toaddtag/.style={tag, fill=orange!15, text=orange!70!black, draw=orange!40},
  mirror/.style={<->, >=Stealth, orange!60!black, semithick, dashed},
  meta/.style={font=\footnotesize\sffamily, text=gray!60},
]
  \node[dot, label={[font=\footnotesize\ttfamily, red!70!black]left:mul\_one}]
    (mo) at (0, 4.0) {};
  \node[simptag, anchor=west] at (0.25, 4.0) {simp};
  \node[meta, anchor=west] at (1.1, 4.0) {$\kappa=12$\enspace$\upsilon=1$};
  \node[dot, label={[font=\footnotesize\ttfamily, red!70!black]left:mul\_comm}]
    (mc) at (0, 2.0) {};
  \node[toaddtag, anchor=west] at (0.25, 1.45) {to\_additive};
  \node[meta, anchor=west] at (1.8, 1.45) {$\kappa=34$\enspace$\upsilon=1$};
  \node[dot, fill=orange!50!red, label={[font=\footnotesize\ttfamily, orange!60!black]right:add\_comm}]
    (ac) at (5.0, 2.0) {};
  \node[meta, anchor=west] at (5.25, 1.45) {$\kappa=34$\enspace$\upsilon=1$};
  \node[dot, label={[font=\footnotesize\ttfamily, red!70!black]left:Prod.mk.eta}]
    (pe) at (0, 0) {};
  \node[simptag, anchor=west] at (0.25, 0) {simp};
  \node[exttag, anchor=west] at (1.05, 0) {ext};
  \node[meta, anchor=west] at (1.75, 0) {$\kappa=8$\enspace$\upsilon=2$};
  \draw[mirror] (mc) -- node[above, font=\footnotesize\sffamily, text=orange!60!black] {$\sim_{\mathrm{add}}$} (ac);
  \draw[depedge] (mo) -- (mc);
  \draw[depedge] (mc) -- (pe);
\end{tikzpicture}
\caption{Nodes of $G_{\mathrm{thm}}$ enriched with attribute tags, proof complexity~$\kappa$, and universe degree~$\upsilon$. The \textcolor{orange!60!black}{dashed} arrow marks a \texttt{to\_additive} mirror pair.}
\label{fig:attr-graph}
\end{subfigure}
\caption{Declaration metadata as node attributes (Definition~\ref{def:decl-attributes}). (a)~Lean source with attribute annotations. (b)~The declaration graph enriched with per-node metadata: colored tags indicate attributes ($\textcolor{teal!70!black}{\texttt{simp}}$, $\textcolor{violet!70!black}{\texttt{ext}}$, $\textcolor{orange!70!black}{\texttt{to\_additive}}$); $\kappa$ and $\upsilon$ are numeric features. The \textcolor{orange!60!black}{dashed} bidirectional arrow marks the automatically generated additive mirror \decl{add\_comm} $\sim_{\mathrm{add}}$ \decl{mul\_comm}.}
\label{fig:decl-attributes}
\end{figure}
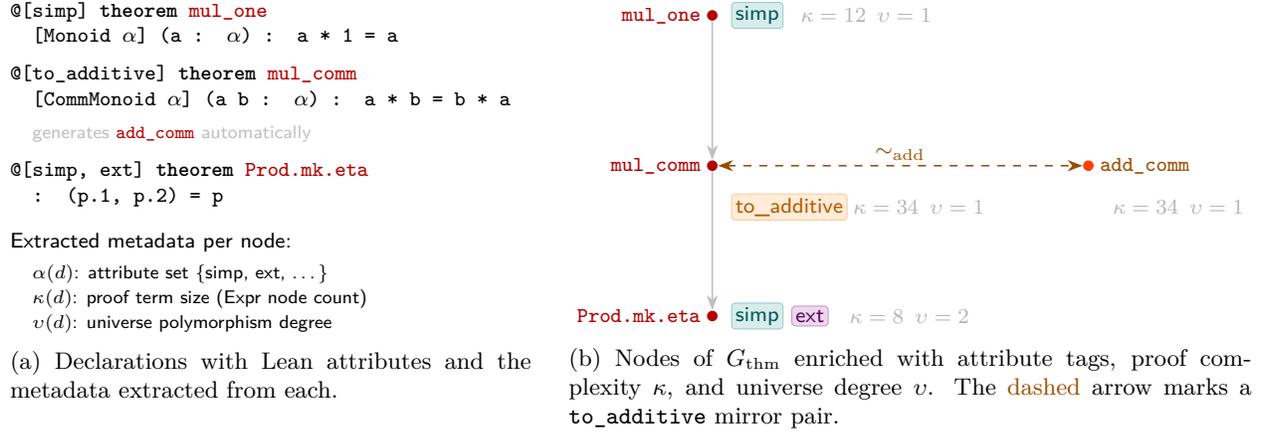

\subsubsection{Import utilization: a cross-level metric}
\label{sec:import-util}
With the declaration graph in hand, we can now define a metric that bridges the module and declaration levels, measuring how efficiently each module consumes its imports.

\begin{definition}[Import utilization]\label{def:import-util}
For each import edge $(m_i, m_j) \in E_{\mathrm{module}}$, the \emph{import utilization} is
\[
  \mathrm{util}(m_i, m_j) = \frac{|\mathrm{refs}(m_i) \cap \mathcal{D}_{m_j}|}{|\mathcal{D}_{m_j}|},
\]
where $\mathcal{D}_{m_j} \subset \mathcal{D}$ is the set of declarations defined in module~$m_j$ and $\mathrm{refs}(m_i) = \bigcup_{d \in \mathcal{D}_{m_i}} \mathrm{refs}(d.\pi)$ is the set of all premises invoked by declarations in~$m_i$. A low utilization indicates that $m_i$ imports $m_j$ for a small fraction of its exports, suggesting the module boundary is too coarse.
\end{definition}

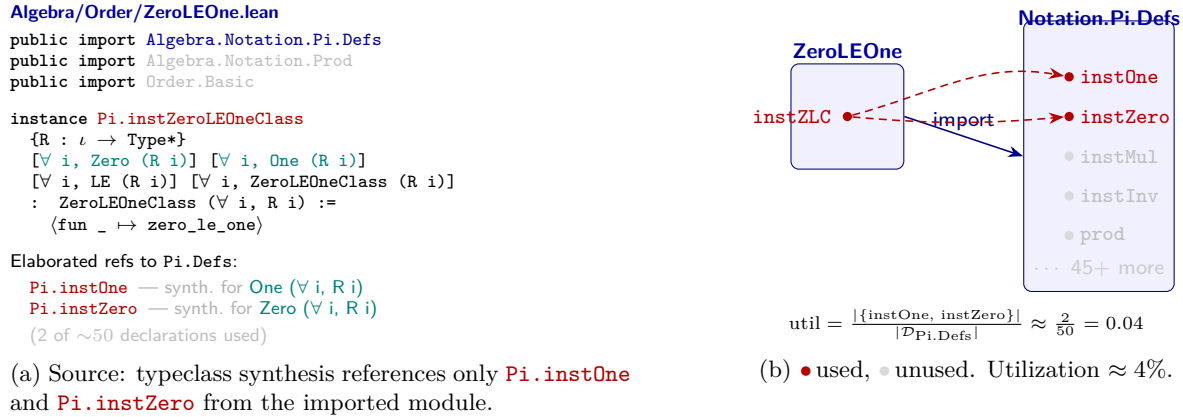
\begin{figure}[!htb]
\centering
\begin{subfigure}[t]{0.50\textwidth}
\centering
\vspace{0pt}
\raggedright\scriptsize\ttfamily
\textsf{\bfseries\color{blue!60!black}Algebra/Order/ZeroLEOne.lean}\\[2pt]
\textbf{public import} \textcolor{blue!60!black}{Algebra.Notation.Pi.Defs}\\
\textbf{public import} \textcolor{gray!50}{Algebra.Notation.Prod}\\
\textbf{public import} \textcolor{gray!50}{Order.Basic}\\[6pt]
\textbf{instance} \textcolor{red!70!black}{Pi.instZeroLEOneClass}\\
\hspace{1em}\{R : $\iota$ \textnormal{$\to$} Type*\}\\
\hspace{1em}[{\textcolor{teal}{\textnormal{$\forall$} i, Zero (R i)}}]
  [{\textcolor{teal}{\textnormal{$\forall$} i, One (R i)}}]\\
\hspace{1em}[\textnormal{$\forall$} i, LE (R i)]
  [\textnormal{$\forall$} i, ZeroLEOneClass (R i)]\\
\hspace{1em}: ZeroLEOneClass (\textnormal{$\forall$} i, R i) :=\\
\hspace{2em}\textnormal{$\langle$}fun \_ \textnormal{$\mapsto$} zero\_le\_one\textnormal{$\rangle$}\\[6pt]
\textnormal{\sffamily Elaborated refs to \texttt{Pi.Defs}:}\\[2pt]
\hspace{1em}\textcolor{red!70!black}{Pi.instOne}%
  \textnormal{\sffamily\scriptsize\textcolor{gray!60}{\hspace{4pt}--- synth.\ for \textcolor{teal}{One ($\forall$ i, R i)}}}\\
\hspace{1em}\textcolor{red!70!black}{Pi.instZero}%
  \textnormal{\sffamily\scriptsize\textcolor{gray!60}{\hspace{4pt}--- synth.\ for \textcolor{teal}{Zero ($\forall$ i, R i)}}}\\[2pt]
\hspace{1em}\textnormal{\sffamily\scriptsize\textcolor{gray!50}{(2 of ${\sim}50$ declarations used)}}
\caption{Source: typeclass synthesis references only \decl{Pi.instOne} and \decl{Pi.instZero} from the imported module.}
\label{fig:util-source}
\end{subfigure}%
\hfill
\begin{subfigure}[t]{0.46\textwidth}
\centering
\vspace{0pt}
\begin{tikzpicture}[scale=0.88, every node/.append style={transform shape},
  every node/.style={font=\scriptsize, inner sep=2pt},
  mod/.style={rounded corners=3pt, draw=blue!50!black, fill=blue!6,
    font=\footnotesize\sffamily\bfseries, text=blue!60!black},
  used/.style={circle, fill=red!70!black, minimum size=3.5pt, inner sep=0pt},
  unused/.style={circle, fill=gray!30, minimum size=3.5pt, inner sep=0pt},
  imp/.style={->, >=Stealth, blue!50!black, semithick},
  ref/.style={->, >=Stealth, red!70!black, semithick, densely dashed},
]
  \node[mod, minimum height=1.4cm, minimum width=1.5cm] (mi) at (0, 1.0) {};
  \node[mod, draw=none, fill=none] at (0, 2.0) {ZeroLEOne};
  \node[used, label={[font=\footnotesize\ttfamily, red!70!black, xshift=-2pt]left:instZLC}]
    (zlc) at (0, 1.0) {};

  \node[mod, minimum height=3.6cm, minimum width=2.0cm] (mj) at (3.8, 0.4) {};
  \node[mod, draw=none, fill=none] at (3.8, 2.5) {Notation.Pi.Defs};
  \node[used, label={[font=\footnotesize\ttfamily, red!70!black]right:instOne}]
    (io) at (3.35, 1.6) {};
  \node[used, label={[font=\footnotesize\ttfamily, red!70!black]right:instZero}]
    (iz) at (3.35, 1.0) {};
  \node[unused, label={[font=\footnotesize\ttfamily, gray!50]right:instMul}]
    at (3.35, 0.4) {};
  \node[unused, label={[font=\footnotesize\ttfamily, gray!50]right:instInv}]
    at (3.35, -0.2) {};
  \node[unused, label={[font=\footnotesize\ttfamily, gray!50]right:prod}]
    at (3.35, -0.8) {};
  \node[font=\footnotesize\sffamily, gray!40] at (3.8, -1.3) {$\cdots$ 45+ more};

  \draw[imp] (mi.east) -- node[above, font=\footnotesize\sffamily, text=blue!50!black] {import} (mj.west);
  \draw[ref] (zlc) to[out=15, in=170] (io);
  \draw[ref] (zlc) to[out=-5, in=185] (iz);
\end{tikzpicture}

\vspace{2pt}
{\scriptsize $\mathrm{util} = \frac{|\{\text{instOne, instZero}\}|}{|\mathcal{D}_{\text{Pi.Defs}}|} \approx \frac{2}{50} = 0.04$}
\caption{\textcolor{red!70!black}{$\bullet$}\,used, \textcolor{gray!30}{$\bullet$}\,unused. Utilization $\approx 4\%$.}
\label{fig:util-graph}
\end{subfigure}
\caption{Import utilization (Definition~\ref{def:import-util}). (a)~\module{Mathlib.Algebra.Order.ZeroLEOne} (line 43) imports \module{Mathlib.Algebra.Notation.Pi.Defs}, which defines ${\sim}50$ declarations (pointwise instances for \texttt{One}, \texttt{Mul}, \texttt{Inv}, \texttt{Div}, etc.\ and their \texttt{to\_additive} mirrors). Typeclass synthesis resolves only \decl{Pi.instOne} and \decl{Pi.instZero} (\textcolor{red!70!black}{solid}); the remaining ${\sim}48$ declarations (\textcolor{gray!30}{gray}) are unused. (b)~The utilization is $2/50 = 0.04$, below the median of $1.6\%$ across all import edges. All names have the \module{Mathlib.}\ prefix removed.}
\label{fig:import-util}
\end{figure}

\subsection{The Namespace Graph: Definitions and Examples}
\label{sec:namespace-graph}

This appendix collects the formal definitions, worked examples, and illustrative figures for the namespace graphs. Extended statistical analysis (degree distributions, centrality, community detection, robustness curves) appears in Appendix~\ref{app:ns-detail}.

\subsubsection{Namespaces versus modules}
Each declaration carries a fully qualified name (e.g., \decl{Nat.Prime.dvd\_mul}) whose dot-separated components form a hierarchy determined by \texttt{namespace} blocks in the source code, not by the file path. The following example illustrates how Lean's naming mechanisms determine fully qualified names:

\begin{codebox}
\textit{\color{gray}-- \textbf{Namespace blocks} determine the declaration's qualified name:}\\
\textbf{namespace} \textcolor{green!50!black}{Nat}\\
\hspace{1.5em}\textbf{theorem} \textcolor{red!70!black}{add\_comm} (a b : Nat) : a + b = b + a := \ldots\\
\hspace{1.5em}\textit{\color{gray}-- fully qualified name: \textcolor{red!70!black}{Nat.add\_comm} \quad(depth 1)}\\[3pt]
\hspace{1.5em}\textbf{namespace} \textcolor{green!50!black}{Prime}\\
\hspace{3em}\textbf{theorem} \textcolor{red!70!black}{dvd\_mul} (hp : Prime p) : p | a * b $\to$ \ldots\ := \ldots\\
\hspace{3em}\textit{\color{gray}-- fully qualified name: \textcolor{red!70!black}{Nat.Prime.dvd\_mul} \quad(depth 2)}\\
\hspace{1.5em}\textbf{end} \textcolor{green!50!black}{Prime}\\
\textbf{end} \textcolor{green!50!black}{Nat}\\[6pt]
\textit{\color{gray}-- \textbf{File path} determines the module name (a separate hierarchy):}\\
\textit{\color{gray}-- File:}\quad \textcolor{blue!60!black}{Mathlib/Data/Nat/Prime/Basic.lean}\\
\textit{\color{gray}-- Module:}\quad \textcolor{blue!60!black}{Mathlib.Data.Nat.Prime.Basic}
\end{codebox}

\noindent
Each declaration's fully qualified name is a dot-separated sequence determined by its enclosing \texttt{namespace} blocks (\textcolor{green!50!black}{green}), not by the file path. The file path determines the \emph{module} name (\textcolor{blue!60!black}{blue}), which governs \texttt{import} visibility but does not affect qualified names. In practice, \module{Mathlib} conventions strongly correlate the two hierarchies (declarations in \ns{Nat.Prime} typically live in modules under \module{Mathlib.Data.Nat.Prime}), but the correlation is a social convention, not a language rule. A single namespace may span many files, and a single file may contribute declarations to multiple namespaces (\S\ref{sec:ns-mod-cross}).

Truncating names to varying depths yields an intermediate family of graphs between the coarse module level and the fine declaration level.

The $308{,}129$ declarations in \module{Mathlib} span six depth levels. Table~\ref{tab:ns-depth-count} shows how the number of distinct namespaces grows with truncation depth.
\begin{table}[ht]
\centering
\caption{Number of distinct namespaces at each truncation depth~$k$.}
\label{tab:ns-depth-count}
\begin{tabular}{cr}
\toprule
Depth $k$ & $|\mathcal{N}_k|$ \\
\midrule
1          & $3{,}184$ \\
2          & $10{,}097$ \\
3          & $13{,}785$ \\
4          & $15{,}198$ \\
5          & $15{,}443$ \\
6 (max)    & $15{,}456$ \\
\bottomrule
\end{tabular}
\end{table}

\noindent
Growth saturates rapidly: $90\%$ of all namespaces are distinguished by depth~$3$. Declarations with names of the form $X.y$ (a single namespace component plus a short name) account for $67\%$ of all declarations; only $10\%$ have depth $\ge 4$. The naming convention of \module{Mathlib} is, in practice, shallow.

\begin{definition}[Namespace graph at depth~$k$]
\label{def:ns-graph}
For each depth $k \ge 1$, let $\mathrm{ns}_k(d)$ denote the namespace of declaration~$d$ truncated to depth~$k$: the first $k$ dot-separated components of $d$'s fully qualified name. If $d$'s name has fewer than $k+1$ components, $\mathrm{ns}_k(d)$ defaults to the full parent namespace; if $d$'s name has no dot, $\mathrm{ns}_k(d) = \ns{\_root\_}$. The \emph{namespace graph at depth~$k$} is the weighted directed graph
\[
  G_{\mathrm{ns}}^{(k)} = (\mathcal{N}_k,\, E_k),
\]
where $\mathcal{N}_k = \{\mathrm{ns}_k(d) \mid d \in \mathcal{D}\}$ is the set of all distinct depth-$k$ namespaces, and each edge $(n_1, n_2) \in E_k$ carries a weight equal to the number of edges $(d_1, d_2) \in E_{\mathrm{thm}}$ such that $\mathrm{ns}_k(d_1) = n_1 \neq n_2 = \mathrm{ns}_k(d_2)$.
\end{definition}

\begin{figure}[!htb]
\centering
\begin{tikzpicture}[
  every node/.style={font=\scriptsize, inner sep=2pt},
  dot/.style={circle, fill=red!70!black, minimum size=3pt, inner sep=0pt},
  dep/.style={->, >=Stealth, red!70!black, semithick},
  depx/.style={->, >=Stealth, red!70!black, semithick, dashed},
  nsbox/.style={rounded corners=5pt, draw=green!50!black, thick, fill=green!10},
  agg/.style={->, >=Stealth, gray!50, very thick, dashed, shorten <=2pt, shorten >=2pt},
]
  \fill[blue!6, rounded corners=3pt] (-0.6,-2.5) rectangle (3.0,2.2);
  \fill[blue!6, rounded corners=3pt] (3.4,-1.0) rectangle (6.4,2.2);
  \fill[green!8, rounded corners=3pt] (-0.2,-0.4) rectangle (6.0,1.6);
  \fill[green!8, rounded corners=2pt] (-0.2,-2.1) rectangle (2.7,-0.7);
  \draw[blue!50!black, rounded corners=3pt, thick, dashed]
    (-0.6,-2.5) rectangle (3.0,2.2);
  \node[font=\scriptsize\sffamily\bfseries, blue!60!black, anchor=north west]
    at (-0.5,2.15) {\textnormal{\module{Data.Nat.Basic}}};
  \draw[blue!50!black, rounded corners=3pt, thick, dashed]
    (3.4,-1.0) rectangle (6.4,2.2);
  \node[font=\scriptsize\sffamily\bfseries, blue!60!black, anchor=north west]
    at (3.5,2.15) {\textnormal{\module{Data.Nat.Defs}}};
  \draw[green!50!black, rounded corners=3pt, thick, dashed]
    (-0.2,-0.4) rectangle (6.0,1.6);
  \node[font=\scriptsize\sffamily\bfseries, green!40!black, anchor=south west]
    at (-0.1,1.65) {\ns{Nat}};
  \draw[green!50!black, rounded corners=2pt, thick, dashed]
    (-0.2,-2.1) rectangle (2.7,-0.7);
  \node[font=\scriptsize\sffamily\bfseries, green!40!black, anchor=south west]
    at (-0.1,-0.65) {\ns{Nat.Primrec}};
  \node[dot, label={[font=\scriptsize\ttfamily, red!70!black]left:add\_comm}]
    (ac) at (2.0,1.1) {};
  \node[dot, label={[font=\scriptsize\ttfamily, red!70!black]left:add\_left\_comm}]
    (alc) at (0.5,0.1) {};
  \node[dot, label={[font=\scriptsize\ttfamily, red!70!black]right:Primrec.zero}]
    (prz) at (0.7,-1.4) {};
  \node[dot, label={[font=\scriptsize\ttfamily, red!70!black]right:succ\_eq\_one}]
    (sea) at (4.8,1.0) {};
  \node[dot, label={[font=\scriptsize\ttfamily, red!70!black]right:zero\_add}]
    (za) at (4.8,0.2) {};
  \draw[dep] (alc) -- (ac);
  \draw[dep, dashed] (sea) to[out=175,in=10] (ac);
  \draw[dep, dashed] (za) to[out=175,in=-10] (alc);
  \draw[dep, densely dotted, thick] (prz) to[out=70,in=-100] (ac);
  \draw[->, >=Stealth, blue!60!black, thick]
    (3.4,1.9) -- node[above, font=\scriptsize\sffamily, text=blue!60!black] {\texttt{import}} (3.0,1.9);
  \node[font=\small\sffamily, anchor=north] at (2.9,-2.9) {(a) $G_{\mathrm{thm}}$};
  \draw[agg] (6.7,-0.1) -- node[above, font=\scriptsize\sffamily, text=gray!60] {aggregate}
    node[below, font=\scriptsize\sffamily, text=gray!60] {depth $2$} (8.1,-0.1);
  \node[nsbox, minimum width=2.5cm, minimum height=1.3cm]
    (nat2) at (9.6,0.55) {};
  \node[font=\small\sffamily\bfseries, green!50!black] at (9.6,0.8) {\ns{Nat}};
  \node[font=\scriptsize\sffamily, text=gray] at (9.6,0.3) {4 declarations};
  \node[nsbox, minimum width=2.5cm, minimum height=1.0cm]
    (npr2) at (9.6,-1.15) {};
  \node[font=\scriptsize\sffamily\bfseries, green!50!black] at (9.6,-0.95) {\ns{Nat.Primrec}};
  \node[font=\scriptsize\sffamily, text=gray] at (9.6,-1.35) {1 declaration};
  \draw[->, >=Stealth, green!50!black, thick] (npr2) --
    node[right, font=\scriptsize, text=green!50!black] {$w\!=\!1$} (nat2);
  \node[font=\scriptsize\itshape, text=gray, align=center] at (9.6,-2.1)
    {\decl{Primrec.zero} $\to$ \decl{add\_comm}};
  \draw[green!50!black, ->, >=Stealth, thin, opacity=0.35]
    ([xshift=-4pt]nat2.north) to[out=130,in=50,looseness=6]
    ([xshift=4pt]nat2.north);
  \node[font=\scriptsize, text=green!40!black] at (9.6,1.85) {$3$ internal};
  \node[font=\small\sffamily, anchor=north] at (9.6,-2.9) {(b) $G_{\mathrm{ns}}^{(2)}$};
  \draw[agg] (11.2,-0.1) -- node[above, font=\scriptsize\sffamily, text=gray!60] {aggregate}
    node[below, font=\scriptsize\sffamily, text=gray!60] {depth $1$} (12.6,-0.1);
  \node[nsbox, minimum width=2.5cm, minimum height=2.5cm]
    (nat1) at (14.0,-0.1) {};
  \node[font=\small\sffamily\bfseries, green!50!black] at (14.0,0.5) {\ns{Nat}};
  \node[font=\scriptsize\sffamily, text=gray] at (14.0,0.1) {5 declarations};
  \node[font=\scriptsize\sffamily, text=gray] at (14.0,-0.25) {4 internal edges};
  \node[font=\scriptsize\itshape, text=red!50!black] at (14.0,-0.65) {contain.\,$=100\%$};
  \node[font=\small\sffamily, anchor=north] at (14.0,-2.9) {(c) $G_{\mathrm{ns}}^{(1)}$};
\end{tikzpicture}
\caption{Aggregation from $G_{\mathrm{thm}}$ to $G_{\mathrm{ns}}^{(k)}$. (a)~Declarations in~$G_{\mathrm{thm}}$; \textcolor{green!40!black}{green} = namespace boundaries, \textcolor{blue!60!black}{blue} = module boundaries. (b)~Depth-$2$ namespaces. (c)~Depth~$1$. All names have the \texttt{Nat.}\ prefix removed.}
\label{fig:ns-aggregation}
\label{fig:ns-aggregation-app}
\end{figure}
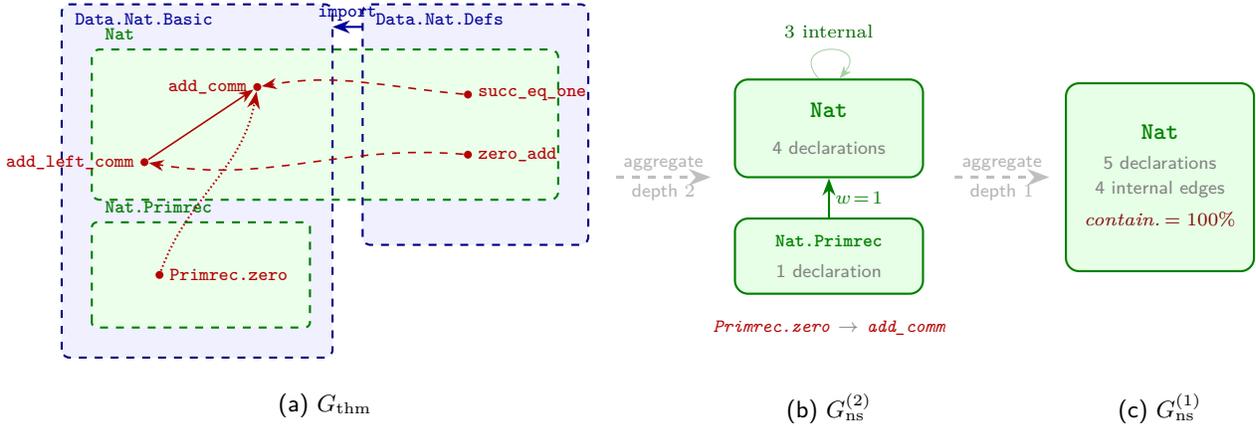

\noindent At $k=1$, the $3{,}184$ coarse namespaces correspond roughly to mathematical topics; at $k=6$, the $15{,}456$ fine-grained namespaces approach the resolution of individual naming scopes. The aggregation process is illustrated in Figure~\ref{fig:ns-aggregation}.

\noindent\textbf{Cycles in the namespace graph.}
\label{rem:ns-cycles}
Unlike $G_{\mathrm{module}}$ and $G_{\mathrm{thm}}$, which are directed acyclic graphs (the compiler and the type-checker respectively forbid circular dependencies), the namespace graph $G_{\mathrm{ns}}^{(k)}$ can contain directed cycles. Two namespaces may depend on each other: namespace~$A$ contains a declaration that cites a declaration in namespace~$B$, while $B$ contains a different declaration that cites one in~$A$. The underlying declaration-level graph remains acyclic, but the aggregation into namespaces collapses the strict topological ordering.

At depth~$2$, $G_{\mathrm{ns}}^{(2)}$ contains $38$ strongly connected components with more than one node, the largest comprising $5{,}899$ namespaces ($58\%$ of all depth-$2$ namespaces). This giant component reflects the deeply interconnected core of Mathlib: most mathematical sub-disciplines at this granularity depend mutually on one another. This structural difference (DAGs at the module and declaration levels, cyclic graphs at the namespace level) reflects the fact that namespace boundaries, unlike file boundaries, are not constrained by a compiler-enforced import ordering.

\begin{figure}[ht]
\centering
\begin{tikzpicture}[
  every node/.style={font=\scriptsize, inner sep=2pt},
  dot/.style={circle, fill=red!70!black, minimum size=3pt, inner sep=0pt},
  dep/.style={->, >=Stealth, red!70!black, semithick},
  nsbox/.style={rounded corners=5pt, draw=green!50!black, thick, fill=green!10,
                minimum width=2.8cm, minimum height=1.6cm},
  agg/.style={->, >=Stealth, gray!50, very thick, dashed,
              shorten <=2pt, shorten >=2pt},
]
  \node[font=\small\sffamily, anchor=south] at (2.0,2.5) {(a) $G_{\mathrm{thm}}$ (acyclic)};
  \fill[green!8, rounded corners=3pt] (-0.3,-0.3) rectangle (2.0,2.2);
  \draw[green!50!black, rounded corners=3pt, thick, dashed]
    (-0.3,-0.3) rectangle (2.0,2.2);
  \node[font=\scriptsize\sffamily\bfseries, green!40!black, anchor=north west]
    at (-0.2,2.15) {\ns{Nat}};

  \fill[green!8, rounded corners=3pt] (2.5,-0.3) rectangle (4.8,2.2);
  \draw[green!50!black, rounded corners=3pt, thick, dashed]
    (2.5,-0.3) rectangle (4.8,2.2);
  \node[font=\scriptsize\sffamily\bfseries, green!40!black, anchor=north west]
    at (2.6,2.15) {\ns{Int}};

  \node[dot, label={[font=\scriptsize\ttfamily, red!70!black, align=center]below:prime\_iff\_\\[-1pt]prime\_int}]
    (npip) at (1.0,1.4) {};
  \node[dot, label={[font=\scriptsize\ttfamily, red!70!black]below:lcm}]
    (nlcm) at (1.0,0.3) {};

  \node[dot, label={[font=\scriptsize\ttfamily, red!70!black]below:dvd\_natAbs}]
    (idna) at (3.5,1.4) {};
  \node[dot, label={[font=\scriptsize\ttfamily, red!70!black]below:lcm\_neg}]
    (ilcm) at (3.5,0.3) {};

  \draw[dep] (npip) -- (idna);
  \draw[dep] (ilcm) -- (nlcm);

  \draw[agg] (5.3,0.95) -- node[above, font=\scriptsize\sffamily, text=gray!60] {aggregate}
    (7.0,0.95);

  \node[font=\small\sffamily, anchor=south] at (9.5,2.5) {(b) $G_{\mathrm{ns}}^{(2)}$ (cyclic)};
  \node[nsbox, minimum width=1.6cm, minimum height=1.3cm] (ns_nat) at (8.0,0.95) {};
  \node[font=\scriptsize\sffamily\bfseries, green!50!black] at (8.0,1.1) {\ns{Nat}};
  \node[font=\scriptsize\sffamily, text=gray] at (8.0,0.7) {2 decls};

  \node[nsbox, minimum width=1.6cm, minimum height=1.3cm] (ns_int) at (11.0,0.95) {};
  \node[font=\scriptsize\sffamily\bfseries, green!50!black] at (11.0,1.1) {\ns{Int}};
  \node[font=\scriptsize\sffamily, text=gray] at (11.0,0.7) {2 decls};

  \draw[->, >=Stealth, green!50!black, thick]
    ([yshift=2pt]ns_nat.north east) to[out=30,in=150]
    node[above, font=\scriptsize, text=green!40!black] {$w\!=\!1$}
    ([yshift=2pt]ns_int.north west);
  \draw[->, >=Stealth, green!50!black, thick]
    ([yshift=-2pt]ns_int.south west) to[out=210,in=330]
    node[below, font=\scriptsize, text=green!40!black] {$w\!=\!1$}
    ([yshift=-2pt]ns_nat.south east);
\end{tikzpicture}
\caption{How aggregation creates cycles. (a)~At the declaration level, two edges cross the \ns{Nat}--\ns{Int} boundary in opposite directions: \decl{Nat.prime\_iff\_prime\_int} cites \decl{Int.dvd\_natAbs}, and \decl{Int.lcm\_neg} cites \decl{Nat.lcm}. No cycle exists among these four declarations. (b)~After aggregation to depth-$2$ namespaces, both edges collapse to namespace-level edges, creating a directed cycle \ns{Nat} $\to$ \ns{Int} $\to$ \ns{Nat}. The declaration-level DAG property is lost.}
\label{fig:ns-cycle-example}
\end{figure}
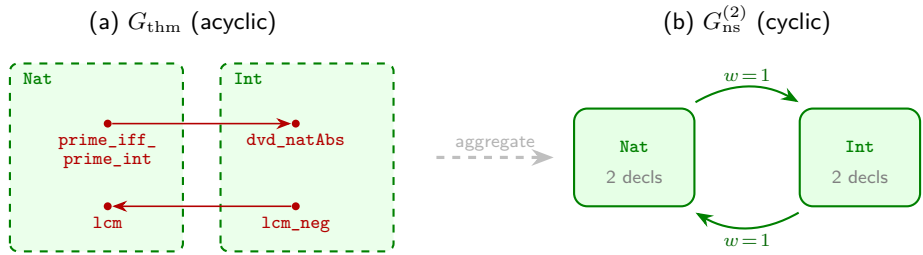

\paragraph{Containment Decay}
\label{sec:containment-decay}

We now quantify how namespace containment erodes as the grouping becomes finer.

\begin{definition}[Containment ratio]
\label{def:containment}
The \emph{containment ratio at depth~$k$} is the proportion of edges in $G_{\mathrm{thm}}$ whose endpoints share the same depth-$k$ namespace:
\[
  \mathrm{contain}(k) \;=\; \frac{|E_{\mathrm{intra}}^{(k)}|}{|E_{\mathrm{thm}}|},
  \qquad
  E_{\mathrm{intra}}^{(k)} = \{(d_1,d_2) \in E_{\mathrm{thm}} \mid \mathrm{ns}_k(d_1) = \mathrm{ns}_k(d_2)\}.
\]
\end{definition}

\noindent
This definition parallels the module containment ratio (Definition~\ref{def:module-containment}), which measures the same concept at a coarser granularity: there, the denominator is $|E_{\mathrm{module}}|$ (import edges between modules) and the grouping is by dot-separated module path; here, the denominator is $|E_{\mathrm{thm}}|$ (all declaration-level dependency edges) and the grouping is by namespace truncation. The two metrics are thus not directly comparable in magnitude, since they operate on different edge sets.

A containment ratio of $1$ would mean all dependencies stay within namespace boundaries; $0$ would mean every edge crosses a boundary. The decay is steep: from $48.9\%$ at the top level to ${\sim}14\%$ at depth~$2$, with the steepest drop at the discipline--sub-discipline transition. Full data appear in Appendix~\ref{app:ns-detail} (Table~\ref{tab:containment-decay}, Figure~\ref{fig:containment-curve}).

Degree distribution, centrality, community detection, and robustness analyses for the namespace graph are reported in Appendix~\ref{app:ns-detail}.

\subsection{Additional Graph Definitions}
\label{sec:future-graphs}

This section defines two graph constructions that require data sources beyond a single commit snapshot: the co-modification graph (\S\ref{sec:co-edit-graph}), which captures logical coupling from version control history, and the temporal graph sequence (\S\ref{sec:temporal-graphs}), which enables longitudinal structural analysis.

\subsubsection{The File Co-Modification Graph}
\label{sec:co-edit-graph}

The module graph captures \emph{structural coupling} (explicit import dependencies), but software engineering research has long recognized a complementary form of coupling: \emph{logical coupling}, measured by co-modification frequency in version control~\cite{Decan_2019}. Two modules that are frequently modified together, even without a direct import relationship, likely share a hidden dependency that the import graph does not capture.

\begin{definition}[Co-modification graph]\label{def:co-edit-graph}
Let $\{p_1, \ldots, p_R\}$ be the set of merged pull requests to \module{Mathlib} over a given time window. For each PR~$p_r$, let $\mathrm{files}(p_r) \subseteq \mathcal{M}$ be the set of modules modified. The \emph{co-modification graph} is the weighted undirected graph
\[
  G_{\mathrm{co}} = (\mathcal{M},\, E_{\mathrm{co}},\, w_{\mathrm{co}}), \quad w_{\mathrm{co}}(m_i, m_j) = |\{p_r : m_i, m_j \in \mathrm{files}(p_r)\}|.
\]
An edge exists between $m_i$ and $m_j$ if $w_{\mathrm{co}}(m_i, m_j) \ge 1$.
\end{definition}

\begin{figure}[!htb]
\centering
\begin{subfigure}[t]{0.46\textwidth}
\centering
\vspace{0pt}
\raggedright\scriptsize\ttfamily
\textsf{\bfseries\color{gray!70}PR \#18042}\textsf{\color{gray!50}\enspace``refactor Nat order lemmas''}\\[2pt]
\hspace{1em}\textcolor{blue!60!black}{Data/Nat/Defs.lean}\\
\hspace{1em}\textcolor{blue!60!black}{Data/Nat/Order.lean}\\
\hspace{1em}\textcolor{blue!60!black}{Data/Int/Defs.lean}\\[8pt]
\textsf{\bfseries\color{gray!70}PR \#18107}\textsf{\color{gray!50}\enspace``add Int.cast lemmas''}\\[2pt]
\hspace{1em}\textcolor{blue!60!black}{Data/Nat/Defs.lean}\\
\hspace{1em}\textcolor{blue!60!black}{Data/Int/Defs.lean}\\[8pt]
\textsf{\bfseries\color{gray!70}PR \#18253}\textsf{\color{gray!50}\enspace``move group lemmas''}\\[2pt]
\hspace{1em}\textcolor{blue!60!black}{Algebra/Group/Defs.lean}\\
\hspace{1em}\textcolor{blue!60!black}{Data/Nat/Order.lean}\\[8pt]
\textnormal{\sffamily Co-modification counts:}\\[2pt]
\hspace{1em}\textnormal{\sffamily\scriptsize%
Nat.Defs $\leftrightarrow$ Int.Defs: 2 PRs,\quad
Nat.Defs $\leftrightarrow$ Nat.Order: 1,}\\
\hspace{1em}\textnormal{\sffamily\scriptsize%
Int.Defs $\leftrightarrow$ Nat.Order: 1,\quad
Group.Defs $\leftrightarrow$ Nat.Order: 1}
\caption{Three PRs and the modules they touch. Pairs co-modified in multiple PRs get higher weight.}
\label{fig:co-edit-source}
\end{subfigure}%
\hfill
\begin{subfigure}[t]{0.50\textwidth}
\centering
\vspace{0pt}
\begin{tikzpicture}[
  every node/.style={font=\small, inner sep=2pt},
  mod/.style={rounded corners=2pt, draw=blue!50!black, fill=blue!6,
    font=\small\ttfamily, text=blue!60!black, inner sep=4pt, minimum height=18pt},
  coedge/.style={-, blue!50!black, semithick},
  strongedge/.style={-, purple!70!black, very thick},
  hiddenedge/.style={-, red!60!black, semithick, dashed},
]
  \node[mod] (natd) at (0, 2.8) {Data.Nat.Defs};
  \node[mod] (nato) at (4.2, 2.8) {Data.Nat.Order};
  \node[mod] (intd) at (0, 0) {Data.Int.Defs};
  \node[mod] (grpd) at (4.2, 0) {Algebra.Group.Defs};
  \draw[strongedge] (natd) -- node[left, font=\small\sffamily, text=purple!70!black] {$w=2$} (intd);
  \draw[coedge] (natd) -- node[above, font=\small\sffamily, text=blue!50!black] {$w=1$} (nato);
  \draw[coedge] (intd) -- node[below, font=\small\sffamily, text=blue!50!black, pos=0.55, yshift=-2pt, xshift=3pt] {$w=1$} (nato);
  \draw[hiddenedge] (grpd) -- node[right, font=\small\sffamily, text=red!60!black] {$w=1$} (nato);
  \node[font=\small\sffamily, anchor=north west] at (-0.5, -1.2)
    {\textcolor{blue!50!black}{\rule{8pt}{1.5pt}} in $G_{\mathrm{co}} \cap G_{\mathrm{module}}$};
  \node[font=\small\sffamily, anchor=north west] at (-0.5, -1.8)
    {\textcolor{red!60!black}{- - -} in $G_{\mathrm{co}} \setminus G_{\mathrm{module}}$ (hidden dep.)};
  \node[font=\small\sffamily, anchor=north west] at (-0.5, -2.4)
    {\textcolor{purple!70!black}{\rule{8pt}{2.5pt}} high weight ($w \ge 2$)};
\end{tikzpicture}
\caption{$G_{\mathrm{co}}$: edge weight = number of PRs co-modifying both modules. The \textcolor{red!60!black}{dashed edge} has no import in $G_{\mathrm{module}}$, indicating a hidden dependency.}
\label{fig:co-edit-graph}
\end{subfigure}
\caption{Co-modification graph (Definition~\ref{def:co-edit-graph}). (a)~Three merged PRs and the files each touches. (b)~The resulting weighted undirected graph $G_{\mathrm{co}}$: \textcolor{blue!50!black}{solid} edges exist in both $G_{\mathrm{co}}$ and $G_{\mathrm{module}}$ (structural + logical coupling); the \textcolor{red!60!black}{dashed} edge between \module{Algebra.Group.Defs} and \module{Data.Nat.Order} appears only in $G_{\mathrm{co}}$, revealing a hidden dependency invisible to the import graph. The \textcolor{purple!70!black}{thick} edge ($w=2$) indicates a stronger coupling. All names have the \module{Mathlib.}\ prefix removed.}
\label{fig:co-modification}
\end{figure}

This graph captures traces of \emph{human organizational process} from a data source entirely independent of Lean's compilation model: the Git history. Comparing $G_{\mathrm{co}}$ with $G_{\mathrm{module}}$ tests whether structural coupling predicts logical coupling. Module pairs with high co-modification weight but no direct import edge reveal hidden dependencies, candidates for refactoring or explicit import addition. Conversely, direct import edges with zero co-modification suggest stable, well-encapsulated interfaces.

\subsubsection{Temporal Graph Sequences}
\label{sec:temporal-graphs}

All analyses in this paper are computed from a single snapshot. A longitudinal extension would test whether the structural properties we observe are stable invariants of mathematical content or transient features of a particular development phase.

\begin{definition}[Temporal graph sequence]\label{def:temporal-sequence}
Let $\{c_1, c_2, \ldots, c_T\}$ be a sequence of $T$ historical \module{Mathlib} commits. Each commit $c_t$ yields an environment $\mathcal{E}_t$ and corresponding graphs $G_{\mathrm{thm}}^{(t)} = (\mathcal{D}_t, E_t)$ and $G_{\mathrm{module}}^{(t)} = (\mathcal{M}_t, F_t)$. Define:
\begin{itemize}
  \item \emph{Growth indicators}: $|\mathcal{D}_t|$, $|\mathcal{M}_t|$, and edge density $|E_t| / |\mathcal{D}_t|$.
  \item \emph{Structural stability}: community persistence $\mathrm{NMI}(\mathcal{L}^{(t)}, \mathcal{L}^{(t+1)})$, hub turnover (rank correlation of in-degree rankings between consecutive snapshots), and redundancy trend $r(G_{\mathrm{module}}^{(t)})$ over time.
  \item \emph{Structural events}: discontinuities in the time series coinciding with known events (e.g., the Lean~3$\to$4 port via Mathport, 2023; the module system introduction, November 2025).
\end{itemize}
\end{definition}

\begin{figure}[!htb]
\centering
\begin{subfigure}[t]{0.38\textwidth}
\centering
\vspace{0pt}
\raggedright\scriptsize
\textsf{\bfseries Timeline of snapshots}\\[6pt]
\begin{tikzpicture}[
  every node/.style={font=\small\sffamily},
  dot/.style={circle, fill=blue!60!black, minimum size=3pt, inner sep=0pt},
  event/.style={font=\small\sffamily\itshape, text=red!60!black},
]
  \draw[->, >=Stealth, gray!60, thick] (0, 0) -- (0, -5.2);
  \node[font=\small\sffamily, gray!50, anchor=east] at (-0.15, 0) {$t$};
  \node[dot] (c1) at (0, -0.3) {};
  \node[anchor=west] at (0.2, -0.3) {$c_1$: Jan 2023};
  \node[dot] (c2) at (0, -1.2) {};
  \node[anchor=west] at (0.2, -1.2) {$c_2$: Jul 2023};
  \node[event, anchor=west] at (0.2, -1.55) {Lean 3$\to$4 port};
  \node[dot] (c3) at (0, -2.4) {};
  \node[anchor=west] at (0.2, -2.4) {$c_3$: Jan 2024};
  \node[dot] (c4) at (0, -3.3) {};
  \node[anchor=west] at (0.2, -3.3) {$c_4$: Jan 2025};
  \node[dot] (c5) at (0, -4.2) {};
  \node[anchor=west] at (0.2, -4.2) {$c_5$: Nov 2025};
  \node[event, anchor=west] at (0.2, -4.55) {module system};
  \node[dot, fill=orange!80!red] (c6) at (0, -5.0) {};
  \node[anchor=west, text=orange!80!red] at (0.2, -5.0) {$c_T$: Feb 2026 (this paper)};
\end{tikzpicture}
\caption{Monthly snapshots $\{c_1, \ldots, c_T\}$ with structural events marked.}
\label{fig:temporal-timeline}
\end{subfigure}%
\hfill
\begin{subfigure}[t]{0.58\textwidth}
\centering
\vspace{0pt}
\begin{tikzpicture}[
  every node/.style={font=\small, inner sep=2pt},
  mod/.style={circle, fill=blue!60!black, minimum size=2.5pt, inner sep=0pt},
  edge/.style={-, blue!30!black, semithick, opacity=0.5},
  frame/.style={rounded corners=3pt, draw=gray!40, fill=gray!3, inner sep=4pt},
]
  \node[frame, minimum width=50pt, minimum height=40pt] (f1) at (0, -0.5) {};
  \node[font=\small\sffamily, gray!60, above=0pt of f1] {$G^{(c_1)}$};
  \node[mod] (a1) at (-0.4, -0.2) {};
  \node[mod] (b1) at (0.3, -0.1) {};
  \node[mod] (c1n) at (-0.1, -0.7) {};
  \node[mod] (d1) at (0.5, -0.8) {};
  \draw[edge] (a1) -- (b1);
  \draw[edge] (b1) -- (c1n);
  \draw[edge] (c1n) -- (d1);
  \node[font=\small\sffamily, text=gray!50, below=2pt of f1] {$|\mathcal{D}|{=}50\text{k}$};
  \node[frame, minimum width=55pt, minimum height=45pt] (f3) at (2.8, -0.5) {};
  \node[font=\small\sffamily, gray!60, above=0pt of f3] {$G^{(c_3)}$};
  \node[mod] (a3) at (2.3, -0.1) {};
  \node[mod] (b3) at (3.0, 0.0) {};
  \node[mod] (c3n) at (2.5, -0.5) {};
  \node[mod] (d3) at (3.2, -0.4) {};
  \node[mod] (e3) at (2.8, -0.9) {};
  \node[mod] (f3n) at (2.4, -0.9) {};
  \draw[edge] (a3) -- (b3);
  \draw[edge] (a3) -- (c3n);
  \draw[edge] (b3) -- (d3);
  \draw[edge] (c3n) -- (e3);
  \draw[edge] (d3) -- (e3);
  \draw[edge] (c3n) -- (f3n);
  \node[font=\small\sffamily, text=gray!50, below=2pt of f3] {$|\mathcal{D}|{=}150\text{k}$};
  \node[frame, minimum width=62pt, minimum height=50pt, draw=orange!60!red] (fT) at (5.8, -0.5) {};
  \node[font=\small\sffamily, orange!80!red, above=0pt of fT] {$G^{(c_T)}$};
  \node[mod, fill=orange!80!red] (aT) at (5.3, 0.0) {};
  \node[mod, fill=orange!80!red] (bT) at (6.0, 0.1) {};
  \node[mod, fill=orange!80!red] (cT) at (5.5, -0.3) {};
  \node[mod, fill=orange!80!red] (dT) at (6.2, -0.3) {};
  \node[mod, fill=orange!80!red] (eT) at (5.3, -0.7) {};
  \node[mod, fill=orange!80!red] (fT2) at (5.8, -1.0) {};
  \node[mod, fill=orange!80!red] (gT) at (6.3, -0.7) {};
  \node[mod, fill=orange!80!red] (hT) at (5.6, -0.5) {};
  \draw[-, orange!40!red, semithick, opacity=0.5] (aT) -- (bT);
  \draw[-, orange!40!red, semithick, opacity=0.5] (aT) -- (cT);
  \draw[-, orange!40!red, semithick, opacity=0.5] (bT) -- (dT);
  \draw[-, orange!40!red, semithick, opacity=0.5] (cT) -- (eT);
  \draw[-, orange!40!red, semithick, opacity=0.5] (dT) -- (gT);
  \draw[-, orange!40!red, semithick, opacity=0.5] (eT) -- (fT2);
  \draw[-, orange!40!red, semithick, opacity=0.5] (hT) -- (fT2);
  \draw[-, orange!40!red, semithick, opacity=0.5] (cT) -- (hT);
  \draw[-, orange!40!red, semithick, opacity=0.5] (gT) -- (fT2);
  \node[font=\small\sffamily, text=orange!80!red, below=2pt of fT] {$|\mathcal{D}|{=}318\text{k}$};
  \draw[->, >=Stealth, gray!40, thick] (f1.east) -- (f3.west);
  \draw[->, >=Stealth, gray!40, thick] (f3.east) -- (fT.west);
  \node[font=\small\sffamily, anchor=north west] at (-0.8, -2.0)
    {Track: $|\mathcal{D}_t|$, edge density, community NMI, hub turnover, redundancy $r(G_{\mathrm{module}}^{(t)})$};
\end{tikzpicture}
\caption{Graph snapshots growing over time; structural metrics tracked across the sequence detect events and trends.}
\label{fig:temporal-graph}
\end{subfigure}
\caption{Temporal graph sequence (Definition~\ref{def:temporal-sequence}). (a)~Monthly snapshots of \module{Mathlib} commits, with two structural events: the Lean~3$\to$4 port (2023) and the module system introduction (November 2025). (b)~The declaration graph $G_{\mathrm{thm}}^{(t)}$ at three time points, growing from ${\sim}50$k to $318$k declarations. Tracking structural indicators across the sequence reveals whether properties like the $17.5\%$ redundancy rate are stable invariants or transient features of a particular development phase.}
\label{fig:temporal-sequence}
\end{figure}

The temporal dimension directly captures the \emph{process} of organizational restructuring. If \module{Mathlib}'s dependency structure is primarily determined by inherited mathematical logic, structural indicators should be stable under refactoring; if organizational process dominates, we expect significant structural change at refactoring events while mathematical content remains constant. The Lean~3$\to$4 port is a natural experiment: identical mathematical content, entirely rewritten proof infrastructure.

\section{Network Analysis Details}\label{app:supplementary}

This section presents the detailed network statistics underlying the findings of \S\ref{sec:contribution}. The three dependency graphs defined in Appendix~\ref{sec:graph-definitions}---the module graph $G_{\mathrm{module}}$ (\S\ref{app:module-detail}), the declaration graph $G_{\mathrm{thm}}$ (\S\ref{app:decl-detail}), and the namespace graph $G_{\mathrm{ns}}$ (\S\ref{app:ns-detail})---are each examined through degree distribution, DAG depth, centrality rankings, community structure, and robustness analysis, supplemented by graph-specific measurements such as containment decay, cross-namespace heatmaps, and theorem/lemma validation. Section~\ref{sec:cross_level} then overlays the three levels, quantifying how file boundaries, namespace boundaries, and logical dependencies align and diverge.

\subsection{Module Graph}\label{app:module-detail}

The module graph $G_{\mathrm{module}}$ (Definition~\ref{def:module-graph}; \S\ref{sec:module-import}) captures file-level import dependencies among \module{Mathlib}'s $7{,}563$ source files.

\subsubsection{Containment Decay}

Table~\ref{tab:module-containment} reports the fraction of import edges that remain within the same group at successive depths of the module tree.

\begin{table}[ht]
\centering
\caption{Module-level containment decay by module depth. At depth~$1$ ($13$ top-level packages), $97.5\%$ of import edges stay within the same package; by depth~$4$, containment has fallen to~$8.8\%$.}
\label{tab:module-containment}
\begin{tabular}{crrr}
\toprule
Depth $k$ & Groups & Containment & $\Delta$ (pp) \\
\midrule
$1$ & $13$      & $97.5\%$  & ---     \\
$2$ & $99$      & $60.2\%$  & $-37.3$ \\
$3$ & $1{,}481$ & $34.1\%$  & $-26.1$ \\
$4$ & $5{,}641$ & $8.8\%$   & $-25.3$ \\
$5$ & $7{,}434$ & $0.8\%$   & $-8.0$  \\
\bottomrule
\end{tabular}
\end{table}

At depth~$1$, the $13$ top-level packages (\module{Mathlib}, \module{Batteries}, \module{Init}, etc.) contain $97.5\%$ of all import edges. The steepest single drop occurs at depth~$1 \to 2$ ($-37.3$ pp), when packages split into their subdirectories. The $60.2\%$ containment at depth~$2$ is consistent with the $61.4\%$ intra-directory rate reported in \S\ref{sec:cross-namespace}.\footnote{The small discrepancy ($60.2\%$ vs.\ $61.4\%$) reflects a difference in edge counts between our import parser and the \texttt{importGraph} tool.} By depth~$4$ (the modal depth of modules), containment has fallen to $8.8\%$, and by depth~$5$ it is negligible.

\label{sec:structural-analysis}
\subsubsection{Degree Distribution}
\label{sec:degree-distribution}
\label{sec:sa-degree}

In $G_{\mathrm{module}}$, the out-degree $\deg^{+}(m) = |\mathrm{imports}(m)|$: it counts the number of modules that $m$ directly imports. The in-degree $\deg^{-}(m)$ counts how many other modules depend on~$m$.
Structurally, in-degree measures how foundational a module is (how many others build upon it), while out-degree measures how much of the library a module assembles. A high in-degree module is \emph{infrastructure}; a high out-degree module is an \emph{integrator}.

\begin{table}[H]
\centering
\caption{Degree statistics for $G_{\mathrm{module}}$ and its transitive reduction~$G_{\mathrm{module}}^{-}$.}
\label{tab:degree-stats}
\begin{tabular}{lrrrr}
\toprule
& \multicolumn{2}{c}{$G_{\mathrm{module}}$} & \multicolumn{2}{c}{$G_{\mathrm{module}}^{-}$} \\
\cmidrule(lr){2-3} \cmidrule(lr){4-5}
& In-degree & Out-degree & In-degree & Out-degree \\
\midrule
Mean   & $3.12$ & $3.12$ & $2.57$ & $2.57$ \\
Median & $2$    & $3$    & $1$    & $2$    \\
Std    & $4.69$ & $4.12$ & $3.91$ & $1.83$ \\
Max    & $167$  & $315$  & $151$  & $70$   \\
\bottomrule
\end{tabular}
\end{table}

The module with the highest in-degree in both graphs is \module{Mathlib.Init} ($\deg^{-} = 167$ in $G_{\mathrm{module}}$, $151$ in $G_{\mathrm{module}}^{-}$), the foundational module that most files depend on. Other high in-degree modules include \module{Tactic.Common} ($74$), \module{Analysis.Normed.Group.Basic} ($57$), and \module{Algebra.Ring.Defs} ($42$). The highest out-degree belongs to \module{Mathlib.Tactic} ($315$ in $G_{\mathrm{module}}$), which serves as an umbrella file aggregating tactic imports; transitive reduction reduces this to~$60$.

\begin{figure}[H]
\centering
\includegraphics[width=\textwidth]{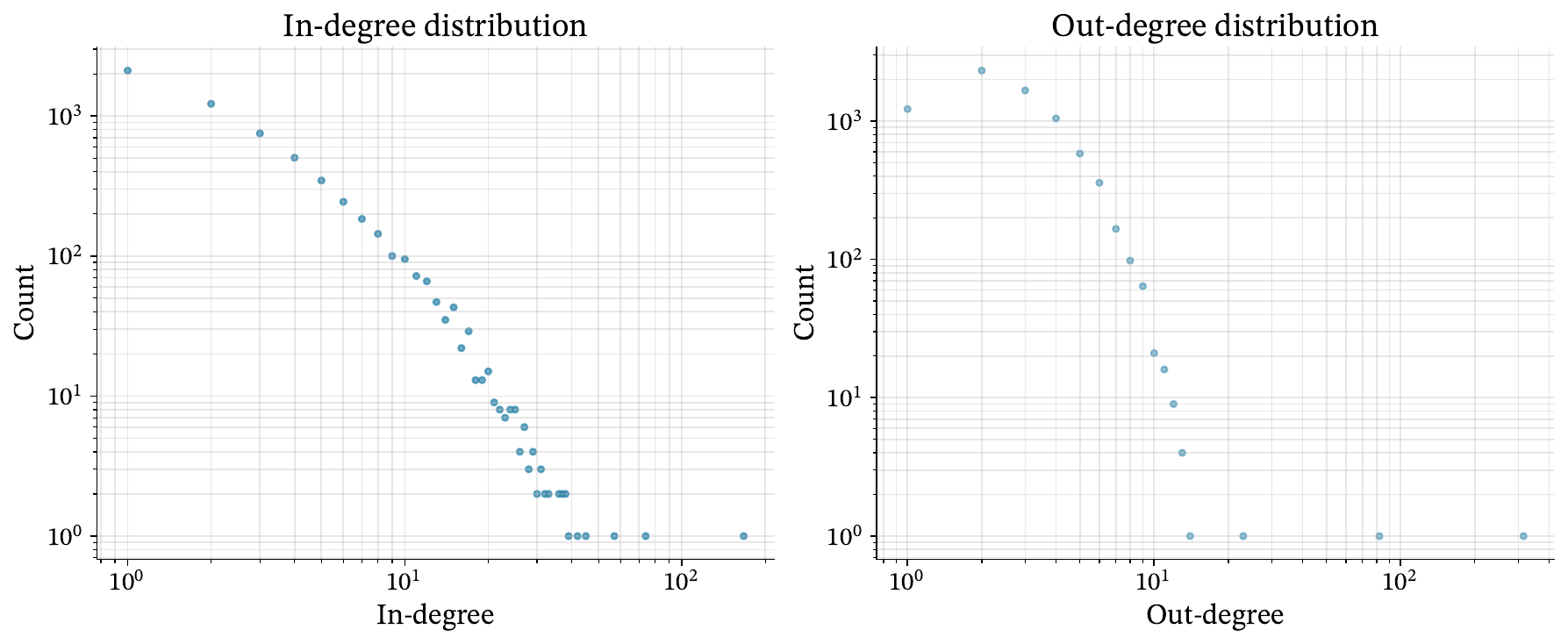}
\caption{In-degree (blue) and out-degree (orange) distributions on log--log axes for $G_{\mathrm{module}}$ (left) and $G_{\mathrm{module}}^{-}$ (right). Dashed lines show power-law references $k^{-\gamma}$. Two features are visually salient: (i)~the in-degree distributions are nearly identical across the two panels, confirming that transitive reduction barely affects how many modules depend on a given file; (ii)~the out-degree tail contracts sharply under transitive reduction (maximum dropping from $315$ to $70$), reflecting the removal of redundant umbrella imports.}
\label{fig:degree-distribution}
\end{figure}

We plot the degree distributions on log--log axes (Figure~\ref{fig:degree-distribution}).

Both degree distributions are heavy-tailed but do not cleanly fit a power law. The out-degree distribution, in particular, becomes much more concentrated after transitive reduction (maximum dropping from $315$ to $70$, standard deviation from $4.12$ to $1.83$), indicating that many high-out-degree imports are transitively redundant.
In contrast, the in-degree distribution is barely affected by transitive reduction (maximum: $167 \to 151$; standard deviation: $4.69 \to 3.91$). This asymmetry has a structural explanation: redundant edges arise primarily from developers importing umbrella modules, an act that inflates the importer's out-degree without changing the importee's in-degree. The development-ergonomic redundancy identified in \S\ref{sec:transitive-reduction} is \emph{directional}: it lives in the ``I depend on others'' direction, not the ``others depend on me'' direction.
The redundancy is not random; it follows the contours of the module tree~$T$, because developers import umbrella modules that aggregate entire sub-hierarchies for refactoring convenience.

\subsubsection{DAG Depth}
\label{sec:dag-depth}

All sources (modules with no incoming edges) sit at level~$0$, while the unique sink \module{Tactic.Linter.DirectoryDependency} sits at level~$153$.

The depth of $G_{\mathrm{module}}$ is $153$, attained by a path from \module{NumberTheory.LSeries.PrimesInAP} (a leaf-level number theory result) down to \module{Tactic.Linter.DirectoryDependency} (a low-level linter utility). This path traverses number theory, special functions, measure theory, topology, order theory, logic, and tactic infrastructure, reflecting the deep dependency chain required to formalize analytic number theory.

$G_{\mathrm{module}}$ has $1{,}433$ source modules (files that no other \module{Mathlib} file imports) and exactly $1$~sink (\module{Tactic.Linter.DirectoryDependency}, the unique module with no imports within Mathlib). The topological levels partition $\mathcal{M}$ into $154$ layers, with a maximum layer width of $1{,}433$ (the source layer) and a median width of~$25$.

Notably, the layer-width profiles of $G_{\mathrm{module}}$ and $G_{\mathrm{module}}^{-}$ are nearly identical: the number of layers, the position of the peak, and the overall funnel shape are preserved under transitive reduction. This means that the vertical structure of the dependency chain reflects genuine logical necessity rather than cognitive redundancy; the 17.5\% redundant edges (post-\texttt{shake}) identified in \S\ref{sec:transitive-reduction} add lateral shortcuts within layers but do not alter the depth of the graph.

A secondary peak is visible near level~$126$, where approximately $59$ modules concentrate, roughly $6$ to $15$ times the width of surrounding layers. This anomaly warrants further investigation; it may reflect a layer of tactic infrastructure or a mathematical theory that branches into many parallel lemmas at a specific depth.

\begin{figure}[H]
\centering
\includegraphics[width=\textwidth]{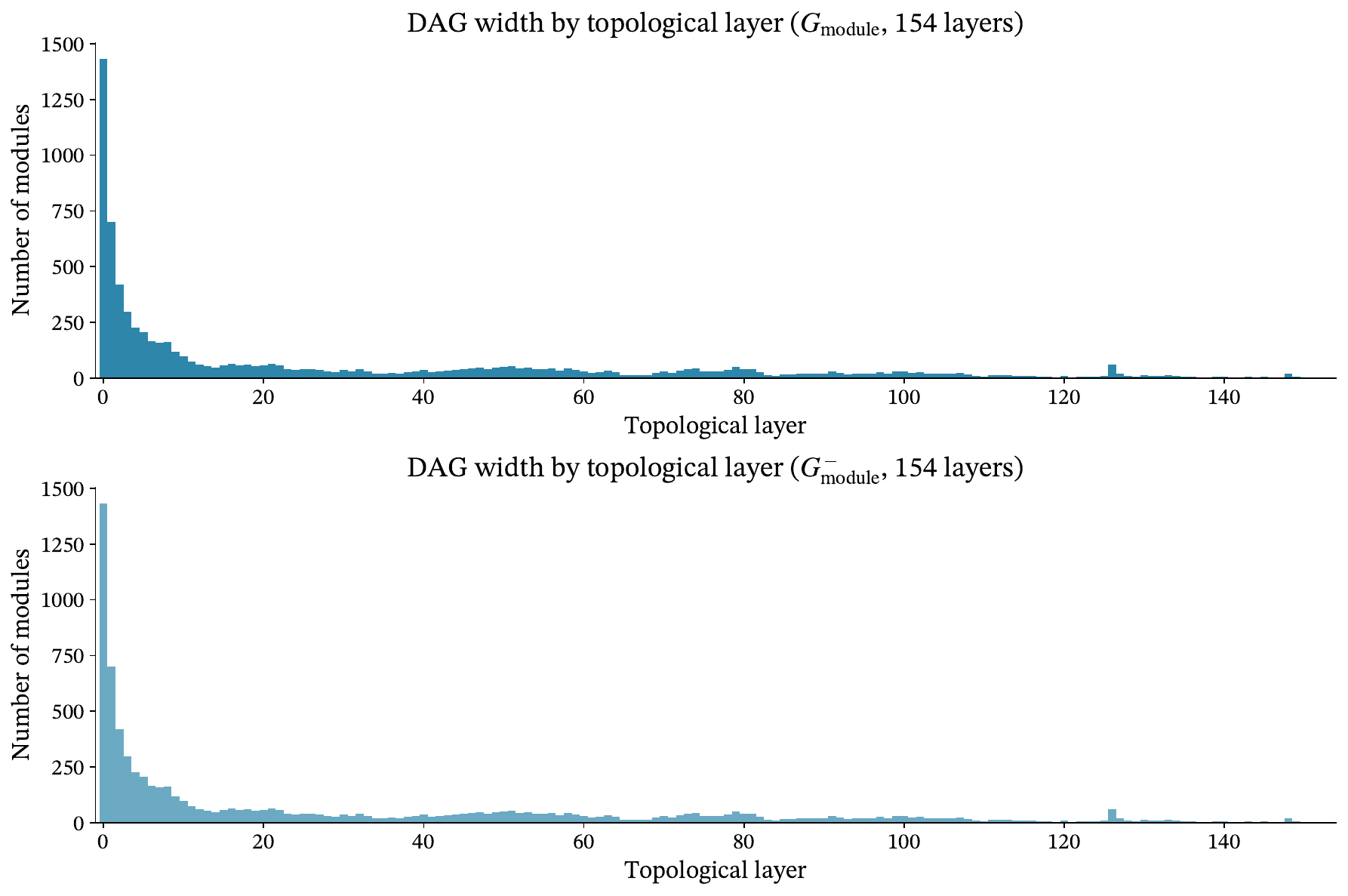}
\caption{DAG width by topological layer for $G_{\mathrm{module}}$ (top) and $G_{\mathrm{module}}^{-}$ (bottom). The distributions share a characteristic funnel shape: a sharp peak at level~$0$ (over $1{,}400$ leaf modules), a rapid decay through the first $20$ layers, a long tail extending to level~$153$, and a secondary peak near level~$126$. The near-identical profiles confirm that the funnel topology arises from logical necessity, not from redundant imports.}
\label{fig:dag-structure}
\end{figure}

The funnel shape (wide at the top, tapering to a single sink) reflects the fact that advanced mathematics is built on a narrow foundation of shared definitions and tactics. Any change to a module in the bottleneck layers propagates upward through a large fraction of the library, creating recompilation cascades. Lean's module system, by enabling private imports, is an existing engineering response to this narrow-waist problem.

\label{rem:import-depth-symmetry}
The DAG depth of~$153$ layers describes the topological extent of the dependency chain. A different notion of depth, the \emph{module depth} (the number of dot-separated components of a module's fully qualified name), describes how deeply a module sits in the naming hierarchy~$T$. Module depth is concentrated: $55.7\%$ of modules have depth~$4$ (e.g., \module{Mathlib.Data.Nat.Basic}), with a mean of~$4.22$. Import edges exhibit near-perfect depth symmetry: $55.4\%$ connect modules at the same naming depth, $23.3\%$ point from deeper to shallower, and $21.3\%$ from shallower to deeper (mean difference~$+0.03$). Only~$1.3\%$ of imports target modules at depth~$\le 2$. Developers import \emph{laterally}, peer to peer within the same naming stratum, rather than vertically across naming layers. As we shall see in \S\ref{sec:cross_level}, this symmetry contrasts sharply with the declaration-level dependency graph, where $37.6\%$ of edges flow from deeper to shallower namespaces and the mean depth difference rises to~$+0.36$: the module graph compresses and hides the strong infrastructure pull that the declaration graph reveals.

\subsubsection{Centrality}
\label{sec:centrality}
\label{sec:sa-centrality}

We compare three centrality measures (in-degree, PageRank, and betweenness; Definition~\ref{def:pagerank}--\ref{def:betweenness}) to identify the most structurally important modules.

\begin{table}[H]
\centering
\caption{Top~10 modules by in-degree, PageRank, and betweenness centrality in $G_{\mathrm{module}}$.}
\label{tab:centrality}
\small
\begin{tabular}{rlrlrl}
\toprule
\multicolumn{2}{c}{In-degree} & \multicolumn{2}{c}{PageRank} & \multicolumn{2}{c}{Betweenness} \\
\cmidrule(lr){1-2} \cmidrule(lr){3-4} \cmidrule(lr){5-6}
$\deg^{-}$ & Module & PR & Module & $c_B$ & Module \\
\midrule
167 & \module{Init}               & .0435 & \module{Init}              & .0113 & \module{Tactic.Common} \\
 74 & \module{Tactic.Common}      & .0316 & \module{Tactic.Linter.Header} & .0037 & \module{Init} \\
 57 & \module{AN.Group.Basic}     & .0285 & \module{Tactic.Linter.DD}  & .0032 & \module{LA.Span.Basic} \\
 45 & \module{Util.CompileInd.}   & .0058 & \module{Tactic.TypeStar}   & .0030 & \module{AN.Module.FD} \\
 42 & \module{Algebra.Ring.Defs}  & .0048 & \module{Algebra.Group.Defs} & .0024 & \module{Analysis.RCLike} \\
 39 & \module{ABO.Finset.Basic}   & .0038 & \module{Tactic.Basic}      & .0021 & \module{AS.Limits.Basic} \\
 38 & \module{Algebra.Field.Defs} & .0037 & \module{CT.Equivalence}    & .0020 & \module{AS.Limits.Normed} \\
 38 & \module{Algebra.Group.Basic}& .0033 & \module{Logic.Function}    & .0019 & \module{LA.Determinant} \\
 37 & \module{Tactic.Fin.Attr}    & .0030 & \module{Logic.Equiv.Defs}  & .0019 & \module{CT.Category.Basic} \\
 37 & \module{Tactic.TypeStar}    & .0030 & \module{CT.Opposites}      & .0017 & \module{Tactic.NormNum} \\
\bottomrule
\end{tabular}
\end{table}

\begin{figure}[H]
\centering
\begin{subfigure}[t]{0.32\textwidth}
\centering
\includegraphics[width=\textwidth]{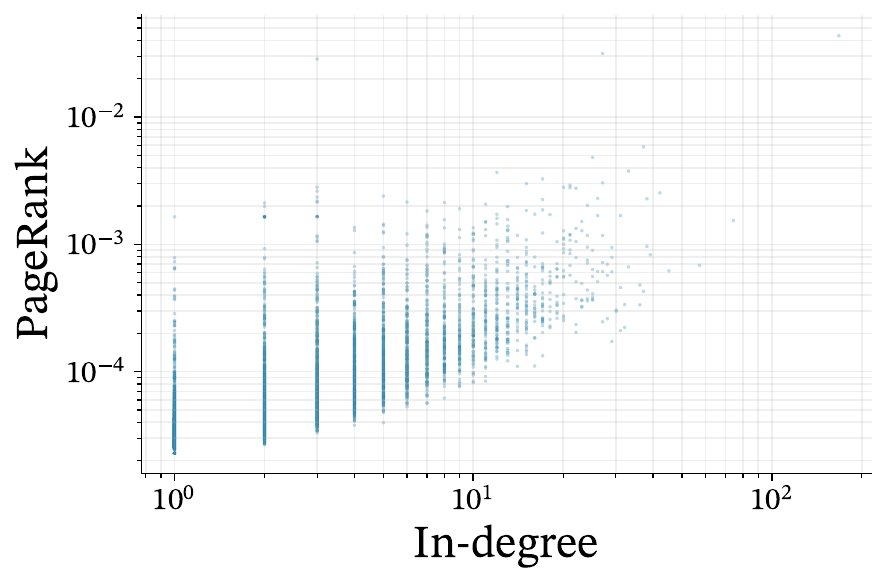}
\caption{In-degree vs.\ PageRank.}
\label{fig:centrality-indeg-pr}
\end{subfigure}%
\hfill
\begin{subfigure}[t]{0.32\textwidth}
\centering
\includegraphics[width=\textwidth]{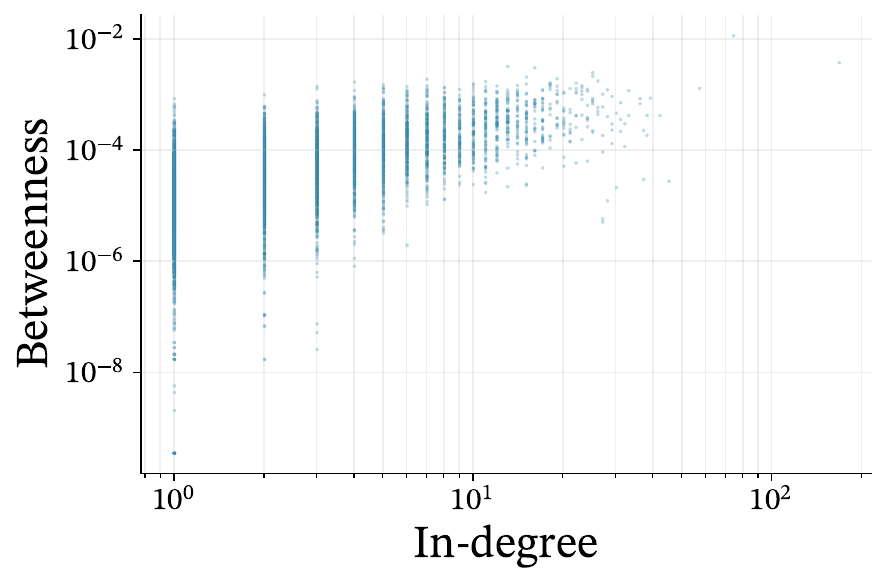}
\caption{In-degree vs.\ Betweenness.}
\label{fig:centrality-indeg-betw}
\end{subfigure}%
\hfill
\begin{subfigure}[t]{0.32\textwidth}
\centering
\includegraphics[width=\textwidth]{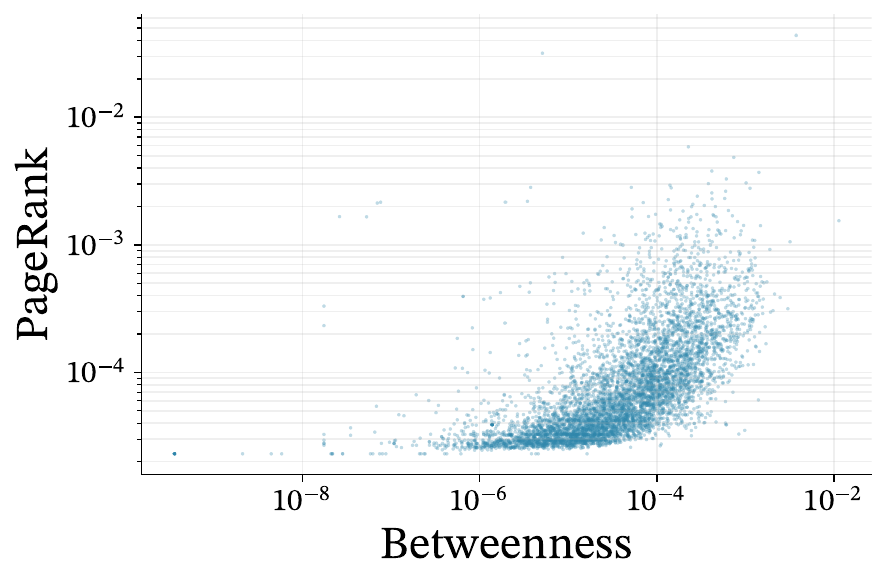}
\caption{Betweenness vs.\ PageRank.}
\label{fig:centrality-betw-pr}
\end{subfigure}
\caption{Pairwise centrality scatter plots for $G_{\mathrm{module}}$. Each point is one module. The wide scatter across all three panels confirms that in-degree, PageRank, and betweenness capture distinct structural roles. The three-way divergence visible in Table~\ref{tab:centrality} pervades the entire distribution, not just the top extremes.}
\label{fig:module-centrality}
\end{figure}

\noindent
\textit{Abbreviations:} AN = \module{Analysis.Normed}, LA = \module{LinearAlgebra}, CT = \module{CategoryTheory},\\
ABO = \module{Algebra.BigOperators.Group}, AS = \module{Analysis.SpecificLimits},\\
FD = \module{FiniteDimension}, DD = \module{DirectoryDependency}, CompileInd.\ = \module{CompileInductive}, Fin.\ = \module{Finiteness}.
We compare in-degree, PageRank, and betweenness pairwise in Figures~\ref{fig:centrality-indeg-pr}--\ref{fig:centrality-betw-pr}. Out-degree is omitted from this comparison because it measures a module's role as an \emph{importer}, how many prerequisites it gathers, rather than its structural position within the network. Moreover, as shown in \S\ref{sec:degree-distribution}, high out-degree values largely reflect redundant umbrella imports and collapse under transitive reduction, making out-degree a poor proxy for structural importance.

The three rankings largely differ: the only module appearing in all three top-20 lists is \module{Mathlib.Init}. This divergence is not a technical detail; it is one of our central findings. It demonstrates that \emph{importance} in a formal mathematical library is irreducibly multidimensional: a module can be heavily used (high in-degree) without being foundational (high PageRank), and foundational without being a bridge (high betweenness). No single centrality measure captures the full structural role of a module; the three measures together define a taxonomy that we develop further in \S\ref{sec:cross-level-summary}.

In particular, high-betweenness modules acquire their bridging role by connecting regions of $G_{\mathrm{module}}$ that the tree $T$ separates. The top betweenness modules in Table~\ref{tab:centrality} typically sit at the boundary between top-level directories, the points where mathematical logic forces a passage across $T$'s cognitive borders.

\subsubsection{Community Structure}
\label{sec:module-community}
\label{sec:sa-community}

We apply Louvain community detection (\S\ref{sec:prelim-community}) to the undirected projection of $G_{\mathrm{module}}$ ($7{,}563$ nodes, $23{,}570$ edges).

\begin{table}[ht]
\centering
\caption{Community detection summary for $G_{\mathrm{module}}$.}
\label{tab:module-community-summary}
\begin{tabular}{lr}
\toprule
Metric & Value \\
\midrule
Number of communities & $16$ \\
Modularity & $0.6395$ \\
Communities with ${>}1{,}000$ nodes & $3$ \\
Communities with ${>}100$ nodes & $9$ \\
\bottomrule
\end{tabular}
\end{table}

The modularity of $0.64$ is the highest among the three graph levels, reflecting the fact that modules are the unit at which human organization is most direct: developers place related files in the same directory, creating dense within-directory import clusters.

\begin{table}[ht]
\centering
\caption{The nine major communities of $G_{\mathrm{module}}$ (Louvain, ${>}100$ nodes), with dominant top-level directories.}
\label{tab:module-communities}
\small
\begin{tabular}{rrl}
\toprule
ID & Size & Dominant directories \\
\midrule
2  & $1{,}445$ & \module{RingTheory} (531), \module{Algebra} (377), \module{LinearAlgebra} (253) \\
7  & $1{,}398$ & \module{CategoryTheory} (893), \module{Algebra} (147), \module{Topology} (92) \\
3  & $1{,}366$ & \module{Algebra} (382), \module{Data} (297), \module{Tactic} (295) \\
1  & $948$     & \module{Analysis} (533), \module{Topology} (124), \module{Geometry} (97) \\
5  & $668$     & \module{Data} (184), \module{Order} (181), \module{Topology} (58) \\
6  & $504$     & \module{MeasureTheory} (232), \module{Probability} (112), \module{Analysis} (88) \\
4  & $455$     & \module{Topology} (233), \module{Analysis} (78), \module{Data} (41) \\
0  & $343$     & \module{Algebra} (111), \module{GroupTheory} (99), \module{Combinatorics} (52) \\
8  & $207$     & \module{Algebra} (102), \module{CategoryTheory} (38) \\
\bottomrule
\end{tabular}
\end{table}

The communities map recognizably onto mathematical domains. \module{CategoryTheory} again dominates a single community ($64\%$ of Community~$7$), mirroring the exceptional autonomy observed at the declaration level (\S\ref{sec:thm-community}). The NMI between Louvain communities and top-level directories is $0.42$ (ARI~$= 0.28$), higher than the declaration-level NMI of $0.34$ (\S\ref{sec:thm-community}). The stronger alignment reflects the fact that modules, unlike declarations, are explicitly organized into a directory tree; the import graph inherits much of that tree structure.

\subsubsection{Robustness}
\label{sec:robustness}
\label{sec:sa-robustness}

$G_{\mathrm{module}}$ is weakly connected: all $7{,}563$ modules belong to a single weakly connected component. We assess robustness by measuring the impact of targeted node removal.

\begin{table}[H]
\centering
\caption{Effect of removing the top-5 modules (by in-degree or betweenness) on the number of weakly connected components.}
\label{tab:robustness}
\begin{tabular}{lrr}
\toprule
Removal strategy & \multicolumn{1}{c}{$G_{\mathrm{module}}$} & \multicolumn{1}{c}{$G_{\mathrm{module}}^{-}$} \\
\midrule
None (baseline)          & $1$  & $1$  \\
Top-5 by in-degree       & $29$ & $56$ \\
Top-5 by betweenness     & $20$ & $48$ \\
\bottomrule
\end{tabular}
\end{table}

Removing the five highest in-degree modules (\module{Init}, \module{Tactic.Common}, \module{Analysis.Normed.Group.Basic}, \module{Util.CompileInductive}, and \module{Algebra.Ring.Defs}) fragments the raw graph into $29$ components, though the largest component still contains $7{,}530$ of the original $7{,}563$ nodes. The effect is more pronounced on $G_{\mathrm{module}}^{-}$ ($56$ components), since transitive reduction removes the alternative paths that would otherwise keep the graph connected.

\label{rem:robustness}
Even after targeted removal, over $99\%$ of modules remain in the giant component. The transitive reduction is more fragile ($56$ components vs.\ $29$) precisely because it removes the redundant connectivity that doubles as structural insurance.

\begin{figure}[ht]
\centering
\includegraphics[width=\textwidth]{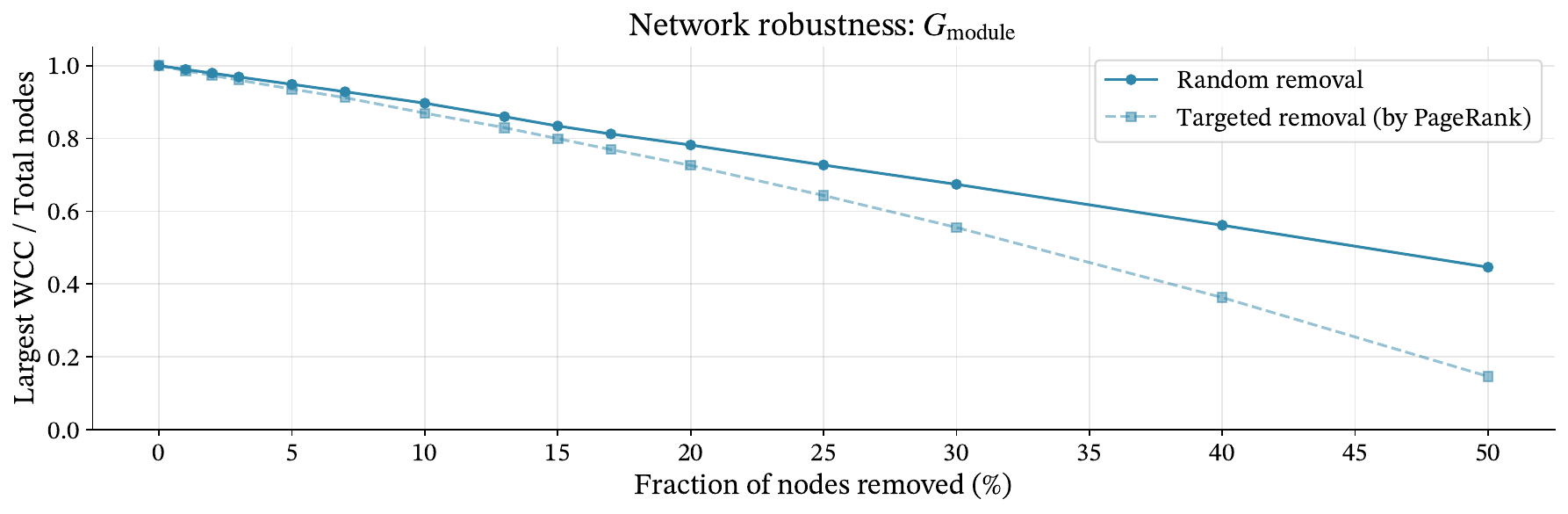}
\caption{Robustness curves for $G_{\mathrm{module}}$: fraction of nodes in the largest WCC as a function of the fraction removed, under random removal (blue) and targeted removal by PageRank (coral).}
\label{fig:module-robustness}
\end{figure}

Figure~\ref{fig:module-robustness} shows the progressive removal curves. Under random removal, the giant component degrades gracefully: removing $20\%$ of modules leaves $80\%$ connected. Under targeted attack on the highest-PageRank modules, the same $20\%$ removal reduces the giant component to $58\%$, a $22$-percentage-point gap that grows with the removal fraction. The curve confirms the hub-dominated vulnerability signature: the module graph's resilience depends on a small number of high-centrality hubs whose removal fragments the network far more effectively than chance would predict.

\subsubsection{Extended Analysis}\label{app:module-analysis}

The $17.5\%$ post-\texttt{shake} edge redundancy (\S\ref{sec:transitive-reduction}) represents a \textit{development ergonomics layer}: developers import umbrella modules to bring tactics, notation, and typeclass instances into scope, sacrificing logical minimality for usability. The fact that \texttt{shake} intentionally preserves many of these imports confirms they serve a structural purpose, suggesting that future proof assistants could benefit from explicitly modeling two dependency layers: a minimal graph for the build system and a richer graph for the interactive environment.

The layer-width profile (\S\ref{sec:dag-depth}) reveals a \textit{funnel topology}: a massive leaf layer (${>}1{,}400$ modules) resting on a deep, narrow foundation. Changes in the narrow ``waist'' (high-betweenness modules) trigger recompilation cascades, a scalability challenge that Lean's module system (November 2025) addresses through private imports. Unlike informal mathematics, where foundational texts are static, \module{Mathlib}'s foundational modules remain mutable code; the maintenance cost of a module is a function of its topological depth. One possible trajectory is that monolithic libraries will hit a \textit{critical depth threshold} requiring \textit{federated tiers}, freezing foundational layers into pre-compiled artifacts.

The three centrality measures (\S\ref{sec:centrality}) admit distinct operational interpretations: in-degree identifies frequently used \textit{vocabulary}; PageRank identifies the \textit{axiomatic core}; betweenness identifies \textit{structural bridges} between subfields. Refactoring a high-betweenness module carries more systemic risk than refactoring a high-in-degree module, since it disrupts cross-theory connectivity.

We hypothesize that \module{Mathlib} has evolved into a \textit{curated small-world network}, growing via \textit{semantic attachment} rather than preferential attachment: new modules cluster within directories while bridge modules maintain global connectivity. The module tree~$T$ and the import graph~$G_{\mathrm{module}}$ continuously reshape each other: developers import along cognitive contours of~$T$, while cross-directory dependencies force periodic reorganization, as in the ongoing absorption of \module{RingTheory} into \module{Algebra}.

\subsection{Declaration Graph}\label{app:decl-detail}

The declaration graph $G_{\mathrm{thm}}$ (Definition~\ref{def:thm-graph}; \S\ref{sec:theorem-premise}) dissolves file boundaries, exposing individual logical dependencies among \module{Mathlib}'s $308{,}129$ declarations.

\subsubsection{Declaration-Level Summary Tables}
\label{sec:decl-summary-tables}

Tables~\ref{tab:kind-overview}--\ref{tab:thm-cross-ns-pairs} collect the basic statistics of $G_{\mathrm{thm}}$: node composition by declaration kind, inter-kind edge flow, per-kind degree statistics, zero-citation rates, edge classification by boundary type, and the heaviest cross-namespace pairs. These tables are referenced throughout the Extended Analysis below.

\begin{table}[ht]
\centering
\caption{Declaration kinds in $G_{\mathrm{thm}}$: role, prevalence, and representative example.}
\label{tab:kind-overview}
\small
\begin{tabular}{lp{5.2cm}rrl}
\toprule
Kind & Role & Count & \% & Example \\
\midrule
theorem     & Proved proposition; proof erased at runtime   & $243{,}725$ & $79.1$ & \decl{Nat.add\_comm} \\
definition  & Computable function; body retained            & $48{,}664$  & $15.8$ & \decl{List.length} \\
abbrev      & Transparent shorthand; always unfolded        & $6{,}667$   & $2.2$  & \decl{OfNat.ofNat} \\
constructor & Introduction rule of an \texttt{inductive}    & $4{,}755$   & $1.5$  & \decl{Nat.succ} \\
inductive   & New type with introduction rules              & $3{,}813$   & $1.2$  & \decl{Nat} \\
opaque      & Sealed body; cannot be unfolded               & $499$       & $0.2$  & \decl{Float.add} \\
quotient    & Kernel primitives for quotient types          & $3$       & ${<}0.1$ & \decl{Quot.mk} \\
axiom       & Accepted without proof; irreducible           & $3$       & ${<}0.1$ & \decl{propext} \\
\bottomrule
\end{tabular}
\end{table}

\begin{table}[H]
\centering
\caption{Inter-kind edge flow in $G_{\mathrm{thm}}$. Each row is a source kind; each column is a target kind. Entries are thousands of edges; blanks denote fewer than $500$.}
\label{tab:inter-kind-flow}
\small
\begin{tabular}{l*{5}{r}}
\toprule
Source $\downarrow$ \textbackslash\ Target $\rightarrow$ & theorem & definition & abbrev & inductive & \textit{other} \\
\midrule
theorem      & $1{,}332$k & $2{,}498$k & $2{,}590$k & $835$k & $241$k \\
definition   & $18$k      & $285$k     & $284$k     & $163$k & $50$k \\
abbrev       & $2$k       & $17$k      & $26$k      & $20$k  & $3$k \\
constructor  & $1$k       & $12$k      & $20$k      & $16$k  & $6$k \\
inductive    &            & $2$k       & $3$k       & $6$k   & $5$k \\
axiom/opaque/quotient &   & $1$k       &            &        & \\
\bottomrule
\end{tabular}
\end{table}

\begin{table}[ht]
\centering
\caption{Degree statistics by declaration type. In-degree measures how often a declaration is cited; out-degree measures how many premises it invokes.}
\label{tab:thm-kind-degree}
\small
\begin{tabular}{lrrrrrrrrr}
\toprule
& & \multicolumn{4}{c}{In-degree} & \multicolumn{4}{c}{Out-degree} \\
\cmidrule(lr){3-6} \cmidrule(lr){7-10}
Kind & Count & Mean & Med.\ & Max & Std & Mean & Med.\ & Max & Std \\
\midrule
axiom       & $3$       & $7{,}823$ & $169$ & $23{,}226$ & $10{,}892$ & $0.7$ & $1$ & $1$ & $0.5$ \\
abbrev      & $6{,}667$ & $438$     & $11$  & $89{,}936$ & $3{,}033$  & $10.0$ & $7$ & $106$ & $10.8$ \\
inductive   & $3{,}813$ & $273$     & $31$  & $44{,}487$ & $1{,}541$  & $4.1$ & $3$ & $54$ & $4.4$ \\
constructor & $4{,}755$ & $59$      & $7$   & $69{,}580$ & $1{,}104$  & $11.7$ & $8$ & $127$ & $10.9$ \\
definition  & $48{,}664$& $58$      & $4$   & $62{,}942$ & $739$      & $16.4$ & $13$ & $146$ & $13.7$ \\
theorem     & $243{,}725$& $5.5$   & $1$   & $58{,}686$ & $237$      & $30.8$ & $21$ & $522$ & $31.2$ \\
quotient    & $3$       & $305$     & $226$ & $596$      & $213$      & $0.3$ & $0$ & $1$ & $0.5$ \\
opaque      & $499$     & $0.5$     & $0$   & $21$       & $1.4$      & $3.9$ & $3$ & $33$ & $3.1$ \\
\bottomrule
\end{tabular}
\end{table}

\begin{table}[ht]
\centering
\caption{Zero-citation rates by namespace (extremes among namespaces with $\ge 100$ theorems).}
\label{tab:thm-leaves}
\small
\begin{tabular}{lrrrlrrr}
\toprule
\multicolumn{4}{c}{Highest zero-citation rate} & \multicolumn{4}{c}{Lowest zero-citation rate} \\
\cmidrule(lr){1-4} \cmidrule(lr){5-8}
Namespace & Total & Zero & Rate & Namespace & Total & Zero & Rate \\
\midrule
\ns{AddEquiv}           & $129$  & $119$ & $92.2\%$ & \ns{Primrec}            & $133$ & $17$ & $12.8\%$ \\
\ns{EMetric}            & $192$  & $163$ & $84.9\%$ & \ns{AnalyticAt}         & $102$ & $15$ & $14.7\%$ \\
\ns{DomMulAct}          & $105$  & $89$  & $84.8\%$ & \ns{AkraBazziRecurrence}& $129$ & $21$ & $16.3\%$ \\
\ns{Ordering}           & $127$  & $105$ & $82.7\%$ & \ns{ContinuousWithinAt} & $119$ & $20$ & $16.8\%$ \\
\ns{LocallyConstant}    & $113$  & $93$  & $82.3\%$ & \ns{HasFDerivWithinAt}  & $120$ & $21$ & $17.5\%$ \\
\ns{MulEquiv}           & $243$  & $194$ & $79.8\%$ & \ns{ConvexOn}           & $111$ & $20$ & $18.0\%$ \\
\ns{AddSubgroup}        & $111$  & $88$  & $79.3\%$ & \ns{LipschitzWith}      & $113$ & $21$ & $18.6\%$ \\
\ns{AddMonoidHom}       & $138$  & $108$ & $78.3\%$ & \ns{DifferentiableAt}   & $113$ & $21$ & $18.6\%$ \\
\bottomrule
\end{tabular}
\end{table}

\begin{table}[ht]
\centering
\caption{Edge classification in $G_{\mathrm{thm}}$ by module and namespace relationship.}
\label{tab:thm-edge-breakdown}
\begin{tabular}{lrr}
\toprule
Category & Edges & Percentage \\
\midrule
Same module                    & $811{,}649$   & $9.6\%$ \\
Cross-module, same namespace   & $805{,}772$   & $9.6\%$ \\
Cross-namespace                & $4{,}296{,}412$ & $50.9\%$ \\
Missing module info            & $2{,}522{,}533$ & $29.9\%$ \\
\bottomrule
\end{tabular}
\end{table}

\begin{table}[ht]
\centering
\caption{Top~10 cross-namespace edge pairs in $G_{\mathrm{thm}}$, ordered by edge count.}
\label{tab:thm-cross-ns-pairs}
\small
\begin{tabular}{llr}
\toprule
Namespace A & Namespace B & Edges \\
\midrule
\ns{AlgebraicGeometry}  & \ns{CategoryTheory}    & $38{,}042$ \\
\ns{CategoryTheory}     & \ns{Quiver}            & $29{,}286$ \\
\ns{CategoryTheory}     & \ns{Eq}                & $27{,}927$ \\
\ns{CategoryTheory}     & \ns{HomologicalComplex} & $15{,}829$ \\
\ns{MeasureTheory}      & \ns{Real}              & $15{,}668$ \\
\ns{MeasureTheory}      & \ns{Set}               & $11{,}954$ \\
\ns{MeasureTheory}      & \ns{ProbabilityTheory} & $11{,}734$ \\
\ns{Eq}                 & \ns{MeasureTheory}     & $10{,}385$ \\
\ns{Complex}            & \ns{Real}              & $10{,}370$ \\
\ns{ENNReal}            & \ns{MeasureTheory}     & $10{,}302$ \\
\bottomrule
\end{tabular}
\end{table}

\subsubsection{Empirical Observations from $G_{\mathrm{thm}}$}

\label{sec:thm-types}
The eight declaration kinds occupy distinct structural niches in $G_{\mathrm{thm}}$ (Table~\ref{tab:thm-kind-degree}). The pattern is systematic: axioms sit at the absolute bottom of the dependency chain with the highest average in-degree ($7{,}823$) and near-zero out-degree; theorems occupy the opposite pole with low in-degree (mean~$5.5$) but the highest out-degree (mean~$30.8$, max~$522$). Among the three axioms, \decl{propext} (propositional extensionality) accounts for $23{,}226$ incoming edges; it is invoked in roughly one out of every thirteen declarations. \decl{Classical.choice} ($169$) and \decl{Quot.sound} ($73$) are invoked far less frequently, suggesting that the vast majority of \module{Mathlib}'s reasoning is constructive at the proof-term level, with classical logic concentrated in specific branches.

Inductive types have high in-degree (mean~$273$, median~$31$) and very low out-degree (mean~$4.1$). These are the structural skeletons of mathematics: type definitions that many theorems depend on but that themselves depend on little. Their in-degree top~$10$ reads as a catalogue of core mathematical concepts: \decl{CategoryTheory.Category} ($44{,}487$), \decl{Real} ($26{,}637$), \decl{TopologicalSpace} ($25{,}310$), \decl{CategoryTheory.Functor} ($24{,}295$), \decl{Module} ($23{,}404$), \decl{CommRing} ($20{,}837$).

Theorems occupy the opposite pole: low in-degree (mean~$5.5$, median~$1$) and the highest out-degree (mean~$30.8$, max~$522$). Theorems are \emph{consumers} of dependencies, not providers. The few high-in-degree theorems are not deep mathematical results but equality reasoning lemmas: \decl{Eq.trans} ($58{,}686$), \decl{of\_eq\_true} ($48{,}801$), \decl{congrFun'} ($45{,}516$), \decl{Eq.symm} ($42{,}542$). These are workhorses of Lean's elaborator and tactic framework: proof infrastructure, not mathematical content.

Abbreviations (\texttt{abbrev}) have the most extreme in-degree distribution among non-axiom types: mean~$438$, maximum~$89{,}936$ (\decl{OfNat.ofNat}). These are type-class coercion paths that the elaborator inserts automatically during type-checking. Their high citation counts reflect the machinery of the proof assistant, not the structure of mathematical reasoning. As Baanen et al.~\cite{Baanen_2022} analyze, typeclass design choices (bundled vs.\ unbundled) shape \module{Mathlib}'s dependency structure; the same authors~\cite{baanen2025growing} report that unbundling gave $6$--$19\%$ compilation speedups, explaining why these language-infrastructure nodes dominate our hub structure.

\label{rem:two-layers}
The degree-by-kind analysis reveals that the hub structure of $G_{\mathrm{thm}}$ decomposes into two layers. The \emph{language infrastructure} (\decl{OfNat.ofNat} ($89{,}936$), \decl{Eq.refl} ($69{,}580$), \decl{DFunLike.coe} ($65{,}437$), \decl{Eq.mpr} ($62{,}942$)) consists of Lean type-system artifacts whose prominence reflects proof-assistant design, not mathematics. The \emph{mathematical infrastructure} (\decl{CategoryTheory.Category}, \decl{Real}, \decl{TopologicalSpace}, \decl{Module}, \decl{CommRing}) consists of core concepts whose prominence reflects genuine logical centrality. The module graph conflates these two layers; the declaration graph separates them.

\label{sec:thm-leaves}
Of the $243{,}725$ theorems in $G_{\mathrm{thm}}$, $109{,}088$ ($44.8\%$) have in-degree zero: they are never cited by any other declaration. These \emph{leaf theorems} form the outermost surface of the library, the boundary where formalized knowledge meets the external world. The zero-citation rate varies dramatically across namespaces (Table~\ref{tab:thm-leaves}): namespaces generated by \texttt{@[to\_additive]} reach $80$--$92\%$, while analysis workbench namespaces (\ns{Primrec}, \ns{AnalyticAt}, \ns{HasFDerivWithinAt}) stay below $18\%$, and nearly every theorem in these feeds into downstream proofs.

\label{rem:leaf-theorems}
These leaf theorems form the outer boundary of the dependency graph: any consumption must come from downstream projects (FLT~\cite{flt_lean}, PFR~\cite{Anderson_Formalization_of_the_2023}, Carleson's theorem~\cite{becker2025carleson_blueprint}, sphere eversion~\cite{vandoorn2023sphereeversion}, etc.), scientific libraries (Physlib~\cite{physlib}, SciLean~\cite{skrivan_scilean}), or AI benchmarks (Formal Conjectures~\cite{formal_conjectures_2025}, Equational Theories~\cite{bolan2025equationaltheoriesprojectadvancing}). The dichotomy between high-zero-rate namespaces (systematically generated \texttt{@[to\_additive]} mirrors) and low-zero-rate namespaces (hand-crafted building blocks) reveals a process-trace: the mirroring machinery inflates the graph's surface area without deepening its logical content.

\label{sec:thm-cross-namespace}
Decomposing the $8{,}436{,}366$ edges by boundary type (Table~\ref{tab:thm-edge-breakdown}), only $9.6\%$ remain within the same module file; $50.9\%$ cross namespace boundaries entirely. This is far higher than the $37.1\%$ cross-directory rate in $G_{\mathrm{module}}^{-}$ (\S\ref{sec:cross-namespace}), because module imports package many individual dependencies into a single edge. The strongest cross-namespace coupling is \ns{AlgebraicGeometry} $\leftrightarrow$ \ns{CategoryTheory} ($38{,}042$ edges); a full ranking is given in Table~\ref{tab:thm-cross-ns-pairs}.

\subsubsection{Degree Distribution}
\label{sec:thm-degree}

The degree statistics of $G_{\mathrm{thm}}$ reveal a striking asymmetry between the two directions of dependency.

\begin{table}[ht]
\centering
\caption{Degree statistics for $G_{\mathrm{thm}}$.}
\label{tab:thm-degree}
\begin{tabular}{lrr}
\toprule
Statistic & In-degree & Out-degree \\
\midrule
Mean    & $27.38$    & $27.38$ \\
Median  & $1$        & $19$ \\
Std Dev & $620.10$   & $29.18$ \\
Max     & $89{,}936$ & $522$ \\
Zero-degree nodes & $119{,}633$ & $367$ \\
\bottomrule
\end{tabular}
\end{table}

The in-degree distribution has median~$1$ but mean~$27$ and standard deviation~$620$, a ratio of standard deviation to mean exceeding~$22$, indicating extreme concentration. A handful of declarations accumulate tens of thousands of citations while the majority are referenced at most once. The out-degree distribution, by contrast, has median~$19$, standard deviation~$29$, and maximum~$522$, a far more concentrated profile.

To test whether these distributions follow a power law, we apply the Clauset--Shalizi--Newman method (\S\ref{sec:prelim-powerlaw}).

\begin{table}[ht]
\centering
\caption{Power law fit and distribution comparisons for $G_{\mathrm{thm}}$. The likelihood ratio $R > 0$ favors power law; $R < 0$ favors the alternative. A $p$-value below $0.05$ indicates a statistically significant comparison.}
\label{tab:thm-powerlaw}
\begin{tabular}{lrrrr}
\toprule
& \multicolumn{2}{c}{In-degree} & \multicolumn{2}{c}{Out-degree} \\
\cmidrule(lr){2-3} \cmidrule(lr){4-5}
Metric & Value & $p$ & Value & $p$ \\
\midrule
$\alpha$ (exponent)     & $1.781$ & --- & $2.936$ & --- \\
$x_{\min}$              & $20$    & --- & $38$    & --- \\
$\sigma$ (std.\ error)  & $0.005$ & --- & $0.008$ & --- \\
Tail fraction           & $11.1\%$ & --- & $20.6\%$ & --- \\
\midrule
vs.\ Lognormal ($R$)          & $-2.48$   & $0.174$   & $-1{,}452.5$ & $\approx 0$ \\
vs.\ Exponential ($R$)        & $+28{,}766$ & $\approx 0$ & $-239.9$ & $0.014$ \\
vs.\ Stretched Exp.\ ($R$)    & $+105.8$  & $\approx 0$ & $-1{,}495.9$ & $\approx 0$ \\
vs.\ Truncated Power Law ($R$) & $-21.5$  & $\approx 0$ & $-1{,}509.5$ & $\approx 0$ \\
\bottomrule
\end{tabular}
\end{table}

For in-degree, the tail ($k \ge 20$, comprising $11.1\%$ of non-zero values) is well described by a heavy-tailed distribution with exponent $\alpha \approx 1.78$. Pure power law decisively beats exponential and stretched exponential alternatives, but a \emph{truncated power law} provides a significantly better fit ($R = -21.5$, $p \approx 0$), and the comparison with lognormal is inconclusive ($p = 0.17$). This pattern (heavy tail, uncertain between power law and lognormal, best fit by a truncated power law) is characteristic of many real-world networks~\cite{Clauset_2009}. Across all three dependency layers, we find that truncated power law or lognormal consistently outperforms pure power law, contradicting the casual assumption that code dependency graphs are scale-free.

For out-degree, the picture is categorically different: \emph{every} alternative distribution significantly outperforms pure power law. The best fits are lognormal and truncated power law, both with overwhelmingly negative likelihood ratios.

\label{rem:thm-degree-asymmetry}
This asymmetry has a structural explanation that connects directly to the central tension of the paper. In-degree measures how often a declaration is \emph{cited}, a cumulative, open-ended quantity determined by the logical structure of the completed library (the product). There is no upper bound: \decl{OfNat.ofNat} can be cited by any proof that involves a numeral, and its in-degree ($89{,}936$) reflects a form of preferential attachment in which foundational declarations attract ever more dependents. Out-degree, by contrast, measures how many premises a single proof \emph{invokes}, a quantity bounded by the cognitive and logical complexity of that proof (the process). No human-written proof cites $90{,}000$ lemmas; the maximum out-degree of~$522$ reflects the practical ceiling of proof complexity. The in-degree distribution records the product; the out-degree distribution records the process.

\begin{figure}[ht]
\centering
\includegraphics[width=\textwidth]{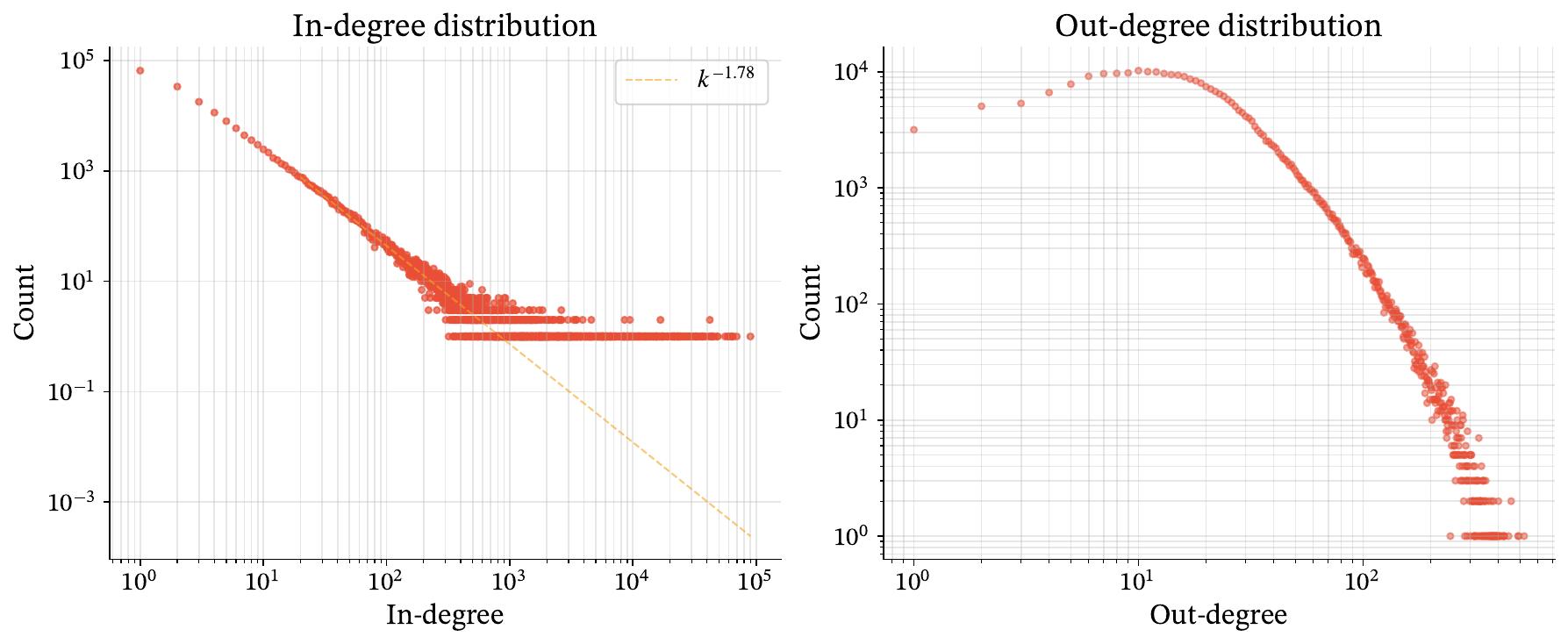}
\caption{Degree distributions of $G_{\mathrm{thm}}$ on log--log axes. Left: in-degree, with power law fit (gold line, $\alpha = 1.78$, $x_{\min} = 20$). Right: out-degree. The in-degree tail extends over four orders of magnitude; the out-degree distribution is sharply bounded.}
\label{fig:thm-degree}
\end{figure}

\label{rem:thm-transitive-reduction}
Unlike $G_{\mathrm{module}}$, where transitive reduction removes $17.5\%$ of edges (\S\ref{sec:transitive-reduction}), $G_{\mathrm{thm}}$ does not admit a meaningful transitive reduction: every edge is extracted mechanically from the proof term, so the data records exactly what the compiler sees. Module-level redundancy measures the gap between compiler requirements and human import choices; at the declaration level, this gap vanishes.

\subsubsection{DAG Depth}
\label{sec:thm-dag-depth}

$G_{\mathrm{thm}}$ is a DAG with $84$ topological layers, roughly half the depth of $G_{\mathrm{module}}$'s $154$ layers. This compression is counterintuitive: the finer-grained graph has \emph{fewer} vertical layers. The explanation lies in transitivity. Each module-level layer spans multiple files that import each other; at the declaration level, the dependencies within those files are resolved in parallel rather than sequentially. The result is a shallower but vastly wider structure: the source layer alone contains $119{,}633$ declarations ($38.8\%$ of all nodes), that is, declarations with zero in-degree, i.e., declarations that no other declaration cites. By contrast, only $401$ declarations are sinks (zero out-degree).

\begin{figure}[ht]
\centering
\includegraphics[width=\textwidth]{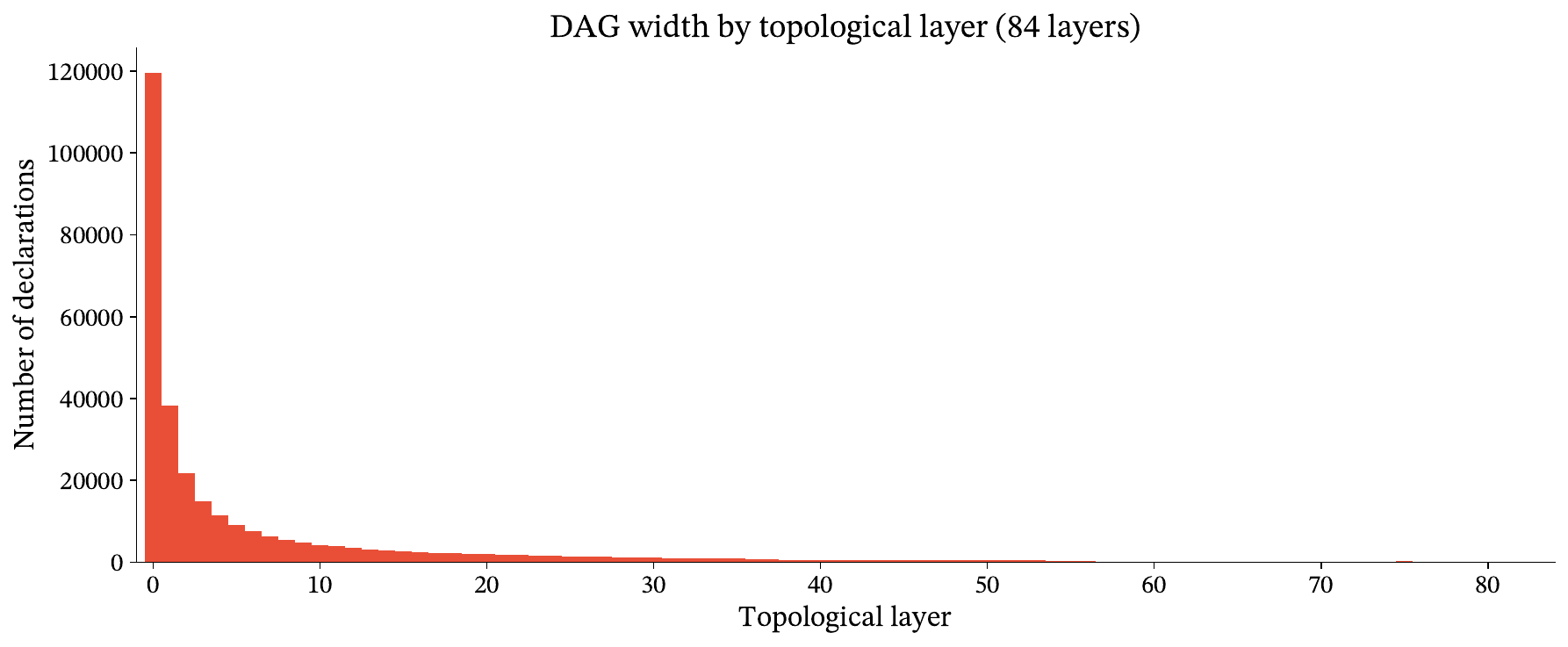}
\caption{DAG width by topological layer for $G_{\mathrm{thm}}$ ($84$ layers). The source layer ($119{,}633$ zero-citation declarations) dominates; the remaining layers decay rapidly, reaching single digits by layer~$60$.}
\label{fig:thm-dag-structure}
\end{figure}

The $119{,}633$ source nodes correspond to the zero-citation declarations of \S\ref{sec:thm-leaves}, giving the declaration-level DAG its characteristic top-heavy profile.

\subsubsection{Centrality}
\label{sec:thm-centrality}

In \S\ref{sec:centrality}, we found that in-degree, PageRank, and betweenness centrality identify different structural roles at the module level. We now compute the same three measures on $G_{\mathrm{thm}}$ and find that the divergence between them is, if anything, more extreme.

\begin{table}[ht]
\centering
\caption{Top~10 declarations by in-degree, PageRank ($\alpha = 0.85$), and betweenness centrality (sampled, $k = 500$) in $G_{\mathrm{thm}}$.}
\label{tab:thm-centrality}
\small
\begin{tabular}{rlrlrl}
\toprule
\multicolumn{2}{c}{In-degree} & \multicolumn{2}{c}{PageRank} & \multicolumn{2}{c}{Betweenness} \\
\cmidrule(lr){1-2} \cmidrule(lr){3-4} \cmidrule(lr){5-6}
$\deg^{-}$ & Declaration & PR & Declaration & $c_B$ & Declaration \\
\midrule
$89{,}936$ & \decl{OfNat.ofNat}       & $.0151$ & \decl{CT.Category.mk}    & $.000338$ & \decl{Real} \\
$69{,}580$ & \decl{Eq.refl}           & $.0145$ & \decl{OfNat.mk}           & $.000338$ & \decl{Real.ofCauchy} \\
$65{,}437$ & \decl{DFunLike.coe}      & $.0140$ & \decl{outParam}           & $.000289$ & \decl{Rat.linearOrder} \\
$63{,}530$ & \decl{Membership.mem}    & $.0125$ & \decl{OfNat}              & $.000256$ & \decl{Real.pi} \\
$62{,}942$ & \decl{Eq.mpr}            & $.0123$ & \decl{CT.Category}        & $.000253$ & \decl{Real.$\exists$cos\_eq\_0} \\
$59{,}997$ & \decl{Set}               & $.0106$ & \decl{Eq.refl}            & $.000183$ & \decl{Rat.instIsStrictOrd.} \\
$58{,}686$ & \decl{Eq.trans}          & $.0097$ & \decl{Quiver}             & $.000177$ & \decl{Rat.le\_refl} \\
$56{,}019$ & \decl{Semiring.toNAR}    & $.0095$ & \decl{Quiver.mk}          & $.000164$ & \decl{Real.zero\_lt\_one} \\
$48{,}801$ & \decl{of\_eq\_true}      & $.0084$ & \decl{LE}                 & $.000157$ & \decl{Complex.log} \\
$48{,}295$ & \decl{PO.toPreorder}     & $.0084$ & \decl{LE.mk}              & $.000155$ & \decl{instSemiringNNReal} \\
\bottomrule
\end{tabular}
\end{table}

\noindent
\textit{Abbreviations:} CT = CategoryTheory, NAR = NonAssocSemiring, PO = PartialOrder, instIsStrictOrd.\ = instIsStrictOrderedRing, $\exists$cos\_eq\_0 = exists\_cos\_eq\_zero.

\medskip

The three columns of Table~\ref{tab:thm-centrality} have almost no overlap, and each tells a different story about the structure of \module{Mathlib}.

\emph{In-degree} is dominated by language infrastructure. The top entries (\decl{OfNat.ofNat}, \decl{Eq.refl}, \decl{DFunLike.coe}, \decl{Membership.mem}, \decl{Eq.mpr}) are type-class coercions and equality constructors that the elaborator inserts into virtually every proof term. Their prominence reflects the design of Lean~4, not the structure of mathematics. This contrasts with the module-level in-degree ranking (Table~\ref{tab:centrality}), where the top entry (\module{Mathlib.Init}) is a file that every module must import; the underlying mechanism (implicit foundational dependency) is analogous, but the theorem level reveals the specific \emph{declarations} responsible.

\emph{PageRank} promotes a different set of nodes: the constructors and types of category theory (\decl{CategoryTheory.Category.mk}, \decl{Quiver}, \decl{Quiver.mk}) and core algebraic infrastructure (\decl{OfNat}, \decl{LE}, \decl{Zero}). The key divergence from in-degree is \decl{CategoryTheory.Category.mk}: it ranks first in PageRank but does not appear in the in-degree top~$20$. Its PageRank reflects not its direct citation count but the fact that the entire $59{,}724$-node category theory subgraph depends on it transitively. PageRank captures \emph{recursive} importance: it identifies the foundations on which large, internally coherent subgraphs rest.

\emph{Betweenness centrality} reveals a third, entirely disjoint set of structurally critical declarations. The top entries (\decl{Real}, \decl{Real.ofCauchy}, \decl{Rat.linearOrder}, \decl{Real.pi}, \decl{Real.exists\_cos\_eq\_zero}) trace the construction chain of the real numbers. These nodes sit at the narrow passage connecting the algebraic world (rational numbers, ordered rings, number fields) to the analytic world (measure theory, complex analysis, probability). \decl{Real.pi} occupies rank~$4$ because $\pi$'s definition threads through number theory, trigonometry, and analysis, making it a crossroads of multiple mathematical disciplines. Only $34{,}564$ of $308{,}129$ nodes ($11.2\%$) have nonzero betweenness, confirming that shortest paths through $G_{\mathrm{thm}}$ are concentrated through a small number of critical bridges. (Note: betweenness centrality is approximated via $k = 500$ pivot nodes; the resulting rankings are stable for high-betweenness nodes but may be noisy in the tail.)

\begin{figure}[H]
\centering
\begin{subfigure}[t]{0.32\textwidth}
\centering
\includegraphics[width=\textwidth]{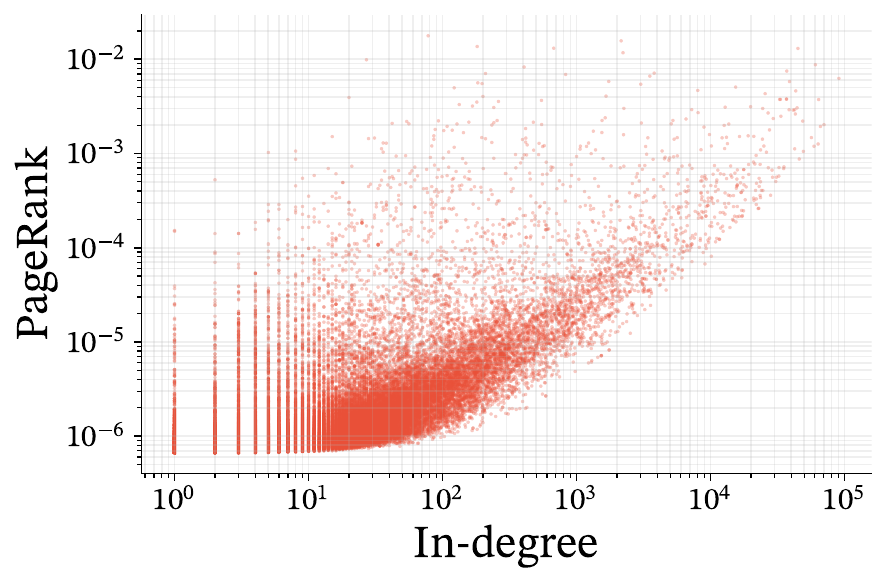}
\caption{In-degree vs.\ PageRank.}
\label{fig:thm-centrality-indeg-pr}
\end{subfigure}%
\hfill
\begin{subfigure}[t]{0.32\textwidth}
\centering
\includegraphics[width=\textwidth]{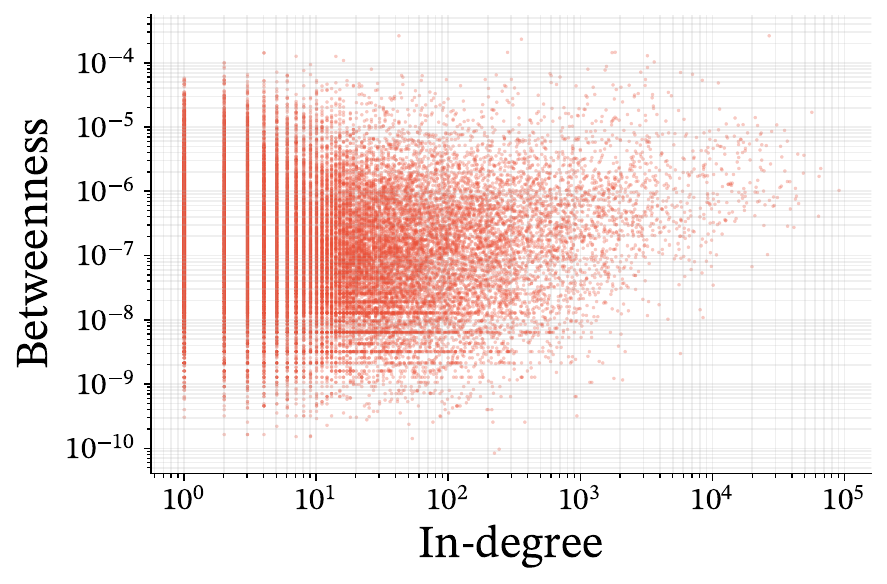}
\caption{In-degree vs.\ Betweenness.}
\label{fig:thm-centrality-indeg-betw}
\end{subfigure}%
\hfill
\begin{subfigure}[t]{0.32\textwidth}
\centering
\includegraphics[width=\textwidth]{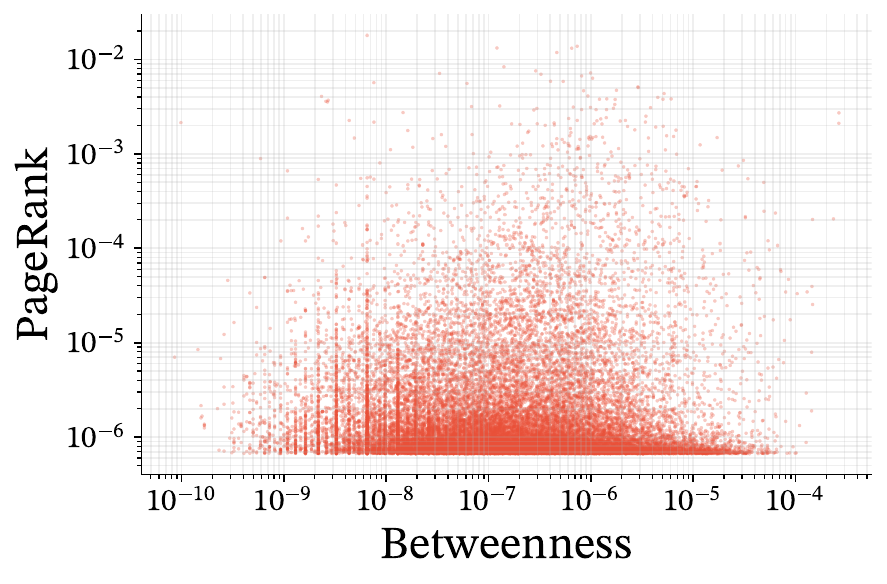}
\caption{Betweenness vs.\ PageRank.}
\label{fig:thm-centrality-betw-pr}
\end{subfigure}
\caption{Pairwise centrality scatter plots for $G_{\mathrm{thm}}$. Each point is one declaration. The wide scatter across all three panels confirms that in-degree, PageRank, and betweenness capture distinct structural roles; the three-way divergence pervades the entire distribution, not just the top-$10$ extremes.}
\label{fig:thm-centrality}
\end{figure}

Figure~\ref{fig:thm-centrality} visualizes the pairwise relationships between the three centrality measures.

\label{rem:thm-centrality-vs-module}
The three-way divergence of centrality measures observed at the module level (\S\ref{sec:centrality}) intensifies at the theorem level: no declaration appears in all three top-$10$ lists of Table~\ref{tab:thm-centrality}. The taxonomy of importance proposed in \S\ref{sec:cross-level-summary} (\emph{vocabulary} (in-degree), \emph{axiomatic core} (PageRank), \emph{structural bridges} (betweenness)) receives sharper instantiation here. At the module level, the centrality rankings are partially confounded by file-organization artifacts (e.g., \module{Mathlib.Init} ranks high in all three measures simply because it is imported by every file). At the theorem level, these artifacts are stripped away, and the three measures separate cleanly into language infrastructure (in-degree), mathematical foundations (PageRank), and inter-disciplinary bridges (betweenness).

\subsubsection{Community Structure}
\label{sec:thm-community}

We apply Louvain community detection (\S\ref{sec:prelim-community}) to the largest weakly connected component of $G_{\mathrm{thm}}$ (converted to an undirected graph, $308{,}054$ nodes, $8{,}431{,}648$ undirected edges).

\begin{table}[ht]
\centering
\caption{Community detection summary for $G_{\mathrm{thm}}$.}
\label{tab:thm-community-summary}
\begin{tabular}{lr}
\toprule
Metric & Value \\
\midrule
Number of communities & $22$ \\
Modularity & $0.4757$ \\
Communities with ${>}1{,}000$ nodes & $7$ \\
Communities with ${>}100$ nodes & $11$ \\
\bottomrule
\end{tabular}
\end{table}

The modularity of $0.48$ indicates a moderately strong community structure, significantly above random ($0$) but below perfectly modular ($1$). If logical dependency alone determined community structure (pure product), modularity would be higher; if cognitive organization were independent of logic (pure process), it would approach zero. The intermediate value quantifies the partial, but incomplete, alignment between mathematical necessity and human practice. The $22$ communities are highly unequal: seven communities of more than $1{,}000$ nodes account for $99.1\%$ of the graph; the remaining fifteen communities contain fewer than $700$ nodes each.

\begin{table}[ht]
\centering
\caption{The seven major communities of $G_{\mathrm{thm}}$ (Louvain, $> 1{,}000$ nodes), with size, dominant mathematical theme, and top representatives by PageRank.}
\label{tab:thm-communities}
\small
\begin{tabular}{rrllp{4cm}}
\toprule
ID & Size & Theme & Top namespace(s) & Representatives \\
\midrule
0  & $63{,}763$ & Order / Sets / Filters & \ns{Set}, \ns{Filter}, \ns{Matroid} & \decl{LE}, \decl{Set}, \decl{Iff.intro}, \decl{Zero} \\
5  & $59{,}724$ & Category Theory & \ns{CategoryTheory} (69\%) & \decl{CT.Category.mk}, \decl{Quiver}, \decl{CT.CategoryStruct} \\
2  & $54{,}649$ & Algebra & \ns{LinearMap}, \ns{Matrix}, \ns{Module} & \decl{outParam}, \decl{Zero.mk}, \decl{Semiring} \\
1  & $49{,}739$ & Combinatorics / Discrete & \ns{List}, \ns{Finset}, \ns{SimpleGraph} & \decl{OfNat.mk}, \decl{Eq.refl}, \decl{List.nil} \\
11 & $38{,}245$ & Number Theory / Polynomials & \ns{Nat}, \ns{Int}, \ns{Polynomial} & \decl{OfNat}, \decl{Mul}, \decl{Semiring.mk} \\
7  & $37{,}368$ & Analysis / Probability & \ns{MeasureTheory}, \ns{Real} & \decl{Real}, \decl{MeasurableSpace} \\
8  & $2{,}773$ & Model Theory / Set Theory & \ns{FirstOrder}, \ns{SetTheory} & \decl{FirstOrder.Language.mk} \\
\bottomrule
\end{tabular}
\end{table}

These seven communities map recognizably onto traditional mathematical subdisciplines. The alignment is not imposed by the analysis; it emerges from the dependency topology alone. Louvain knows nothing about mathematical content; it optimizes modularity on the undirected edge structure, yet it recovers divisions (algebra vs.\ analysis, category theory vs.\ combinatorics) that correspond to centuries-old disciplinary boundaries.

To quantify the alignment between community structure and the namespace hierarchy, we compute the Normalized Mutual Information (NMI) and Adjusted Rand Index (ARI, Definition~\ref{def:nmi}--\ref{def:ari}) between two labelings of the $308{,}054$ nodes: the Louvain community assignment and the top-level namespace extracted from each declaration's module path.

\begin{table}[ht]
\centering
\caption{Alignment between Louvain communities and top-level namespaces.}
\label{tab:thm-nmi}
\begin{tabular}{lrl}
\toprule
Metric & Value & Interpretation \\
\midrule
NMI & $0.3432$ & Moderate association \\
ARI & $0.1817$ & Weak-to-moderate agreement \\
\bottomrule
\end{tabular}
\end{table}

The moderate NMI and low ARI confirm that dependency topology and namespace organization are \emph{related but far from identical}. The most striking case is \ns{CategoryTheory}: $97.4\%$ of its $42{,}094$ declarations fall into Community~$5$, and conversely, $68.7\%$ of Community~$5$ belongs to the \ns{CategoryTheory} namespace. This near-perfect correspondence is unique among the major communities. By contrast, the largest community (Community~$0$, $63{,}763$ nodes) draws its largest namespace contribution from root-level declarations ($17.1\%$), followed by \ns{Set} ($8.1\%$) and \ns{MeasureTheory} ($4.6\%$); no single namespace dominates.

\label{rem:nmi-tension}
The NMI of $0.34$ provides a theorem-level metric for the tension between the module tree $T$ and the logical dependency graph $G$, extending the cross-namespace analysis of \S\ref{sec:cross-namespace}. At the module level, $37.1\%$ of reduced import edges cross namespace boundaries; at the theorem level, the NMI of $0.34$ quantifies the same phenomenon from a complementary angle: the dependency-based community structure aligns only moderately with the human-imposed namespace hierarchy. These two measures converge on the same conclusion: cognitive classification and logical structure are correlated but distinct, and the gap between them widens as granularity increases.

The exceptional case of category theory deserves emphasis. Its near-complete community--namespace alignment ($97.4\%$) suggests that category theory is not merely a namespace convention but a genuinely \emph{autonomous} mathematical discipline within \module{Mathlib}: its internal dependency structure is dense enough to form a self-contained community with minimal need for external foundations beyond the shared core. No other major namespace achieves this degree of self-containment.

\subsubsection{Robustness}
\label{sec:thm-robustness}

We assess the resilience of $G_{\mathrm{thm}}$ under two node-removal strategies: random removal and targeted removal in order of decreasing PageRank.

\paragraph{Single-node removal.} Removing any one of the top-$30$ PageRank nodes has negligible impact on connectivity. The largest effect is the removal of \decl{Eq.refl}, which disconnects exactly $6$ of $308{,}054$ nodes from the giant component ($0.002\%$). No single declaration is a critical point of failure. This contrasts with the module-level finding (Table~\ref{tab:robustness}), where removing the top-$5$ modules fragments $G_{\mathrm{module}}$ into $29$ components. The difference arises because module-level removal is coarser: removing a module removes all declarations it contains, severing many edges simultaneously. At the theorem level, each declaration is individually expendable.

Under progressive node removal, the asymmetry between random and targeted attack becomes apparent.

\begin{table}[ht]
\centering
\caption{Robustness of $G_{\mathrm{thm}}$ under progressive node removal. The table reports the fraction of nodes remaining in the largest weakly connected component.}
\label{tab:thm-robustness}
\begin{tabular}{rrrl}
\toprule
Removal \% & Random & Targeted & Gap \\
\midrule
$1\%$  & $0.990$ & $0.975$ & $0.015$ \\
$5\%$  & $0.950$ & $0.868$ & $0.082$ \\
$10\%$ & $0.899$ & $0.722$ & $0.177$ \\
$15\%$ & $0.849$ & $0.583$ & $0.266$ \\
$20\%$ & $0.799$ & $0.461$ & $0.338$ \\
\bottomrule
\end{tabular}
\end{table}

\begin{figure}[ht]
\centering
\includegraphics[width=\textwidth]{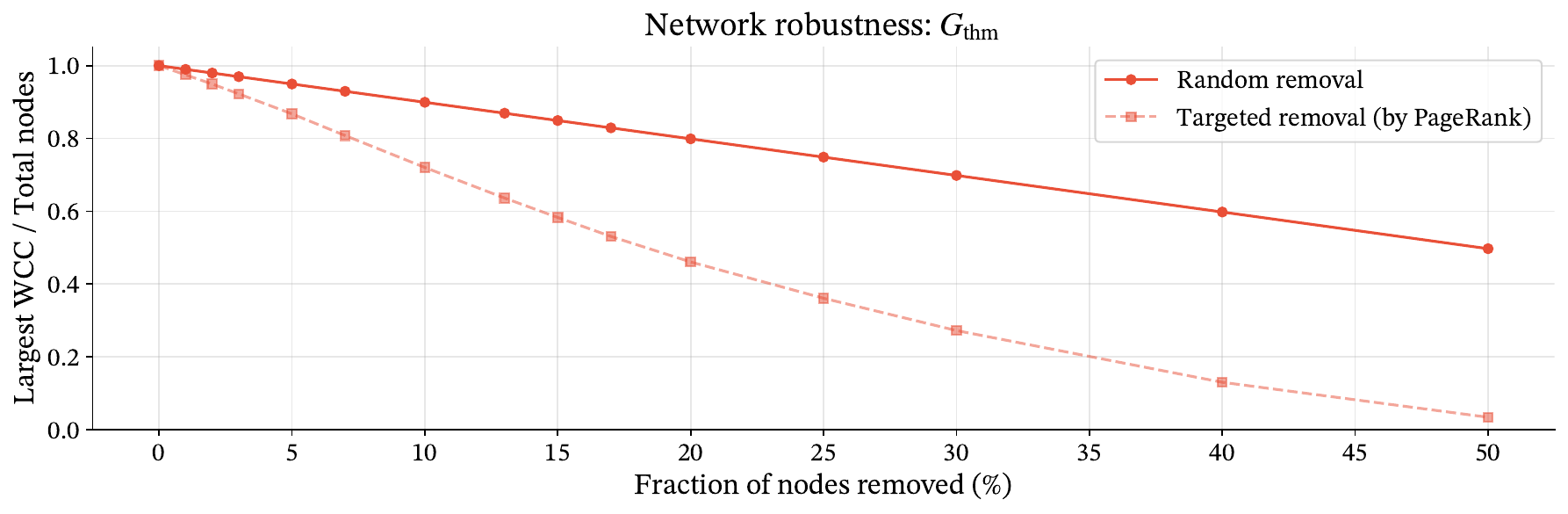}
\caption{Robustness curves for $G_{\mathrm{thm}}$: fraction of nodes in the largest WCC as a function of the fraction removed, under random removal (indigo) and targeted removal by PageRank (gold).}
\label{fig:thm-robustness}
\end{figure}

Under random removal, the giant component shrinks almost linearly: removing $20\%$ of nodes leaves $80\%$ connected. Under targeted attack (removing the highest-PageRank nodes first), $20\%$ removal reduces connectivity to $46\%$, a $34$-percentage-point gap. This is the classic signature of a hub-dominated network~\cite{Albert_2000,Baraba_si_1999}: resilient to random failure, vulnerable to targeted attack on hubs.

\label{rem:thm-robustness-redundancy}
The extreme resilience to single-node removal at the theorem level, where no single declaration is a critical bridge, has a deeper explanation than mere edge density. It reflects the inherent \emph{logical redundancy} of mathematical proof. The same lemma is typically reachable through multiple independent proof paths: \decl{Eq.refl} can be circumvented via \texttt{rfl}, \decl{Eq.mpr}, or tactic-generated proof terms; a group-theoretic fact can be accessed through the group axioms or through the ring axioms that subsume them. This redundancy is not engineered (unlike the module-level redundancy of umbrella imports discussed in \S\ref{sec:transitive-reduction}); it is an intrinsic property of mathematical logic. The product carries its own structural insurance.

\subsubsection{Theorem vs.\ Lemma: Validating Human Importance Judgments}
\label{sec:thm-vs-lemma}

In Lean~4 source code, developers choose between \texttt{theorem} and \texttt{lemma} to declare propositions. While the kernel treats both identically; \texttt{lemma} is syntactic sugar for \texttt{theorem}, and the choice carries cognitive significance: \texttt{theorem} typically marks a major result, while \texttt{lemma} signals an auxiliary stepping stone. Our extraction pipeline records both as kind \texttt{theorem}, erasing this distinction.

To recover it, we parsed the \module{Mathlib} source code, tracking \texttt{namespace} blocks to reconstruct fully qualified names, and cross-referenced with~$G_{\mathrm{thm}}$. Of the $251{,}642$ propositions in the graph, $175{,}567$ ($69.8\%$) were matched to a source declaration: $129{,}444$ marked \texttt{theorem} and $46{,}123$ marked \texttt{lemma}.

\begin{table}[ht]
\centering
\begin{tabular}{lccc}
\toprule
Metric & \texttt{theorem} & \texttt{lemma} & Ratio \\
\midrule
Mean in-degree        & $5.16$ & $3.52$ & $1.47\times$ \\
Zero-citation rate    & $37.2\%$ & $39.4\%$ & --- \\
Mean PageRank         & $7.52 \times 10^{-7}$ & $6.83 \times 10^{-7}$ & $1.10\times$ \\
\bottomrule
\end{tabular}
\caption{Structural comparison of declarations marked \texttt{theorem} vs.\ \texttt{lemma} in source code. All three metrics confirm that human importance judgments align with topological prominence.}
\label{tab:thm-vs-lemma}
\end{table}

All three metrics confirm that human importance judgments are validated by the logical structure: declarations that developers considered major results do occupy more prominent positions in the dependency network. The effect is moderate, a $1.47\times$ ratio in citation count, not an order-of-magnitude difference, suggesting that the theorem/lemma boundary is a noisy but genuine signal of structural importance.

\label{rem:misjudged}
The alignment is imperfect, however. Among the notable outliers: \decl{funext} (function extensionality) is marked \texttt{lemma} in source code but has in-degree $15{,}574$, making it one of the most cited propositions in \module{Mathlib}. Conversely, $48{,}177$ declarations marked \texttt{theorem}, over a third of all matched theorems, have zero citations. The former represents underestimation of structural importance by the author; the latter represents overestimation, or more precisely, the distinction between importance as a mathematical result (worthy of the name ``theorem'') and importance as a dependency hub (actually reused by other proofs). In the language of our framework, the theorem/lemma distinction is a trace of the process, a record of the author's cognitive judgment at the time of writing, that the product partially but not fully corroborates.

\subsubsection{Additional Empirical Observations}
\label{sec:additional-empirical}

The graph definitions of \S\ref{sec:graph-definitions} assign rich metadata to each declaration node. This section reports two completed empirical analyses that characterise the \emph{process} layer of the dependency graph: definitional height (a kernel-level measure of implicit coupling) and tactic usage (a proof-strategy fingerprint).

\label{sec:definitional-height}
Lean's type checker relies on \emph{definitional equality}: to verify that two types match, the kernel may need to unfold a chain of definitions. For each definition, Lean computes a \emph{definitional height} $\delta(d)$, the number of regular (semireducible) definitions in its transitive dependency chain, stored in the \texttt{ReducibilityHints} field. This quantity is invisible in the explicit dependency graph $G_{\mathrm{thm}}$ (it does not appear in $\mathrm{refs}(d.\pi)$) yet it constitutes a form of \emph{implicit coupling}: declarations with high~$\delta$ are brittle, because changes to any definition along the unfolding chain can break type-checking even if the explicit premise set is unchanged.

We extracted $\delta(d)$ for all $101{,}021$ Mathlib definitions via the \texttt{DefinitionVal.hints} API. Of these, $53{,}962$ are \emph{regular} (semireducible) with a numeric height, $42{,}786$ are \emph{abbreviations} (always unfolded, no height), and $4{,}273$ are \emph{opaque} (never unfolded). Among regular definitions, the height distribution has median~$7$, mean~$10.1$, IQR~$[4, 14]$, and a maximum of~$60$. The distribution is roughly log-normal: $58\%$ of definitions have height $< 10$, while a tail of $13.5\%$ has height $\ge 20$. This confirms that most definitions sit at shallow depth, but a non-trivial fraction involves long unfolding chains that may contribute to elaboration and compilation bottlenecks.

\label{sec:tactic-usage}
To extract per-declaration tactic usage, we employ \texttt{jixia}~\cite{jixia2024}, a static analysis tool for Lean~4 developed at BICMR, Peking University (see also~\cite{gao2025herald}). Running \texttt{jixia}'s declaration plugin over all $7{,}563$ Mathlib source files yields the pretty-printed proof term for each declaration, from which we parse tactic names.

Of $235{,}586$ declarations extracted, $170{,}985$ are theorems ($72.6\%$), $35{,}635$ definitions ($15.1\%$), $24{,}947$ instances ($10.6\%$), and $4{,}019$ other (abbreviations, examples, opaques, inductives).
Among theorems, $76{,}708$ ($44.9\%$) use tactic mode (proof begins with \texttt{by}) and $94{,}277$ ($55.1\%$) use term mode. An additional $3{,}202$ non-theorem declarations (primarily instances) also employ tactic proofs, bringing the total tactic-mode count to $79{,}910$.

Tactic-mode proofs contain $325{,}454$ tactic steps in total. The distribution of steps per proof has median~$2$, mean~$4.1$, and IQR~$[1, 4]$, with a maximum of~$173$. Single-step proofs (\texttt{by simp}, \texttt{by exact \ldots}, etc.) account for $39.2\%$ of all tactic proofs, and $68.7\%$ use at most three steps. Only $9.4\%$ require ten or more steps. This confirms that most Mathlib proofs are short: the library favors fine-grained decomposition into small lemmas over long monolithic proof scripts.

Table~\ref{tab:tactic-freq} lists the fifteen most frequently invoked tactics. The top three (\texttt{rw} ($16.9\%$), \texttt{simp} ($11.2\%$), and \texttt{exact} ($10.4\%$)) together account for $38.5\%$ of all tactic steps. The top ten account for $67.0\%$. The dominance of \texttt{rw} (rewriting) reflects Mathlib's equational style: the most common proof strategy is to transform the goal by substituting known equalities. The prominence of \texttt{simp} (simplification) and \texttt{exact} (term-mode embedding) indicates that automation and direct proof-term construction are the next most common strategies.

\begin{table}[ht]
\centering
\small
\caption{Top 15 tactics by invocation count across all $79{,}910$ tactic-mode proofs ($325{,}454$ total steps).}
\label{tab:tactic-freq}
\begin{tabular}{l r r}
\toprule
\textbf{Tactic} & \textbf{Count} & \textbf{\%} \\
\midrule
\texttt{rw}       & 55,074 & 16.9 \\
\texttt{simp}     & 36,380 & 11.2 \\
\texttt{exact}    & 33,767 & 10.4 \\
\texttt{have}     & 23,939 &  7.4 \\
\texttt{refine}   & 18,158 &  5.6 \\
\texttt{apply}    & 12,639 &  3.9 \\
\texttt{obtain}   & 12,020 &  3.7 \\
\texttt{intro}    & 10,321 &  3.2 \\
\texttt{simpa}    &  8,372 &  2.6 \\
\texttt{ext}      &  7,408 &  2.3 \\
\texttt{rcases}   &  6,519 &  2.0 \\
\texttt{simp\_rw} &  5,933 &  1.8 \\
\texttt{let}      &  5,439 &  1.7 \\
\texttt{rintro}   &  5,304 &  1.6 \\
\texttt{by\_cases} & 3,959 &  1.2 \\
\midrule
\emph{Top 3}      & 125,221 & 38.5 \\
\emph{Top 10}     & 218,070 & 67.0 \\
\bottomrule
\end{tabular}
\end{table}

Tactic usage varies significantly across mathematical domains. Table~\ref{tab:tactic-ns} shows the top tactic for each of the ten largest namespaces. Notably, \texttt{rw} is the most common tactic in eight of ten namespaces, but \texttt{simp} leads in \module{CategoryTheory} (where diagrammatic reasoning is heavily automated) and \texttt{exact} ties with \texttt{rw} in \module{Topology} (where point-set arguments are often resolved by direct term construction).

To quantify these differences, we compute the Jensen--Shannon divergence (JSD) between the tactic frequency distributions of namespace pairs. The most divergent pair is \module{CategoryTheory} vs.\ \module{MeasureTheory} ($\mathrm{JSD} = 0.180$), reflecting fundamentally different proof styles: categorical proofs rely heavily on \texttt{simp}, \texttt{aesop}, and \texttt{ext}, while measure-theoretic proofs favor \texttt{have}, \texttt{exact}, and \texttt{filter\_upwards}. The most similar pairs are \module{Algebra} vs.\ \module{LinearAlgebra} ($\mathrm{JSD} = 0.052$) and \module{RingTheory} vs.\ \module{LinearAlgebra} ($\mathrm{JSD} = 0.052$), confirming that algebraic subfields share closely related proof methodologies.

\begin{table}[ht]
\centering
\small
\caption{Tactic usage by top-level namespace (top 10 by tactic-proof count). The three most frequent tactics in each namespace are listed with their share of that namespace's total tactic steps.}
\label{tab:tactic-ns}
\begin{tabular}{l r l}
\toprule
\textbf{Namespace} & \textbf{Proofs} & \textbf{Top 3 tactics (\%)} \\
\midrule
\module{Algebra}         & 11,390 & \texttt{rw} (20), \texttt{simp} (13), \texttt{exact} (9) \\
\module{Analysis}        &  9,677 & \texttt{rw} (16), \texttt{simp} (11), \texttt{exact} (10) \\
\module{Data}            &  8,598 & \texttt{rw} (18), \texttt{simp} (14), \texttt{exact} (9) \\
\module{RingTheory}      &  6,633 & \texttt{rw} (18), \texttt{simp} (10), \texttt{exact} (10) \\
\module{Topology}        &  5,741 & \texttt{rw} (14), \texttt{exact} (14), \texttt{simp} (9) \\
\module{CategoryTheory}  &  5,625 & \texttt{simp} (13), \texttt{rw} (13), \texttt{exact} (9) \\
\module{MeasureTheory}   &  4,631 & \texttt{rw} (15), \texttt{exact} (13), \texttt{have} (10) \\
\module{LinearAlgebra}   &  4,323 & \texttt{rw} (18), \texttt{simp} (12), \texttt{exact} (8) \\
\module{Order}           &  3,780 & \texttt{rw} (16), \texttt{simp} (14), \texttt{exact} (14) \\
\module{NumberTheory}    &  3,049 & \texttt{rw} (18), \texttt{have} (10), \texttt{exact} (10) \\
\bottomrule
\end{tabular}
\end{table}

\subsubsection{Extended Analysis}\label{app:decl-analysis}

The two-layer hub decomposition (Remark~\ref{rem:two-layers}) has a key implication: the separation between language infrastructure and mathematical infrastructure is invisible at the module level, where both layers cohabit the same source files. The language layer is a trace of the process (the specific proof assistant chosen) embedded in the product. The mathematical infrastructure layer plausibly reflects genuine mathematical centrality, but the language infrastructure layer is an artifact of Lean's type system whose prominence would differ under a different proof assistant. Distinguishing intrinsic mathematical centrality from path-dependent formalization artifacts remains an open problem.

Louvain community detection recovers seven major mathematical disciplines from the dependency topology alone (Table~\ref{tab:thm-communities}). Among the major namespaces, only \ns{CategoryTheory} achieves near-complete community correspondence ($97.4\%$), suggesting that it is genuinely autonomous in its dependency structure. All other communities are cross-namespace mixtures, confirming that the dependency topology discovers a structure related to, but distinct from, the human-imposed namespace hierarchy.

\medskip

Table~\ref{tab:thm-vs-module} summarizes the key quantitative contrasts between the two levels of analysis.

\begin{table}[ht]
\centering
\caption{Module-level ($G_{\mathrm{module}}$) vs.\ theorem-level ($G_{\mathrm{thm}}$) comparison.}
\label{tab:thm-vs-module}
\begin{tabular}{lrr}
\toprule
Metric & $G_{\mathrm{module}}$ / $G_{\mathrm{module}}^{-}$ & $G_{\mathrm{thm}}$ \\
\midrule
Nodes                        & $7{,}563$          & $308{,}129$ \\
Edges                        & $23{,}570$ / $19{,}448$ & $8{,}436{,}366$ \\
Cross-namespace edges         & $37.1\%$           & $50.9\%$ \\
In-degree max                & $167$              & $89{,}936$ \\
Community--namespace NMI     & ---                & $0.34$ \\
Single-node removal impact   & $29$ components    & $\le 6$ nodes disconnected \\
Zero-citation nodes          & $1{,}433$ sources  & $109{,}088$ leaf theorems ($44.8\%$) \\
\bottomrule
\end{tabular}
\end{table}

The declaration graph is a more direct projection of the logical product than the module graph, yet the process remains encoded: the language infrastructure layer reflects Lean's type system design, the $44.8\%$ leaf theorem rate includes \texttt{@[to\_additive]} mirrors that inflate surface area without deepening content, and the bounded out-degree distribution records the cognitive complexity ceiling of human-written proofs.

\subsection{Namespace Graph}\label{app:ns-detail}

The namespace graph $G_{\mathrm{ns}}^{(k)}$ (Definition~\ref{def:ns-graph}; \S\ref{sec:namespace-graph}) aggregates declarations by their position in the naming hierarchy, yielding a family of graphs parameterized by truncation depth~$k$. All statistics below use $k = 2$ unless noted otherwise.

\subsubsection{Containment Decay}

Table~\ref{tab:containment-decay} reports the containment ratio at each level of the namespace and module hierarchies.

\begin{table}[ht]
\centering
\caption[Containment across granularity levels]{Containment across granularity levels. Namespace-depth rows are computed on all $8{,}436{,}366$ edges of $G_{\mathrm{thm}}$. The top-level directory and file-level rows are computed on the $2{,}506{,}738$ edges whose both endpoints appear in the file-module mapping.\protect\footnotemark}
\label{tab:containment-decay}
\begin{tabular}{llrr}
\toprule
Granularity & Units & Cross-boundary & Containment \\
\midrule
Top-level directory             & $32$          & $51.1\%$ & $48.9\%$ \\
Namespace depth $1$             & $3{,}184$     & $77.8\%$ & $22.2\%$ \\
Namespace depth $2$             & $10{,}097$    & $85.8\%$ & $14.2\%$ \\
File-level                      & $7{,}225$     & $84.4\%$ & $15.6\%$ \\
Namespace depth $\ge 3$         & ${\sim}15$K   & $87.4\%$ & $12.6\%$ \\
\bottomrule
\end{tabular}
\end{table}
\footnotetext{The file-module mapping covers $217{,}647$ of $308{,}129$ declarations; unmapped declarations are primarily Lean core infrastructure residing outside the \texttt{Mathlib/} source tree.}

\begin{figure}[ht]
\centering
\includegraphics[width=\textwidth]{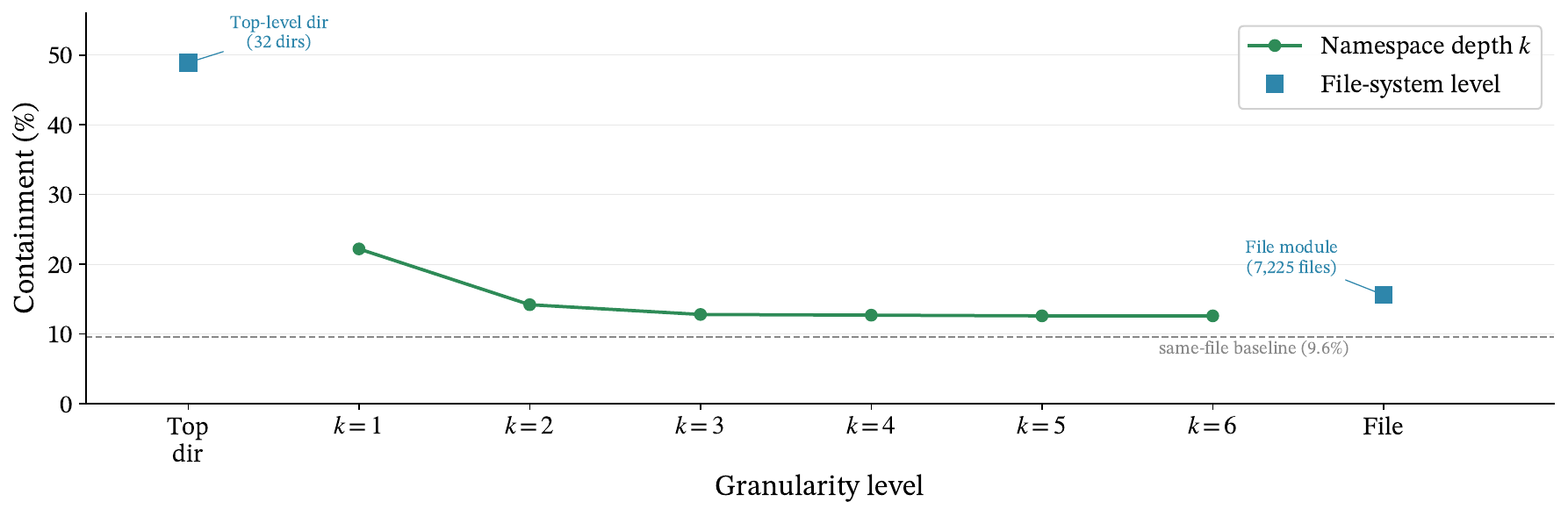}
\caption{Containment decay curve. Green circles connected by the solid line show containment at namespace depth $k = 1, \ldots, 6$; blue squares mark file-system levels (top-level directory and file module). The gray dashed line marks the same-file baseline from Table~\ref{tab:thm-edge-breakdown} ($9.6\%$). Containment drops steeply from the top-level directory to depth~$1$ and saturates by depth~$3$.}
\label{fig:containment-curve}
\end{figure}

The decay exhibits three regimes. At the coarsest granularity (the $32$ top-level directories: \module{Algebra}, \module{CategoryTheory}, \module{Topology}, etc.), nearly half of all declaration-level dependencies ($48.9\%$) stay within the same directory, confirming that discipline-level categories retain genuine organizational force. The steepest single drop occurs at the discipline--sub-discipline transition ($48.9\% \to 22.2\%$): moving from $32$ directories to $3{,}184$ depth-$1$ namespaces eliminates more than half of the remaining containment, indicating that within a discipline, boundaries between \ns{Nat} and \ns{Int}, or between \ns{Set} and \ns{Finset}, carry far less force. Beyond depth~$2$, containment barely decreases (from $14.2\%$ to $12.6\%$ at depth~$6$): finer naming distinctions add almost no intra-namespace edges. The file-module level ($7{,}225$ units, $15.6\%$ containment) falls between namespace depth~$1$ and depth~$2$, offering no containment advantage beyond that of a mid-level namespace.

\subsubsection{Degree Distribution}
\label{sec:ns-degree}

We examine the degree distribution of $G_{\mathrm{ns}}^{(2)}$, the namespace graph at depth~$2$. We choose $k = 2$ because it occupies the sweet spot of the containment decay curve (Table~\ref{tab:containment-decay}): depth~$1$ is too coarse ($3{,}184$ namespaces, comparable to the top-level directory partition), while depth~$3$ and beyond add almost no new cross-boundary structure (containment saturates at ${\sim}13\%$). At depth~$2$, the $10{,}097$ namespaces correspond to familiar mathematical groupings (\ns{Nat.Prime}, \ns{CategoryTheory.Functor}, \ns{MeasureTheory.Measure}) and provide enough nodes for meaningful network analysis. The graph has $332{,}081$ edges (unique namespace pairs), carrying a total weight of $7{,}234{,}844$ declaration-level dependencies.

\begin{table}[ht]
\centering
\caption{Degree statistics for $G_{\mathrm{ns}}^{(2)}$ (unweighted).}
\label{tab:ns-degree-stats}
\begin{tabular}{lrr}
\toprule
& In-degree & Out-degree \\
\midrule
Mean        & $32.9$    & $32.9$ \\
Median      & $1$       & $14$ \\
Std         & $223.2$   & $59.3$ \\
Max         & $9{,}846$ & $3{,}075$ \\
Zero-degree & $3{,}920$ & $6$ \\
\bottomrule
\end{tabular}
\end{table}

The degree distribution is extremely skewed (Table~\ref{tab:ns-degree-stats}). The median in-degree of~$1$ versus a mean of~$33$ reveals a long tail: $3{,}920$ namespaces ($38.8\%$) have zero in-degree (they are referenced by no other namespace), while a handful attract thousands of incoming edges. The top namespaces by in-degree are \ns{\_root\_} ($9{,}846$), \ns{Eq} ($7{,}686$), \ns{OfNat} ($4{,}390$), \ns{Membership} ($3{,}407$), and \ns{Semiring} ($3{,}252$). By out-degree, the leaders are \ns{\_root\_} ($3{,}075$), \ns{MeasureTheory} ($687$), \ns{CategoryTheory} ($656$), and \ns{Polynomial} ($615$).

Power-law fitting (\S\ref{sec:prelim-powerlaw}) yields $\alpha = 1.60$ ($x_{\min} = 4$) for in-degree and $\alpha = 1.90$ ($x_{\min} = 13$) for out-degree. In both cases, a lognormal distribution provides a significantly better fit than a pure power law (likelihood ratio $R = -10.6$ and $R = -289.7$, respectively; $p < 0.01$). This mirrors the pattern observed for $G_{\mathrm{module}}$ (\S\ref{sec:degree-distribution}): heavy-tailed distributions that are not cleanly power-law.

\begin{figure}[ht]
\centering
\includegraphics[width=\textwidth]{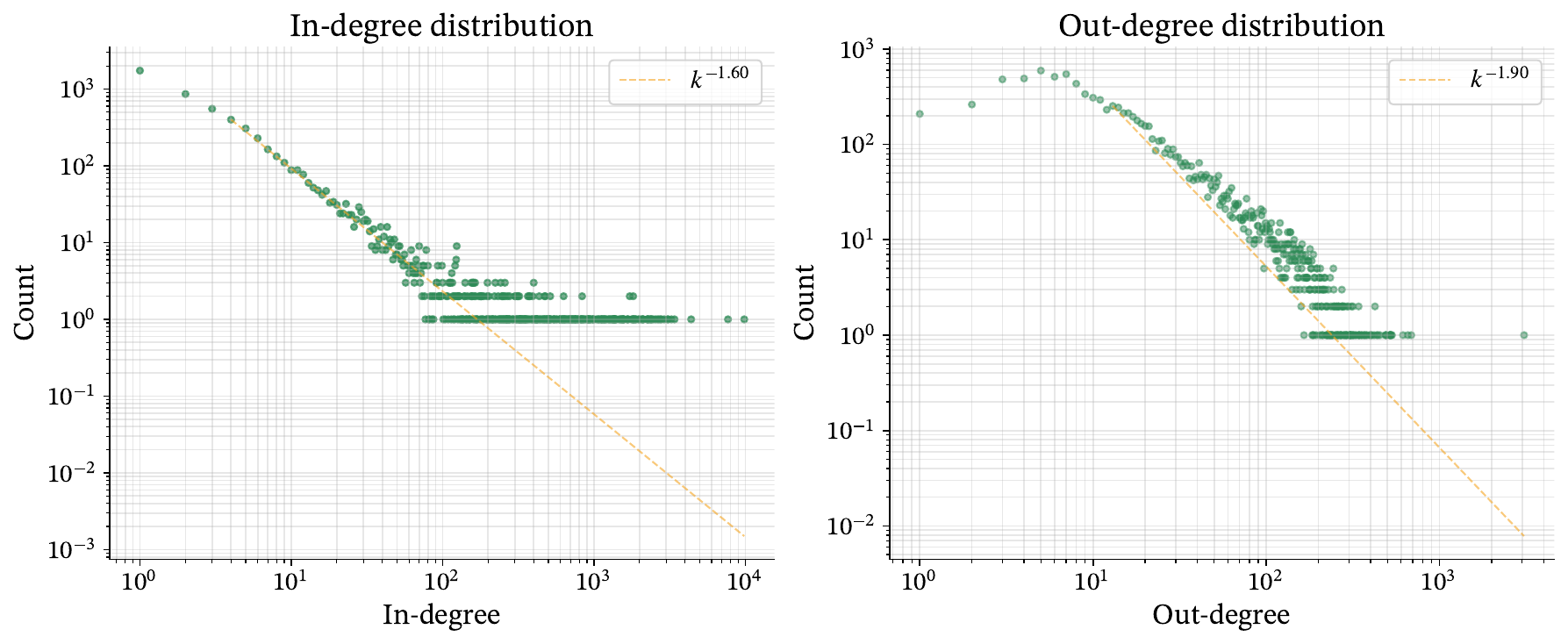}
\caption{Degree distributions of $G_{\mathrm{ns}}^{(2)}$ on log--log axes. Left: in-degree (\textcolor{blue!60!black}{blue}), with power-law reference (gold dashed, $\alpha = 1.60$, $x_{\min} = 4$). Right: out-degree (\textcolor{red!60!black}{coral}), with power-law reference ($\alpha = 1.90$, $x_{\min} = 13$). Both tails are heavy but better fit by a lognormal than a pure power law. The extreme outlier in the in-degree panel is \ns{\_root\_} ($\deg^{-} = 9{,}846$; see Remark~\ref{rem:root-artifact}).}
\label{fig:ns-degree}
\end{figure}

\label{rem:root-artifact}
The namespace \ns{\_root\_} dominates both in-degree and out-degree at depth~$2$. This is an artifact of the truncation rule in Definition~\ref{def:ns-graph}: declarations whose fully qualified names have fewer than three components, including foundational definitions such as \decl{Eq.refl}, \decl{Nat.succ}, and \decl{OfNat.ofNat}, are assigned to \ns{\_root\_}. This namespace is not a coherent conceptual domain; it is a catch-all for the shallow infrastructure of the Lean type system. Its extreme centrality reflects naming-depth convention rather than mathematical importance: the most foundational constructs carry the shortest names. In the analysis that follows, we note where \ns{\_root\_} distorts aggregate statistics.

As at the module (\S\ref{sec:degree-distribution}) and declaration (\S\ref{sec:thm-degree}) levels, the degree distributions are heavy-tailed but better fit by lognormal than pure power law. In-degree skewness is extreme (median~$1$ vs.\ mean~$33$), distorted by the \ns{\_root\_} artifact (Remark~\ref{rem:root-artifact}). The in-degree/out-degree asymmetry mirrors the pattern at the other two levels: open-ended in-degree (product) versus bounded out-degree (process).

\subsubsection{DAG Depth}
\label{sec:ns-dag-depth}

Unlike $G_{\mathrm{module}}$ and $G_{\mathrm{thm}}$, the namespace graph $G_{\mathrm{ns}}^{(2)}$ is \emph{not} a DAG: aggregating declaration-level dependencies to the namespace level creates cycles (e.g., \ns{Nat} $\to$ \ns{Int} $\to$ \ns{Nat}), because declarations in different namespaces can depend on each other bidirectionally (Remark~\ref{rem:ns-cycles}). To recover a layered structure, we condense strongly connected components (SCCs), yielding a DAG of $4{,}080$ super-nodes in $8$ topological layers. The structure is bimodal: layer~$0$ contains $3{,}941$ super-nodes (mostly trivial, single-namespace SCCs that depend on nothing else), while layer~$5$ contains a single super-node, the giant SCC of $5{,}899$ namespaces ($58\%$ of all depth-$2$ namespaces). The intermediate layers ($1$--$4$) hold only $134$ super-nodes that bridge the leaf namespaces to the giant core. Beyond the giant SCC, layers~$6$ and~$7$ contain just $4$ namespaces.

\begin{figure}[ht]
\centering
\includegraphics[width=\textwidth]{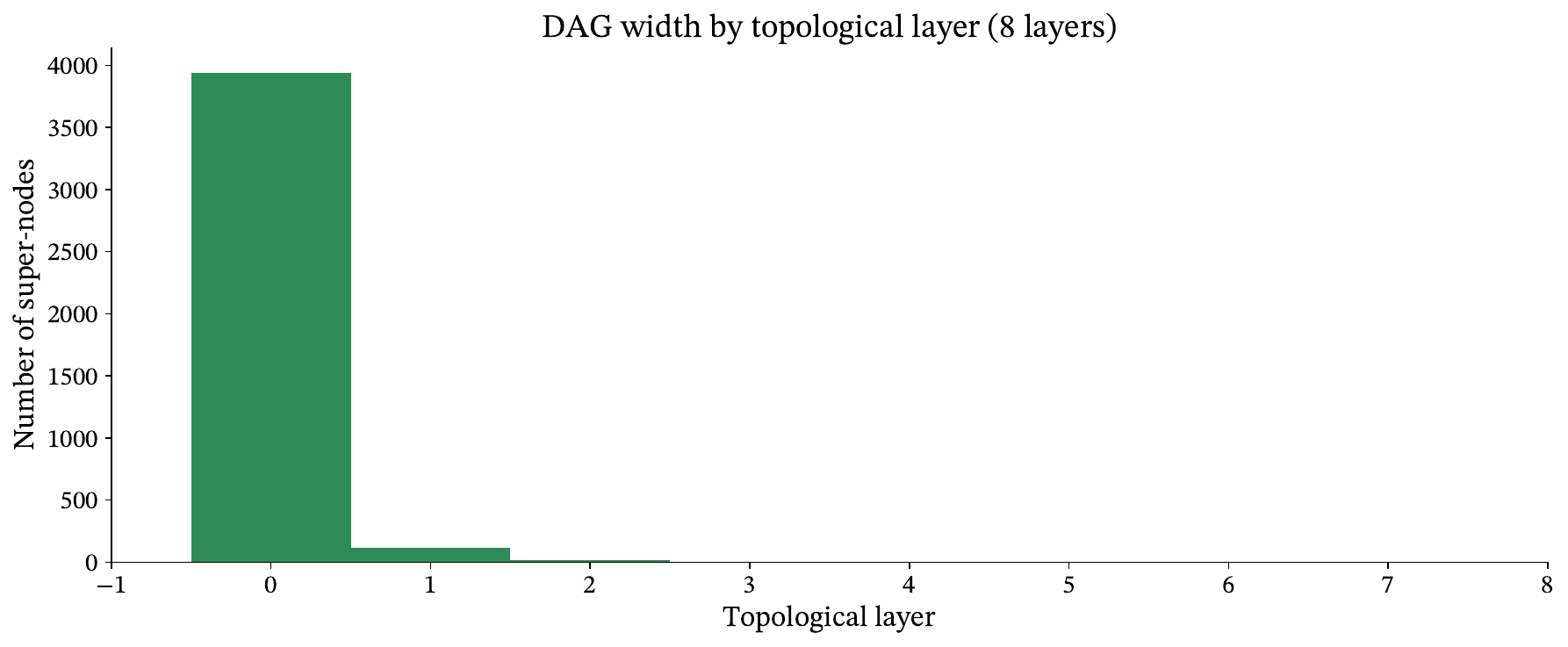}
\caption{Condensed DAG of $G_{\mathrm{ns}}^{(2)}$ by topological layer ($4{,}080$ super-nodes, $8$ layers). Each bar counts the number of super-nodes (SCCs) at that layer. Layer~$0$ holds $3{,}941$ leaf SCCs; layer~$5$ holds a single giant SCC comprising $5{,}899$ namespaces. The shallow, bimodal structure reflects pervasive mutual dependency: most namespaces either sit at the periphery or belong to one densely interconnected core.}
\label{fig:ns-dag-structure}
\end{figure}

The $8$-layer condensed DAG is far shallower than the module ($154$) or declaration ($84$) graphs: at the namespace level, pervasive mutual dependency produces a nearly flat structure rather than vertical hierarchy.

\subsubsection{Centrality}
\label{sec:ns-centrality}

As in \S\ref{sec:centrality} and \S\ref{sec:thm-centrality}, we compute PageRank and betweenness centrality (Definition~\ref{def:pagerank}--\ref{def:betweenness}; $\alpha = 0.85$, weighted; $k = 300$ samples) to identify structurally important namespaces.

\begin{table}[ht]
\centering
\caption{Top~10 namespaces by PageRank and betweenness centrality in $G_{\mathrm{ns}}^{(2)}$.}
\label{tab:ns-centrality}
\small
\begin{tabular}{rlrl}
\toprule
\multicolumn{2}{c}{PageRank} & \multicolumn{2}{c}{Betweenness} \\
\cmidrule(lr){1-2} \cmidrule(lr){3-4}
PR & Namespace & $c_B$ & Namespace \\
\midrule
.247 & \ns{\_root\_}          & .248 & \ns{\_root\_} \\
.035 & \ns{Eq}                & .065 & \ns{And} \\
.017 & \ns{Set}               & .043 & \ns{Eq} \\
.015 & \ns{Iff}               & .036 & \ns{Exists} \\
.014 & \ns{OfNat}             & .031 & \ns{Function} \\
.012 & \ns{Real}              & .030 & \ns{Nat} \\
.012 & \ns{Nat}               & .029 & \ns{CategoryTheory} \\
.012 & \ns{CategoryTheory}    & .019 & \ns{MeasureTheory} \\
.010 & \ns{Preorder}          & .018 & \ns{Set} \\
.008 & \ns{Semiring}          & .017 & \ns{Module} \\
\bottomrule
\end{tabular}
\end{table}

Excluding the \ns{\_root\_} artifact (Remark~\ref{rem:root-artifact}), the two rankings reveal a clear separation (Table~\ref{tab:ns-centrality}). PageRank is dominated by mathematical infrastructure: \ns{Eq}, \ns{Set}, \ns{OfNat}, \ns{Real}, and \ns{Preorder}, the foundational types and structures on which proofs ultimately rest. Betweenness centrality, by contrast, promotes logical connectives and bridging concepts: \ns{And}, \ns{Exists}, and \ns{Function} occupy positions~$2$--$5$ by betweenness but do not appear in the PageRank top~$10$. These namespaces serve as structural bridges: proofs in distant mathematical theories must pass through logical connectives to compose their arguments.

\begin{figure}[H]
\centering
\begin{subfigure}[t]{0.32\textwidth}
\centering
\includegraphics[width=\textwidth]{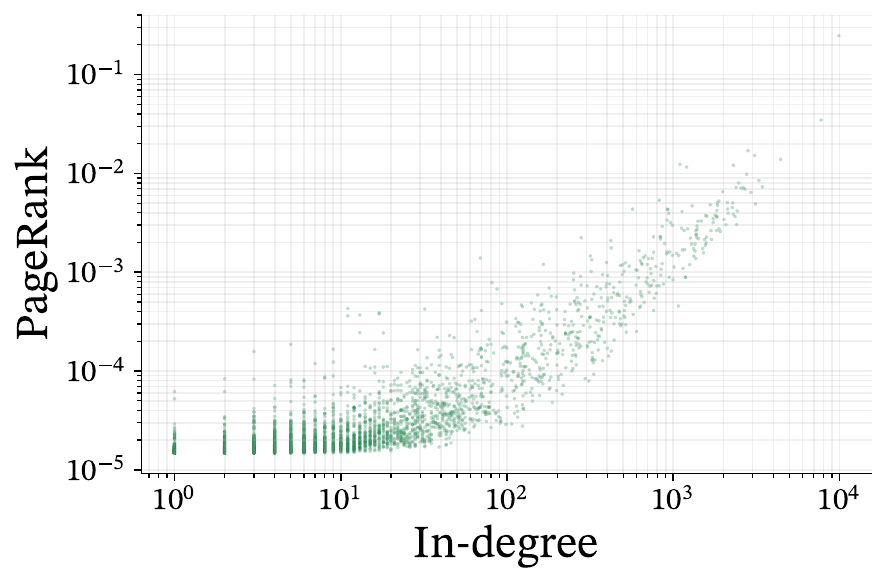}
\caption{In-degree vs.\ PageRank.}
\label{fig:ns-centrality-indeg-pr}
\end{subfigure}%
\hfill
\begin{subfigure}[t]{0.32\textwidth}
\centering
\includegraphics[width=\textwidth]{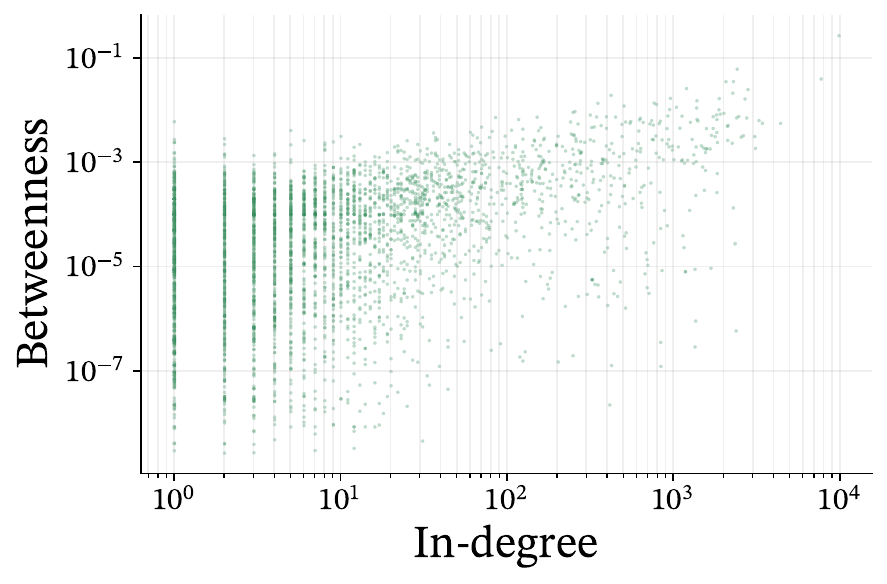}
\caption{In-degree vs.\ Betweenness.}
\label{fig:ns-centrality-indeg-betw}
\end{subfigure}%
\hfill
\begin{subfigure}[t]{0.32\textwidth}
\centering
\includegraphics[width=\textwidth]{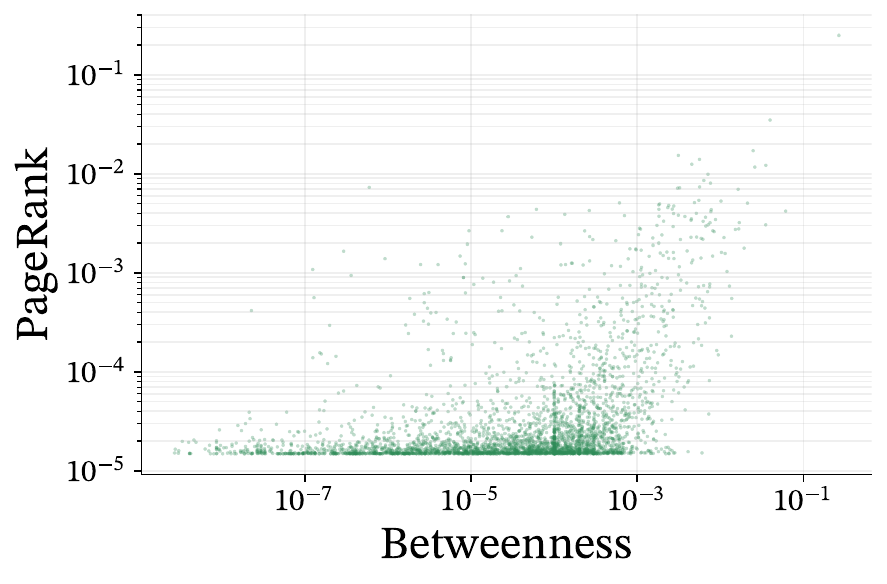}
\caption{Betweenness vs.\ PageRank.}
\label{fig:ns-centrality-betw-pr}
\end{subfigure}
\caption{Pairwise centrality scatter plots for $G_{\mathrm{ns}}^{(2)}$. Each point is one namespace. The extreme outlier in (a) is \ns{\_root\_} (Remark~\ref{rem:root-artifact}). The scatter confirms that PageRank and betweenness capture distinct structural roles at the namespace level.}
\label{fig:ns-centrality}
\end{figure}

This separation echoes the module-level finding of \S\ref{sec:centrality}, where no single module dominates all three centrality measures, and the declaration-level finding of \S\ref{sec:thm-centrality}, where in-degree, PageRank, and betweenness identify disjoint categories of importance. At the namespace level, the two available measures already diverge sharply, confirming that the multidimensionality of structural importance persists across all three granularities. Figure~\ref{fig:ns-centrality} shows the pairwise scatter plots.

\subsubsection{Community Structure}
\label{sec:ns-community}

Louvain community detection (\S\ref{sec:prelim-community}) on the undirected, weighted projection of $G_{\mathrm{ns}}^{(2)}$ yields $12$ communities with modularity $Q = 0.271$.

\begin{table}[ht]
\centering
\caption{Top five communities in $G_{\mathrm{ns}}^{(2)}$ by size, with representative namespaces (highest weighted in-degree within each community).}
\label{tab:ns-communities}
\begin{tabular}{rrl}
\toprule
ID & Size & Representative namespaces \\
\midrule
2 & $5{,}112$ & \ns{\_root\_}, \ns{Eq}, \ns{Set} \\
3 & $2{,}355$ & \ns{Semiring}, \ns{CommRing}, \ns{NonUnitalNonAssocSemiring} \\
0 & $1{,}227$ & \ns{CategoryTheory}, \ns{CategoryTheory.Functor}, \ns{CategoryTheory.CategoryStruct} \\
4 & $927$     & \ns{Real}, \ns{Complex}, \ns{NormedField} \\
1 & $462$     & \ns{Monoid}, \ns{Subgroup}, \ns{MulOneClass} \\
\bottomrule
\end{tabular}
\end{table}

The namespace-level community structure differs markedly from the declaration-level structure (\S\ref{sec:thm-community}). At the declaration level, Louvain detection produces $22$ communities with modularity $0.48$ and NMI~$= 0.34$ against the top-level namespace hierarchy. At the namespace level, the number of communities is smaller ($12$ vs.\ $22$), the modularity is lower ($0.271$ vs.\ $0.48$), and the NMI with the top-level directory falls to~$0.261$ (Table~\ref{tab:ns-communities}).

The lower modularity reflects the coarsening inherent in namespace aggregation. When thousands of declarations are collapsed into a single namespace node, the fine-grained community boundaries that Louvain detects at the declaration level are blurred. Edges between namespaces are weighted sums of many declaration-level edges, and the averaging effect smooths out the modular structure that exists at finer granularity. The largest community (ID~$2$, $5{,}112$ nodes) spans half the graph and lacks a coherent mathematical identity: its dominant members (\ns{\_root\_}, \ns{Eq}, \ns{Set}) are foundational constructs used across all disciplines. This contrasts with the declaration-level communities, where the largest communities map onto recognizable mathematical disciplines (\S\ref{sec:thm-community}).

\subsubsection{Robustness}
\label{sec:ns-robustness}

We assess the robustness of $G_{\mathrm{ns}}^{(2)}$ by progressively removing nodes and measuring the fraction of the original graph remaining in the largest weakly connected component (GCC). The original graph has $4$ weakly connected components, with the largest containing $10{,}094$ of $10{,}097$ nodes.

\begin{table}[ht]
\centering
\caption{Robustness of $G_{\mathrm{ns}}^{(2)}$: GCC fraction after random vs.\ targeted (PageRank-ordered) node removal.}
\label{tab:ns-robustness}
\begin{tabular}{rrrr}
\toprule
Removal \% & Random GCC & Targeted GCC & Gap \\
\midrule
$5\%$  & $0.950$ & $0.726$ & $0.223$ \\
$10\%$ & $0.899$ & $0.509$ & $0.391$ \\
$20\%$ & $0.798$ & $0.208$ & $0.590$ \\
$30\%$ & $0.688$ & $0.054$ & $0.634$ \\
$50\%$ & $0.477$ & $0.000$ & $0.477$ \\
\bottomrule
\end{tabular}
\end{table}

\begin{figure}[ht]
\centering
\includegraphics[width=\textwidth]{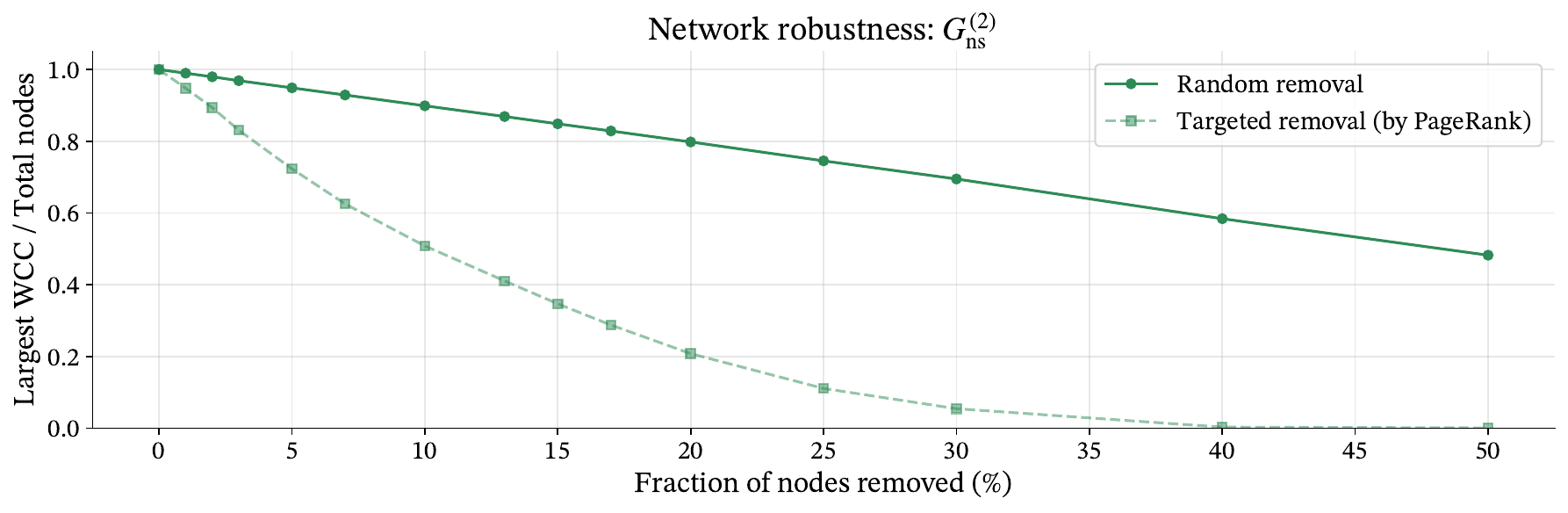}
\caption{Robustness curves for $G_{\mathrm{ns}}^{(2)}$: fraction of nodes in the largest WCC as a function of the fraction removed, under random removal (blue) and targeted removal by PageRank (coral).}
\label{fig:ns-robustness}
\end{figure}

The pattern matches the classic hub-dominated vulnerability signature observed at both the module level (\S\ref{sec:robustness}) and the declaration level (\S\ref{sec:thm-robustness}): high resilience under random failure, extreme fragility under targeted attack (Table~\ref{tab:ns-robustness}, Figure~\ref{fig:ns-robustness}). At $20\%$ targeted removal, the GCC collapses to $20.8\%$ of its original size, more severe than the declaration-level graph, where the same removal fraction leaves $46.1\%$ (\S\ref{sec:thm-robustness}). The namespace graph's greater fragility reflects the concentration of connectivity in the \ns{\_root\_} super-node and a small number of infrastructure namespaces (Remark~\ref{rem:root-artifact}); once these are removed, the remaining mathematical namespaces fragment into isolated clusters with no shared foundation.

\subsubsection{Extended Analysis}\label{app:ns-analysis}

The containment decay curve (Figure~\ref{fig:containment-curve}) measures how well the namespace hierarchy captures dependency structure. The disciplinary taxonomy (Algebra, Analysis, Topology, Number Theory) retains genuine organizational force at the broadest scale, but this force dissipates rapidly at the sub-discipline transition ($48.9\% \to 22.2\%$). Within a discipline, boundaries between \ns{Nat} and \ns{Int} carry far less constraining power than the boundary between \ns{Algebra} and \ns{Topology}. The saturation beyond depth~$2$ (${\sim}13\%$) suggests the namespace tree is ``informationally shallow'': its first two levels encode most dependency structure that human naming captures, and deeper levels serve primarily as disambiguation.

The heightened fragility of the namespace graph under targeted attack (GCC reduced to $20.8\%$ at $20\%$ removal, versus $46.1\%$ at the declaration level) arises because namespace aggregation concentrates dependencies: a handful of infrastructure namespaces (\ns{Eq}, \ns{OfNat}, \ns{DFunLike}) absorb thousands of declaration-level dependencies into single namespace-level hubs. Refactoring such bottleneck namespaces carries disproportionate systemic risk and should be guided by centrality rankings.

\subsection{Cross-Level Analysis}
\label{sec:cross_level}
\label{sec:module-declaration}

Sections~\ref{sec:module-import}--\ref{sec:namespace-graph} analyzed the library at three distinct granularities: source files (\S\ref{sec:module-import}), declarations (\S\ref{sec:theorem-premise}), and namespaces (\S\ref{sec:namespace-graph}). Each level reveals a different facet of the tension between human organizational process and inherited mathematical structure. We now overlay these levels, measuring how the three organizational structures (file boundaries, namespace boundaries, and logical dependencies) align and diverge.

The overlay yields three cross-level measurements. Namespace--module alignment (\S\ref{sec:ns-mod-cross}) measures the agreement between the two human organizational systems (naming and filing). Module cohesion (\S\ref{sec:cohesion}) quantifies how much declaration-level dependency traffic stays within file boundaries. Namespace--declaration interaction (\S\ref{sec:ns-decl-interaction}) compares the constraining power of the two systems at the declaration level and reveals a directional depth asymmetry between the import and declaration graphs. Assembling these measurements (\S\ref{sec:three-layers}) reveals a consistent pattern: human organizational process is internally coherent but collectively divergent from inherited mathematical structure.

\subsubsection{Namespace--Module Alignment}
\label{sec:ns-mod-cross}

The containment analysis of \S\ref{sec:containment-decay} and the cohesion analysis below each compare one human organizational system against the inherited mathematical structure. We first ask a different question: how well do the two \emph{human} systems (naming and filing) agree with each other? If both reflect the same cognitive model, their alignment should be high even when both diverge from logic.

Table~\ref{tab:ns-mod-cross} cross-tabulates the namespace and module partitions for the $217{,}487$ declarations with file mappings. The normalized mutual information (NMI) between the two partitions is $0.708$ at depth~$1$ and $0.766$ at depth~$2$, substantially higher than the NMI of~$0.34$ between Louvain communities and the namespace hierarchy (\S\ref{sec:thm-community}).

\begin{table}[ht]
\centering
\caption{Namespace--module cross-tabulation for $217{,}487$ declarations with file mappings.}
\label{tab:ns-mod-cross}
\begin{tabular}{lrr}
\toprule
Metric & Depth $1$ & Depth $2$ \\
\midrule
Unique namespaces              & $2{,}241$  & $4{,}943$ \\
Unique modules (files)         & \multicolumn{2}{c}{$6{,}993$} \\
NMI(namespace, module)         & $0.708$    & $0.766$ \\
Single-file namespaces         & $50\%$     & $64\%$ \\
Single-namespace files         & $58\%$     & $39\%$ \\
Max files per namespace        & \multicolumn{2}{c}{$2{,}169$ (\ns{\_root\_})} \\
Max namespaces per file        & $31$       & $34$ \\
\bottomrule
\end{tabular}
\end{table}

At depth~$1$, half of all namespaces appear in exactly one file; for these, naming and filing are perfectly aligned. But the tail is heavy: \ns{CategoryTheory} spans $977$ files, and~$10\%$ of namespaces span more than $10$ files (Table~\ref{tab:ns-mod-cross}). In the reverse direction, $58\%$ of files contain declarations from only one namespace; the most namespace-diverse files are concentrated in topology: \module{Topology.Constructions} and \module{Topology.Separation.Basic} each host over $30$ depth-$1$ namespaces, reflecting their role as bridge files that simultaneously manipulate declarations from many conceptual domains.

The two directions of misalignment reflect distinct engineering pressures. When a namespace spans many files (namespace~$>$~module), a single conceptual domain has been partitioned into multiple compilation units, typically to manage build times. When a file contains many namespaces (module~$>$~namespace), related concepts from different naming domains have been gathered in one place for cognitive convenience. Both patterns are rational organizational decisions; they produce NMI~$\approx 0.7$ rather than~$1.0$ because the two pressures (compilation efficiency and cognitive proximity) do not always agree.

\label{rem:two-pressures}
The NMI of~$0.71$ between namespaces and modules is \emph{twice} the NMI of~$0.34$ between namespaces and dependency-based Louvain communities (\S\ref{sec:thm-community}). Human organizational decisions (naming and filing) agree with each other far more than either agrees with the logical structure of the dependency graph. This is unsurprising: both naming and filing are human decisions, often made by the same contributor, while the dependency structure is determined by mathematical necessity.

\subsubsection{Module--Declaration Interaction}
\label{sec:cohesion}

\begin{figure}[H]
\centering
\begin{subfigure}[t]{0.36\textwidth}
\centering
\vspace{0pt}
\begin{forest}
  for tree={
    font=\scriptsize\ttfamily,
    text=blue!60!black,
    grow'=0,
    child anchor=west,
    parent anchor=south,
    anchor=west,
    calign=first,
    edge path={
      \noexpand\path [draw, black, \forestoption{edge}]
        (!u.south west) +(7.5pt,0) |- (.child anchor)\forestoption{edge label};
    },
    before typesetting nodes={
      if n=1 {insert before={[,phantom]}} {}
    },
    fit=band,
    before computing xy={l=12pt},
  }
[Data/
  [Nat/
    [Basic.lean
      [\textcolor{red!70!black}{add\_comm}]
      [\textcolor{red!70!black}{add\_left\_comm}]
    ]
    [Defs.lean
      [\textcolor{red!70!black}{succ\_eq\_add\_one}]
      [\textcolor{red!70!black}{zero\_add}]
    ]
  ]
]
\end{forest}
\caption{File-system hierarchy~$T$ (\textcolor{blue!60!black}{blue}), extended to the declaration level.}
\label{fig:tree-view-app}
\end{subfigure}%
\hfill
\begin{subfigure}[t]{0.60\textwidth}
\centering
\vspace{0pt}
\raggedright\scriptsize
\textsf{\bfseries\color{blue!60!black}Basic.lean}\\[2pt]
\ttfamily
\textbf{import} \textcolor{blue!60!black}{Init.Prelude}\\[3pt]
\textbf{theorem} \textcolor{red!70!black}{add\_comm} (n m : Nat)\\
\hspace{1.5em}: n + m = m + n := \textbf{by}\\
\hspace{1.5em}induction n <;> simp [\textcolor{red!70!black}{Eq.refl}]\\[3pt]
\textbf{theorem} \textcolor{red!70!black}{add\_left\_comm} (n m k : Nat)\\
\hspace{1.5em}: n + (m + k) = m + (n + k) := \textbf{by}\\
\hspace{1.5em}rw [\textcolor{red!70!black}{add\_comm}]; simp [\textcolor{red!70!black}{Eq.refl}]\\[6pt]
\textsf{\bfseries\color{blue!60!black}Defs.lean}\\[2pt]
\ttfamily
\textbf{import} \textcolor{blue!60!black}{Data.Nat.Basic}\\
\textbf{import} \textcolor{blue!60!black}{Init.Prelude}\\[3pt]
\textbf{theorem} \textcolor{red!70!black}{succ\_eq\_add\_one} (n : Nat)\\
\hspace{1.5em}: succ n = n + 1 := \textbf{by}\\
\hspace{1.5em}rw [\textcolor{red!70!black}{add\_comm}]; exact \textcolor{red!70!black}{Eq.refl} \_\\[3pt]
\textbf{theorem} \textcolor{red!70!black}{zero\_add} (n : Nat)\\
\hspace{1.5em}: 0 + n = n := \textbf{by}\\
\hspace{1.5em}induction n <;> exact \textcolor{red!70!black}{Eq.refl} \_
\caption{Proof source code: each \textcolor{red!70!black}{red identifier} in a proof term generates a dependency edge in~(c).}
\label{fig:source-view-app}
\end{subfigure}

\par\vspace{4pt}

\begin{subfigure}[b]{\textwidth}
\centering
\begin{tikzpicture}[
  every node/.style={font=\scriptsize\ttfamily, inner sep=2pt},
  dot/.style={circle, fill=red!70!black, minimum size=3.5pt, inner sep=0pt},
  dep/.style={->, >=Stealth, red!70!black, semithick},
  depext/.style={->, >=Stealth, red!70!black, semithick, dashed},
  modimp/.style={->, >=Stealth, blue!60!black, thick},
]
  \filldraw[fill=blue!8, draw=blue!50!black, semithick, line join=round]
    (0.2,4.7) -- (10.2,4.7) -- (11.4,5.09) -- (1.4,5.09) -- cycle;
  \filldraw[fill=blue!18, draw=blue!50!black, semithick, line join=round]
    (0.2,4.7) -- (10.2,4.7) -- (10.2,4.57) -- (0.2,4.57) -- cycle;
  \node[font=\scriptsize\sffamily, blue!60!black, anchor=west] at (10.4,4.63)
    {\textnormal{\module{Data.Nat.Defs}}};
  \node[dot] (sea) at (3.5,4.88) {};
  \node[font=\scriptsize\ttfamily, red!70!black, anchor=west] at (3.6,4.88) {succ\_eq\_add\_one};
  \node[dot] (za)  at (7.5,4.88) {};
  \node[font=\scriptsize\ttfamily, red!70!black, anchor=west] at (7.6,4.88) {zero\_add};

  \filldraw[fill=blue!8, draw=blue!50!black, semithick, line join=round]
    (0,2.5) -- (10.0,2.5) -- (11.2,2.89) -- (1.2,2.89) -- cycle;
  \filldraw[fill=blue!18, draw=blue!50!black, semithick, line join=round]
    (0,2.5) -- (10.0,2.5) -- (10.0,2.37) -- (0,2.37) -- cycle;
  \node[font=\scriptsize\sffamily, blue!60!black, anchor=west] at (10.2,2.43)
    {\textnormal{\module{Data.Nat.Basic}}};
  \node[dot] (ac)  at (3.3,2.68) {};
  \node[font=\scriptsize\ttfamily, red!70!black, anchor=east] at (3.2,2.68) {add\_comm};
  \node[dot] (alc) at (7.3,2.68) {};
  \node[font=\scriptsize\ttfamily, red!70!black, anchor=west] at (7.4,2.68) {add\_left\_comm};

  \filldraw[fill=blue!8, draw=blue!50!black, semithick, line join=round]
    (1.0,0.15) -- (8.0,0.15) -- (8.9,0.46) -- (1.9,0.46) -- cycle;
  \filldraw[fill=blue!18, draw=blue!50!black, semithick, line join=round]
    (1.0,0.15) -- (8.0,0.15) -- (8.0,0.03) -- (1.0,0.03) -- cycle;
  \node[font=\scriptsize\sffamily, blue!60!black, anchor=west] at (8.2,0.09)
    {\textnormal{\module{Init.Prelude}}};
  \node[dot] (er)  at (4.8,0.3) {};
  \node[font=\scriptsize\ttfamily, red!70!black, anchor=west] at (4.9,0.3) {Eq.refl};

  \draw[blue!50, semithick]
    (-0.2,2.37) -- (-0.5,2.37) -- (-0.5,5.09) -- (-0.2,5.09);
  \node[font=\scriptsize\sffamily, blue!60!black, rotate=90, anchor=south]
    at (-0.75,3.73) {namespace \textnormal{\ns{Data.Nat}}};

  \node[circle, fill=blue!60!black, minimum size=5.5pt, inner sep=0pt] (md) at (9.8, 4.88) {};
  \node[circle, fill=blue!60!black, minimum size=5.5pt, inner sep=0pt] (mb) at (10.0, 2.68) {};
  \node[circle, fill=blue!60!black, minimum size=5.5pt, inner sep=0pt] (mo) at (7.2, 0.3) {};
  \draw[modimp] (md) to[out=-85,in=85] (mb);                         
  \draw[modimp] (mb) to[out=-70,in=40] (mo);                         
  \draw[modimp] (md) to[out=-50,in=55] (mo);                         

  \draw[dep] (alc) -- (ac);
  \draw[depext] (sea) to[out=-88,in=92] (ac);
  \draw[depext] (ac)  to[out=-88,in=105] (er);
  \draw[depext] (alc) to[out=-88,in=60] (er);
  \draw[depext] (sea) to[out=-85,in=120] (er);
  \draw[depext] (za)  to[out=-85,in=45] (er);

\end{tikzpicture}
\caption{An isometric view of $G_{\mathrm{thm}}$ layered by module. Each \textcolor{blue!60!black}{blue card} represents a module file; \textcolor{red!70!black}{red nodes} are declarations. Solid \textcolor{red!70!black}{red arrows} show intra-module edges ($E_{\mathrm{int}}$); dashed \textcolor{red!70!black}{red arrows} show external edges ($E_{\mathrm{ext}}$) that pierce module boundaries. \textcolor{blue!60!black}{Blue arrows} on the right show the corresponding module-level imports from $G_{\mathrm{module}}$.}
\label{fig:dep-view-app}
\end{subfigure}
\caption{Three views of the same declarations (cf.\ Figure~\ref{fig:three-views-main} for the introductory version). (a)~The file-system hierarchy~$T$ (\textcolor{blue!60!black}{blue}), extended to the declaration level. (b)~Proof source code: each \textcolor{red!70!black}{red identifier} creates a dependency edge in~(c). (c)~An isometric view of~$G_{\mathrm{thm}}$ layered by module; of six dependency edges, only one stays within its module, visualizing the low cohesion of~$0.107$ (\S\ref{sec:cohesion}). The \textcolor{blue!60!black}{blue} boundaries represent source files, in contrast to the \textcolor{green!40!black}{green} namespace boundaries of Figure~\ref{fig:ns-aggregation-app}; the two layers partially overlap (NMI~$\approx 0.71$; \S\ref{sec:ns-mod-cross}). All names have the \texttt{Nat.}\ prefix removed.}
\label{fig:module-region-app}
\end{figure}
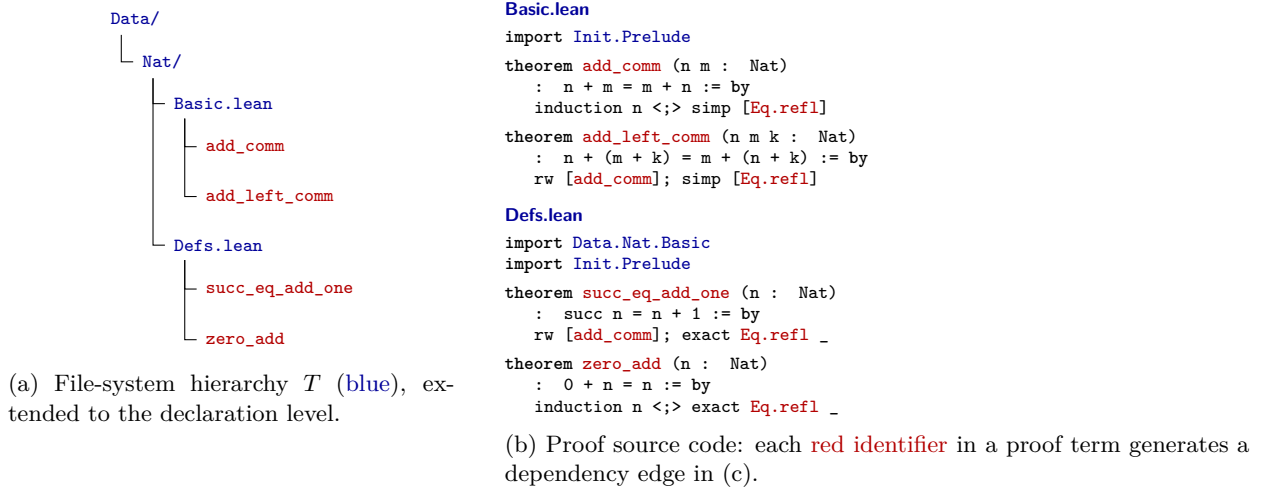
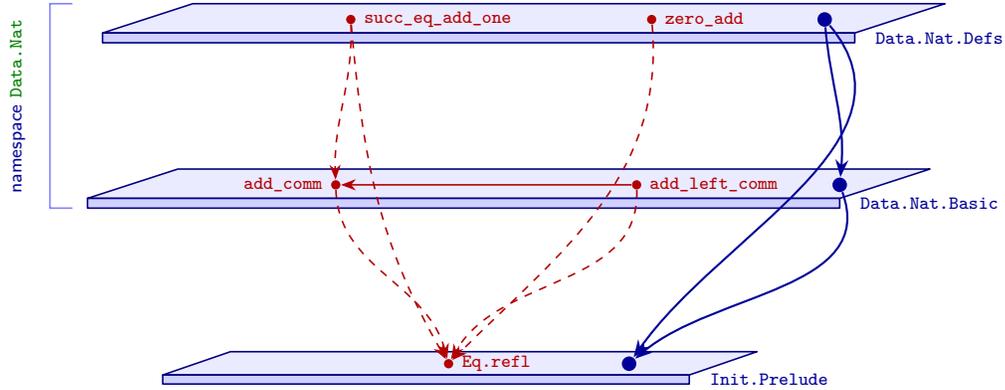

In $G_{\mathrm{thm}}$, each module becomes a \emph{region} (a set of declaration-vertices sharing a common source file). For each module~$m$, let $\mathcal{D}_m \subseteq \mathcal{D}$ denote its declarations. We partition the incident edges into two classes:
\begin{align*}
  E_{\mathrm{int}}(m) &= \{(d_1, d_2) \in E_{\mathrm{thm}} \mid d_1, d_2 \in \mathcal{D}_m\}, \\
  E_{\mathrm{ext}}(m) &= \{(d_1, d_2) \in E_{\mathrm{thm}} \mid d_1 \in \mathcal{D}_m \oplus d_2 \in \mathcal{D}_m\},
\end{align*}
where $\oplus$ denotes exclusive or. Figure~\ref{fig:module-region-app} illustrates this partition: of six dependency edges among declarations in \module{Data.Nat.Basic}, \module{Data.Nat.Defs}, and \module{Init.Prelude}, only one stays within its module; the rest pierce through file boundaries.

\begin{definition}[Module cohesion]\label{def:cohesion}
The \emph{cohesion} of a module~$m$ is
\[
  \mathrm{coh}(m) = \frac{|E_{\mathrm{int}}(m)|}{|E_{\mathrm{int}}(m)| + |E_{\mathrm{ext}}(m)|},
\]
with the convention that $\mathrm{coh}(m) = 0$ when $|E_{\mathrm{int}}(m)| + |E_{\mathrm{ext}}(m)| = 0$.
\end{definition}

\begin{table}[ht]
\centering
\caption{Global distribution of module cohesion across $6{,}993$ file-level modules.}
\label{tab:cohesion-global}
\begin{tabular}{lr}
\toprule
Statistic & Value \\
\midrule
Mean       & $0.107$ \\
Median     & $0.074$ \\
Std Dev    & $0.121$ \\
Max        & $1.000$ \\
Modules with $\mathrm{coh} = 0$ & $584$ / $6{,}993$ ($8.4\%$) \\
\bottomrule
\end{tabular}
\end{table}

Module cohesion is low throughout: the mean is~$0.107$ and the median is~$0.074$ (Table~\ref{tab:cohesion-global}). Internal edges constitute only about $10\%$ of a typical module's total connectivity. If module boundaries reflected logical structure, cohesion would cluster near~$1$; the observed value indicates that file boundaries and dependency structure are only weakly aligned.

\label{rem:zero-cohesion}
Among the $584$ modules ($8.4\%$) with zero cohesion, most are \emph{lemma collections}: declarations grouped by concept rather than by logical interdependence. Developers place ``all lemmas about~$X$'' in one file for discoverability, even when these lemmas depend on definitions across the library. Zero cohesion is not a defect but the structural signature of cognitive organization.

\label{sec:import_vs_declaration}
Aggregating $G_{\mathrm{thm}}$ to the module level yields the \emph{file-aggregated graph}~$G_{\mathrm{file}}$, in which an edge $m_1 \to m_2$ exists whenever some declaration in~$m_1$ cites one in~$m_2$ ($m_1 \neq m_2$). The gap is striking: $G_{\mathrm{file}}$ has $215{,}211$ edges versus $23{,}235$ in~$G_{\mathrm{module}}$, a $9.1\times$ amplification.

\begin{table}[ht]
\centering
\caption{Edge-level comparison of $G_{\mathrm{module}}$ (human-written imports) and $G_{\mathrm{file}}$ (declaration-aggregated dependencies).}
\label{tab:import-vs-file}
\begin{tabular}{lrr}
\toprule
Category & Edges & Percentage \\
\midrule
\multicolumn{3}{l}{\emph{$G_{\mathrm{module}}$ breakdown ($23{,}235$ edges):}} \\
\quad Active imports (in both $G_{\mathrm{module}}$ and $G_{\mathrm{file}}$) & $16{,}742$ & $72.1\%$ \\
\quad Unused imports ($G_{\mathrm{module}}$ only) & $6{,}493$ & $27.9\%$ \\
\midrule
\multicolumn{3}{l}{\emph{$G_{\mathrm{file}}$ breakdown ($215{,}211$ edges):}} \\
\quad Direct import exists & $16{,}742$ & $7.8\%$ \\
\quad Indirect, transitively reachable & $197{,}639$ & $91.8\%$ \\
\quad Indirect, not reachable & $830$ & $0.4\%$ \\
\bottomrule
\end{tabular}
\end{table}

From the import side, $72.1\%$ of import edges are \emph{active}: the importing file's declarations actually cite the imported module (Table~\ref{tab:import-vs-file}). The remaining $27.9\%$ provide transitive visibility only. (These ``unused'' imports overlap with what Mathlib's \texttt{shake} linter targets for removal; some may be intentionally preserved by \texttt{shake} for instance or notation availability, while others represent residual redundancy.) From the declaration side, only $7.8\%$ of cross-file dependencies correspond to a direct import; the remaining $92.2\%$ are reached through transitive chains. The module graph functions as a sparse gateway: each file declares a handful of direct imports (mean out-degree $3.1$ in $G_{\mathrm{module}}$), but through transitivity these few edges open access to a vast dependency space (mean out-degree $30.8$ in $G_{\mathrm{file}}$).

\label{rem:unused-vs-redundant}
Section~\ref{sec:transitive-reduction} found $17.5\%$ of import edges transitively redundant; here, $27.9\%$ are unused. Both figures are measured after the \texttt{shake} linter. The gap ($27.9\% > 17.5\%$) has a precise interpretation: $78.4\%$ of unused imports are graph-theoretically essential (the sole path for transitive visibility), while $21.6\%$ are both unused and redundant.

The binary active/unused classification above treats all active imports equally. Import utilization (Definition~\ref{def:import-util}) refines this to a continuous measure: for each import edge $(m_i, m_j)$, what fraction of $m_j$'s declarations does $m_i$ actually reference? We computed $\mathrm{util}(m_i, m_j)$ for all $30{,}793$ import edges between modules that define at least one declaration, cross-referencing the $499{,}732$ declaration-to-module mappings from Lean's \texttt{Environment} API with the $8.4$M declaration-level edges. The distribution is heavily right-skewed: the median utilization is $1.6\%$, the mean is $5.1\%$, and the inter-quartile range spans $[0\%, 5.7\%]$. Strikingly, $11{,}410$ import edges ($37\%$) have \emph{zero} utilization: the importing module does not directly reference any declaration from the imported module, relying entirely on transitive re-exports. These findings complement the cohesion analysis: cohesion measures how much traffic stays \emph{inside} a module boundary ($10.7\%$), while utilization measures how much of what crosses the boundary is actually \emph{consumed} ($1.6\%$ median). Both are low, confirming from opposite directions that module boundaries are a coarse organizational overlay on the fine-grained dependency structure.

The preceding edge-count comparison treats all cross-module dependencies uniformly. We now refine the analysis by comparing the \emph{depth} of modules that a file directly imports in $G_{\mathrm{module}}$ with the depth of modules whose declarations it actually cites in $G_{\mathrm{thm}}$. For each module~$A$, let $\overline{d}_{\mathrm{imp}}(A)$ be the mean module depth of $A$'s direct imports and $\overline{d}_{\mathrm{use}}(A)$ the mean depth of the modules containing declarations cited by $A$'s own declarations. The \emph{depth difference} is $\Delta(A) = \overline{d}_{\mathrm{use}}(A) - \overline{d}_{\mathrm{imp}}(A)$.

Across the $6{,}861$ modules for which both quantities are defined, the mean depth difference is $\overline{\Delta} = -0.098$ (median $-0.069$, $\sigma = 0.445$). The distribution is roughly symmetric but slightly left-skewed: $54.2\%$ of modules have $\Delta < 0$ (usage shallower than imports), $40.3\%$ have $\Delta > 0$ (usage deeper), and $5.5\%$ have $\Delta = 0$. Actual usage thus reaches, on average, \emph{shallower} modules than those directly imported: through deep import chains, declarations access shallow algebraic and order-theoretic infrastructure.

\begin{table}[ht]
\centering
\caption{Mean depth difference $\overline{\Delta}$ by top-level directory (five most negative and five most positive among directories with ${\geq}30$ modules).}
\label{tab:depth-diff-by-dir}
\begin{tabular}{lrrrrr}
\toprule
Directory & $N$ & $\overline{d}_{\mathrm{imp}}$ & $\overline{d}_{\mathrm{use}}$ & $\overline{\Delta}$ & Median~$\Delta$ \\
\midrule
\multicolumn{6}{l}{\emph{Usage shallower than imports ($\Delta < 0$):}} \\
\module{CategoryTheory}      & 919 & 4.339 & 3.979 & $-0.360$ & $-0.333$ \\
\module{Geometry}             & 119 & 4.658 & 4.319 & $-0.339$ & $-0.364$ \\
\module{Algebra}              & 1210 & 4.487 & 4.309 & $-0.178$ & $-0.151$ \\
\module{Analysis}             & 702 & 4.522 & 4.358 & $-0.164$ & $-0.141$ \\
\module{MeasureTheory}        & 284 & 4.380 & 4.218 & $-0.162$ & $-0.141$ \\
\midrule
\multicolumn{6}{l}{\emph{Usage deeper than imports ($\Delta > 0$):}} \\
\module{LinearAlgebra}        & 324 & 4.054 & 4.234 & $+0.180$ & $+0.193$ \\
\module{NumberTheory}         & 207 & 4.233 & 4.375 & $+0.142$ & $+0.179$ \\
\module{RingTheory}           & 621 & 4.103 & 4.247 & $+0.145$ & $+0.177$ \\
\module{FieldTheory}          &  73 & 3.917 & 4.231 & $+0.314$ & $+0.307$ \\
\module{RepresentationTheory} &  32 & 3.949 & 4.262 & $+0.313$ & $+0.195$ \\
\bottomrule
\end{tabular}
\end{table}

Table~\ref{tab:depth-diff-by-dir} reveals a clear split. \module{CategoryTheory} ($\overline{\Delta} = -0.360$) stands out: its files import other category-theoretic modules at depth~${\sim}4.3$ but actually use declarations from modules at depth~${\sim}4.0$. This reflects the highly layered, self-contained nature of categorical formalization: deep categorical imports serve as portals to shallow definitional infrastructure. At the opposite extreme, \module{FieldTheory} ($\overline{\Delta} = +0.314$) imports modules at depth~${\sim}3.9$ but uses declarations at depth~${\sim}4.2$: its imports serve as cognitive shortcuts to deeper algebraic and ring-theoretic foundations. The positive-$\Delta$ directories (\module{RingTheory}, \module{NumberTheory}, \module{LinearAlgebra}, \module{FieldTheory}) are precisely those that build on deep algebraic infrastructure, confirming that the module graph functions as a sparse gateway to deeper dependency chains.

\subsubsection{Namespace--Declaration Interaction}
\label{sec:ns-decl-interaction}

The containment analysis of \S\ref{sec:containment-decay} measured how much declaration-level dependency traffic stays within namespace boundaries (see also Figure~\ref{fig:ns-aggregation-app}). We can now compare this with module cohesion to assess the relative constraining power of the two human organizational systems at the declaration level.

At depth~$1$, namespace containment is $22.2\%$: roughly one in five edges stays within the same depth-$1$ namespace. By depth~$2$, containment drops to $14.2\%$ and saturates near $12.6\%$ by depth~$3$ (Table~\ref{tab:containment-decay}). Module cohesion, the mean proportion of a module's edges that stay internal, is $0.107$, close to the file-level containment of $15.6\%$ reported in the same table.

The comparison reveals a consistent ordering: namespace containment at depth~$1$ ($22.2\%$) exceeds file-level containment ($15.6\%$), which in turn exceeds module cohesion ($10.7\%$). Namespaces, as conceptual categories, retain slightly more dependency traffic than files; a naming domain like \ns{Nat} captures more intra-group dependencies than the file \module{Data.Nat.Basic}, because the namespace spans multiple files and encloses a larger set of related declarations. But all three figures are far below~$50\%$: regardless of which human boundary one draws (naming hierarchy or file system), the majority of mathematical dependencies cross it.

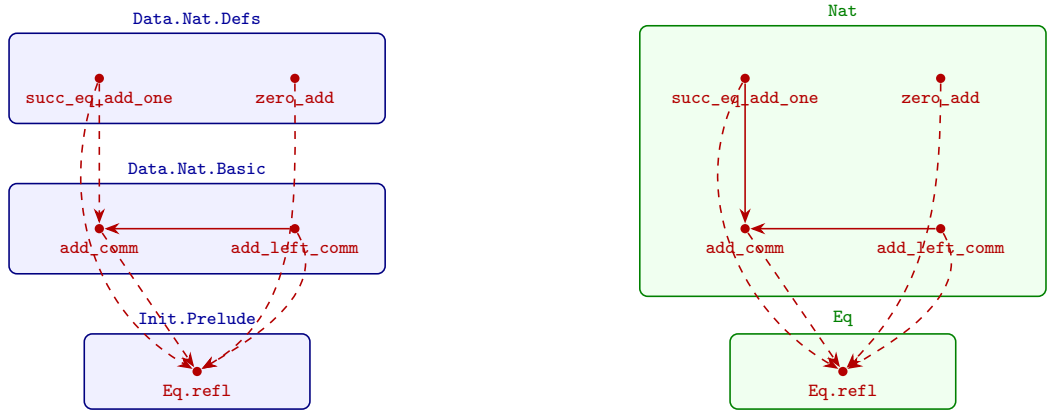
\begin{figure}[ht]
\centering
\begin{subfigure}[b]{0.48\textwidth}
\centering
\begin{tikzpicture}[
  every node/.style={font=\scriptsize, inner sep=2pt},
  dot/.style={circle, fill=red!70!black, minimum size=3.5pt, inner sep=0pt},
  dep/.style={->, >=Stealth, red!70!black, semithick},
  depext/.style={->, >=Stealth, red!70!black, semithick, dashed},
  modbox/.style={draw=blue!50!black, fill=blue!6, rounded corners=3pt, semithick},
]
  \node[modbox, minimum width=5.0cm, minimum height=1.2cm] (defs) at (2.3,3.2) {};
  \node[font=\scriptsize\sffamily, text=blue!60!black] at (2.3,4.0) {\module{Data.Nat.Defs}};
  \node[dot] (sea) at (1.0,3.2) {};
  \node[font=\scriptsize\ttfamily, red!70!black, below] at (1.0,3.1) {\decl{succ\_eq\_add\_one}};
  \node[dot] (za) at (3.6,3.2) {};
  \node[font=\scriptsize\ttfamily, red!70!black, below] at (3.6,3.1) {\decl{zero\_add}};

  \node[modbox, minimum width=5.0cm, minimum height=1.2cm] (basic) at (2.3,1.2) {};
  \node[font=\scriptsize\sffamily, text=blue!60!black] at (2.3,2.0) {\module{Data.Nat.Basic}};
  \node[dot] (ac) at (1.0,1.2) {};
  \node[font=\scriptsize\ttfamily, red!70!black, below] at (1.0,1.1) {\decl{add\_comm}};
  \node[dot] (alc) at (3.6,1.2) {};
  \node[font=\scriptsize\ttfamily, red!70!black, below] at (3.6,1.1) {\decl{add\_left\_comm}};

  \node[modbox, minimum width=3.0cm, minimum height=1.0cm] (init) at (2.3,-0.7) {};
  \node[font=\scriptsize\sffamily, text=blue!60!black] at (2.3,0.0) {\module{Init.Prelude}};
  \node[dot] (er) at (2.3,-0.7) {};
  \node[font=\scriptsize\ttfamily, red!70!black, below] at (2.3,-0.8) {\decl{Eq.refl}};

  \draw[dep] (alc) -- (ac);

  \draw[depext] (sea) -- (ac);
  \draw[depext] (ac) -- (er);
  \draw[depext] (alc) to[out=-60,in=30] (er);
  \draw[depext] (sea) to[out=-110,in=150] (er);
  \draw[depext] (za) to[out=-90,in=30] (er);
\end{tikzpicture}
\caption{Grouped by \textcolor{blue!60!black}{module}. Only $1$ of $6$ edges is intra-module (solid); $5$ are cross-module (dashed). Cohesion~$\approx 10.7\%$.}
\label{fig:ns-vs-mod-module-cross}
\end{subfigure}%
\hfill
\begin{subfigure}[b]{0.48\textwidth}
\centering
\begin{tikzpicture}[
  every node/.style={font=\scriptsize, inner sep=2pt},
  dot/.style={circle, fill=red!70!black, minimum size=3.5pt, inner sep=0pt},
  dep/.style={->, >=Stealth, red!70!black, semithick},
  depext/.style={->, >=Stealth, red!70!black, semithick, dashed},
  nsbox/.style={draw=green!50!black, fill=green!6, rounded corners=3pt, semithick},
]
  \node[nsbox, minimum width=5.4cm, minimum height=3.6cm] (nat) at (2.3,2.1) {};
  \node[font=\scriptsize\sffamily, text=green!50!black] at (2.3,4.1) {\ns{Nat}};

  \node[dot] (sea) at (1.0,3.2) {};
  \node[font=\scriptsize\ttfamily, red!70!black, below] at (1.0,3.1) {\decl{succ\_eq\_add\_one}};
  \node[dot] (za) at (3.6,3.2) {};
  \node[font=\scriptsize\ttfamily, red!70!black, below] at (3.6,3.1) {\decl{zero\_add}};

  \node[dot] (ac) at (1.0,1.2) {};
  \node[font=\scriptsize\ttfamily, red!70!black, below] at (1.0,1.1) {\decl{add\_comm}};
  \node[dot] (alc) at (3.6,1.2) {};
  \node[font=\scriptsize\ttfamily, red!70!black, below] at (3.6,1.1) {\decl{add\_left\_comm}};

  \node[nsbox, minimum width=3.0cm, minimum height=1.0cm] (eq) at (2.3,-0.7) {};
  \node[font=\scriptsize\sffamily, text=green!50!black] at (2.3,0.0) {\ns{Eq}};
  \node[dot] (er) at (2.3,-0.7) {};
  \node[font=\scriptsize\ttfamily, red!70!black, below] at (2.3,-0.8) {\decl{Eq.refl}};

  \draw[dep] (alc) -- (ac);
  \draw[dep] (sea) -- (ac);

  \draw[depext] (ac) -- (er);
  \draw[depext] (alc) to[out=-60,in=30] (er);
  \draw[depext] (sea) to[out=-120,in=150] (er);
  \draw[depext] (za) to[out=-90,in=45] (er);
\end{tikzpicture}
\caption{Grouped by \textcolor{green!50!black}{namespace}. $2$ of $6$ edges are intra-namespace (solid); $4$ are cross-namespace (dashed). Containment~$\approx 22.2\%$.}
\label{fig:ns-vs-mod-namespace-cross}
\end{subfigure}
\caption{The same five declarations and six dependency edges, grouped by module (a) vs.\ namespace~(b). The edge \decl{succ\_eq\_add\_one}~$\to$~\decl{add\_comm} crosses a \textcolor{blue!60!black}{module boundary} (dashed in~a) but stays within the \textcolor{green!50!black}{\ns{Nat}} namespace (solid in~b). Namespace boundaries, spanning multiple files, capture more intra-group dependencies than file boundaries, explaining why containment ($22.2\%$) exceeds cohesion ($10.7\%$).}
\label{fig:ns-vs-mod-cross}
\end{figure}

\label{sec:depth-asymmetry}
\noindent The $9.1\times$ edge amplification measures the \emph{quantitative} gap between $G_{\mathrm{module}}$ and declaration-level dependencies. A complementary analysis of naming depth reveals a \emph{directional} gap. For each module in $G_{\mathrm{module}}$, the \emph{module depth} is the number of dot-separated components (e.g., \module{Mathlib.Data.Nat.Basic} has depth~$4$). For each declaration in $G_{\mathrm{thm}}$, the \emph{declaration depth} is the number of namespace components (e.g., \decl{Nat.add\_comm} has depth~$1$). Table~\ref{tab:depth-asymmetry} compares the depth patterns of edges in the two graphs.

\begin{table}[ht]
\centering
\caption{Depth asymmetry: import edges vs.\ declaration-level dependencies.}
\label{tab:depth-asymmetry}
\begin{tabular}{lrr}
\toprule
& $G_{\mathrm{module}}$ & $G_{\mathrm{thm}}$ \\
\midrule
Same depth                         & $55.4\%$ & $51.4\%$ \\
Source deeper (deep $\to$ shallow) & $23.3\%$ & $37.6\%$ \\
Target deeper (shallow $\to$ deep) & $21.3\%$ & $11.0\%$ \\
Mean depth difference              & $+0.03$  & $+0.36$  \\
\bottomrule
\end{tabular}
\end{table}

Import edges are nearly symmetric: the mean depth difference is~$+0.03$, and the deep$\,\to\,$shallow and shallow$\,\to\,$deep proportions are balanced ($23.3\%$ vs.\ $21.3\%$). Developers import \emph{laterally}, from one depth-$4$ module to another, rather than vertically across naming layers (Remark~\ref{rem:import-depth-symmetry}). At the declaration level, the pattern inverts: the mean depth difference rises to~$+0.36$, and deep$\,\to\,$shallow edges ($37.6\%$) outnumber shallow$\,\to\,$deep edges ($11.0\%$) by more than $3{\times}$. Among declarations at depth~$\ge 2$, fully $69.3\%$ of outgoing edges target depth~$\le 1$, the shallow infrastructure layer of definitions like \decl{OfNat.ofNat}, \decl{Eq.refl}, and \decl{DFunLike.coe} that appeared as extreme hubs in \S\ref{sec:thm-centrality}.

The contrast is a structural consequence of transitivity. A single \texttt{import} brings an entire module into scope, providing transitive access to all upstream declarations. This compresses the vertical depth asymmetry into a lateral peer-to-peer pattern: importing \module{Data.Nat.Basic} implicitly conveys access to all the shallow infrastructure in \module{Init.Prelude}, without the importing module needing to reference depth-$0$ declarations directly. The module graph is thus not merely a sparse summary of declaration-level dependencies; it is a \emph{depth-compressed} summary that masks the gravitational pull of shallow infrastructure.

\subsubsection{Three Layers Combined}
\label{sec:three-layers}

Table~\ref{tab:three-layers-app} assembles the cross-level measurements into a single comparison.

\begin{table}[ht]
\centering
\caption{Cross-level alignment summary. Each row compares two organizational layers.}
\label{tab:three-layers-app}
\begin{tabular}{llrl}
\toprule
Comparison & Measure & Value & Interpretation \\
\midrule
Namespace vs.\ module            & NMI          & $0.71$  & High: human--human agreement \\
Namespace vs.\ Louvain community & NMI          & $0.34$  & Moderate: human $\neq$ logic \\
Namespace vs.\ declaration       & Containment  & $22.2\%$ & Low: naming $\neq$ dependency \\
Module vs.\ declaration          & Cohesion     & $0.107$ & Low: files $\neq$ logical units \\
\bottomrule
\end{tabular}
\end{table}

Table~\ref{tab:three-layers-app} reveals a striking contrast: the two human organizational systems (naming and filing) largely agree with each other (NMI~$0.71$), but neither aligns well with the logical dependency structure. The problem is not that humans organize the library poorly (their internal consistency is high) but that mathematical logic does not respect human cognitive categories. The $99.1\%$ of same-namespace declaration pairs that reside in different files illustrates this tension: large namespaces like \ns{CategoryTheory} are partitioned across hundreds of compilation units.

The depth asymmetry analysis (\S\ref{sec:depth-asymmetry}) reveals a further subtlety. At the declaration level, $37.6\%$ of edges flow from deeper to shallower namespaces (mean depth difference~$+0.36$), reflecting the gravitational pull of infrastructure definitions. At the module level, import edges are nearly symmetric (mean depth difference~$+0.03$). Transitivity compresses the vertical dependency structure into a lateral peer-to-peer pattern: $69.3\%$ of edges from deep declarations ($\mathrm{depth} \ge 2$) target shallow infrastructure ($\mathrm{depth} \le 1$), but the module graph hides this asymmetry behind a few peer-level \texttt{import} statements. The module graph is not merely sparse; it is \emph{depth-compressed}.

The cross-level analysis also implicitly rests on a design parameter that is itself a process-trace: the granularity of modules. \module{Mathlib}'s $6{,}993$ modules contain $308{,}129$ declarations, yielding a mean of approximately $44$ declarations per module. This ratio is not determined by mathematics; it is determined by the cognitive capacity of human developers to manage a coherent unit of code in a single file. Two thought experiments illuminate the point.

\begin{figure}[ht]
\centering
\begin{subfigure}[b]{0.30\textwidth}
\centering
\begin{tikzpicture}[
  dot/.style={circle, fill=red!70!black, minimum size=3.5pt, inner sep=0pt},
  depx/.style={->, >=Stealth, red!70!black, thin, dashed},
  modbox/.style={draw=blue!50!black, fill=blue!6, rounded corners=2pt, semithick,
                 minimum size=0.6cm},
]
  \node[modbox] (m1) at (0,1.2) {};
  \node[modbox] (m2) at (1.1,1.2) {};
  \node[modbox] (m3) at (2.2,1.2) {};
  \node[modbox] (m4) at (0.55,0) {};
  \node[modbox] (m5) at (1.65,0) {};
  \draw[depx] (0.55,0) -- (0,1.2);
  \draw[depx] (0.55,0) -- (1.1,1.2);
  \draw[depx] (1.65,0) -- (1.1,1.2);
  \draw[depx] (1.65,0) -- (2.2,1.2);
  \draw[depx] (0,1.2) -- (1.1,1.2);
  \node[dot] (d1) at (0,1.2) {};
  \node[dot] (d2) at (1.1,1.2) {};
  \node[dot] (d3) at (2.2,1.2) {};
  \node[dot] (d4) at (0.55,0) {};
  \node[dot] (d5) at (1.65,0) {};
  \node[font=\scriptsize, text=gray] at (2.85,0.6) {$\cdots$};
\end{tikzpicture}
\caption{1 decl/module}
\label{fig:gran-fine}
\end{subfigure}%
\hfill
\begin{subfigure}[b]{0.36\textwidth}
\centering
\begin{tikzpicture}[
  dot/.style={circle, fill=red!70!black, minimum size=3.5pt, inner sep=0pt},
  dep/.style={->, >=Stealth, red!70!black, thin},
  depx/.style={->, >=Stealth, red!70!black, thin, dashed},
  modimp/.style={->, >=Stealth, blue!60!black, thick},
  modbox/.style={draw=blue!50!black, fill=blue!6, rounded corners=3pt, semithick},
]
  \node[modbox, minimum width=1.6cm, minimum height=1.6cm] at (0,0.6) {};
  \node[font=\scriptsize\sffamily, text=blue!60!black] at (0,1.65) {module $A$};
  \node[modbox, minimum width=1.6cm, minimum height=1.6cm] at (2.8,0.6) {};
  \node[font=\scriptsize\sffamily, text=blue!60!black] at (2.8,1.65) {module $B$};
  \draw[dep] (0,0.05) -- (-0.35,1.0);
  \draw[dep] (0,0.05) -- (0.35,1.0);
  \draw[dep] (-0.35,1.0) -- (0.35,1.0);
  \draw[dep] (2.45,0.2) -- (2.8,1.0);
  \draw[depx] (2.45,0.2) -- (0.35,1.0);
  \draw[modimp] (2.0,0.6) -- (0.8,0.6);
  \node[dot] (d1) at (-0.35,1.0) {};
  \node[dot] (d2) at (0.35,1.0) {};
  \node[dot] (d3) at (2.8,1.0) {};
  \node[dot] (d4) at (0,0.05) {};
  \node[dot] (d5) at (2.45,0.2) {};
\end{tikzpicture}
\caption{${\sim}44$ decl/module (\module{Mathlib})}
\label{fig:gran-current}
\end{subfigure}%
\hfill
\begin{subfigure}[b]{0.30\textwidth}
\centering
\begin{tikzpicture}[
  dot/.style={circle, fill=red!70!black, minimum size=3.5pt, inner sep=0pt},
  dep/.style={->, >=Stealth, red!70!black, thin},
  modbox/.style={draw=blue!50!black, fill=blue!6, rounded corners=3pt, semithick},
]
  \node[modbox, minimum width=2.4cm, minimum height=1.6cm] at (1.1,0.6) {};
  \node[dot] (d1) at (0.35,1.0) {};
  \node[dot] (d2) at (1.1,1.0) {};
  \node[dot] (d3) at (1.85,1.0) {};
  \node[dot] (d4) at (0.7,0.25) {};
  \node[dot] (d5) at (1.5,0.25) {};
  \draw[dep] (d4) -- (d1);
  \draw[dep] (d4) -- (d2);
  \draw[dep] (d5) -- (d2);
  \draw[dep] (d5) -- (d3);
  \draw[dep] (d1) -- (d2);
\end{tikzpicture}
\caption{All decl/1 module}
\label{fig:gran-coarse}
\end{subfigure}
\caption{The granularity spectrum. \textcolor{blue!60!black}{Blue boxes} are modules; \textcolor{red!70!black}{red dots} are declarations; \textcolor{red!70!black}{solid red arrows} are intra-module dependencies; \textcolor{red!70!black}{dashed red arrows} are cross-module dependencies. (a)~One declaration per module: $G_{\mathrm{module}} \cong G_{\mathrm{thm}}$ ($308$K modules, unnavigable). (b)~Current \module{Mathlib}: multiple cross-module declaration edges collapse into a single \textcolor{blue!60!black}{blue import} arrow; internal edges are invisible at the module level. (c)~All declarations in one module: $G_{\mathrm{module}}$ collapses to a single vertex with no edges.}
\label{fig:granularity-spectrum}
\end{figure}
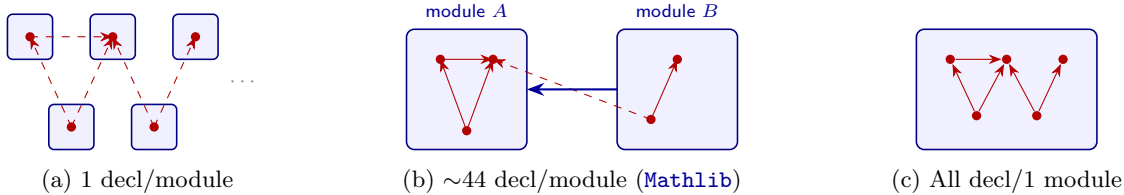

At one extreme, if each declaration occupied its own module, $G_{\mathrm{module}} \cong G_{\mathrm{thm}}$ ($308{,}129$ nodes, unnavigable). At the other extreme, a single module collapses the graph to one vertex with no organizational signal. The current equilibrium (${\sim}44$ declarations per module) represents the largest unit a developer can hold in working memory while writing and reviewing proofs. Figure~\ref{fig:granularity-spectrum} illustrates these three regimes.

As AI systems become primary contributors to formal libraries, this equilibrium may shift. An AI author is not subject to the same working-memory constraint; it could operate productively with finer-grained modules optimized for compilation parallelism, or with coarser modules optimized for logical coherence. The direction of the shift (toward machine-optimal granularity rather than human-optimal granularity) would alter every cross-level metric reported in this section: cohesion, containment, edge amplification, and depth compression would all change, not because the mathematics changed, but because the production process changed. Tracking these metrics longitudinally as AI contributions grow will distinguish intrinsic mathematical structure from artifacts of the human production mode.

A parallel thought experiment applies to the namespace axis. \module{Mathlib}'s $15{,}456$ leaf namespaces contain $308{,}129$ declarations, yielding a mean of approximately $20$ declarations per namespace. Again, this ratio is a product of human naming conventions, not of mathematical necessity.

\begin{figure}[ht]
\centering
\begin{subfigure}[b]{0.30\textwidth}
\centering
\begin{tikzpicture}[
  dot/.style={circle, fill=red!70!black, minimum size=3.5pt, inner sep=0pt},
  depx/.style={->, >=Stealth, red!70!black, thin, dashed},
  nsbox/.style={draw=green!50!black, fill=green!6, rounded corners=2pt, semithick,
                minimum size=0.6cm},
]
  \node[nsbox] (n1) at (0,1.2) {};
  \node[nsbox] (n2) at (1.1,1.2) {};
  \node[nsbox] (n3) at (2.2,1.2) {};
  \node[nsbox] (n4) at (0.55,0) {};
  \node[nsbox] (n5) at (1.65,0) {};
  \draw[depx] (0.55,0) -- (0,1.2);
  \draw[depx] (0.55,0) -- (1.1,1.2);
  \draw[depx] (1.65,0) -- (1.1,1.2);
  \draw[depx] (1.65,0) -- (2.2,1.2);
  \draw[depx] (0,1.2) -- (1.1,1.2);
  \node[dot] (d1) at (0,1.2) {};
  \node[dot] (d2) at (1.1,1.2) {};
  \node[dot] (d3) at (2.2,1.2) {};
  \node[dot] (d4) at (0.55,0) {};
  \node[dot] (d5) at (1.65,0) {};
  \node[font=\scriptsize, text=gray] at (2.85,0.6) {$\cdots$};
\end{tikzpicture}
\caption{1 decl/namespace}
\label{fig:ns-gran-flat}
\end{subfigure}%
\hfill
\begin{subfigure}[b]{0.36\textwidth}
\centering
\begin{tikzpicture}[
  dot/.style={circle, fill=red!70!black, minimum size=3.5pt, inner sep=0pt},
  dep/.style={->, >=Stealth, red!70!black, thin},
  depx/.style={->, >=Stealth, red!70!black, thin, dashed},
  nsbox/.style={draw=green!50!black, fill=green!6, rounded corners=3pt, semithick},
  nsinr/.style={draw=green!50!black, fill=green!3, rounded corners=2pt, thin, dashed},
]
  \node[nsbox, minimum width=1.6cm, minimum height=1.6cm] at (0,0.6) {};
  \node[font=\scriptsize\sffamily, text=green!50!black] at (0,1.65) {\ns{Nat}};
  \node[nsbox, minimum width=1.6cm, minimum height=1.6cm] at (2.8,0.6) {};
  \node[font=\scriptsize\sffamily, text=green!50!black] at (2.8,1.65) {\ns{Nat.Prime}};
  \draw[dep] (0,0.05) -- (-0.35,1.0);
  \draw[dep] (0,0.05) -- (0.35,1.0);
  \draw[dep] (-0.35,1.0) -- (0.35,1.0);
  \draw[dep] (2.45,0.2) -- (2.8,1.0);
  \draw[depx] (2.45,0.2) -- (0.35,1.0);
  \draw[->, >=Stealth, green!50!black, thick] (2.0,0.6) -- (0.8,0.6);
  \node[dot] (d1) at (-0.35,1.0) {};
  \node[dot] (d2) at (0.35,1.0) {};
  \node[dot] (d3) at (2.8,1.0) {};
  \node[dot] (d4) at (0,0.05) {};
  \node[dot] (d5) at (2.45,0.2) {};
\end{tikzpicture}
\caption{Hierarchical (\module{Mathlib})}
\label{fig:ns-gran-current}
\end{subfigure}%
\hfill
\begin{subfigure}[b]{0.30\textwidth}
\centering
\begin{tikzpicture}[
  dot/.style={circle, fill=red!70!black, minimum size=3.5pt, inner sep=0pt},
  dep/.style={->, >=Stealth, red!70!black, thin},
  nsbox/.style={draw=green!50!black, fill=green!6, rounded corners=3pt, semithick},
]
  \node[nsbox, minimum width=2.4cm, minimum height=1.6cm] at (1.1,0.6) {};
  \node[dot] (d1) at (0.35,1.0) {};
  \node[dot] (d2) at (1.1,1.0) {};
  \node[dot] (d3) at (1.85,1.0) {};
  \node[dot] (d4) at (0.7,0.25) {};
  \node[dot] (d5) at (1.5,0.25) {};
  \draw[dep] (d4) -- (d1);
  \draw[dep] (d4) -- (d2);
  \draw[dep] (d5) -- (d2);
  \draw[dep] (d5) -- (d3);
  \draw[dep] (d1) -- (d2);
\end{tikzpicture}
\caption{All decl/1 namespace}
\label{fig:ns-gran-single}
\end{subfigure}
\caption{The namespace granularity spectrum. \textcolor{green!50!black}{Green boxes} are namespaces; \textcolor{red!70!black}{red dots} are declarations; \textcolor{red!70!black}{solid red arrows} are intra-namespace dependencies; \textcolor{red!70!black}{dashed red arrows} cross namespace boundaries. (a)~One declaration per namespace: every edge crosses a boundary (containment~$= 0\%$). (b)~Current \module{Mathlib}: nested namespaces with partial containment. (c)~All declarations in one namespace: every edge is internal (containment~$= 100\%$, no organizational signal).}
\label{fig:ns-granularity-spectrum}
\end{figure}
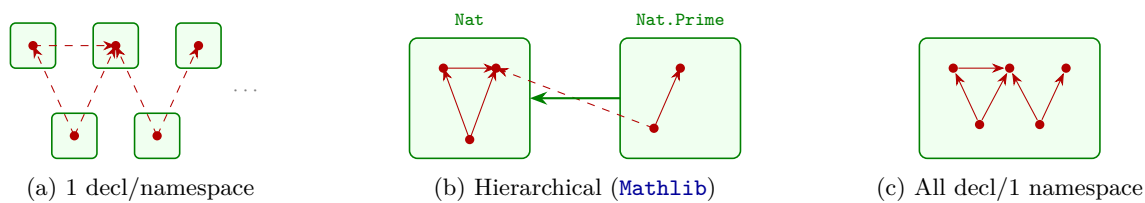

At one extreme, if each declaration occupied its own namespace (flat naming: \ns{add\_comm}, \ns{dvd\_mul}, \ldots), containment at depth~$1$ would be trivially $0\%$ (no two declarations could share a namespace boundary) and the naming hierarchy would carry no organizational information. At the other extreme, if all declarations shared a single namespace, containment would be trivially $100\%$ but the hierarchy would provide no finer-than-global grouping. The current Mathlib equilibrium ($6$ depth levels, $22.2\%$ containment at depth~$1$, rising to $86.9\%$ at depth~$3$) reflects a naming convention that balances logical grouping against navigational convenience. Figure~\ref{fig:ns-granularity-spectrum} illustrates these three regimes.

Figures~\ref{fig:granularity-spectrum} and~\ref{fig:ns-granularity-spectrum} together reveal that library organization is a \emph{two-dimensional} design choice: module granularity (how many declarations per file) and namespace granularity (how many declarations per naming group) are independently adjustable parameters. \module{Mathlib} currently occupies one point in this design space (${\sim}44$ declarations per module, ${\sim}20$ per leaf namespace), but alternative design points are entirely feasible, and each would produce different cross-level statistics. The product (mathematical dependency) constrains neither axis; only the process (human or machine authorship) does.
\label{sec:structural-overview}
\label{sec:cross-level-summary}

\end{document}